\documentclass[final,3p,times,twocolumn,authoryear,toc]{elsarticle}

\usepackage{hyperref}

\usepackage{graphicx}

\usepackage{amssymb}

\bibpunct{(}{)}{;}{a}{}{,}

\newcommand{\msun}{{M_\odot}}
\newcommand{\eff}{{\epsilon_{\rm ff}}}
\newcommand{\tdep}{{t_{\rm dep}}}
\newcommand{\tff}{{t_{\rm ff}}}

\newcommand{\red}[1]{{{#1}}}
\newcommand{\blue}[1]{{{#1}}}

\journal{Physics Reports}

\begin{document}

\begin{frontmatter}



\title{The Big Problems in Star Formation: the Star Formation Rate, Stellar Clustering, and the Initial Mass Function}


\author{Mark R. Krumholz}

\address{Department of Astronomy \& Astrophysics, University of California, Santa Cruz, 95064, USA}

\begin{abstract}
Star formation lies at the center of a web of processes that drive cosmic evolution: generation of radiant energy, synthesis of elements, formation of planets, and development of life. Decades of observations have yielded a variety of empirical rules about how it operates, but at present we have no comprehensive, quantitative theory. In this review I discuss the current state of the field of star formation, focusing on three central questions: what controls the rate at which gas in a galaxy converts to stars? What determines how those stars are clustered, and what fraction of the stellar population ends up in gravitationally-bound structures? What determines the stellar initial mass function, and does it vary with star-forming environment? I use these three question as a lens to introduce the basics of star formation, beginning with a review of the observational phenomenology and the basic physical processes. I then review the status of current theories that attempt to solve each of the three problems, pointing out links between them and opportunities for theoretical and numerical work that crosses the scale between them. I conclude with a discussion of prospects for theoretical progress in the coming years.
\end{abstract}


\end{frontmatter}

\onecolumn
\tableofcontents
\twocolumn


\section{Introduction}
\label{sec:intro}

Star formation is one of the least understood phenomena in cosmic evolution. It is difficult to formulate a general theory for star formation in part because of the wide range of physical processes involved. The interstellar gas out of which stars form is a supersonically-turbulent, weakly-ionized plasma governed by non-ideal magnetohydrodynamics (MHD). This by itself would make star formation a difficult problem, since we have at best a partial understanding of subsonic hydrodynamic turbulence, let alone supersonic non-ideal MHD turbulence. The behavior of star-forming gas is obviously influenced by gravity, which adds complexity, and the dynamics of the interstellar medium (ISM) is also strongly affected by both continuum and line radiative processes. Finally, its behavior is influenced by a wide variety of chemical processes, including formation and destruction of molecules and dust grains (which changes the thermodynamic behavior of the gas) and changes in ionization state (which alter how strongly the gas couples to magnetic fields). As a result of these complexities, there is nothing like a generally agreed-upon theory of star formation as there is for stellar structure. Instead, we are forced to take a much more phenomenological approach.

Before diving into this phenomenology, however, it is worth pausing to consider the motivation for this review. Star formation has been the subject of a number of recent reviews, focusing on theory \citep{mckee07a}, numerical simulations \citep{klessen11a}, observations in the Milky Way and nearby galaxies \citep{kennicutt12a}, and a number of other much more detailed topics \citep[e.g.,][]{hennebelle12a, kruijssen14a, dobbs14a, offner14a, padoan14a, krumholz14a, tan14a}. Each of these reviews provides a valuable description of one or more aspects of the star formation process, and some aim at a much more comprehensive overview of the field. Replicating either the scope or the detail of this previous work is neither a useful exercise, nor, given the limitations of space and reader attention, a viable goal.

Part of the motivation for this review is simply to provide an update. While there are a number of very recent reviews of specialized topics within star formation, the last comprehensive review of star formation theory as whole, by \citet{mckee07a}, is now six years old. This is a very long time in a fast-moving field like star formation. However, the aim of this review also differs from that of previous reviews in two ways.

First, in this review I provide a significantly more pedagogic introduction to the subject, particularly on the basic physics background that is often assumed or skipped in higher-level reviews. While there are several textbooks on star formation \citep{stahler05a, ward-thompson11a, bodenheimer11a, schulz12a}, this material is not part of the standard graduate curriculum at most institutions, and there is little material available that bridges the gap between a textbook-level introduction and a specialist review. The discussion I provide here is intended to occupy that middle ground, and is aimed at readers who are not experts in either the observational phenomenology or background theory of star formation or the molecular ISM, but want a much shorter and higher-level introduction than is provided by a textbook. I divide this introduction into a review of the observational phenomenology (Section \ref{sec:phenomenology}) and an introduction to the physics of the star-forming interstellar medium (Section \ref{sec:theorybackground}). The latter in particular focuses on basic theory as much as recent results. Both these sections are intended to bring students and other non-experts up to speed, and may safely be skimmed or skipped by readers who already possess a deep familiarity with star formation.

The second goal of this review is to provide a (necessarily biased) perspective focused on what I consider the minimal elements required for a predictive theory of star formation, and to suggest a common approach to tackling them. This is an important difference from most of the more narrowly-focused reviews listed above, which strive to cover one aspect of star formation in great detail. My goal here is instead to step back and identify those questions that will need to be answered before star formation theory becomes like the theory of stellar evolution: a field that continues to hold unsolved problems, but one for which enough basic results have been established that researchers in other areas of astrophysics can make use of them with some confidence, and without the need for continual worry if even the zeroth-order results are robust. Reaching this point is necessary if we are ever to have confidence in extrapolating the results of star formation studies to galactic environments far-removed from those familiar to us from the nearby Universe.

Any predictive theory must provide a statistical description of the star formation process, and while many statistics can be defined, I focus on three in this review. First, given the large-scale properties of a galaxy or some subsection of it (e.g., gas and stellar distributions and kinematics, metal content), what will be the \textit{star formation rate} (SFR) in that galaxy? Second, what will be the \textit{spatial and temporal distribution} of the star formation, i.e.~how will the newborn stars be clustered together in space and time? Third, what will be the \textit{mass distribution} of the resulting stars, the initial mass function (IMF)? Each of these questions could be (and has been) the basis of its own review, but as I argue below, they are inextricably linked, and must be solved together.

The remainder of this review is organized as follows. As mentioned above, I begin in Section \ref{sec:phenomenology} with a review of some of the necessary observational background, and provide a similar introduction to the theoretical background in Section \ref{sec:theorybackground}. I then review the three questions of the star formation rate, the clustering of star formation, and the mass distribution of stars in Sections \ref{sec:sfr}, \ref{sec:clustering}, and \ref{sec:imf}, respectively, before attempting a rough synthesis and pointing the way for future work in Section \ref{sec:discussion}.

\section{Observational Phenomenology}
\label{sec:phenomenology}

\subsection{The Star Formation Rate}
\label{ssec:sfrobs}

\subsubsection{Galactic Scales}
\label{sssec:sfrobsgalactic}

\begin{figure*}[ht!]
\centerline{
\includegraphics[width=6.5in]{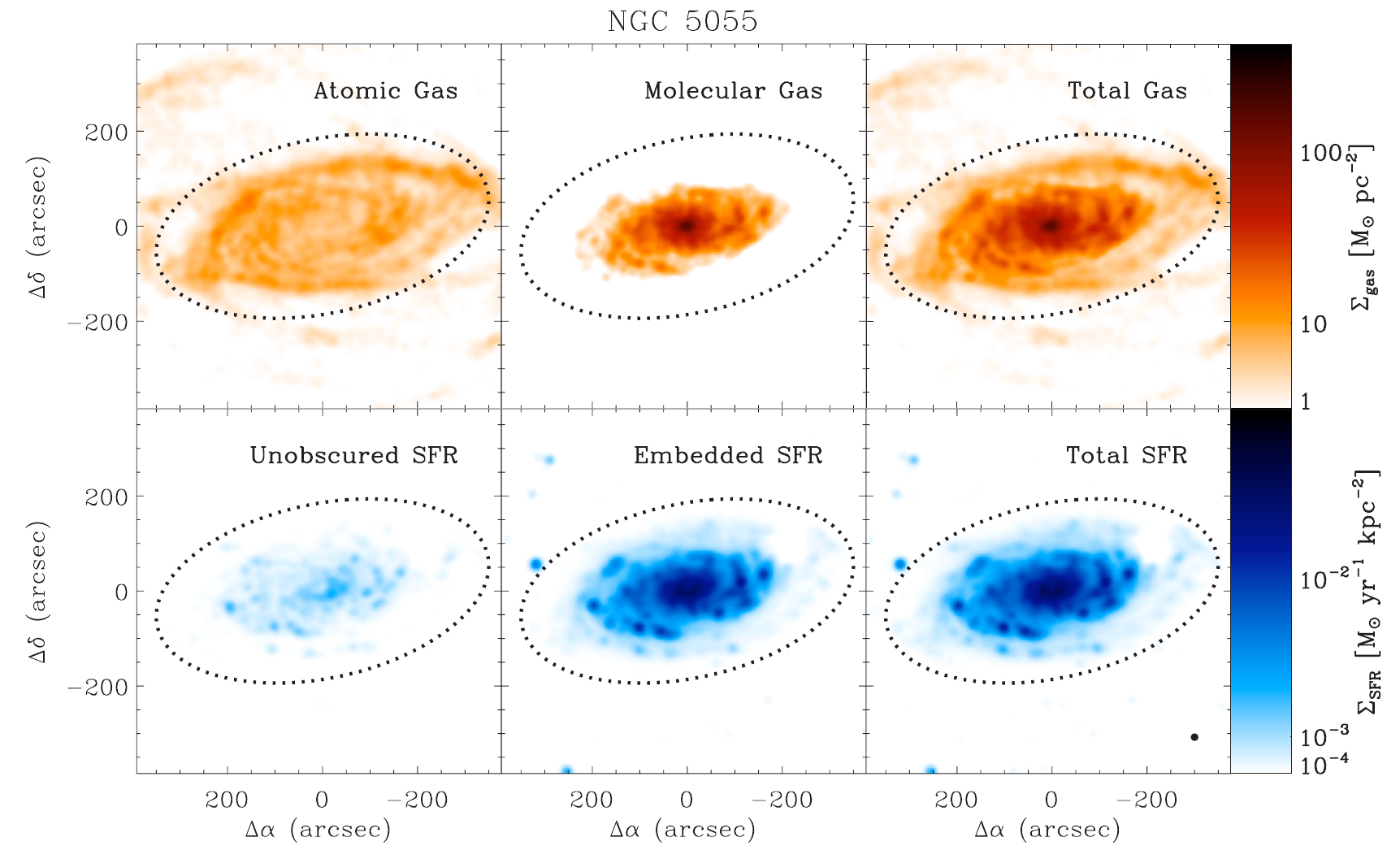}
}
\caption{
\label{fig:leroy08}
Maps the the distribution of gas (top row) and star formation (bottom row) in the nearby galaxy NGC 5055, reprinted from \citet{leroy08a} and reproduced by permission of the AAS. The top three panels show, from left to right, the surface densities of atomic gas inferred from the H~\textsc{i} 21 cm line, molecular gas inferred from the CO $J=2\rightarrow 1$ line, and the sum of the two. The bottom three panels show, also from left to right, the star formation rate per unit area inferred from far ultraviolet emission (``unobscured"), the star formation rate per unit area inferred from infrared emission (``embedded"), and the sum of the two. \red{Here ``embedded" means that the light we are seeing does not come directly from the stars themselves, but instead arises from re-emission by warm dust grains that have been heated by the light of young stars. The luminosity of the warm dust is a proxy for the bolometric output of the stellar population, and since this is dominated by young stars, it is a reasonable proxy for the star formation rate.}
}
\end{figure*}

\begin{figure*}[ht!]
\centerline{
\includegraphics[width=6in]{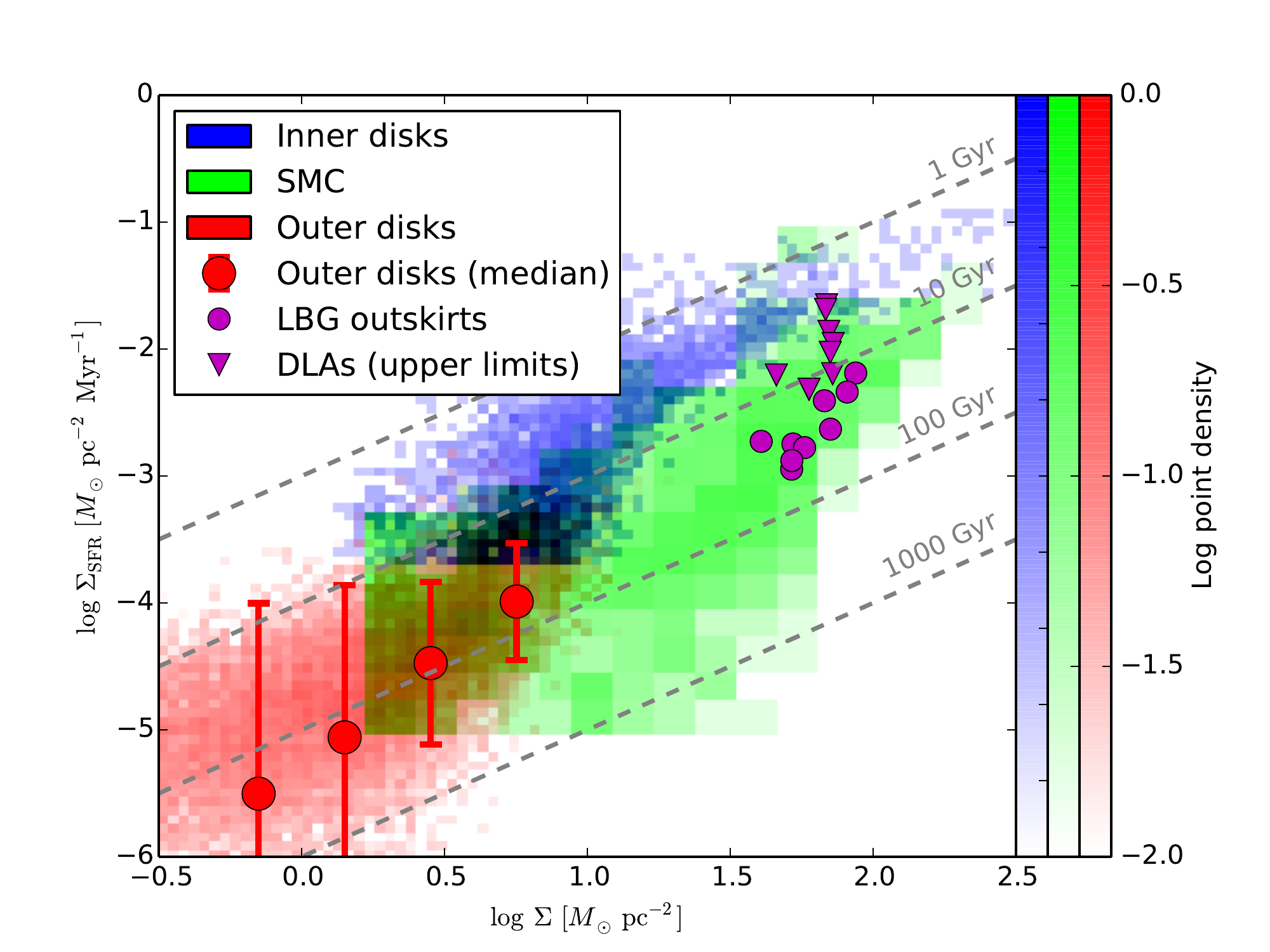}
}
\caption{
\label{fig:sflawtot}
Four data sets showing the relationship between surface density of gas $\Sigma$ (including all neutral phases) and surface density of star formation $\Sigma_{\rm SFR}$. Gray dashed lines are lines of constant depletion time, from $1-1000$ Gyr as indicated. Magenta circles show the outskirts of Lyman Break Galaxies at $z\approx 3$, from \citet{rafelski11a}, and magenta downward triangles show upper limits on star formation rates in damped Lyman $\alpha$ absorbers (DLAs) from \citet{wolfe06a}. Blue pixels show the inner parts of galaxies, defined as the region inside $R_{25}$, in local Universe galaxies, from \citet{bigiel08a}\red{, using CO as a proxy for H$_2$ and 21 cm emission to measure H~\textsc{i}.}. Red pixels show the portions of local Universe spiral and dwarf galaxies outside $R_{25}$, from \citet{bigiel10a}\red{, measured in the same way}. Green pixels show the Small Magellanic Cloud, from \citet{bolatto11a}\red{, using cold dust emission as a proxy for total gas column density}. For the blue, red, and green pixels, the intensity of each pixel in the image is proportional to the log of the density of points in that bin of $(\Sigma, \Sigma_{\rm SFR})$, where each point is an independent region of the target galaxy. For the inner and outer disk data, regions are $\approx 750$ pc in size, while for the SMC regions are 200 pc in size. The color scale for each data set has been independently normalized so that the most populated pixel has a value of unity, and sharp edges to the data at particular values of $\Sigma$ and $\Sigma_{\rm SFR}$ are due to the sensitivity limits of the individual surveys. The red circles with error bars show the median and $1\sigma$ scatter of the outer galaxy data indicated by the red pixels. The median and scatter lie below most of the red pixels shown because they are computed properly accounting for observational errors that can lead to negative values of $\Sigma_{\rm SFR}$, which are masked by the logarithmic $y$ axis.
}
\end{figure*}

\begin{figure*}[ht!]
\centerline{
\includegraphics[width=6in]{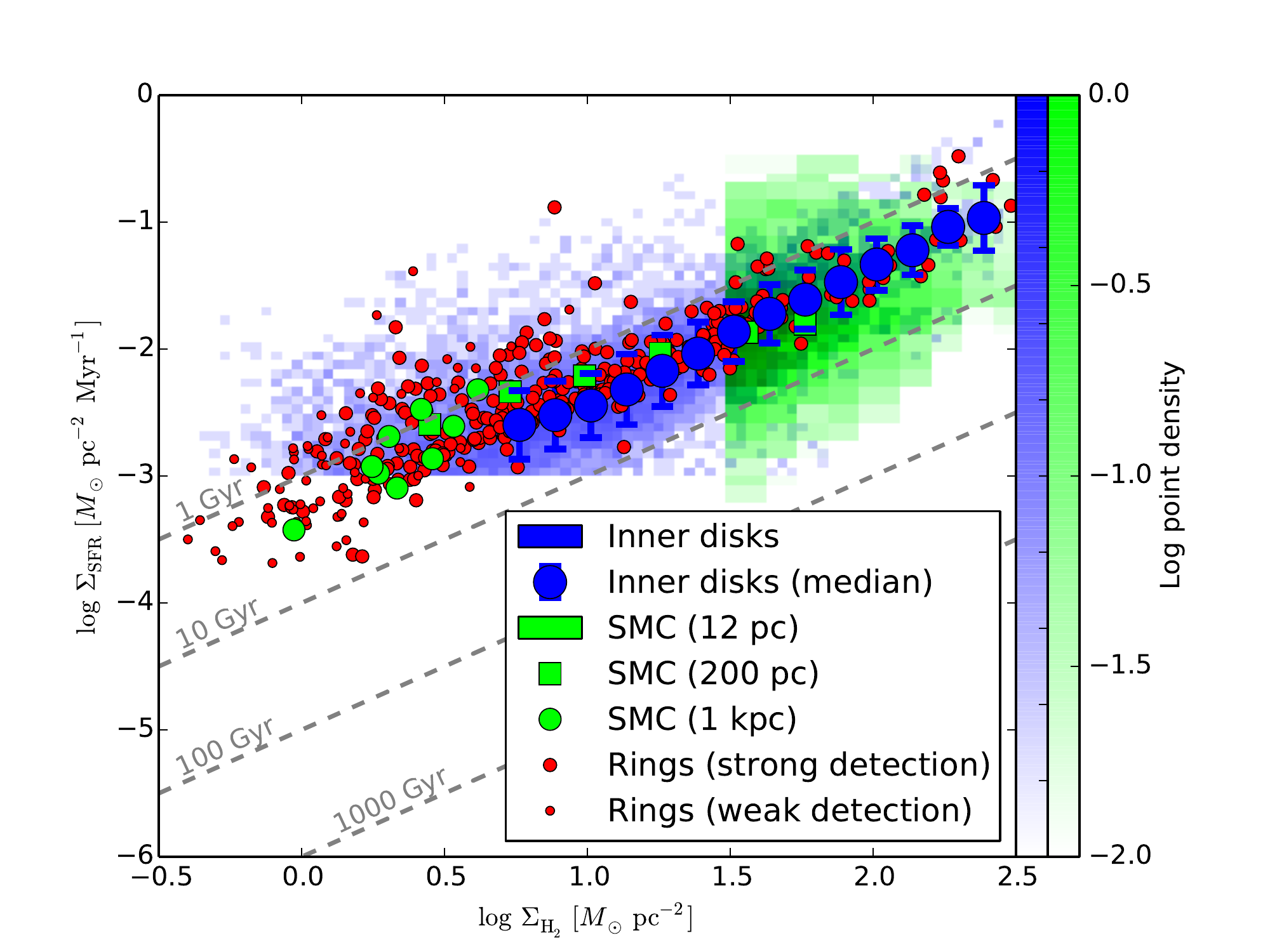}
}
\caption{
\label{fig:sflawh2}
Three data sets showing the relationship between surface density of molecular gas $\Sigma_{\rm H_2}$ and surface density of star formation $\Sigma_{\rm SFR}$. Lines and axes are the same as in Figure \ref{fig:sflawtot}. Blue pixels show the inner parts of galaxies in the local Universe, from \citet{leroy13a}; blue circles with error bars show the median and scatter of this data set. (Note that this is an extended version of data set from \citet{bigiel08a} shown in blue in Figure \ref{fig:sflawtot}.) Green pixels are the data of \citet{bolatto11a} for the SMC, but with each pixel representing a $12$ pc aperture; green squares and circles are the same data set, but averaged over 200 pc and 1 kpc apertures instead. Red points are averages over azimuthal rings, with widths from $220 - 1800$ pc depending on the distance of the target, in nearby spiral galaxies from \citet{schruba11a}. The size of the symbol indicates whether Schruba et al.~classify the detection as strong or marginal. \red{The inner disk and ring data sets are based on CO emission as a proxy for H$_2$, while the SMC data set uses dust emission as a proxy because, for reasons discussed in Section \ref{sssec:carbon}, CO is an unreliable tracer of molecular gas in low-metallicity galaxies like the SMC.}
}
\end{figure*}

\red{\citet{schmidt59a} was the first to conjecture a powerlaw relationship between galaxies' gas content and their star formation, but the first large, multi-galaxy data sets testing this conjecture were assembled in the 1990s. These revealed a clear} correlation between the gas surface density of galaxies and their rates of star formation \citep{kennicutt98a, kennicutt98b}. In the past decade\red{, however,} our knowledge of star formation at the galactic scale has improved \red{still further} thanks to the advent of spatially-resolved surveys. These surveys have allowed us to map out the spatial distribution of star formation within galaxies, and to correlate it with the spatial distribution of both atomic and molecular gas. One can use the correlation between star formation and gas (molecular, atomic, or without regard to phase) to define a timescale, called the depletion time, which is simply the ratio of the gas mass to the SFR: $\tdep \equiv M_{\rm gas} / \dot{M}_*$. This characterizes the rate of star formation in that gas. Figure \ref{fig:leroy08} shows an example of this type of data for the galaxy NGC 5055, and Figures \ref{fig:sflawtot} and \ref{fig:sflawh2} summarize the currently-observed correlation between star formation and total gas, and between star formation and molecular gas. 

\paragraph{Phase Dependence}

Examining Figure \ref{fig:leroy08}, one is immediately struck by the strong correlation between the maps of H$_2$ and star formation, and the correspondingly weaker correlation between total gas and star formation. One can see this effect quantitatively by studying the red and blue pixels in Figure \ref{fig:sflawtot}, which show correlations measured in a collection of $\sim 20$ nearby galaxies, including NGC 5055. Each of these galaxies is pixelized into $\sim 1$ kpc-sized regions, and the data plotted show the distribution of these pixels in gas surface density  $\Sigma$ versus star formation surface density $\Sigma_{\rm SFR}$. These data should be thought of as describing the typical state of star formation in roughly Solar-metallicity galaxies in the nearby Universe. At gas surface densities above $\Sigma \approx 10$ $\msun$ pc$^{-2}$, \blue{mostly represented by the blue pixels,} there is a close to linear relationship between star formation rate and gas surface density, corresponding to a nearly constant depletion time \blue{of a few Gyr. At gas surface densities below $\Sigma \approx 10$ $\msun$ pc$^{-2}$, mostly illustrated by the red pixels, there is again a roughly constant depletion time of $\sim 100$ Gyr. Thus there is a factor of $\sim 50$ change in the gas depletion time near $\Sigma \approx 10$ $M_\odot$ pc$^{-2}$, where the red and blue pixels meet.}

If one examines Figure \ref{fig:sflawh2}, it is clear that there is no corresponding feature in the relationship between molecular gas surface density $\Sigma_{\rm H_2}$ and $\Sigma_{\rm SFR}$. The data instead appear consistent with a roughly constant depletion time in the H$_2$. The feature in Figure \ref{fig:sflawtot} seen at a total gas surface density of 10 $\msun$ pc$^{-2}$ corresponds to a sharp transition between the ISM being H$_2$-dominated at large surface densities and being H~\textsc{i}-dominated at smaller surface densities. In the H~\textsc{i}-dominated region, the dispersion of star formation rates at fixed gas surface density is extremely large, and ``second parameters" such as metallicity \citep{bolatto11a, krumholz13c} and stellar surface density \citep{blitz04a, blitz06b, leroy08a} appear to become important. Thus the first lesson we can extract from the observations shown in Figures \ref{fig:leroy08} -- \ref{fig:sflawh2}, and from numerous other surveys \citep[e.g.,][]{wong02a, kennicutt07a, blanc09a}, is that star formation is much more strongly and directly correlated with H$_2$ than with H~\textsc{i}.

Next consider the green pixels, which show data on the Small Magellanic Cloud (SMC). A second interesting feature apparent in Figures \ref{fig:sflawtot} and \ref{fig:sflawh2} is that there is a very clear offset between the SMC data and the spiral galaxy data in the plot for total gas, but not in the corresponding relation using molecular gas only. Similarly, the data for the outskirts of Lyman Break Galaxies (LBGs) and for damped Lyman $\alpha$ absorbers (DLAs) (shown in white in Figure \ref{fig:sflawtot}) are systematically below the $\Sigma-\Sigma_{\rm SFR}$ correlation that applies to nearby spiral galaxies, but agree roughly with the SMC. The SMC, DLAs, and LBG outskirts all seem to have much longer depletion times at fixed gas surface density than most local galaxies. The physical reason for this is something to be discussed below, but an obvious candidate is that the SMC, LBG outskirts, and DLAs all have much lower metallicities than the other galaxies plotted -- roughly $20\%$ of Solar for the SMC \citep{bolatto11a}, and likely between $1\%$ and $10\%$ of Solar for DLAs and LBG outskirts\footnote{It is important here that the regions being measured are the outskirts of LBGs, not the central star-forming disks, which likely have higher metallicities.} \citep{prochaska03a, rafelski12a}.

On the other hand, examining Figure \ref{fig:sflawh2}, there is no corresponding change in the relationship between molecular gas and star formation in the SMC. The depletion time for H$_2$ is the same in the SMC as in other galaxies. (We lack corresponding data on the H$_2$ content of the LBG outskirts.) Interestingly, the fixed depletion time is seen only if one consider the H$_2$, and not the CO. In low-metallicity galaxies like the SMC, there are large regions of H$_2$ without any CO (for reasons I discuss in Section \ref{ssec:gmcchemistry}), and in this case one obtains a constant depletion time only if one considers all the H$_2$, not just the H$_2$ where CO is also present \citep{krumholz11b, bolatto11a}. This strongly suggests that the correlation between star formation and H$_2$ is the fundamental one.

\paragraph{Depletion Times and Star Formation Efficiency}

The next thing to consider about the observed star formation-gas correlation, beyond its dependence on phase, is the quantitative value of the depletion time. Figure \ref{fig:sflawh2} shows that, averaged over scales of $\sim 1$ kpc or more, the observed depletion time of molecular gas in nearby galaxies is $\tdep \approx 1-2$ Gyr. There is some uncertainty as to whether this value is in fact constant \citep{bigiel08a, blanc09a, rahman11a, rahman12a, leroy13a}, or varies slightly with the gas surface density or other large-scale properties of a galaxy \citep{kennicutt07a, liu11a, saintonge11b, calzetti12a, meidt13a, momose13a, shetty13a}. Much of the uncertainty stems from technical issues of how to convert between CO luminosity and molecular gas mass, and from various tracers of star formation activity to the physical star formation rate. However, the range of variation in the disks of nearby spiral galaxies is at most a factor of a few; averaged over kpc scales, we do not see molecular gas with a depletion time much below 1 Gyr \blue{(except perhaps in the very centers of galaxies)} or much above $\sim 10$ Gyr. \red{This limited range is not a fundamental physical limitation so much as a statement about the demographics of disk galaxies in the local Universe.} The depletion time can be an order of magnitude shorter in nuclear regions of spirals and in compact irregular galaxies that appear to have experienced recent mergers or disturbances, both locally and at high redshift \citep[e.g.][]{kennicutt98a, daddi10a, genzel10a}. \red{Such galaxies are simply rare today, and the nearest examples are beyond the Local Group.} Conversely, Figure \ref{fig:sflawtot} shows that H$_2$-poor regions can have depletion times for the neutral ISM as a whole (i.e., including the H~\textsc{i} as well as molecular gas) up to $\sim 100$ Gyr \citep{wyder09a, bigiel10a, rafelski11a, bolatto11a, cantalupo12a, huang12a}. 

These depletion times can be compared to some other natural timescales. One is the Hubble time. H~\textsc{i}-dominated galaxies have depletion times comparable to or longer than the Hubble time, and this suggests that these systems have not yet reached any sort of equilibrium between cosmological infall of gas and star formation. Instead, their time-averaged accretion rate from the intergalactic medium up to this point in cosmic time has exceeded the rate at which they are capable of processing that gas into stars. This is not true in present-day spirals with $\tdep\sim 1$ Gyr,\blue{\footnote{\blue{However, \citet{kennicutt12a} point out that even the Milky Way has a depletion time of $\approx 5.5$ Gyr if one includes all the H~\textsc{i} in the far outer disk. Thus the outskirts of the Milky Way are likely out of equilibrium even if the inner regions are not. This is consistent with the models of \citet{forbes14a}, where galaxies equilibrate inside-out.}}} or even for large star-forming galaxies up to $z\sim 3$ \citep{saintonge13a}, though it might have been true for their progenitors at even higher redshift \citep{krumholz12d}.

Even in galaxies where $t_{\rm dep} < t_{\rm Hubble}$, the depletion time is still a factor of $\sim 10$ longer than the galactic orbital period, and a factor of $\sim 100$ longer than the dynamical times in the molecular clouds where stars form. The natural timescale for a self-gravitating gas cloud is the free-fall time,
\begin{equation}
\tff = \sqrt{\frac{3\pi}{32G\rho}} = 4.3 n_{\rm H,2}^{-1/2}\mbox{ Myr},
\end{equation}
where $\rho$ is the density, $n_{\rm H}$ is the number density of H nuclei, $n_{\rm H,2} = n_{\rm H}/100$ cm$^{-3}$, and the calculation assumes a mean mass per H nucleus of $2.3\times 10^{-24}$ g, as expected for gas with the standard cosmological He mass faction. In the Milky Way and similar spirals, molecular clouds have mean volume densities of $n_{\rm H} \approx 50 - 1000$ cm$^{-3}$ \citep[e.g.,][]{bolatto08a, roman-duval10a}, implying free-fall times of $\sim 1-10$ Myr. Thus the observed depletion times are $\sim 100$ times longer than the free-fall timescale. The depletion times are smaller in starbursts, but the gas densities are also generally much higher \citep[e.g.,][]{downes98a}, so the offset between free-fall and depletion timescales remains large. The dimensionless ratio of the free-fall and depletion times (first introduced by \citet{krumholz05c}) is conventionally denoted
\begin{equation}
\eff \equiv \frac{\tff}{\tdep}.
\end{equation}

\begin{figure*}[ht!]
\centerline{
\includegraphics[width=6in]{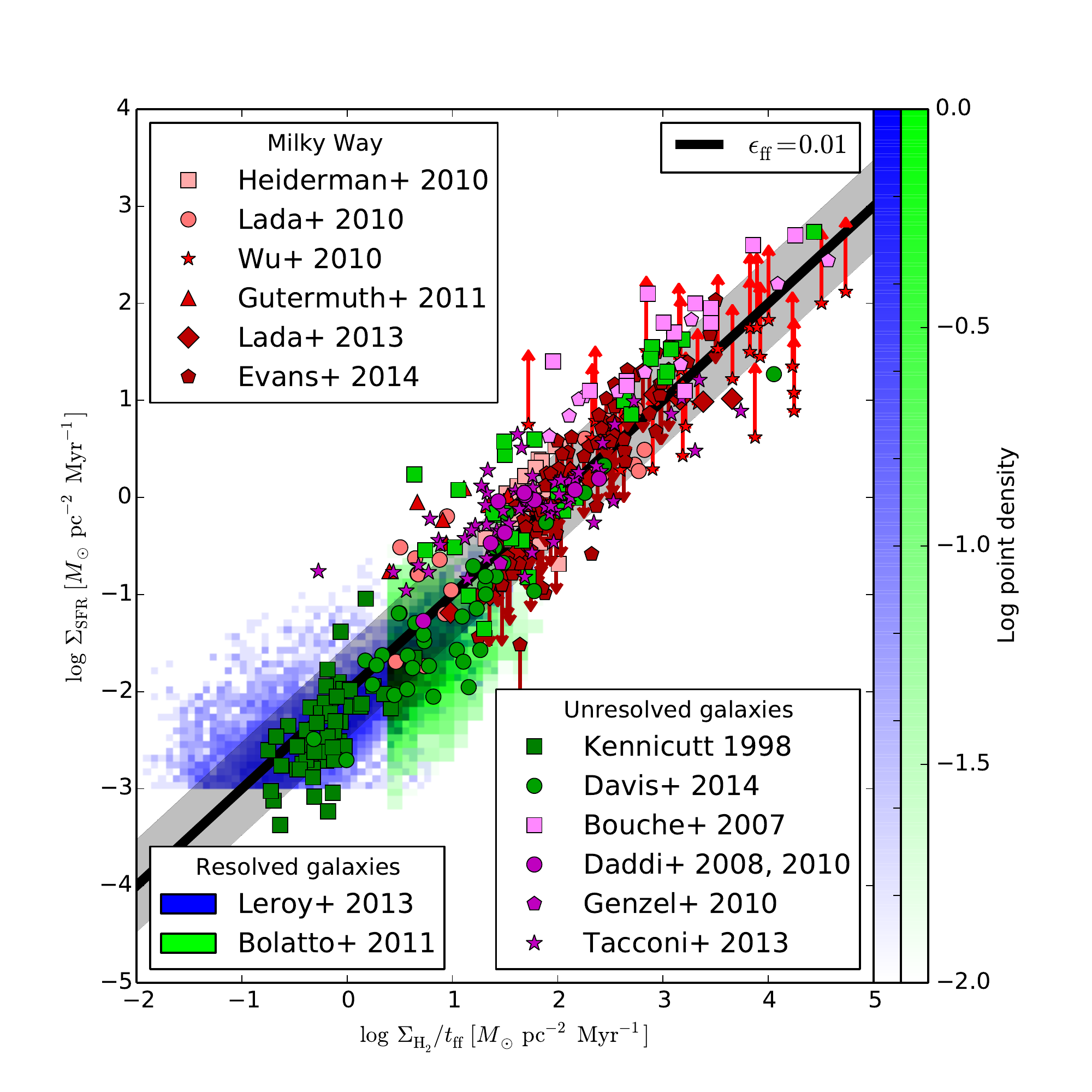}
}
\caption{
\label{fig:eff}
Surface density of star formation versus surface density of molecular gas normalized by estimated free-fall time $\Sigma/\tff$. 
\blue{
The free-fall time for all objects has been estimated following the method of \citet{krumholz12a}. The black thick line shows $\epsilon_{\rm ff} = 0.01$; the gray band indicates a factor of 3 scatter about this value. The data shown in the plot are as follows: individual molecular clouds in the Milky Way (red-hued points) are from \citet[red squares]{heiderman10a}, \citet[red circles]{lada10a}, \citet[red stars, upward arrows indicate lower limits]{wu10a}, \citet[red diamonds]{lada13a}, and \citet[red pentagons; downward arrows indicate upper limits]{evans14a}; resolved observations of nearby galaxies (rasters, same data as shown in Figure \ref{fig:sflawh2}) are from a sample of the inner disks of spirals \citep[blue raster]{leroy13a} and the 12 pc resolution data form the Small Magellanic Cloud \citep[green raster]{bolatto11a}; unresolved observations of $z=0$ galaxies (green points) are spirals and starbursts from \citet[green squares]{kennicutt98b}, and the molecular disks of early-type galaxies from \citet{davis14a}; unresolved observations of $z>0$ galaxies (magenta points) are from \citet[magenta squares]{bouche07a}, \citet[magneta circles]{daddi08a, daddi10b}, \citet[magenta pentagons]{genzel10a}, and \citet[magenta stars]{tacconi13a}. All CO-to-H$_2$ conversion factors have been standardized to the fiducial values of \citet{daddi10a}: $\alpha_{\rm CO} = 0.8$ $M_\odot / (\mbox{K km s}^{-1}\mbox{ pc}^{-2})$ in starbursts at all redshifts, $\alpha_{\rm CO} = 4.6$ $M_\odot / (\mbox{K km s}^{-1}\mbox{ pc}^{-2})$ in $z=0$ disks, and $\alpha_{\rm CO} = 3.6$ $M_\odot / (\mbox{K km s}^{-1}\mbox{ pc}^{-2})$ in $z>0$ disks. Within each data set, lighter colored points are those for which a starburst-like $\alpha_{\rm CO}$ value was adopted, while darker points are those using a disk-like $\alpha_{\rm CO}$. The exception is the early-type galaxy sample of \citet{davis14a}, where it is not clear which to use, and I have therefore deferred to their recommended, intermediate value $\alpha_{\rm CO}=3.4$ $M_\odot / (\mbox{K km s}^{-1}\mbox{ pc}^{-2})$.}
}
\end{figure*}

\citet{krumholz07a} and \citet{krumholz12a} collect a large sample of observations, including both resolved regions of galaxies and entire galaxies, and conclude that all the data are consistent with a universal value $\eff \sim 0.01$. \blue{Figure \ref{fig:eff} shows an updated compilation, analyzed following the same method as described in \citet{krumholz12a}, that includes more recent observations as well.}

Contrary to this, when averaging over whole galaxies (but \textit{not} considering resolved regions), \citet{faucher-giguere13a} argue that some galaxies have $\eff \sim 0.1-0.3$, and that there is a significant correlation between $\eff$ and gas fraction in the galaxy. The origin of the difference appears to be in the way the two groups define the gas free-fall time. All values of $\eff>0.1$ that \citeauthor{faucher-giguere13a}~find come from marginally-resolved, large disk galaxies at $z\approx 1-3$\blue{, taken from the sample of \citet{tacconi13a}.\footnote{\blue{The \citet{tacconi13a} galaxies were not included in the original \citet{krumholz12a} paper because they were published later, but they \textit{are} included in Figure \ref{fig:eff}, where they appear as magenta stars.}}} In these galaxies, \citeauthor{krumholz12a}'s method of estimating the free-fall time assumes that star formation takes place in discrete molecular clouds like in the Milky Way, and would assign these clouds free-fall times $t_{\rm ff}\lesssim 10$ Myr, comparable to what is seen in Milky Way clouds. In contrast, \citeauthor{faucher-giguere13a}'s method simply computes the mean density in the galactic disk, ignoring any cloud structure. This leads to lower densities, and longer free-fall times of $\gtrsim 30$ Myr, accounting for the factor of 10 difference in the typical value of $\eff$ deduced for them. This disagreement only affects the high-redshift sample; both teams conclude that $\eff \approx 0.01$ for galaxies in the local Universe.

\subsubsection{Sub-Galactic Scales}
\label{sssec:sfrsubgalactic}

One can also examine the star formation rates of sub-kpc scales, with the usual tradeoff between the resolution that one can reach and the distance out to which the observation is possible. Between $\sim 100$ pc and 1 kpc, the observed correlation between SFR and molecular gas progressively worsens \blue{as one moves to smaller scales}, reaching multiple order of magnitude-level scatter at $\sim 100$ pc scales \citep{onodera10a, schruba10a}. Similarly, within the Milky Way, the amount of infrared or ionizing luminosity per unit molecular mass varies by several orders of magnitude from one giant molecular cloud to another \citep{mooney88a, murray10b, vutisalchavakul13a}. The scatter is not random: samples that select star-forming regions by ionizing luminosity or some other selection based on star formation rate tend to give $\eff \sim 0.1 - 0.2$, while those that select based on tracers of molecular gas mass instead find $\eff \sim 0.001$ or less. Only when the observed region is large enough to average over multiple maxima of both the CO emission and the star-formation rate tracer (usually infrared or H$\alpha$ emission) does one recover $\eff \sim 0.01$.

This variation might plausibly be explained as an evolutionary effect: when clouds first form they begin their lives containing few stars, and so their star formation rates and values of $\eff$ appear low. As stars form, they begin to destroy the cloud with their feedback, reducing $M_{\rm gas}$, and at the same time the cloud and newborn stars begin to drift apart, since stellar orbits through the galaxy are determined only by gravity, while gas is subject to pressure forces as well. As a result, an observational aperture centered on newborn stars and measuring the present-day gas mass (as opposed to the gas mass when the stars formed, which is what we really want) tends to underestimate $M_{\rm gas}$, while one centered on the gas tends to underestimate $\dot{M}_*$ \citep{feldmann11a, kim13b}. While models incorporating these effects appear able to reproduce the small-scale observed variations in $\eff$ even assuming a fixed true $\eff$, it is not at present possible to rule out a model in which there is also true variation in $\eff$ either between clouds, or within a single cloud over its lifetime.

Below $\sim 100$ pc, an observation generally captures only a single molecular cloud, since typical sizes of large molecular clouds are $\sim 10-100$ pc \citep[and references therein]{dobbs14a}. To reach these scales, one must either restrict the survey to the Solar Neighborhood, or one must give up on spatial resolution and instead select dense regions using density-sensitive molecular line observations. In the former category, the recent \textit{Spitzer Cores to Disks} legacy survey of low-mass star-forming clouds near the Sun gives $\eff \sim 0.01 - 0.1$ for clouds with mean densities  $n_{\rm H_2} \sim 10^3$ cm$^{-3}$ \citep{evans09a}. The star formation rate per unit mass within a given cloud also appears to be strongly correlated with the amount of gas it possesses above some threshold volume or column density \citep{heiderman10a, lada10a, lada12a, evans14a}. 

In the latter category, there have been a number of surveys of extragalactic star formation using HCN, HCO$^+$, higher rotational transitions CO, and other molecules that have critical densities ranging from $n_{\rm H_2} \sim 10^4 - 10^8$ cm$^{-3}$ \citep{gao04b, gao04a, narayanan05a, gracia-carpio06a, gao07a, bussmann08a, baan08a, juneau09a, bayet09a, wu10a, garcia-burillo12a}. Their high critical densities mean that these molecules require relatively high densities to be excited, and so, even if observations using these molecules do not spatially resolve the emitting regions \blue{(as is the case in all extragalactic applications)}, presumably the emission arises from compact and dense structures. Converting the observed luminosities to gas masses is non-trivial, and requires the application of a large-velocity gradient, escape probability, or similar approximation, and so the majority of these studies do not attempt to assess absolute values of $\eff$. However, to the extent that such estimates can be made, they also tend to find $\eff\sim 0.01$, albeit with large uncertainties \citep{krumholz07e, garcia-burillo12a}.

These small-scale results should be taken with caution, as they are subject to a number of systematic uncertainties of varying severity. As already mentioned, obtaining gas masses from molecular line observations always carries with it some degree of uncertainty, and that uncertainty is probably larger at small scales. The conversion from CO $J=1-0$ luminosity to mass has received the most attention, and is probably good to within a factor of $\sim 2$ for galaxies with metallicities and surface densities comparable to that of the Milky Way (see the recent review by \citet{bolatto13a}, and references therein), but the conversion for other molecules is certainly less well known, and for all molecules the conversion factor should depend on the abundances, temperatures and velocity dispersions of the emitting clouds \citep[e.g.][]{narayanan11a, narayanan12a, shetty11a, shetty11b, feldmann12a, feldmann12b}, which vary from galaxy to galaxy.

At small scales measuring the star formation rate is also non-trivial. Conventional conversions between tracers of star formation activity (e.g., ionizing or infrared luminosity) and true star formation rate all rely on an assumption that the emitting stellar population samples the full IMF and the full range of stellar evolutionary states, from stars just reaching the zero-age main sequence to stars dying as supernovae and ceasing to emit. The former condition requires that the stellar population have a mass of at least $\sim 10^3-10^4$ $\msun$ \citep{cervino04a, cervino06a, wu05a, fouesneau12a}, while the latter is only satisfied for stellar populations older than a few Myr \citep{krumholz07e, kennicutt12a}, and only if the mean star formation rate within the region under study is larger than $\sim 0.1$ $\msun$ yr$^{-1}$ \citep{da-silva12a}. Many small star-forming regions fail to satisfy these conditions, and the size of the resulting errors in inferred SFR depend on how badly they are violated. Estimates of SFRs from direct star counts are possible if the region being studied is close enough to resolve individual stars, and do not suffer from these problems. However, this method depends on uncertain estimates of the pre-main sequence lifetimes, and tends to give results that differ from those based on integrated light at the factor of $\sim 2$ level \citep{chomiuk11a, vutisalchavakul13a}. See \citet{kennicutt12a} for a thorough discussion of the pitfalls of various methods of measuring star formation rates.

\subsubsection{Combining Scales}

One can also combine the galactic and sub-galactic scales. This is of interest in part because the galactic-scale relationship between gas and star formation must ultimately be the sum of the relationship in numerous sub-galactic regions, but there are numerous plausible ways that this sum could be achieved. For example, the depletion time of $\tdep \approx 2$ Gyr seen for molecular gas on $\sim$kpc scales might be the result of numerous clouds that all have $\tdep\approx 2$ Gyr, or it might be the result of averaging together two distinct populations of clouds, one with $\tdep \ll 2$ Gyr and one with $\tdep \gg 2$ Gyr. Figure \ref{fig:sflawtot} suggests that the latter is certainly possible in outer galaxies, since the scatter in SFR at fixed gas content is more than an order of magnitude. This seems less likely at surface densities above $\sim 10$ $M_\odot$ pc$^{-2}$, where the scatter in $\Sigma_{\rm SFR}$ is far smaller, but it could still be the case that each $\sim 750$ pc pixel contains some actively star-forming and some passive clouds.


\blue{
Figure \ref{fig:eff} shows a combined plot that includes both small- and large-scale measurements of the star formation law. The core data set plotted was compiled by \citet{krumholz12a}, and then extended by \citet{federrath13c}. Figure \ref{fig:eff} further extends the data set by adding several more recent observations as well, including a sample of molecular gas in early-type galaxies from \citet{davis14a}. The data show that} both individual clouds and entire galaxies have roughly the same values of $\eff\approx 0.01$\blue{, and that this conclusion applies to all types of galaxies: dwarfs, disks and both low and high redshift, starbursts and mergers, and early types}. This suggests that the galactic-scale rate of star formation can plausibly be thought of as simply the sum of star formation in numerous clouds that are, to first order, all about the same in their properties.

\begin{figure*}[ht!]
\centerline{
\includegraphics[width=6in]{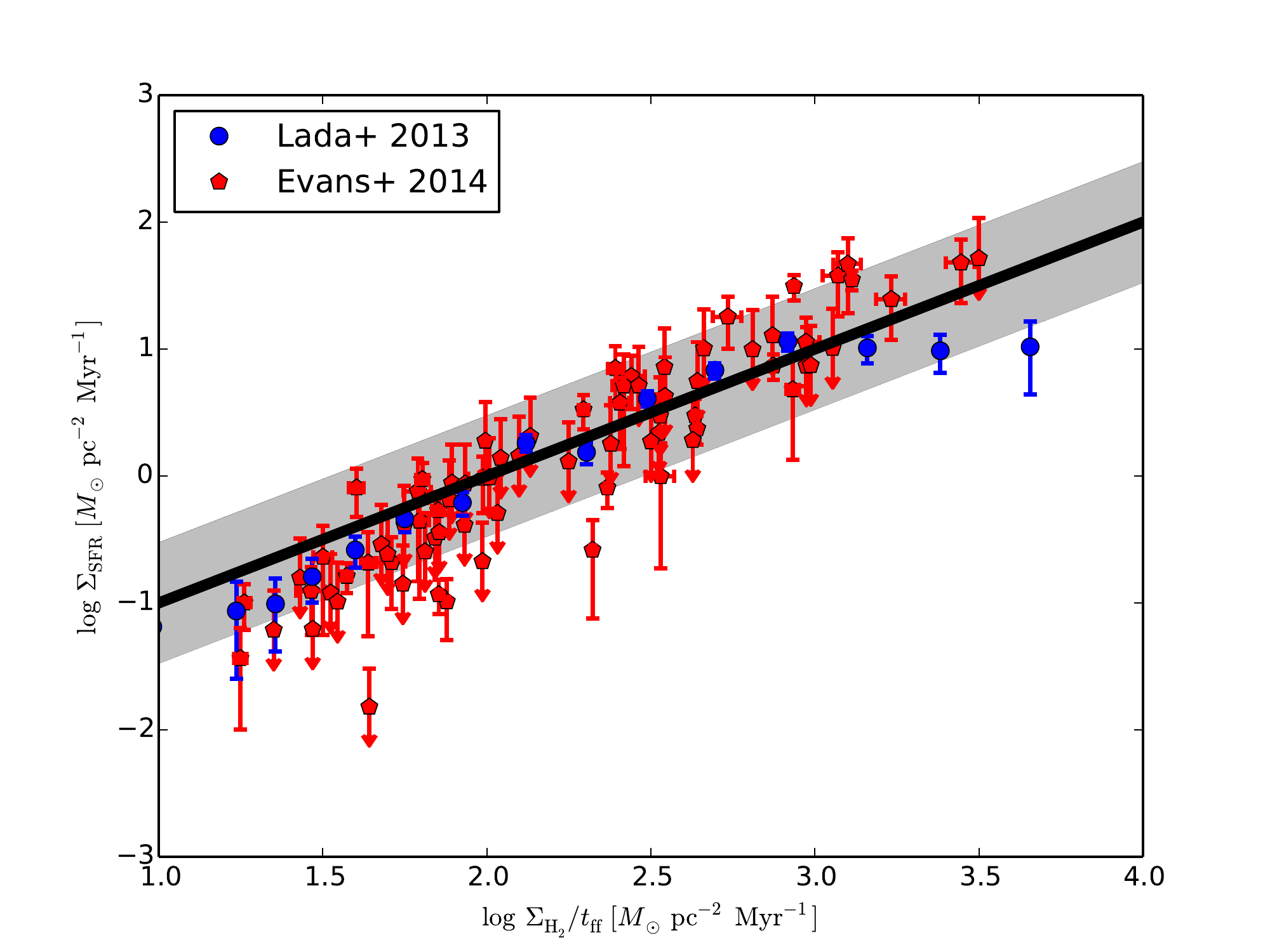}
}
\caption{
\label{fig:eff_intracloud}
\blue{
Surface density of star formation versus surface density of molecular gas normalized by estimated free-fall time $\Sigma/\tff$. This figure differs from Figure \ref{fig:eff} in that the plot shows the relationship plotted for successive contours of column density within individual clouds, rather than comparing multiple clouds (see text for details). The data shown are for the Orion A cloud \citep[blue circles]{lada13a}, and for several clouds selected from the c2d and Gould's Belt surveys \citep[red pentagons]{evans14a}. The black line and gray band show $\eff=0.01$ and a factor of 3 range around it, as in Figure \ref{fig:eff}.
}
}
\end{figure*}

\blue{Figure \ref{fig:eff} shows primarily an \textit{inter}cloud relationship, where for the most part there is one data point per object. However, for some Galactic clouds it is also possible to compute an \textit{intra}cloud relationship.}
This is accomplished by selecting portions of clouds above a given threshold surface density, and asking how quickly stars form within the regions defined by different surface density contours.
\blue{By counting the gas mass and number of young stars that are contained between any two such contours, we can estimate a value of $\eff$ for that gas. Figure \ref{fig:eff_intracloud} shows a result of this computation from two recent sets of observations of nearby clouds.}
The Figure shows that, while $\eff \approx 0.01$ gives roughly the right SFR, \blue{it may not fully describe the relationship between gas and star formation within an individual cloud. Indeed, \citet{evans14a} fit their data and find that, rather than a constant value of $\eff$, the data are best fit by $\eff \propto (\Sigma/\tff)^{0.3-0.5}$, with the best fit slope varying slightly depending on the fitting method used.} 
The systematic concerns regarding SFR estimates on small scales apply to both of the data sets shown in Figure \ref{fig:eff_intracloud}, but leaving these aside, the data suggest that $\eff\sim 0.01$ is a reasonable estimate on scales from entire galaxies to single clouds, but that within individual clouds something more complex may be taking place.

The data shown in Figures \ref{fig:sflawtot} -- \ref{fig:eff_intracloud} represent the first challenge that any predictive theory of star formation must meet: what physical processes are responsible for setting $\eff$, at scales from individual clouds to entire galaxies? On the other hand, observations show that $\eff$ is much less than $0.01$ in the atomic phase of the ISM. Why is that? Are there any molecular environments where $\eff$ deviates significantly from $\sim 0.01$, and if so, why?

\subsection{Stellar Clustering}
\label{ssec:obscluster}

\begin{figure*}[ht!]
\centerline{
\includegraphics[width=4in]{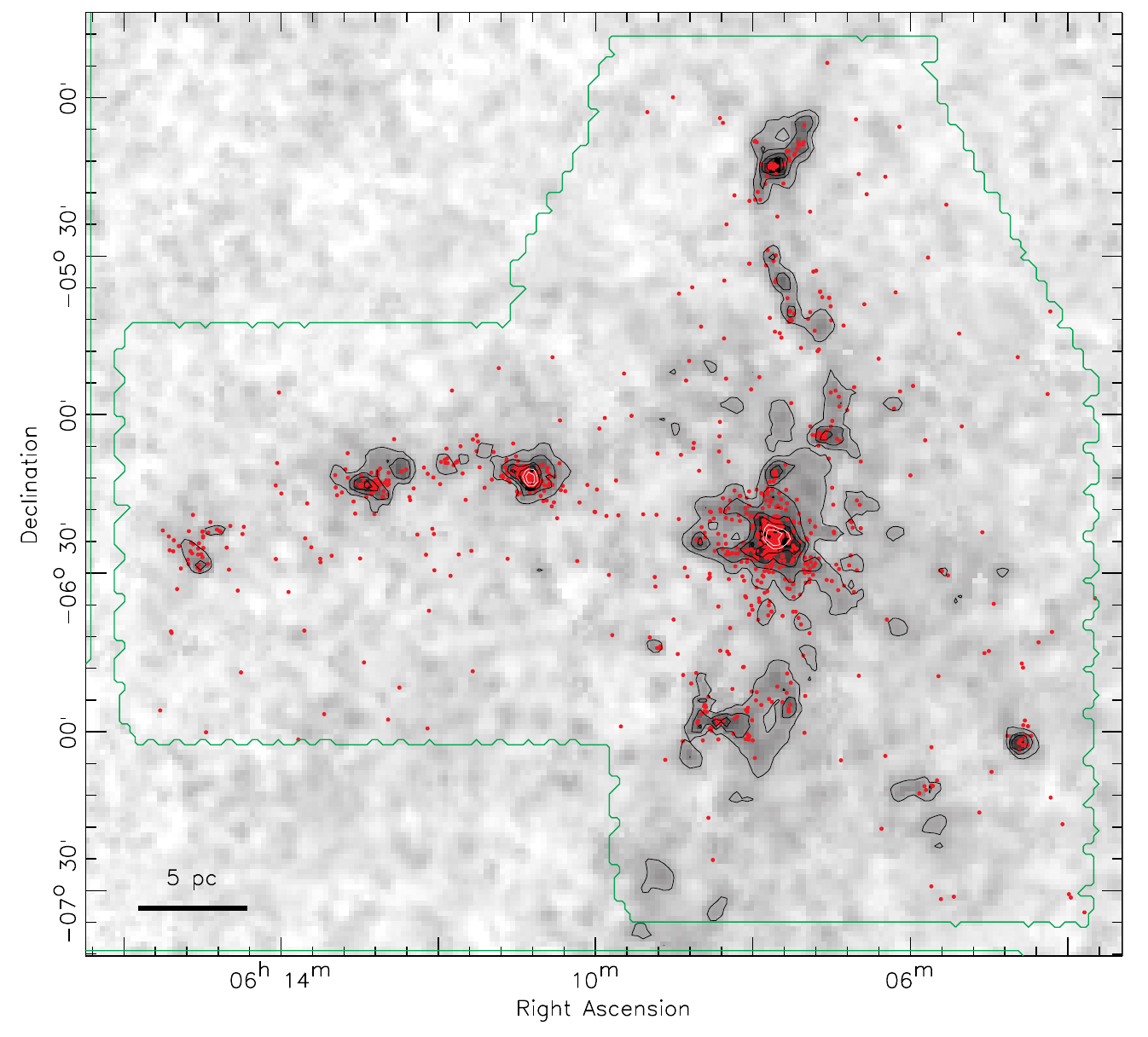}
}
\centerline{
\includegraphics[width=4in]{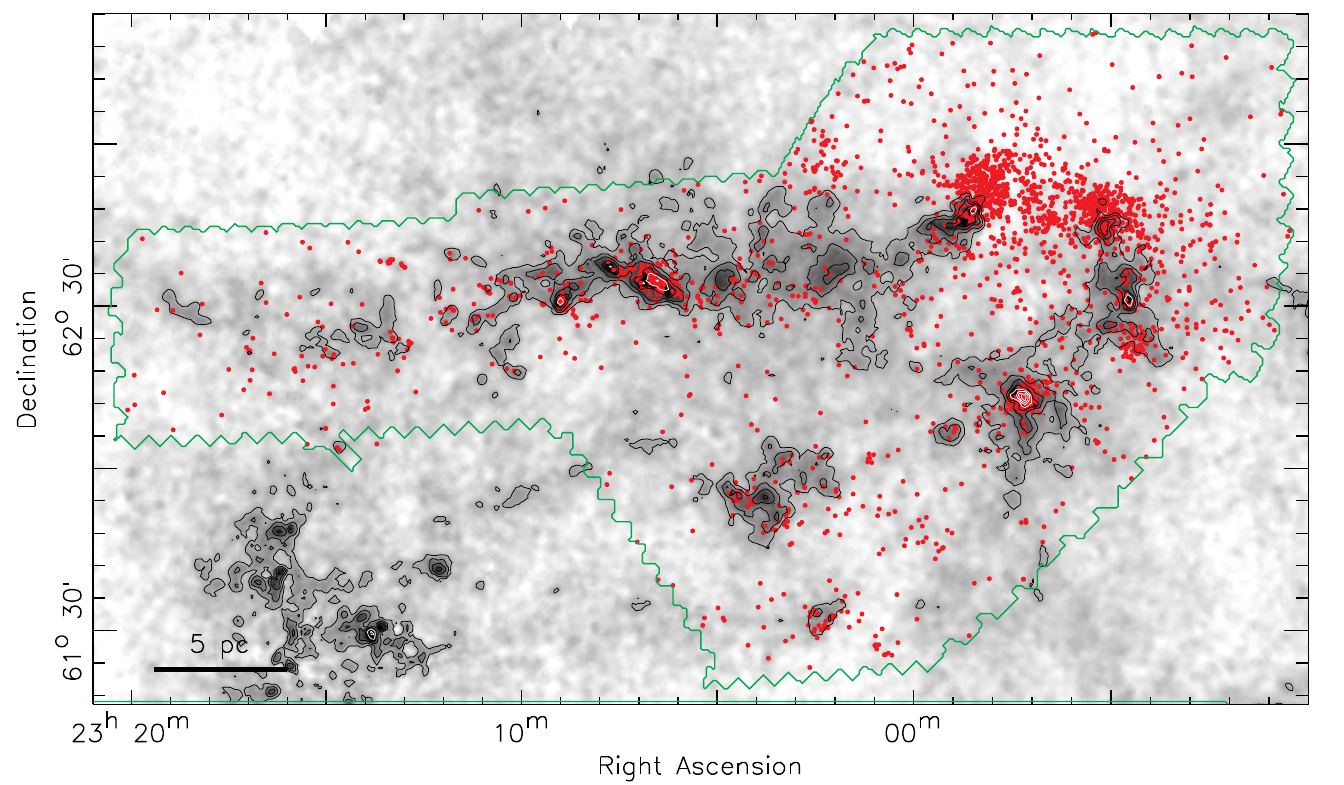}
}
\caption{
\label{fig:gutermuth11}
Distributions of gas and young stars in two star-forming regions near the Sun: MonR2 (top) and CepOB3 (bottom). In both panels, the inverted grayscale shows the gas column density as measured by the dust extinction; the color scale is from $A_V = -1$ to 10 mag, linearly stretched. \blue{(Note that negative $A_V$ is possible due to noise in the observations.)} Contours start at $A_V=3$ mag and increase by $2$ mag thereafter. Red circles indicate projected positions of young stellar objects identified by infrared excess as by \textit{Spitzer}. The green contour marks the outer edge of the \textit{Spitzer} coverage. Reprinted from \citet{gutermuth11a}, reproduced by permission of the AAS.
}
\end{figure*}

The dissection of the relationship between star formation and gas within a single molecular cloud illustrated in Figure \ref{fig:eff_intracloud} naturally points to the second topic of this review: how are young stars spatially arranged, with respect to one another and to the gas clouds from which they form? In nearby clouds, we see that stars form in a highly inhomogeneous fashion, with the stellar surface density varying by orders of magnitude even within the limited range of star-forming environments found within $\sim 1$ kpc of the Sun \citep{gutermuth09a, bressert10a, gutermuth11a}. Figure \ref{fig:gutermuth11} shows examples of the gas and stellar distributions in two nearby star-forming regions, MonR2 and CepOB3.

\subsubsection{Statistical Description of Gas and Stars}

The apparent inhomogeneity can be characterized using wide variety of statistical tools. One of the most commonly-used is the two-point correlation function, or equivalently the mean surface density of companions, which simply measures the excess number of stars around a given star as a function of angular separation, compared to what one would expect for a Poisson distribution. Other quantitative techniques include fractal dimensions (which are closely related to two-point correlation functions) \citep{larson95a}, parameters extracted from minimum spanning trees \citep{cartwright04a}, and dendrograms \citep{rosolowsky08c, gouliermis10a}, to name only a few. When these techniques are applied to the stars in young clusters, the general result is that the stars are structured on a wide range of scales, as indicated by roughly powerlaw behavior in the two-point correlation function. However, there are breaks at both large and small scales, indicating deviations from scale-free behavior \citep[e.g.,][]{gomez93a, larson95a, simon97a, bate98a, nakajima98a, hartmann02a, hennekemper08a, kraus08a, schmeja09a}. 

Observations of the gas in star-forming regions find similar signatures of hierarchical structure, though these are somewhat harder to interpret as the results may depend on the choice of gas tracer used. The relatively low density gas traced by $^{13}$CO shows results similar to those measured for stars: scale-free behavior, as indicated by a powerlaw correlation function or similar statistic, over a broad range of scales, but with breaks at both large and small scales \citep[e.g.,][]{blitz97a, schneider11a}. If one instead focuses on high-density tracers, in nearby regions one can identify individual, dense structures known as cores. The structure of a single core is definitely not hierarchical and scale-free \citep[e.g][]{barranco98a, goodman98b, pineda10a}, but if one treats the dense cores as point particles like stars and analyzes their positions relative to one another, the result is again a powerlaw very similar to that observed for stars  \citep[e.g.,][]{johnstone00a, stanke06a, enoch08a}.

In the case of stars, the small-scale break in the correlation function has been interpreted as the transition between the regime of binary stars and that of correlations between stars in a cluster that are not bound to one another individually, but only to the cluster as a whole. In the case of gas, it has been interpreted as revealing the Jeans length in the cloud \citep{larson95a, blitz97a}, and these interpretations are not necessarily mutually exclusive. The large-scale break has been interpreted as representing the transition between scales where the free-streaming of stars after their birth has erased structure and those where it has not, though it conceivably also represents an edge to star formation associated with the transition from star-forming molecular gas to non-star-forming atomic gas.

For regions in which spectroscopy is available, \blue{one can also examine the velocity structure of the stars and the gas. In general, the velocities are hierarchically-correlated in much the same manner as the position. However, there are some systematic differences between low-density gas, dense gas, and cores. Both dense cores \citep{andre07a, kirk07a, rosolowsky08a} and stars \citep{furesz08a, tobin09a} show systematically smaller velocity dispersions than the diffuse gas in the same region. Despite their lower velocity dispersion, cores \citep{walsh04a} and stars \citep{furesz08a, tobin09a} have mean velocities that are similar to those of the surrounding, low-density gas. This behavior is perhaps easiest to understand when it is expressed in terms of moments of the velocity distribution. Consider observing a star-forming region, and making a map of the the first moment (the mean) and second moment (the dispersion) of the velocity distribution as a function of position. The observational situation is that the first moment map is qualitatively similar for the low-density gas, dense cores, and stars. The second moment map is qualitatively similar for the stars and dense gas, but both  stars and dense cores have significantly smaller second moments of their velocity distribution than does the low-density gas around them.}

 \subsubsection{Time Evolution of Stellar Clustering}
 
The correlations between stellar positions and between gas and stars are noticeably variable from one region to another, as is clear simply from visual inspection of Figure \ref{fig:gutermuth11}. In MonR2, the stellar distribution is well-correlated with the gas, while in CepOB3 the peaks of the gas and stellar distribution are noticeably offset from one another. \red{Quantitative analysis backs up the visual impression: \citet{gutermuth11a} find that the Pearson correlation coefficient between the logarithms of the gas and stellar surface densities is $0.87$ in MonR2, but only $0.17$ in CepOB3.} This is probably an evolutionary effect: CepOB3 contains multiple early-type stars whose feedback has likely dispersed the gas in which they were initially embedded. This is clearly related to the process by which the correlation between gas and  star formation breaks down on sufficiently small scales, as discussed in Section \ref{sssec:sfrsubgalactic}. These images therefore present us with a dual problem: what determines the spatial (and also kinematic) relationship of gas and stars in regions like MonR2 that are still gas-rich, and what processes cause a transition to things that look like CepOB3, where the gas and stars are spatially distinct?

\begin{figure*}[ht!]
\centerline{\includegraphics[width=6in]{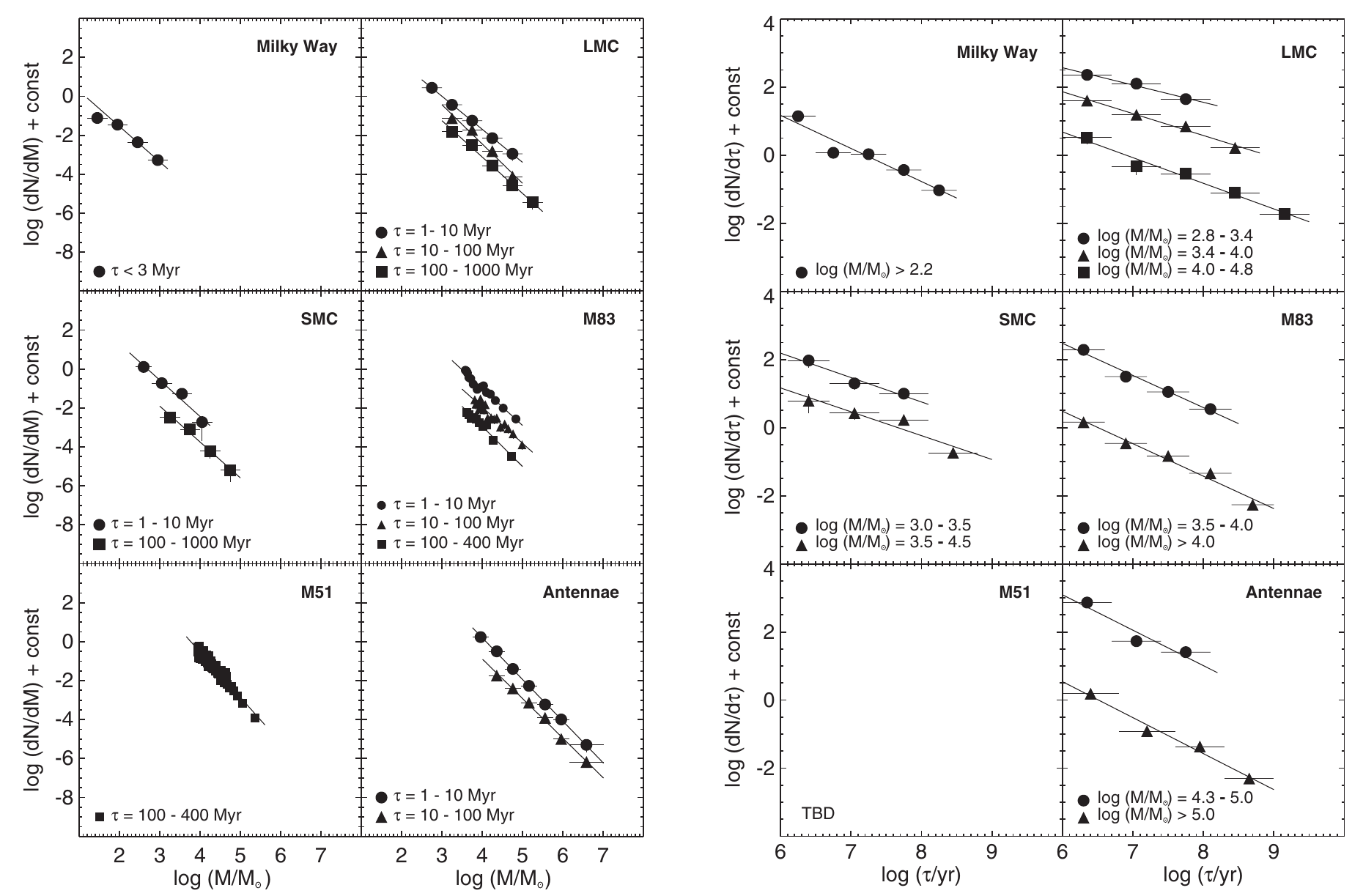}}
\caption{
\label{fig:fall12}
Cluster mass (left) and age (right) distributions in six galaxies. In the left panel, the figures show the number of clusters in each logarithmically-spaced bin in mass, with different symbols corresponding to different cluster ages. The right panel shows number of clusters in logarithmically-spaced age bins with different symbols for different cluster masses. Taken from \citet{fall12a}, reprinted by permission of the AAS.
}
\end{figure*}

The time-evolution of the spatial distribution of stars is also interesting over longer times. The stars shown in Figure \ref{fig:gutermuth11} are detected by their excess infrared emission, an observational feature produced by circumstellar material (almost certainly a disk) that reprocesses starlight into the infrared. Such signatures are present for only several Myr after a star forms \citep[e.g.,][]{haisch01a, hernandez08a}. The stellar density around such young stars is invariably much higher than that found in the Galactic field, but the density drops rapidly with stellar age. Even populations with ages of a few Myr have noticeably lower densities than stars that have just formed \citep{gutermuth09a}, and by the time stellar populations reach an age of $\sim 10 - 100$ Myr their densities have dropped dramatically, with only a small fraction remaining in gravitationally-bound structures \citep{silva-villa11a, fall12a, silva-villa13a}. The right panel of Figure \ref{fig:fall12} illustrates the important result: the number of clusters of a given age declines dramatically from ages below $\sim 3$ Myr to $\sim 300$ Myr.

Before moving on, an important caveat is in order, which is that there is considerable debate in the literature about the exact functional form of this decline, with some authors favoring a power law in age $\tau$ of the form $dN/d\tau \propto \tau^{\gamma}$ with $\gamma \approx -1$ \citep{fall05a, fall09a, chandar10a, chandar10b, chandar11a} while others argue for little or no cluster disruption, but that only a small number of stars are formed in clusters in the first place \citep{boutloukos03a, de-grijs03a, gieles07a, bastian11a, bastian12a, bastian12b, silva-villa13a}. There is also a secondary debate about whether the fraction of stars that remain in clusters over long times is universal, or depends on some galaxy-scale property. Much but not all of this debate is semantic, and has to do with whether one should classify young star systems that still contain gas, or that have just become star-dominated but are not yet dynamically relaxed, as star clusters. Those who define clusters purely as stellar overdensities tend to obtain power law declines with age, while those who limit their samples based on morphology, age relative to crossing time, or other indicators of a relaxed state tend to find that most stars form in unbound structures (referred to as associations) rather than clusters, but that those clusters that do form are likely to survive for many dynamical times.

This semantic debate, however, should not obscure the interesting underlying physics questions, which can be posed independently of the definition of cluster: why does the density of stars begin to drop dramatically as soon as stars emerge from their parent gas clouds? The gas clouds from which star clusters or associations form appear to be gravitationally bound \citep[e.g.,][]{dobbs14a}, so why aren't the stars themselves? Is the process that regulates what fraction of stars remain in clusters governed mainly by processes internal to the star-forming clouds, or is the galactic environment the dominant influence?

A related question has to do with the mass function of those structures that do remain bound. While there is significant dispute about cluster age distributions, observations in a wide variety of galaxies consistently find that the mass function for open clusters is well-described by a power law $dN/dM\propto M^{-\beta}$ with $\beta \approx 2$ over most of its range\blue{; the uncertainty on the value of $\beta$ is roughly $0.1-0.2$} \citep{williams97a, zhang99b, larsen02a, bik03a, de-grijs03b, goddard10a, bastian11a, fall12a}. The left panel of Figure \ref{fig:fall12} provides an example. Some authors also report a non-powerlaw cutoff at the highest masses \citep{bastian08a, larsen09a, bastian12b}, though the reality of this feature is disputed \citep{fall12a}. The index $\beta \approx 2$ is interesting, in that it is noticeably shallower than the index describing the masses of individual stars (roughly $2.3$; see the following section), but slightly steeper than that describing the mass function of individual molecular clouds in the molecule-rich parts of galactic disks (roughly $1.5$ to $2.0$; \citep{solomon87a, heyer01a, rosolowsky05a, roman-duval10a, gratier12a}). \blue{It is about comparable to the mass function one obtains by selecting dense regions within molecular clouds \citep{shirley03a}.} The origin and universality of this cluster mass function has received fairly little theoretical attention, far less than the stellar mass function, but it no less a problem for star formation theory to explain.

\subsection{The Initial Mass Function}
\label{ssec:imfintro}

Zooming in even further from stellar clusters, we reach the scale of individual stars. Numerous properties of stars are important for determining their observable characteristics and evolutionary path, but of course the most important is their mass. The distribution of stellar masses at birth is known as the initial mass function (IMF). The current state of research into the IMF has been the subject of several recent reviews \citep{bastian10a, jeffries12a, kroupa13a, offner14a}, so I here I only provide a short synopsis, and refer readers to those reviews for more details. It should be noted that there is some level of disagreement even on the observational side (for example see \citet{kroupa13a} versus \citet{offner14a} and \citet{bastian10a}), but for the purposes of this review I have mostly limited myself to those issues about which there is some consensus\blue{; where I make controversial claims about the observations that are not universally-accepted, I will attempt to make this clear}. Observational efforts to measure the IMF can be divided into two categories: those that make use of resolved stellar populations, and those that attempt to measure the IMF of unresolved stellar populations. 

\subsubsection{Resolved Stellar Populations}
\label{sssec:imfresolved}

\paragraph{Field Star Surveys}

The most obvious way to measure the IMF is to begin with field stars visible in the Solar neighborhood, an effort that began with \citet{salpeter55a}. Measuring the IMF from the field involves five main steps, the first two of which involve construction of the sample, and the last three of which involve derivation of the IMF from it. First, one must determine the luminosity function for stars in the sample region in some observing band. This requires that apparent magnitudes be combined with distance measurements, which are not trivial to obtain. For stars within $\sim 20$ pc of the Sun, accurate parallax distances from Hipparcos can be used \citep[e.g.,][]{reid02a}, but such a small survey volume provides limited statistics, and so it is more common to use distances estimated from photometry or spectroscopy \citep[e.g.,][]{bochanski10a}, which are subject to significant systematic uncertainties. Thus the design of the survey involves a tradeoff between statistical and systematic errors. Second, the sample must be corrected for Malmquist and Lutz-Kelker \citep{lutz73a} bias; the former describes the tendency of intrinsically-brighter stars to be overrepresented in a magnitude-limited sample because they are visible to large distances, and the latter describes an asymmetry whereby, in constructing a volume-limited sample using parallax distances, it is more likely that stars from outside the sample region to scatter into it due to error than for stars inside the sample volume to scatter out.

Once the sample is constructed, the third step is to convert absolute magnitudes into stellar masses using a mass-magnitude relationship, either empirical or theoretical, which gives the present-day mass function. Fourth, the present-day mass function (PDMF) must be transformed into the IMF, which requires correcting for the effects of stellar evolution. For stars whose main sequence lifetimes are longer than the age of the Galaxy (roughly those with $M<0.8$ $M_\odot$), the IMF and PDMF are identical, but for more massive stars we must correct the observed number by a factor $t_{\rm MS}(M)/t_{\rm gal}$, where $t_{\rm MS}(M)$ is the time for which stars of mass $M$ remain on the main sequence, and $t_{\rm gal}$ is the time over which stars of that mass have been forming in the Galaxy. Thus determining the right evolutionary correction requires knowledge of the star formation history of the region being observed. Fifth and finally, if the volume being observed does not sample the full Galactic scale height, one must correct for the mass-dependence of stellar scale heights, which causes more massive stars to be overrepresented in a sample near the Galactic plane \citep{miller79a}. 

\paragraph{Young Cluster Surveys}

\blue{An} alternative strategy using resolved stars is to study a young star cluster. Compared to observations using field stars, this approach has several important advantages. The correction from the PDMF to the IMF is, depending on the age of the cluster, either much smaller or non-existent. This makes clusters a more reliable means of measuring the massive end of the IMF. Another advantage of clusters is that the stars are coeval or nearly so, and are also of uniform chemical composition. For stars on the main sequence, this removes a major source of error that arises in the conversion between absolute magnitude and mass for field stars. Moreover, the stars are all at nearly the same distance, and for at least some clusters this distance is known quite precisely from interferometry-based parallaxes to radio-flare stars \citep[e.g.,][]{hirota07a, sandstrom07a, menten07a, reid09a}. Low-mass stars and brown dwarfs are also much brighter when they are young, they are far easier to detect in clusters than in the field.

\begin{figure}[ht!]
\centerline{\includegraphics[width=3in]{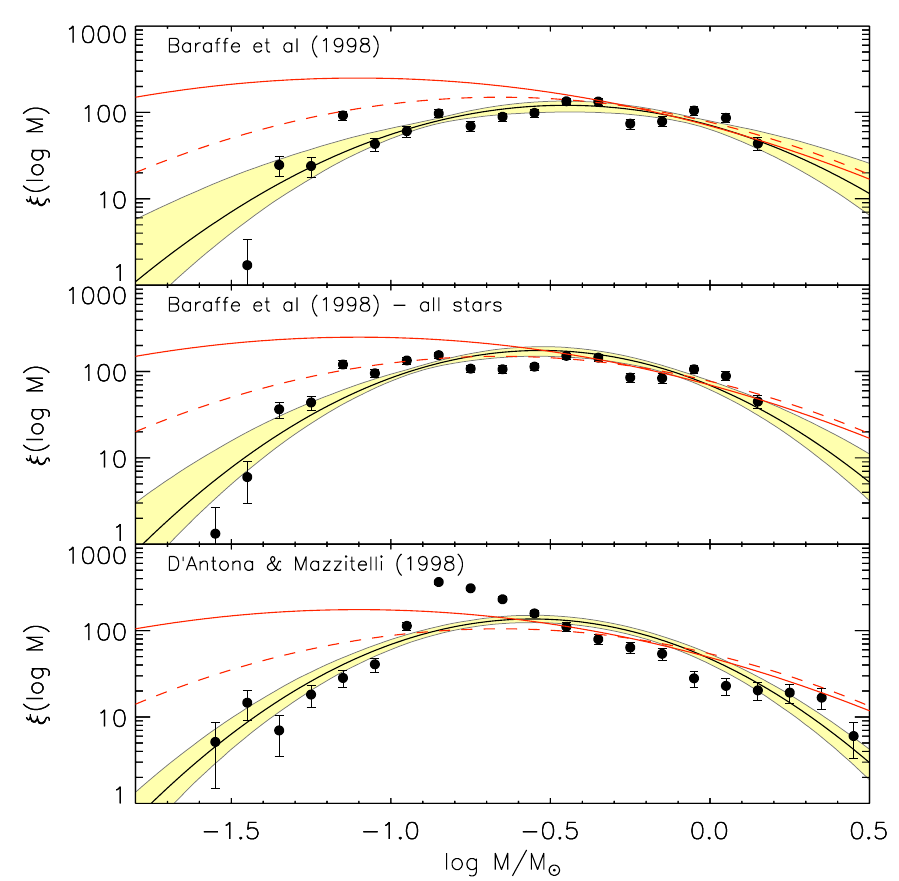}}
\caption{
\label{fig:da_rio12}
The IMF $\xi(\log m) = dn/d\log m$ of stars in the Orion Nebula Cluster as inferred from \textit{Hubble Space Telescope} photometry. In each panel the black points show the data; the error bars are the 1$\sigma$ errors that result from a combination of counting statistics and incompleteness. Although the underlying data in each panel are the same, the three panels show the results of converting the observed colors and magnitudes to stellar masses using three different models. The bottom panel uses the models of \citet{dantona98a}, while the top two panels both use the models of \citet{baraffe98a}, using two different methods for handling stars that fall outside Baraffe et al.'s model grid. The red solid and dashed lines are the single-star and system IMFs of \citet{chabrier03a}, while the black curve is the best fit of the data to a lognormal functional form; the yellow band shows the $1\sigma$ uncertainty on the fit. Taken from \citet{da-rio12a}, reprinted by permission of the AAS.
}
\end{figure}

On the other hand, clusters also have significant downsides compared to the field, particularly for stars near the peak of the IMF. In clusters young enough to contain massive stars, stars with masses $\lesssim 1$ $M_\odot$ will not yet be on the main sequence, which greatly complicates the mapping between luminosity and mass, and introduces a significant source of systematic error. Figure \ref{fig:da_rio12} illustrates an example of this uncertainty, by showing IMFs derived using two different sets of pre-main sequence evolution models. In cluster cores, where stellar densities are high, confusion can become a significant problem \citep{ascenso09a}, and in practice this is what limits the distance from the Sun to which the cluster method can be used to measure the IMF. Young clusters are often still partially shrouded by dust, and this creates difficulties in measuring accurate luminosities in the first place, particularly since the extinctions are not necessarily the same from star to star. They are also often mass-segregated, and this creates problems if the observations only sample the cluster core or envelope \citep[e.g.,][]{pang13a, lim13a}. Mass segregation in combination with either confusion or a radial gradient in dust extinction within the cluster creates particularly difficult-to-remove systematic effects, as they cause the errors associated with either extinction or confusion to correlate systematically with stellar mass \citep{parker12b}. N-body interactions that eject massive stars from clusters entirely can also create systematics that are very difficult to remove \citep{banerjee12c}.

\paragraph{\blue{Globular Cluster Surveys}}

\blue{
A third method for studying the IMF is to use the resolved stellar population in globular clusters. This method shares several of the advantages of the young cluster method, in that the population is at a known, uniform distance, and is close enough to coeval and chemically homogenous that corrections for age and abundance variations are not a major source of uncertainty. Moreover, globular clusters offer the only opportunity to perform resolved studies of low-metallicity, ancient stellar populations, which are otherwise accessible only via techniques for the study of unresolved populations, which have their own pitfalls (see Section \ref{sssec:unresolvedimf}). While these observations obviously provide little or no information about the IMF for massive stars, they provide one of the few ways to explore whether the IMF of low mass stars varies with metallicity or over cosmic time.
}

\blue{
The price for access to these low-metallicity, ancient stars is that one is faced with systematic uncertainties stemming from dynamical evolution. Globular clusters undergo significant mass segregation, and can lose a significant fraction of their low-mass stars through two-body evaporation; depending on the cluster, stellar collisions may also significantly modify the mass function \citep{spitzer87a}. Thus the procedure for deriving an IMF for a globular cluster is not simply a matter of fitting to the observations and then perhaps making a correction for star formation history. One must instead start with a proposed IMF, calculate how the mass function will evolve over the age of the cluster, and then compare the result to the observations. Fortunately calculations of purely N-body evolution are reasonably straightforward computationally, and the processes involved are well-understood analytically, so such corrections can be done with some level of confidence. However, there are still significant uncertainties stemming from poorly known parameters such as the cluster's binary fraction and degree of mass segregation at birth, and the cluster's orbit around the Milky Way; the latter matters because it affects the strength of the tidal potential responsible for stripping off low mass stars.
}

\paragraph{\blue{Chemical Abundance Patterns}}

\blue{
In principle measurements of the chemical abundance patterns in stars can provide a fourth path to measuring the IMF \citep[e.g.,][]{tolstoy03a, mcwilliam13a}. This is because different elements are primarily produced by stars of differing masses; for example, $\alpha$ elements are produced primarily by type II supernovae occurring in stars larger than $\sim 8$ $M_\odot$, while iron peak element production is dominated by type I supernovae whose progenitors are white dwarfs with significantly lower birth masses. Thus measuring element ratios in principle makes it possible to infer the IMF of the stellar population that produced those elements. However, such inferences are subject to a vast number of confounding uncertainties, involving stellar yields, binary stellar evolution, metal mixing in the ISM, and galactic winds. These uncertainties are such that any conclusions drawn from this technique are tentative at best, and for this reason I will not discuss it further.
}

\paragraph{Binarity}

Finally, there is one important limitation that affects \blue{the young cluster, globular cluster,} and field star methods: unresolved binaries. None of the observations used to make these measurements are capable of resolving binaries \red{except for those with the very largest angular separations}, and so the observed magnitudes that are used to estimate masses will in some cases be system rather than single-star magnitudes. Since stellar luminosities are generally steep functions of mass, to first order the effect of this is simply that a number of low-mass stars in multiple systems will be hidden by the light of their more massive companions. However, the extent to which this statement is true depends on the choice of observing band (since the mass-magnitude relationship is steeper in bluer bands than in redder ones) and on the underlying distribution of binary separations and mass ratios.

If the underlying distributions are known at least approximately, as is the case in the field, it is possible to attempt to correct for the bias introduced by unresolved binaries in order to produce separate single-star and system IMFs \citep[e.g.,][]{chabrier05a}. The correction is not huge, because most low-mass stars are single \citep{fischer92a, lada06a, basri06a, allen07b, raghavan10a}. While most massive stars do have companions \citep{preibisch99a, mason09a}, the number of massive stars is relatively small, implying a fairly sharp upper limit to the absolute number of low-mass stars that could be cloaked by companions. Brown dwarfs represent a possible exception to this statement, since they are both intrinsically rarer than stars and easily concealed by a stellar companion. Fortunately there appear to be few brown dwarf-stellar binaries \citep[e.g.,][]{dieterich12a}, but the exact form of the IMF at low masses is quite sensitive to exactly how rare they are, since K and M stars are so numerous that even a small number of brown dwarf companions to them might represent a non-negligible contribution to the total number of brown dwarfs.

For young clusters \blue{and globular clusters}, on the other hand, it is at present not feasible to correct for binarity, because the binary star fraction, as well as the mass ratio and orbital period distributions, appear to be functions of both cluster properties and age \citep[and references therein]{duchene13a}. There have been some theoretical attempts to reverse-engineer the binary populations of embedded clusters based on dynamical modeling \citep{marks11a, marks12a}, but these are still works in progress, and have not yet been used in an attempt to make binarity corrections to IMF measurements in young clusters.

\paragraph{Results}

With the caveat about binaries aside, observations using the field star \citep{kroupa01a, kroupa02c, chabrier03a, chabrier05a, covey08a, deacon08a, bochanski10a, parravano11a}, young cluster \citep{muench02a, chabrier03a, chabrier05a, sabbi08a, andersen09a, sung10a, lodieu11a, lodieu12a, lodieu12b, da-rio12a, habibi13a}\blue{, and globular cluster \citep{de-marchi00a, de-marchi10a, leigh12a}} methods \blue{all} appear to produce roughly consistent results, at least in the stellar regime. Figure \ref{fig:da_rio12} shows a typical result from one of these studies. As shown in the figure, the IMF has a distinct peak in the mass range $0.1 - 1$ $\msun$. It falls off as a powerlaw $dN/dm \propto m^{-\alpha}$ at higher masses, with $\alpha \approx 2.35$, the value originally determined by \citet{salpeter55a}. There are numerous possible functional representations of the IMF: broken powerlaws \citep{kroupa01a, kroupa02c}, lognormals to represent the peak coupled with powerlaws for the tail \citep{chabrier03a, chabrier05a}, and powerlaws with exponential cutoffs at low mass \citep{parravano11a}. The combined lognormal-powerlaw form for the single-star IMF suggested by \citet{chabrier05a} is
\begin{equation}
\label{eq:chabrierimf}
\frac{dn}{d\log m} =
\mathcal{N}
\left\{
\begin{array}{ll}
\exp\left[-\frac{(\log m - \log m_c)^2}{2\sigma^2}\right], & m \leq m_b \\
A (m/m_b)^{1-\alpha}, & m > m_b
\end{array},
\right.
\end{equation}
with $\sigma=0.55$, $m_b = 1$, $\alpha=2.35$, and $A = \exp\{-[\log (m_b/m_c)]^2/(2\sigma^2)\}$ (so as to guarantee continuity across the lognormal-powerlaw break). Here stellar mass $m$ is measured in units of $M_\odot$, and $\mathcal{N}$ is a normalization constant. The alternate functional forms are generally identical within the spread of observational error. The greatest uncertainty is in the brown dwarf regime below $0.08$ $\msun$, where there is clearly a fall-off from the peak, but its exact sharpness and functional form are poorly-determined. Some authors report evidence for a discontinuity between stars and brown dwarfs \citep{thies07a, thies08a}. There may also be an upper cutoff somewhere between $100$ and $150$ $M_\odot$ \citep{elmegreen00a, weidner04a, figer05a}, although this possibility has been challenged by recent observations of stars that appear to exceed the proposed limit \citep{crowther10a, doran13a}. Even if there is a cutoff to the PDMF of massive stars, it is possible that this is a result of a sharp increase in instability and mass loss beyond a certain limiting mass, rather than an aspect of star formation \citep[and references therein]{tan14a}. 

There are only a few convincing cases for deviations from this IMF based on resolved stellar populations, and unfortunately the subject has a long history of disputes over whether results are statistically significant, with the most conservative and careful analyses suggesting that published uncertainties are often significantly underestimated \citep{weisz13a}. For example, \citet{geha13a} measure the IMF in two ultra-faint dwarf satellite galaxies of the Milky Way via direct counting of stars in the mass range $0.52-0.77$ $\msun$, below the main sequence turnoff mass for these stellar populations. They find that, if they fit a powerlaw mass function in this range, their best-fit slope is strongly inconsistent with the Salpeter slope $\alpha=2.35$, and with the slope of $\alpha=2.3$ used in the \citet{kroupa02c} broken-powerlaw functional form for the IMF. They report this inconsistency as evidence for IMF variation. However, if they instead choose to fit a lognormal form to the data, the results are consistent at the $1\sigma$ level with the best-fit values given in equation (\ref{eq:chabrierimf}). Similarly, \citet{kalirai13a} perform star counts in a field in the outskirts of the Small Magellanic Cloud over a mass range  $0.37- 0.93$ $\msun$, and find that the data can be fit by a single powerlaw with no turnover. However, the data are again not capable of excluding the functional form given by equation (\ref{eq:chabrierimf}) even at the $2\sigma$ level \citep{offner14a}. The most convincing cases for IMF variation based on resolved stellar populations are for the clusters forming near the Galactic center, including the Quintuplet cluster \citep{husmann12a} and the nuclear star cluster \citep{lu13a} do appear to have IMFs where the high-mass slope is somewhat flatter than the Salpeter value $\alpha=2.35$.

\blue{
Several authors have also claimed that the IMF varies systematically with the mass of the mass of the star cluster in which the stars formed, with the powerlaw tail at high masses being truncated at a value that depends on the cluster mass, based either on direct comparisons of stellar and cluster masses in the Milky Way \citep{kroupa03a, weidner06a, weidner10a, weidner13a}, or on indirect indicators such as X-ray binary populations \citep{dabringhausen09a, dabringhausen12a} and globular cluster properties \citep{marks12b}. If true, this would imply that the IMF integrated over an entire galaxy is steeper than that of individual massive clusters, since much star formation occurs in low mass clusters, and this in turn would have major implications for models of chemical evolution \citep{koppen07a} and interpretation of star formation rate indicators \citep{pflamm-altenburg08a, pflamm-altenburg09a}. Models based on this \textit{ansatz} are known as integrated galactic IMF (IGIMF) models.}

\blue{
However, the observational basis for IGIMF models is questionable. The most convincing evidence for an IGIMF effect is the direct comparisons of stellar and star cluster masses, as the indirect measures depend strongly on a number of poorly known parameters (e.g., how the star-formation efficiency in a forming globular cluster scales with the number of massive stars). For these direct tests, published claims of statistically-significant cluster-to-cluster variation are sensitively dependent on the values adopted for the errors in the measurement of stellar and cluster mass. Since these are dominated by systematic uncertainties, they are extremely hard to estimate, and quite easy to underestimate. To give an example, for the Orion Nebula Cluster (M42) \citet{weidner06a} and \citet{weidner10a} adopt a stellar mass of $2200\pm 300$ $M_\odot$ from \citet{hillenbrand98a}. However, the much more recent survey of \citet{da-rio12a}, which improves on the original data set significantly by using space-based photometry, star-by-star extinction modeling, and a new distance estimate based on radio parallax, yields a total stellar mass closer to $1000$ $M_\odot$ (da Rio, 2012, priv.~comm.). Using the new, lower mass, the expected maximum mass for a non-truncated IMF is in fact close to the observed mass of the most massive star. More recent IGIMF analyses have used newer estimates for the mass of the ONC and continue to report a significant IGIMF effect \citep{weidner13a}, but that does not obviate the point of the example, which is how easy it is to misestimate the error bar. \citet{weidner06a}'s original error bar of $\pm 300$ $M_\odot$ was clearly too optimistic by a factor of several. This should serve as an important caution about how much weight to give to claims of statistical significance that depend on such error bars.
}

A further, related, concern is that searches for a cluster mass-dependent truncation of the IMF have thus far yielded positive results only using data culled from the literature, where there is no uniform definition for what counts as a cluster, and masses for both clusters and stars have not been derived in the same way from one object to another. All searches using homogeneously-observed and -analyzed data sets have thus far returned null detections \citep{calzetti10b, koda12a, andrews13a}. These studies are based on unresolved photometry, which certainly has its own systematic errors, but these are probably better understood and easier to model than the systematics that affect the inhomogenously-selected Galactic data set, yielding a cleaner measurement.

\blue{
Finally, it is worth noting that a claim that the IMF varies depending on the mass of the star cluster in which the stars formed runs up against a problem that should be clear to any reader who has examined Section \ref{ssec:obscluster}. While star clusters have well-defined masses once they have dynamically relaxed, stars that are still forming out of their parent clouds cannot easily or uniquely be divided into identifiable clusters with definite masses. Depending on how one defines clusters, a large fraction of stars may not form in them at all. Even if one adopts an expansive definition such that most stars are born in clusters, the mass that one assigns to a given cluster cannot be specified independent of the cluster definition. Thus an IGIMF model faces a fundamental problem: it is coherent and predictive only to the extent that one can find a meaningful and physically-motivated way of defining the masses of star clusters when the stars are still embedded in their parent clouds. Thus far no such definition has been proposed.
}

\subsubsection{Unresolved Populations}
\label{sssec:unresolvedimf}

The field and young cluster methods for determining the IMF can be used only in the Milky Way and a few of its closest galactic neighbors; beyond this distance, it is no longer possible to resolve individual stars. As a result, the range of star-forming environments accessible via the resolved population methods is somewhat limited, and it is desirable to push further to check if the IMF might depend on the environment. Doing so requires the use of integrated light measurements, which means that one must resort to stellar population synthesis (SPS) models to interpret the data. Such models can be applied to either spectroscopic data or to data that combines photometry with dynamical modeling.

\paragraph{Bottom-Heavy IMFs in Early Type Galaxies}

On the spectroscopic side, in a series of papers, van Dokkum and Conroy \citep{van-dokkum10a, van-dokkum11a, van-dokkum12a, conroy12a, conroy13a} (also see \citet{spiniello12a}) introduced a method to analyze the IMF in early type galaxies using a variety of spectral features that are sensitive to both stars' effective temperature and surface gravity. The latter sensitivity makes it possible to separate main sequence stars  from giants with similar surface temperatures, and the former picks out a particular mass range, generally $\sim 0.1-1$ $\msun$ depending on the particular spectral feature used. They find that the spectra in these galaxies are strongly inconsistent with a turnover in the IMF in the $0.1-1$ $\msun$ range; instead, the IMF must remain as steep as a powerlaw of slope $\alpha=2.35$, or perhaps even steeper. In a crucial consistency check, the signatures of such a bottom-heavy IMF are \textit{not} found in the similarly-ancient populations of globular clusters, where a steep IMF such as that inferred in early type galaxies is ruled out by dynamical constraints \citep{van-dokkum11a, strader11a}. If anything, \citet{strader11a} conclude that the globular clusters appear to have a shallower IMF than the disk of the Milky Way. Overall, the level of bottom-heaviness in the IMF appears to correlate with the velocity dispersion of the galaxy.

The other available method for measuring the IMF in unresolved stellar populations is to compare observed mass to light ratios with values expected for a given theoretical IMF. This approach has two parts. First, one must measure the stellar mass to light ratio of a target galaxy. This means determining the underlying mass distribution, which can be accomplished using either stellar kinematics \citep{cappellari12a}, constraints from the lensing of background sources by the target galaxy \citep{thomas11a}, or a combination of both \citep{sonnenfeld12a}; with these data, one can fit both the total mass distribution (including dark matter) and the stellar mass distribution. The second step is to compute a theoretical mass to light ratio using an SPS model, and compare the observed and predicted ones. Modeling of this sort shows that, consistent with the spectroscopic method, the most massive early type galaxies have mass to light ratios significantly larger than would be expected for an IMF like that given by equation (\ref{eq:chabrierimf}), and instead are in better agreement with an IMF that is a pure powerlaw of slope $\alpha=2.35$ or steeper in the range $0.1-1$ $M_\odot$ \citep{thomas11a, cappellari12a, sonnenfeld12a}.

There are potential systematic worries with both of the above methods. The spectroscopic method relies on the ability of stellar population synthesis models to reproduce the properties of the ancient, metal-rich stars found in giant early type galaxies, and there is a dearth of similar stars in the Milky Way or similarly nearby locations where the stars could be resolved, and their spectra compared to the models directly. While the models do pass a number of consistency checks in the stars that are available, there is still a possibility that they are missing something important. Similarly, the dynamical models rely on the ability of SPS models to predict the mass to light ratios of these stellar populations. If changes in stellar evolution lead to a much higher mass of dark remnants (neutron stars and black holes) than current models predict, that would explain the elevated mass to light ratios without resort to variations in the IMF. However, the systematics that would be required to explain both sets of observations without varying the IMF are quite different, and the fact that the two methods give consistent results adds significantly to their credibility.

\paragraph{A Cautionary Tale}

Despite the growing evidence for IMF variation in early type galaxies, it seems appropriate to end this section with a cautionary tale about the interpretation of light from unresolved stellar populations as variation in the IMF. Prior to the current generation of observations focusing on early type galaxies, there were similar claims in the literature for variations in the IMF of dwarf galaxies. The primary piece of evidence for this claim came from the H$\alpha$ emission of these galaxies; H$\alpha$ is produced by recombination in ionized gas, and thus H$\alpha$ emission is a (relatively) straightforward proxy for ionizing luminosity. Since ionizing luminosity comes primarily from the most massive stars, a comparison of the H$\alpha$ / ionizing luminosity to luminosities in other bands that are less weighted toward massive stars in principle provides an efficient means of measuring the high-mass slope of the IMF. Galaxies, or regions of galaxies, with low total or areal star formation rates show systematic deficiencies in the amount of H$\alpha$ they produce per unit far ultraviolet (FUV) emission \citep{boissier07a, lee09a, boselli09a, meurer09a}, and systematically low H$\alpha$ equivalent widths \citep{hoversten08a, gunawardhana11a}. Some authors interpreted this as evidence that these galaxies have an IMF systematically deficient in massive stars \citep[e.g.,][]{pflamm-altenburg08a, pflamm-altenburg09a, krumholz08a}. 

However, improvements in stellar population synthesis modeling revealed a more prosaic explanation: stars form in temporally-correlated clusters, and as a result, in regions with low star formation rates, the H$\alpha$ luminosity undergoes very large fluctuations. In a single star cluster, the H$\alpha$ to FUV ratio and H$\alpha$ equivalent width \blue{are initially large,} when the massive stars \blue{that dominate ionizing photon production} are still on the main sequence. These quantities then fall \blue{over a $\sim 10$ Myr time scale} as the stellar population ages \blue{and these stars leave the main sequence}. At high star formation rates a galaxy contains many clusters at different stages of this cycle, and an unresolved observation that combines the light from all the clusters washes out the fluctuations, resulting in a fairly steady H$\alpha$ luminosity \blue{and values of the H$\alpha$ to FUV ratio and H$\alpha$ equivalent width that vary little}. At low star formation rates, however, the number of clusters present at any time is not large, and as result the H$\alpha$ luminosity of the entire galaxy undergoes large excursions about the mean. These excursions are asymmetric, such that galaxies spend most of their time in a state of low H$\alpha$ luminosity compared to the mean, and only briefly undergo periods of very high H$\alpha$ luminosity. Any given observation is much more likely to capture the former phase than the latter.

This stochasticity was ignored in earlier generations of SPS models, but detailed comparisons between the observations and newer SPS codes that include stochasticity \citep{eldridge09a, da-silva12a} show that stochastic star formation plus a normal IMF is fully consistent with the data, and in fact provides a much better match than models with the proposed ``top-light" IMFs \citep{fumagalli11a, eldridge12a, weisz12a}. This theoretical explanation has now received direct observational support from measurements of H$\alpha$ to FUV ratios in individual star clusters in nearby galaxies, which show that individual clusters are indeed likely to be H$\alpha$-deficient, but that this is because a randomly-selected cluster is likely to be at an age when it most massive stars have already left the main sequence \citep{gogarten09a}. However, when the light of many clusters is added together, the summed H$\alpha$ to FUV ratio averages to the value predicted for a normal IMF \citep{calzetti10b, andrews13a}, exactly as predicted in the stochastic models. The lesson of this history is that discrepancies between SPS models and observations that are taken as evidence of IMF variation may in fact be due to some inadequacy in the SPS models that had simply not been considered before. Caution is warranted.

\subsubsection{Summary of the IMF}

This brings us to the third set of questions for star formation theory that animates this review: can we explain the origin of the IMF, and in particular can we explain both the powerlaw slope at the high mass end and the the existence of a characteristic mass in the range $0.1-1$ $\msun$? Can we explain the extent to which these properties vary with star-forming environment, and can we do so with enough confidence to extrapolate to the conditions that prevailed at high redshift, when star-forming galaxies looked very different than they do today?

\section{Theoretical Background}
\label{sec:theorybackground}

Having reviewed the observational background on star formation, I now turn to the physics of the star-forming phase of the ISM. As discussed in Section \ref{ssec:sfrobs}, star formation \blue{in the present-day Univese} appears to occur exclusively in regions where the hydrogen is mostly in the form of H$_2$ and, at least at Solar metallicity, the carbon mostly in the form of CO. If we are to understand star formation, we must therefore understand the dominant physical processes in this gas. The goal of this section is to acquaint the reader with some of the basic theoretical results that will be invoked over the remainder of this review. This section covers the chemistry (Section \ref{ssec:gmcchemistry}), thermodynamics (Section \ref{ssec:gmcthermo}), hydrodynamics (Section \ref{ssec:gmcflows}), and global stability and force balance (Section \ref{ssec:gmcstability}) of molecular gas. It is helpful in this section to keep some basic observationally-defined parameters in mind. The typical star-forming giant molecular cloud in the Milky Way has a mass $M\sim 10^4-10^6$ $M_\odot$, a size $R\sim 10-100$ pc, a surface density $\Sigma\approx 100$ $M_\odot$ pc$^{-2}$, a velocity dispersion $\sigma \approx 1-10$ km s$^{-1}$, a magnetic field strength $B\sim 1-10$ $\mu$G \blue{at densities $n_{\rm H} \lesssim 300$ cm$^{-3}$, rising as $n_{\rm H}^{2/3}$ thereafter}, and a gas temperature $T\approx 10$ K \citep{dobbs14a, crutcher12a}.

\subsection{The Atomic to Molecular Transition}
\label{ssec:gmcchemistry}

The first question to address in understanding the physics of the ISM is to understand what controls the transition between H~\textsc{i} and H$_2$, and the related transition from C$^+$ to CO. This may or may not be relevant to the regulation of star formation, as I discuss below, but the question is important regardless because the observed correlation between star formation and H$_2$, and the corresponding lack of correlation between star formation and H~\textsc{i}, demands an explanation, and this explanation must invoke the physics of the atomic to molecular transition. Although H$_2$ and CO are both lower-energy states than atomic hydrogen and atomic carbon plus oxygen, and the reactions required to form them can proceed spontaneously, the bulk of the matter in the ISM of the Milky Way and similar galaxies is in a chemical state where H and C$^+$ are the dominant repositories of hydrogen and carbon. The reason for this is the interstellar FUV field is capable of dissociating H$_2$ and CO molecules, and of ionizing carbon atoms. The chemical state of the gas is therefore determined by a competition between formation and destruction processes.

\subsubsection{Hydrogen Chemistry}

\paragraph{Formation}
Investigation of the formation and destruction of H$_2$ in the ISM dates back to the seminal work of \citet{gould63a} and \citet{hollenbach71a}. Formation of H$_2$ is less straightforward than one might initially expect, because the most obvious reaction for making it, ${\rm H}+{\rm H}\rightarrow {\rm H}_2$, occurs at a rate so low as to be negligible. The low reaction rate is a product of the symmetry of the system; if the hydrogen atoms are both in the ground state, then there are no allowed radiative transitions that can remove the binding energy of the free hydrogen atoms, and unless the temperature exceeds several thousand K, the population of H atoms in excited states is negligibly small \citep{gould63a, latter91a}. Three-body reactions of the form $3{\rm H}\rightarrow {\rm H}_2+{\rm H}$ are negligible unless the density is $\gtrsim 10^8$ cm$^{-3}$ \citep{palla83a, abel97a}, vastly higher than typical ISM densities. Gas-phase reactions to form H$_2$ therefore require the presence of free electrons and protons, which allow the reactions
\begin{eqnarray}
{\rm H}+e & \rightarrow & {\rm H}^- + h\nu \\
{\rm H}^- + {\rm H} & \rightarrow & {\rm H}_2 + e
\label{eq:h-channel}
\end{eqnarray}
and
\begin{eqnarray}
{\rm H}+{\rm H}^+ & \rightarrow & {\rm H}_2^+ + h\nu \\
{\rm H}_2^+ + {\rm H} & \rightarrow & {\rm H}_2 + {\rm H}^+.
\end{eqnarray}
In these reactions either an electron or a proton acts as a catalyst. In the first step, a free hydrogen undergoes radiative association with the catalyst, which is not forbidden because the system is not symmetric. Then the intermediate product encounters another hydrogen atom and forms H$_2$, while the catalyst particle carries off the remaining binding energy, obviating the need for a radiative transition. The former pair of reactions is generally much faster, because the lower mass of electrons compared to protons produces a much higher radiative association rate with H.  However, the rate of this reaction is sharply limited by two factors. First, the supply of free electrons (and free protons) is quite small in the dense gas where overall reaction rates are highest -- under Milky Way conditions, typical free electron fractions at densities $\gtrsim 1$ cm$^{-3}$ are $\lesssim 10^{-4}$ \citep{wolfire03a}. Second, H$^{-}$ is vulnerable to photodetachment: ${\rm H}^- + h\nu\rightarrow {\rm H}+e$. This reaction turns out to occur far more often than reaction (\ref{eq:h-channel}) \citep{glover03a}. Thus only a small fraction of H$^{-}$-forming reactions go on to catalyze the production of H$_2$. See \citet{lepp02a} and \citet{abel97a} for thorough reviews of the gas-phase chemistry of H$_2$ formation, including discussions of several other, sub-dominant formation channels that I have omitted here.

The inefficiency of H$_2$ formation in the gas phase makes another formation channel dominant, at least in the modern universe: formation on the surfaces of dust grains. Dust grains can catalyze H$_2$ formation for the same reason that free electrons can: the availability of a solid surface connected to a vibrational lattice provides a repository for the energy of formation that does not require the very low-probability emission of forbidden photons. The rate of H$_2$ formation via grain catalysis can be written as
\begin{eqnarray}
\frac{dn_{\rm H_2}}{dt} & = & \frac{1}{2} \left(\frac{8k_BT}{\pi m_{\rm H}}\right)^{1/2} \Sigma_{\rm gr} S(T)  \epsilon_{\rm H_2} n_{\rm H} n_{\rm H_0} 
\nonumber \\
& \equiv & \mathcal{R}_{\rm gr}  n_{\rm H} n_{\rm H_0}
\label{eq:h2formrate}
\end{eqnarray}
where $\Sigma_{\rm gr}$ is the total geometric cross section of dust grains per H nucleon, $S(T)$ is the temperature-dependent probability that a hydrogen atom that strikes a dust grain sticks to it, $\epsilon_{\rm H_2}$ is the probability that a stuck H atom will leave the grain by forming H$_2$ rather than becoming unstuck due to a thermal fluctuation or a photon, $n_{\rm H}$ and $n_{\rm H_0}$ are the number densities of H nucleons and free hydrogen atoms, respectively,  and the quantity in parentheses is the usual factor for collisions of neutral species that arises from integration over the Maxwellian distribution of particle velocities. \blue{The factor of $1/2$ appears because two H atoms must stick to a grain to produce one H$_2$ molecule.}

The quantities $\epsilon_{\rm H_2}$ and $S(T)$ can be calculated from models or measured from laboratory experiments \citep[e.g.][]{hollenbach79a, cazaux02a, cazaux04a, cazaux05a}, while $\Sigma_{\rm gr}$ can be constrained approximately from the level of dust extinction in the UV \citep[and references therein]{draine03a, draine11a}. However, a more common approach is to constrain the entirety of $\mathcal{R}_{\rm gr}$ via observations of C~\textsc{i}, C~\textsc{ii}, H~\textsc{i}, and H$_2$ column densities. In the Milky Way, such analysis points to a total rate coefficient $\mathcal{R}_{\rm gr} \approx 3\times 10^{-17}$ cm$^3$ s$^{-1}$ \citep{jura75a, gry02a, wolfire08a}, with some variation with environment. In the Magellanic Clouds, where the metallicity and dust abundance are smaller, the rate coefficient is correspondingly smaller \citep{browning03a}. Except at very small dust abundance, this rate coefficient is high enough so that dust-mediated H$_2$ formation completely dominates H$_2$ production -- see \citet{glover03a} for a much more thorough comparison of the two channels.

\paragraph{Destruction}
The destruction of H$_2$ is dominated by photodissociation by FUV photons in the interstellar radiation field (ISRF). As with formation, the symmetry of the H$_2$ system means that the process is slightly more complex than one might suppose at first. Although the binding energy of H$_2$ in the ground state is only $4.5$ eV,\footnote{However, the more relevant energy is the energy difference between the attractive and repulsive states at the equilibrium internuclear separation, which is $8-10$ eV \citep{gould63a}.} transitions of the form ${\rm H_2} + h\nu \rightarrow {\rm H}+{\rm H}$ are forbidden unless one of the resulting H atoms is left in an excited state, which requires a photon with a minimum energy of 14.5 eV. However, since photons of this energy are capable of ionizing neutral hydrogen, they are mostly absent from the ISRF. Thus direct dissociation in which both H atoms are left in the ground state is very slow because it is forbidden, and direct dissociation with one of the H atoms in an excited state is slow due to the lack of sufficiently energetic photons in the ISRF.

Instead, the dominant dissociation channel is a two-step process using lower-energy photons \blue{\citep{stecher67a}}. A hydrogen molecule in the ground electronic state $X(v,J)$, where $v$ and $J$ are the vibrational and rotational quantum numbers, can make an allowed transition to the first or second electronic states $B(v',J')$ and $C(v',J')$. Since each $(v,J)\rightarrow (v',J')$ combination has a slightly different energy, the transitions form a series of densely spaced lines; transitions to the first excited state are known as the Lyman band, and those to the second excited state are the Werner band. Both bands have transitions in the energy range $11-13.6$ eV, below the ISRF cutoff from neutral H ionization. An H$_2$ molecule in an excited electronic state will eventually spontaneously decay via photon emission back to the ground electronic state, and there is a finite probability that this state will be an unbound one. The probability of this happening depends on the excited state from which the decay is occurring, and thus the total dissociation rate is the product of the excitation rate into a given state and the dissociation probability from that state, summed over all possible upper states. The latter depends on quantum mechanics alone, while the former also depends on the properties of the radiation field responsible for exciting the H$_2$ molecules.

\red{The unattenuated ISRF of the Milky Way carries an energy density $\approx 6-9\times 10^{-14}$ erg cm$^{-3}$ integrated over the energy range $6-13.6$ eV \citep{draine78a, mathis83a, draine11a}. For a fixed spectral shape, the rate at which H$_2$ molecules are excited into the upper states of the Lyman-Werner bands is proportional to this energy density, with the value given above corresponding to an excitation rate (summed over all upper states) of} $\zeta_{\rm exc}\approx 3\times 10^{-10}$ s$^{-1}$. The mean dissociation probability (weighted by the relative excitation rates of the upper states) is $f_{\rm diss}=0.11-0.13$ depending on the assumed radiation field \citep{draine96a, browning03a, draine11a}, so the net dissociation rate is roughly $\zeta_{\rm diss} \approx 4\times 10^{-11}$ s$^{-1}$ \citep[and references therein]{draine11a}. Equating the formation and dissociation rates, the expected equilibrium H$_2$ fraction in the ISM of the Milky Way is
\begin{eqnarray}
\frac{n_{\rm H_2}}{n_{\rm H}} & = &
\frac{\mathcal{R}_{\rm gr} n_{\rm H^0}}{\zeta_{\rm diss}}
\nonumber\\
& = & 8\times 10^{-6}\left(\frac{4\times 10^{-11}\mbox{ s}^{-1}}{\zeta_{\rm diss}}\right)\left(\frac{n_{\rm H_0}}{10\mbox{ cm}^{-3}}\right).
\end{eqnarray}
Thus H$_2$ will be subdominant over the bulk of the ISM. For H$_2$ to become dominant, the density must be extraordinarily high, or the photodissociation rate (and thus the flux of FUV photons) must fall significantly. The latter can occur in regions where the column density is high enough to attenuate the ISRF. FUV photons can be absorbed or scattered by dust grains, and they can also be absorbed by H$_2$ molecules\footnote{One might think that the resonant absorption by H$_2$ would only remove a photon if the molecule in question were actually dissociated, but even if the molecule survives and returns to the ground state, it generally does so via a series of rotational and vibrational transitions giving rise to infrared photons, rather than via emission of a single photon with energy equal to that of the initially-absorbed one. As a result, to good approximation every absorption removes an FUV photon, even if only $\sim 10\%$ of absorptions lead to dissociation.}. The balance between the two depends on the density and metallicity of the gas, but for gas at the density typical of the cold neutral atomic medium, dust absorption and H$_2$ self-shielding contribute about equally over a broad range of metallicities and radiation fields \citep{krumholz08a, krumholz09a}.

\paragraph{Formation-Destruction Balance}
Detailed calculations of the H~\textsc{i} - H$_2$ transition can be performed in a variety of approximations, with some approaches emphasizing a more faithful treatment of quantum mechanics or radiative transfer, and others using less accuracy in these areas but allowing for more general geometries and non-equilibrium effects. For simple static geometries (generally either slabs or spheres, but in principle for any specified geometry) one can perform a coupled numerical calculation of chemical equilibrium and radiative transfer in order to obtain the position-dependence of both the spectrum of the radiation field and the populations of the various quantum states of H$_2$ \citep{federman79a, van-dishoeck86a, black87a, draine96a, neufeld96a, spaans97a, hollenbach99a, liszt00a, liszt02a, browning03a, allen04a}. Based in part on these results, a number of authors have developed analytic approximations for the attenuation of the radiation field as a function of column density \citep{draine96a}, and for the amount of shielding required for a transition from H~\textsc{i} to H$_2$ in either slab \citep{sternberg88a} or spherical \citep{elmegreen93a, krumholz08c, krumholz09a, mckee10a} geometry.

The basic result of all these computations is relatively straightforward, and can be understood to first order via a simple argument analogous to the Str\"omgren analysis of H~\textsc{ii} regions \citep{mckee10a}. Consider a flux of Lyman-Werner band photons $F_{0}^*$ (measured in photons per unit area per unit time) incident on the surface of a slab of interstellar gas of density $n_{\rm H}$. The FUV photons will keep the surface of the slab in a form dominated by H~\textsc{i}, but at some depth the radiation field will be attenuated sufficiently for the gas to transition to H$_2$-dominated. If we approximate that the transition is sharp and occurs at some depth $\ell_{\rm HI}$ into the slab, then we can solve for the H~\textsc{i} column density $n_{\rm H} \ell_{\rm HI}$ simply by equating the rates of H$_2$ formation and dissociation per unit area:
\begin{equation}
\mathcal{R}_{\rm gr} n_{\rm H}^2 \ell_{\rm HI} = f_{\rm diss} F_0^* \quad\Rightarrow\quad
 n_{\rm H} \ell_{\rm HI} = \frac{f_{\rm diss} F_0^*}{\mathcal{R}_{\rm gr} n_{\rm H}}.
\end{equation}
This approximation ignores dust attenuation, but, as noted above, this is generally not the dominant process. It also assumes that the transition is sharp, which is a reasonable approximation under typical conditions but can fail in cases where the dissociating radiation field is fairly weak \citep{krumholz08c, mckee10a}, though in this case once can still obtain a good estimate using a somewhat more sophisticated approximation \citep{sternberg88a}.

\citet{krumholz09a} point out that the ratio $F_0^*/n_{\rm H}$ is not really free, because atomic gas tends to spontaneously segregate into warm, diffuse, and cold, dense phases, and the latter is likely to dominate shielding around molecular regions. If the warm and cold atomic phases are in pressure balance, then the characteristic density of the cold phase is close to directly proportional to the FUV photon flux. Since the cold phase dominates the absorption, one can approximate the ratio $F_0^*/n_{\rm H}$ using this characteristic value, which can then be inserted into the above equation to derive a characteristic H~\textsc{i} column density that depends only on $\mathcal{R}_{\rm gr}$:
\begin{equation}
\Sigma_{\rm HI} = \mu_{\rm H} n_{\rm H} \ell_{\rm HI} \approx 9 \left(\frac{\mathcal{R}_{\rm gr}}{\mathcal{R}_{\rm gr,MW}}\right)^{-1} \,M_\odot\mbox{ pc}^{-2},
\label{eq:sigmahi}
\end{equation}
where $\mathcal{R}_{\rm gr,MW} = 3\times 10^{-17}$ cm$^3$ s$^{-1}$ is the approximate H$_2$ formation rate coefficient for the Milky Way, and $\mu_{\rm H} = 2.34\times 10^{-24}$ g is the mean mass per H nucleus. This explains why the H~\textsc{i} - H$_2$ transition tends to occur at $\sim 10$ $\msun$ pc$^{-2}$.

\paragraph{Non-Equilibrium Effects}

The calculations mentioned thus far are based on an assumption of chemical equilibrium, but there is an important caveat, which is that it is not at all clear that this assumption is a good one. The characteristic time required for the H~\textsc{i} - H$_2$ balance to reach chemical equilibrium is
\begin{equation}
t_{\rm H_2,eq} = \frac{1}{\mathcal{R}_{\rm gr} \blue{\mathcal{C}} \langle n_{\rm H}\rangle } \approx \blue{(\mathcal{C}/10)^{-1}} \langle n_{\rm H,2}\rangle ^{-1}\mbox{ Myr}
\label{eq:th2eq}
\end{equation}
for the Milky Way value of $\mathcal{R}_{\rm gr}$. \blue{Here $\langle n_{\rm H}\rangle$ is the volume-averaged number density of the gas, and $\langle n_{\rm H,2}\rangle \equiv \langle n_{\rm H}\rangle/100$ cm$^{-3}$. The quantity $\mathcal{C}$ is a clumping factor, defined as $\mathcal{C} = \langle n_{\rm H}^2\rangle / \langle n_{\rm H}\rangle^2$, and it appears because the chemical reaction rate per unit volume varies as the square of the volume density. This means that, if a region has a non-uniform density, reactions within it will proceed faster than they would in a uniform region of the same volume-averaged density. The point to take from this calculation is that the timescale for H$_2$ formation is shorter than} the observationally-estimated lifetime of a molecular cloud \citep{kawamura09a, koda09a}\blue{, but not by a huge margin, suggesting that non-equilibrium effects might be important in some circumstances}. In the last few years, several authors have examined this question at a variety of size scales. The general approach in these models is to add a time-dependent chemical network to a hydrodynamic or MHD simulation \citep{koyama00a, bergin04a, glover07a, glover07b, pelupessy09a, gnedin09a, christensen12a, mac-low12a, clark12a, inoue12a}, which can be either one-dimensional or three-dimensional. In the latter case, the radiative transfer and shielding must be treated in a far more approximate manner than in the static calculations, for reasons of computational cost.

Unfortunately it is difficult thus far to draw general conclusions from this work as to the extent to which the hydrogen chemistry deviates from equilibrium. At $\sim 100$ pc and larger scales the deviation seems to be small \citep{krumholz11a}, but at the resolution achievable in these models, the small-scale density structure is unknown. \blue{Thus one must adopt a clumping factor $\mathcal{C}$ to model the unresolved substructure that accelerates the chemical reactions. The standard practice in large-scale models has been to adopt a value of $\mathcal{C}$} calibrated so that the simulations match observations in the Milky Way and Magellanic Couds \blue{\citep[e.g.,][]{gnedin09a, pelupessy09a}}. There are obviously potential pitfalls in this procedure.

On $\sim 1-10$ pc scales, \citet{mac-low12a} and \citet{clark12b} find that non-equilibrium effects are important, and that their simulations do not reach equilibrium between formation and destruction over timescales of several tens of Myr, long enough that they expect clouds to be dispersed before reaching equilibrium. However, these models focus on a small turbulent region that is supposed to represent the interior of a molecular cloud, and thus do not capture galaxy-scale dynamics accessible to the lower-resolution simulations. Moreover, the results appear on their surface to contradict observations showing that clouds on such small scales do have close-to-equilibrium H~\textsc{i} to H$_2$ ratios \citep{goldsmith05a, lee12a}. At present the resolution to this problem is not clear.

One final remark on hydrogen chemistry is in order, which is that the dependence of both the H~\textsc{i} shielding column (equation \ref{eq:sigmahi}) and the equilibration time (equation \ref{eq:th2eq}) on $\mathcal{R}_{\rm gr}$ has an important but subtle implication. As noted above, $\mathcal{R}_{\rm gr}$ is observed to be substantially smaller in the Magellanic Clouds than in the Milky Way, and this is to be expected given their lower metallicities: an ISM containing fewer heavy elements should also have a smaller total dust surface area available to catalyze chemical reactions. However, this means that, at low metallicity, we expect that both the column of H~\textsc{i} required before the gas transitions to H$_2$ and the time required for the gas to reach chemical equilibrium will rise compared to their values in the Milky Way. There is direct evidence for an increase in the H~\textsc{i} shielding column in low-metallicity galaxies \citep{fumagalli10a, bolatto11a, wong13a}, and as a result of the timescale effect, even those models that predict equilibrium under Milky Way conditions suggest that non-equilibrium effects will become dominant once the metallicity is reduced to $\sim 1\%$ of Solar \citep{krumholz11a, krumholz12e}.

\subsubsection{Carbon Chemistry}
\label{sssec:carbon}

While hydrogen is the dominant species in the ISM, it is extraordinarily difficult to observe directly, due to its large level spacings, the same characteristic that makes it a very poor coolant at low temperatures. As a result, observers are generally forced to use proxies to observe molecular gas, and the most common one is CO. As I discuss in the next section, CO (and to a lesser extent C$^+$) are also the most important coolants in molecular clouds. Thus the chemistry of carbon is nearly as important as that of hydrogen for understanding the behavior of the star-forming ISM. The formation and destruction mechanisms of CO were first explored in detail by \citet{van-dishoeck86a, van-dishoeck88a}, and have subsequently been elucidated by a number of additional authors \citep[e.g.,][]{bergin95a, nelson99a}. As with H$_2$, models for carbon chemistry form a spectrum from those with very detailed treatments of chemistry and radiative transfer but generally simple geometries and/or simple treatments of the time-dependence \citep[e.g.,][]{van-dishoeck86a, van-dishoeck88a, bergin95a, nelson99a, wolfire10a, levrier12a} to those with greatly simplified treatments of the chemistry and radiative transfer but more more sophisticated treatments of the flow \citep{glover10a, glover11b, glover12b, glover12c, clark12b, clark12a}.

\begin{figure}
\centerline{\includegraphics[width=2.5in]{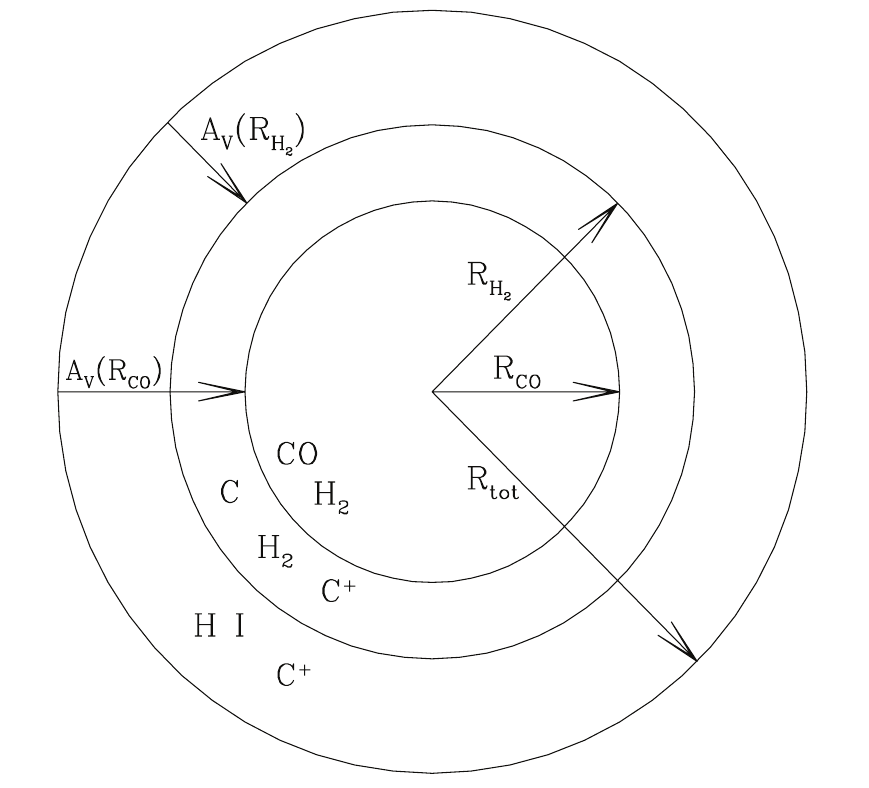}}
\centerline{\includegraphics[width=2.5in]{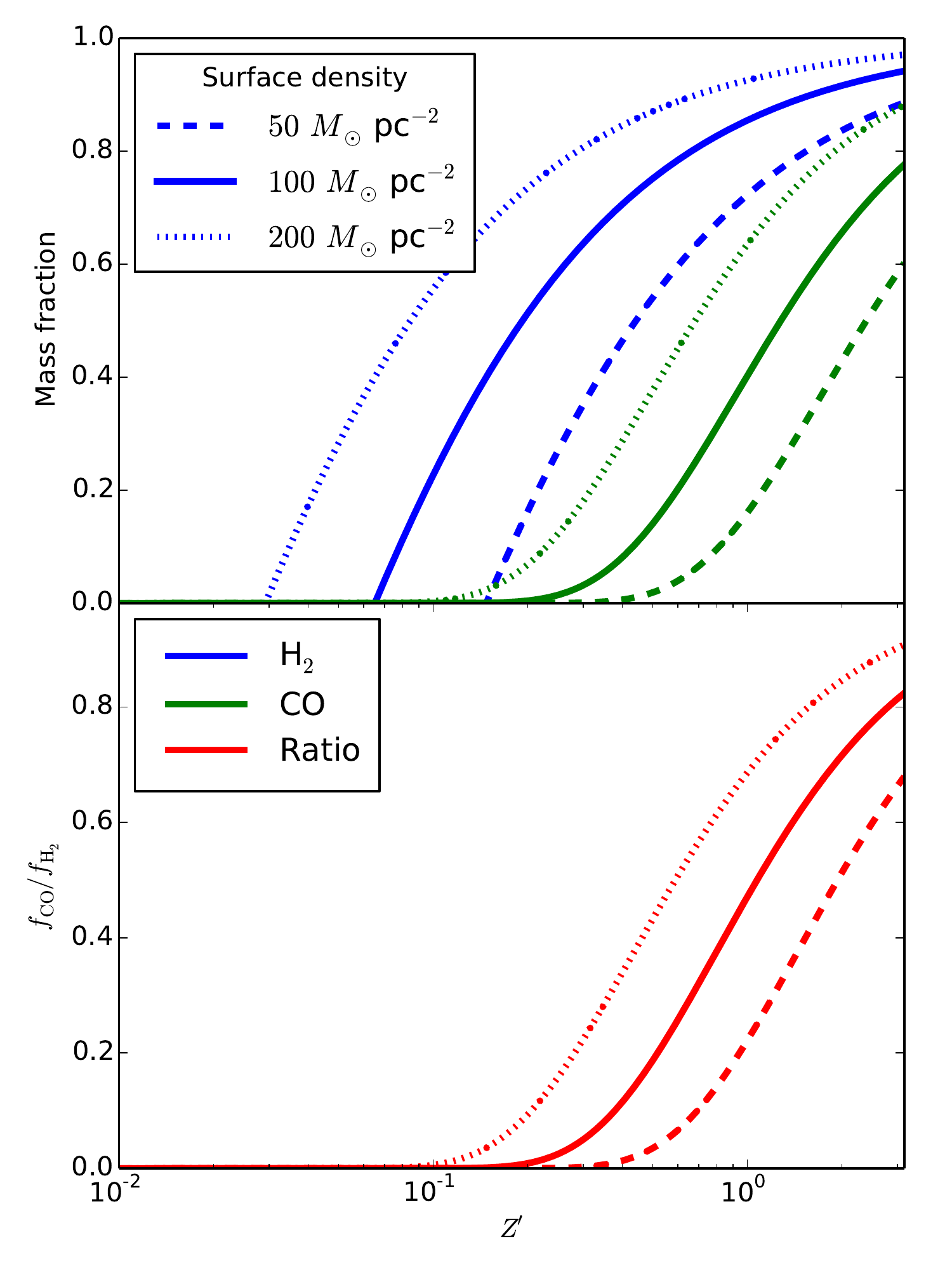}}
\caption{
\label{fig:h2co}
Top: schematic diagram of the chemical structure of a molecular cloud, with H~\textsc{i} and C$^+$ dominant at the surface, H$_2$ and C$^+$ further in, and H$_2$ and CO at the center; reprinted from \citet{wolfire10a}, reproduced by permission of the AAS. Middle: mass fractions of spherical clouds of mean surface density $\Sigma=50$, 100, and 200 $\msun$ pc$^{-2}$ (dashed, solid, dotted lines) where the dominant hydrogen form is H$_2$ (blue) and the dominant carbon form is CO (green), as a function of gas metallicity normalized to Solar, $Z'$. Bottom: ratio of CO and H$_2$ mass fractions. \red{Note that all the mass fractions plotted here are \textit{not} the fractions of the cloud mass comprised of CO and H$_2$ molecules. Rather, they are, respectively, the fraction of the cloud mass within which the majority of the carbon atoms are locked into CO molecules, and the majority of the hydrogen atoms are locked into H$_2$ molecules. Since it is possible for both of these conditions to occur simultaneously, the sum of the two mass fractions can exceed unity.} The H$_2$ mass fraction has been computed using the analytic model of \citet{mckee10a}, and the CO mass fraction using the analytic approximation of \citet{wolfire10a}; these have been combined using the method (and choice of fiducial parameters) described in \citet{krumholz11b}.
}
\end{figure}

The formation of CO is substantially different than that of H$_2$ in that it is dominated by gas-phase rather than grain-surface reactions. Since the temperatures in regions where this reaction is taking place tend to be low, as discussed in the next section, ion-molecule reactions are a key component of this process. Unlike neutral-neutral reactions, they have rate coefficients that are only weak functions of temperature. There are two main pathways that lead to the formation of CO starting from C$^+$, O, and H$_2$. One route passes through the OH molecule, and can proceed in the following chain:
\begin{eqnarray}
{\rm H}_2 + {\rm CR} & \rightarrow & {\rm H}_2^+ + e + {\rm CR} \\
{\rm H}_2^+ + {\rm H}_2 & \rightarrow & {\rm H}_3^+ + {\rm H} \\
{\rm H}_3^+ + {\rm O} & \rightarrow & {\rm OH}^+ + {\rm H}_2 \\
{\rm OH}^+ + {\rm H}_2 & \rightarrow & {\rm OH}_2^+ + {\rm H} \\
{\rm OH}_2^+ + e & \rightarrow & {\rm OH} + {\rm H} \\
{\rm C}^+ + {\rm OH} & \rightarrow & {\rm CO}^+ + {\rm H} \\
{\rm CO}^+ + {\rm H}_2 & \rightarrow & {\rm HCO}^+ + {\rm H} \\
{\rm HCO}^+ + e & \rightarrow & {\rm CO} + {\rm H}.
\end{eqnarray}
Here CR indicates cosmic ray. There are also a number of possible variants (e.g., the OH$_2^+$ could form OH$_3^+$ before proceeding to OH\blue{, or the OH may form on the surface of a dust grain rather than in the gas phase \citep{wolfire10a}}). The second main route is through the CH molecule, where reaction chains tend to follow the general pattern
\begin{eqnarray}
{\rm C}^+ + {\rm H}_2 & \rightarrow & {\rm CH}_2^+ + h\nu \\
{\rm CH}_2^+ + e & \rightarrow & {\rm CH} + {\rm H} \\
{\rm CH} + {\rm O} & \rightarrow & {\rm CO} + {\rm H}.
\end{eqnarray}
The rate at which the first reaction chain manufactures CO is limited by the supply of cosmic rays that initiate the production of H$_2^+$, while the rate at which the second reaction chain proceeds is limited by the rate of the final neutral-neutral reaction. Which chain dominates depends on the cosmic ray ionization rate, density, temperature, and similar details. However, note that both of these reaction chains require the presence of H$_2$. The net result is that clouds tend to have a layered structure, as illustrated in the top panel of Figure \ref{fig:h2co}. In poorly-shielded regions where the FUV has not yet been attenuated, H~\textsc{i} and C$^+$ dominate. Further in, where the FUV has been partly attenuated, H$_2$ and C$^+$ dominate. Finally a transition to H$_2$ and CO as the dominant chemical states occurs at the center.

CO is destroyed via radiative excitation followed by dissociation in essentially the same manner as H$_2$. The shielding process for CO is slightly different however. As with H$_2$, photons that dissociate CO can be absorbed both by dust grains and by CO molecules. However, due to the much lower abundance of CO compared to H$_2$, the balance between these two processes is quite different than it is for hydrogen, with dust shielding generally the more important of the two. Moreover, there is non-trivial overlap between the resonance lines of CO and those of H$_2$, and thus there can be cross-shielding of CO by H$_2$. The process is sufficiently complex that no good analytic approximations exist, only fitting formulae tabulated to the results of numerical calculations \citep[e.g.,][]{van-dishoeck88a}.

The differences in formation process and shielding between H$_2$ and CO have several important implications. First of all, because CO formation proceeds via ion-neutral reactions that are much faster than the grain surface processes that dominate for H$_2$, the time required to reach equilibrium is much shorter. Thus the CO abundance is generally always in equilibrium, though that equilibrium value may change as the H$_2$ abundance does \citep{glover10a, glover11b, glover12b}.

Second, because CO requires H$_2$ to form, and is less effectively shielded from the ISRF, the outer parts of molecular clouds tend to consist of ``dark gas", where the hydrogen is mostly in the form of H$_2$ but the carbon is still dominated by C$^+$ \citep{wolfire10a, glover11b}; this gas is referred to as dark because it is particularly hard to observe, as the \blue{bulk of the H$_2$ is at temperatures too low to emit}, and the lines of C$^+$ are not observable from the ground and tend to be masked by the ubiquitous C$^+$ emission of the atomic ISM. Under Milky Way conditions, models suggest that the dark gas is only a $\sim 30\%$ contribution to the total molecular mass budget. However, because it cannot self-shield effectively, CO is much more sensitive than H$_2$ to the dust abundance. Both models \citep{wolfire10a, glover11b, krumholz11b, narayanan11a, narayanan12a, shetty11a, shetty11b, feldmann12a, feldmann12b} and observations \citep{bolatto11a, leroy11a, genzel12a, leroy13a, bolatto13a} suggest that, at low metallicity dark, CO-free gas can completely dominate the gas budget of molecular clouds, leaving the CO-emitting region as only the tip of a much larger iceberg. Figure \ref{fig:h2co} demonstrates this effect using the theoretical models of \citet{mckee10a} to estimate the H$_2$ content and \citet{wolfire10a} to estimate the CO content of clouds of varying metallicities and column densities. As the plot shows, for idealized spherical clouds the mass in the CO-dominated region falls off significantly faster than the mass in the H$_2$-dominated region, so that the ratio of the masses in these two regions decreases as the metal and dust content do.

\subsection{Thermodynamics of Molecular Clouds}
\label{ssec:gmcthermo}

Having discussed the chemical processes that govern the formation of molecular clouds, I next review the thermal properties of this gas, which are quite different from those of other phases of the ISM. Interstellar gas can be heated by the grain photoelectric effect and by cosmic rays, cooled by line emission, and, depending on the temperature of the dust, can be both heated and cooled by collisional exchange with the dust. Gas can also change its temperature due to adiabatic compression or expansion, or due to dissipation of bulk motions, either in shocks or via any other process. Determining the temperature of a gas cloud requires modeling all of these effects, and there are a number of codes that can perform this calculation numerically at a variety of levels of approximation \citep{ferland98a, goldsmith01a, lesaffre05a, meijerink05a, le-petit06a, narayanan11a, narayanan12a, clark12a, krumholz14b}. However, before proceeding to a numerical solution, it is helpful to make some order of magnitude estimates for the star-forming phase of the ISM.

The rates of grain photoelectric and cosmic ray heating per H nucleus can be approximated by \citep{krumholz14b}
\begin{eqnarray}
\label{eq:gammape}
\Gamma_{\rm PE} & = & 4.0\times 10^{-26} \chi_{\rm FUV} Z_d' e^{-(1/2) \kappa_{d,\rm PE} \Sigma}\mbox{ erg s}^{-1} \\
\Gamma_{\rm CR} & = & 2.4\times 10^{-27} \zeta_{-16} \mbox{ erg s}^{-1}
\end{eqnarray}
where $\chi_{\rm FUV}$ is the strength of the interstellar far ultraviolet (FUV) radiation field normalized to the Solar neighborhood value, $Z'_d$ is the dust abundance normalized to the Solar neighborhood value, $\kappa_{d,\rm PE}$ is the dust opacity per unit gas mass at the $\approx 1000$ \AA~wavelengths that dominate photoelectric heating, $\Sigma$ is the gas column density, and $\zeta_{-16}$ is the cosmic ray primary ionization rate normalized to $10^{-16}$ s$^{-1}$; observationally-estimated values are $\zeta_{-16} \approx 0.2-2$ \citep{wolfire10a, neufeld10a, indriolo12a}, and ionization rates in the well-shielded interiors of molecular clouds are probably toward the low end of this range. \blue{The factor of $1/2$ in the exponent
of equation (\ref{eq:gammape}) amount to approximating that the typical atom inside an optically thick cloud is shielded by an amount equal to half the area-averaged cloud optical depth.}

For Milky Way dust $\kappa_{d,\rm PE} \approx 500$ cm$^2$ g$^{-1}$ \citep{draine03a}, and combining this with the typical surface density $\Sigma\approx 100$ $M_\odot$ pc$^{-2}$ mentioned above, we see that the $\kappa_{d,\rm PE}\Sigma \approx 10$, so the exponential factor in equation (\ref{eq:gammape}) is of order a few times $10^{-3}$. As a result cosmic ray heating is probably dominant over photoelectric heating in molecular cloud interiors, though they are likely competitive at cloud surfaces, in clouds with low surface densities, and in regions with FUV radiation fields significantly stronger than the Solar neighborhood value. \blue{Note that, because much of the mass in molecular clouds lies are relatively low column densities, FUV heating may be dominant for most of the mass, even if it is sub-dominant in the coldest regions \citep{mckee89a}.}

The rate of grain-gas energy exchange is roughly 
\begin{equation}
\Psi_{\rm gd} = \alpha_{\rm gd} n_{\rm H} T_g^{1/2} (T_d-T_g),
\end{equation}
where $T_g$ and $T_d$ are the gas and dust temperatures, and $\alpha_{\rm gd} \approx 1\times 10^{-33}$ erg cm$^3$ K$^{-3/2}$ \citep{goldsmith01a, young04a, krumholz11b, krumholz14b} is the dust-gas coupling coefficient for an H$_2$-dominated medium. \blue{This value is uncertain at the factor of few level, and is dependent on the values adopted for the grain surface area per H atom and on the accommodation coefficient for H$_2$-grain collisions.} At typical molecular cloud gas temperatures $T_g \approx 10$ K and dust-gas temperature differences $T_d-T_g\approx 10$ K, this rate is negligible compared to cosmic ray heating until the density is quite high, $n_{\rm H} \approx 10^4 - 10^5$ cm$^{-3}$. Most material in star-forming clouds is not that dense, and so coupling to dust grains is unimportant \blue{for gas thermodynamics, though it still controls the flow of radiant energy. However, once the density is high enough for grain-gas coupling, as in denser ``clump" regions, dust becomes dominant.}

The rate at which shocks or other forms of dissipation of bulk motion can heat a cloud is determined by the energy content of the bulk motions and by how quickly their kinetic energy can be converted to thermal energy. As discussed in detail in the following section, the time required for this is generally of order a crossing time, so in a cloud of characteristic density $\rho$, velocity dispersion $\sigma$ and radius $R$, the rate of shock heating is expected to be
\begin{equation}
\Gamma_{\rm diss} \approx \frac{\rho \sigma^3}{R} \approx 8\times 10^{-27} n_{\rm H,2} \sigma_0^3 R_1^{-1}\mbox{ erg s}^{-1},
\end{equation}
where $\sigma_0 = \sigma/1$ km s$^{-1}$ and $R_1 = R/10$ pc.   Rates of change of the internal energy due to adiabatic expansion or contraction are at most of this order. This is generally smaller than the rate of cosmic ray heating, though not by a huge margin. The above estimate probably also overstates the importance of dissipation heating, because such heating tends to be highly localized and provide high heating rates over small volumes. The extra heat injected in these small volumes is more efficiently radiated away by the higher temperature gas that it produces, and thus the impact on the median gas temperature is less than one would naively expect. However, there is considerable debate on the subject of whether dissipation heating might in fact be more important than cosmic rays, with some authors arguing that it is \citep{pan09a} and others disagreeing \citep{li12c}.

Rates of cooling are more difficult to estimate, as they depend on the chemical state of the gas. Moreover, in regions where CO is the dominant carbon form, cooling is dominated by optically-thick lines of CO, and \blue{the effects of optical depth and (for higher $J$ levels) sub-thermal excitation} are non-trivial to estimate. CO is a quantum rotor, and the rate of cooling by the various rotational transitions of CO is determined by the competition between optical depth effects, which suppress cooling from low $J$ levels because the photons cannot escape, and excitation effects, which suppress cooling in high-$J$ lines because the relevant levels are sub-thermally populated due to low densities. A very rough rule of thumb is that the cooling will be dominated by the lowest $J$ level that is marginally optically thin, and that this level will have a population that is not too far from its local thermodynamic equilibrium (LTE) value; for CO molecules in Galactic giant molecular clouds, this competition tends to result in cooling being dominated by transitions between $J=2\rightarrow 1$ and $J=5\rightarrow 4$, depending on the conditions in the emitting region.

For an optically thin transition of a quantum rotor where the population is in LTE, the rate of energy emission per H nucleus from transitions between angular momentum quantum numbers $J$ and $J-1$ is given by
\begin{eqnarray}
\label{eq:lambdaco}
\Lambda_{J,J-1} & = & x_{\rm em} \frac{(2J+1)e^{-E_J/k_B T}}{Z(T)} \nonumber \\
& & \qquad {} \cdot A_{J,J-1} (E_J - E_{J-1}) \\
E_J & = & h B J (J+1) \\
A_{J,J-1} & = & \frac{512\pi^4 B^3\mu^2}{3hc^3} \frac{J^4}{2J+1}.
\end{eqnarray}
Here $x_{\rm em}$ is the abundance of the emitting species per H nucleus, $T$ is the gas temperature, $Z(T)$ is the partition function, $A_{J,J-1}$ is the Einstein $A$ coefficient from transitions from state $J$ to state $J-1$, $E_J$ is the energy of state $J$, $B$ is the rotation constant for the emitting molecule, and $\mu$ is the electric dipole moment of the emitting molecule. The first equation is simply the statement that the energy loss rate is given by the abundance of emitters multiplied by the fraction of emitters in the $J$ state in question times the spontaneous emission rate for this state times the energy emitted per transition. \blue{Note that there is no explicit density dependence as a result of our assumption that the level with which we are concerned is in LTE.} The latter two equations are general results for quantum rotors. The CO molecule has $B=57$ GHz and $\mu=0.112$ Debye, and at Solar metallicity its abundance in regions where CO dominates the carbon budget is $x_{\rm CO} \approx 1.1\times 10^{-4}$ \citep{draine11a}. For these values, at $T=10$ K, cooling rates for $J=1-5$ are of order $10^{-27}-10^{-26}$ erg cm$^{-3}$, comparable to the heating rate from cosmic rays, which is why the equilibrium temperature is $\sim 10$ K.

As discussed in the previous section, at sub-Solar metallicity molecular clouds are increasingly-dominated by regions where the dominant form of carbon is C$^+$ rather the CO. Calculation of the cooling rate in this case is much simpler, as C$^+$ cooling is dominated by a single line (at 158 $\mu$m), and this line is generally optically thin \citep{krumholz12e}. The cooling rate through this line, assuming that the chemical composition is dominated by H$_2$ and neglecting the sub-dominant contribution from collisional excitations of C$^+$ by He and by free electrons, is
\begin{equation}
\label{eq:lambdac+}
\Lambda_{\rm C^+} \approx k_{\rm C^+-H_2} x_{\rm C^+} k_B T_{\rm C^+}\left(\frac{n_{\rm H}}{2}\right),
\end{equation}
where $k_{\rm C^+-H_2} \approx 6.6\times 10^{-10} e^{-T_{\rm C^+}/T}$ cm$^3$ s$^{-1}$ is the collisional excitation rate coefficient for C$^+$ by H$_2$\,\footnote{The value given is for an H$_2$ mixture of 0.25 ortho-to-para; this rate coefficient comes from the Leiden Atomic and Molecular Database, 
\texttt{http://home.strw.leidenuniv.nl/$\sim$moldata/} \citep{flower77a, flower88a, schoier05a}.}, $T_{\rm C^+}=91$ K is the energy of the upper state measured in K, $x_{\rm C^+} \approx 1.1\times 10^{-4} Z'$ is the carbon abundance \citep{draine11a}, and $Z'$ is the metallicity relative to Solar. At a density $n_{\rm H} = 100$ cm$^{-3}$ and a temperature of 10 K, this cooling rate is of order $10^{-30}$ erg cm$^{-3}$ s$^{-1}$, far less than the heating rate. However, due to the exponential factor in $k_{\rm C^+-H_2}$ it is extremely sensitive to temperature, so that the C$^+$ cooling rate becomes of order $10^{-27}-10^{-26}$ erg cm$^{-3}$ s$^{-1}$, comparable to the heating rate, at temperatures of $\sim 20-30$ K. We therefore expect equilibrium temperatures of $20-30$ K in regions where C$^+$ rather than CO dominates the cooling.

In addition to providing an estimate of the equilibrium temperature, the above analysis reveals a few other interesting points. First, the temperature will be relatively insensitive to variations in the local heating rate. The cosmic ray and photoelectric heating rates are to good approximation temperature-independent, but the cooling rate is extremely temperature sensitive because, for the dominant cooling lines, either C$^+$ or CO, the level energies are large compared to $k_B T$. \blue{For C$^+$ cooling, equation (\ref{eq:lambdac+}) implies that the cooling rate is exponentially sensitive to temperature.  For CO, equation (\ref{eq:lambdaco}) would seem to suggest the same thing, but in fact the true dependence is somewhat shallower because as the temperature changes the level that dominates the cooling also changes. \citet{hollenbach79a} find that the cooling rate scales as temperature roughly as a powerlaw $\Lambda_{\rm CO} \propto T^p$ with $p\approx 2-3$. Regardless of the exact functional form, the implication is that the temperature will only be a weak function of the gas heating rate, because even a small change in temperature will produce a large change in the heating rate. In regions where cosmic ray heating dominates, this means that there are unlikely to be significant temperature variations. As noted above, there are significant portions of molecular clouds where photoelectric heating dominates, and these will experience somewhat larger temperature variations. Even there, though, simulations suggest that these variations will be limited to factors of a few \citep{glover10a}.}

A second important point is the timescales involved. The thermal energy per H nucleus for a gas of molecular hydrogen is $e = (3/4) f k T$, where the factor $f$ is a number of order unity that depends on the temperature and the division of H$_2$ between ortho- and para-states \citep{boley07a, tomida13a, krumholz14b}. The characteristic time for the gas to return to thermal equilibrium is simply this energy divided by the cooling rate, $t_{\rm eq} \approx e/\Lambda$. Depending on the exact temperature \blue{and density}, this timescale is generally of order $10-100$ kyr. In comparison, the crossing timescale is $t_{\rm cross} = R/\sigma = 10 (R_1/\sigma_0)$ Myr, and the free-fall timescale is similar. Thus the timescale for the gas in a molecular cloud to reach thermal equilibrium is extremely small compared to any reasonable estimate of the mechanical time scale. The timescale is longer at lower metallicity where the CO abundance is smaller, but \blue{at typical molecular cloud densities above 100 cm$^{-3}$, it} remains smaller than the dynamical timescale down to metallicities as low as $\sim 0.1\%$ of Solar \citep{krumholz12e}.

The combination of a temperature that is quite insensitive to variations in the local heating rate due to the stiffness of the cooling function, and a cooling timescale that is very short compared to the mechanical timescale, means that in many circumstances one can regard the gas in molecular clouds as roughly isothermal. This has important consequences for the dynamics, as I discuss in the next section.

\begin{figure}
\centerline{\includegraphics[width=2.9in]{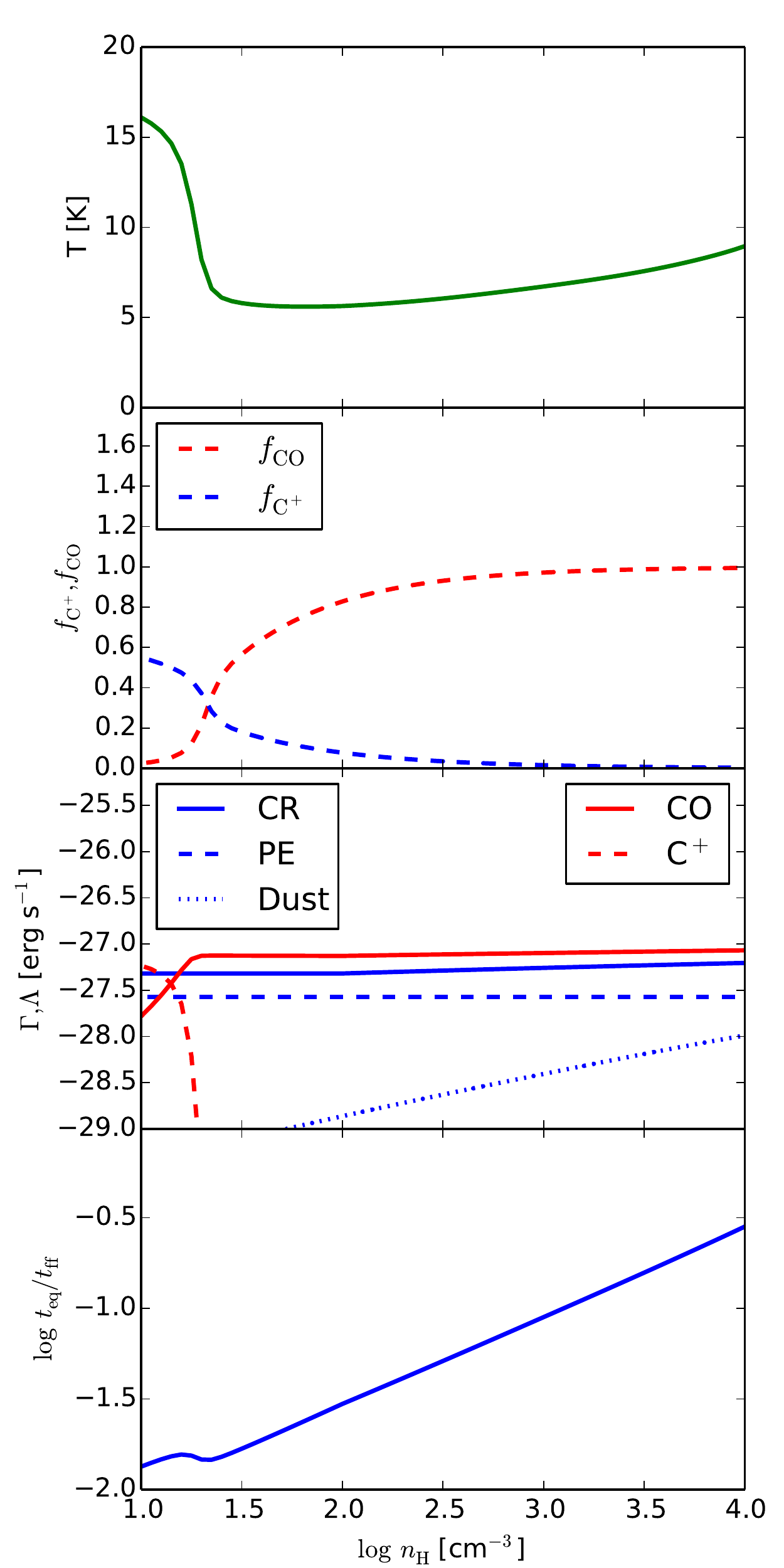}}
\caption{
\label{fig:despotic}
Properties of a cloud \blue{interior} versus mean density computed using the \textsc{despotic} code \citep{krumholz14b}. The first and second panels show the equilibrium gas temperature and the chemical composition, where $f_{\rm CO}$ and $f_{\rm C^+}$ show the fraction of all carbon nuclei in the form of CO and C$^+$, respectively. The third panel shows rates of heating due to cosmic rays, the grain photoelectric effect, and dust-gas energy exchange, and the rates of cooling due to CO and C$^+$ line emission, as indicated by the legend; all heating and cooling rates are per H nucleus. Finally, the bottom panel shows the ratio of the thermal equilibration and free-fall times; the former is defined as $t_{\rm eq} = e_{\rm int}/\Gamma = e_{\rm int}/\Lambda$, where $\Gamma$ and $\Lambda$ are the total heating and cooling rates per H nucleus (equal in equilibrium) and $e_{\rm int}$ is the internal energy per H nucleus. These properties have been computed for a cloud with mean column density $N_{\rm H} = 10^{22}$ H cm$^{-2}$, virial ratio $\alpha_{\rm vir} = 1$, with molecular hydrogen and helium abundances $x_{\rm o-H_2} = 0.1$, $x_{\rm p-H_2} = 0.4$, and $x_{\rm He} = 0.1$ per H nucleus, and using $\chi = 1$, $\zeta_{-16} = 0.3$, an infrared radiation field temperature of 10 K. For more information on these parameters and how they are defined in the \textsc{despotic} code, see \citet{krumholz14b}; the script that generated this figure is included in the \textsc{despotic} package, available at
\texttt{https://code.google.com/p/despotic/}.
}
\end{figure}

To verify the analytic estimates made above, Figure \ref{fig:despotic} shows some sample numerical results computed using the \textsc{despotic} code \citep{krumholz14b}. The code self-consistently computes the chemical state of the gas using a reduced chemical network \citep{nelson99a}, and the temperature of the gas including all of the processes listed above except dissipation heating; it handles optically-thick line cooling using an escape probability approximation. The numerical results confirm the analytic estimates above. In particular, the figure shows that, once the chemical composition becomes dominated by CO, the temperature is nearly independent of density, it is set by a competition mostly between CO line cooling and cosmic ray heating (with a small but non-negligible contribution from photoelectric heating), and that the cooling time is always much shorter than the free-fall time. \blue{(In a region of active star formation, the infrared radiation field would be larger, and the dust heating term would also become a significant contributor at the higher densities.)} Moreover, the numerical results show that the thermodynamics are not very different even when the carbon is mostly in the form of C$^+$, again in accordance with our analytic estimates. In that case, the cooling is dominated by C$^+$ rather than CO line emission, and the temperature is somewhat higher, $15-20$ K rather than $5 - 10$ K, but otherwise the gas remains subject to cooling on a timescale short compared to any mechanical timescale.

\subsection{Flows in Molecular Clouds}
\label{ssec:gmcflows}

\subsubsection{Characteristic Numbers}

The low temperatures of molecular clouds imply that their sound speeds are small: the isothermal sound speed $c_s \approx 0.2$ km s$^{-1}$ for H$_2$-dominated gas at 10 K. However, the observed linewidths of molecular clouds, $\sigma \approx 0.5-10$ km s$^{-1}$, are far larger than this. Indeed, large linewidths seem inevitable simply based on cloud masses. The natural velocity scale in a cloud of mass $M$ and radius $R$ is roughly $\sqrt{GM/R} = 6.6 (M_5/R_1)^{1/2}$ km s$^{-1}$, where $M_5 = M/10^5$ $M_\odot$. Even if a gas cloud initially had a velocity much smaller than this, gravity would accelerate it to a speed of this order in a time comparable to the free-fall time. Thus the gas within molecular clouds is highly supersonic. In terms of the dimensionless numbers routinely used to describe hydrodynamics, the Mach number $\mathcal{M}$ of flows in molecular clouds is $\gg 1$. In contrast, the flows are not necessarily fast compared to the Alfv\'{e}n speed,
\begin{equation}
v_A = \frac{B}{\sqrt{4\pi \rho}} = 1.8 B_1 n_{\rm H,2}^{-1/2}\mbox{ km s}^{-1}
\end{equation}
where $B_1 = B/10$ $\mu$G. The Alfv\'{e}n Mach number $\mathcal{M}_A$, which measures the ratio of the characteristic speed to $v_A$, is closer to unity. Whether it is actually about unity or is somewhat lower or higher is a subject of significant observational and theoretical debate, as the observations required to measure $\mathcal{M}_A$ are highly non-trivial \citep{padoan04b, padoan04c, heyer12a, crutcher12a}. One important implication of $\mathcal{M}\gg 1$ while $\mathcal{M}_A\sim 1$ is that the thermal pressure in molecular clouds is far smaller than the magnetic pressure; we parameterize this through the ratio
\begin{equation}
\beta = \frac{\rho c_s^2}{B^2/8\pi} = 2\left(\frac{\mathcal{M}_A}{\mathcal{M}}\right)^2 \ll 1.
\end{equation}

In addition to being supersonic and about trans-Alfv\'{e}nic, the motions within molecular clouds are also certainly turbulent. The transition to turbulence is an unsolved problem in physics, but is associated with the Reynolds number of the flow, which is the ratio of the size scale of the system to the dissipation scale. Formally, for a system of size $L$, characteristic velocity $V$, and kinematic viscosity $\nu$, the Reynolds number is
\begin{equation}
\mbox{Re} = \frac{LV}{\nu};
\end{equation}
here $\nu/V$ is the characteristic dissipation length. For a \red{non-magnetic} diffuse gas \red{(I discuss magnetic effects below)}, the kinematic viscosity is $\nu=2\overline{u}\lambda$, where $\overline{u}$ is the root mean square particle speed and $\lambda$ is the particle mean free path. The RMS particle speed is comparable to the sound speed, so $\overline{u}\sim 0.2$ km s$^{-1}$ in a molecular cloud. The mean free path $\lambda \sim 1/n\sigma$, where $n$ is the particle number density and $\sigma$ is the cross section, typically $\sim 0.01-1$ nm$^2$ for a neutral particle like H$_2$. At a density $n\sim 100$ cm$^{-2}$, this implies $\lambda\sim 10^{12}$ cm, so $\nu\sim 10^{16}$ cm$^2$ s$^{-1}$. Combining this viscosity with a length scale of 10 pc and a velocity scale of 1 km s$^{-1}$ implies that the typical Reynolds number in molecular clouds is $\mbox{Re} \sim 10^9$. Flows at such high Reynolds numbers are invariably turbulent.

However, it is important to note that ordinary fluid viscosity is not the only, or even the dominant, dissipative process operating inside molecular clouds. \red{While there are a large number of non-ideal effects that become important in exceptional circumstances, for example inside shocks or in very high density regions, the dominant effect in the bulk of molecular clouds is ion-neutral drift, also known as ambipolar diffusion (AD). This, rather than ordinary fluid viscosity, is probably the dominant dissipation mechanism inside molecular clouds, at least for motions that are not purely aligned with the magnetic field \citep{li12c}.} 

AD is a non-ideal MHD process that occurs in weakly-ionized plasmas. Because ionizing photons cannot penetrate molecular clouds, cosmic rays are the only significant source of free ions and electrons in them. As a result, their ionization fractions tend to be very low, ranging from a few times $10^{-4}$ in regions where the carbon is mostly C$^+$ down to as low as $\sim 10^{-9}$ in the most dense, shielded regions. A rough fit is given by \blue{$\chi_i \approx 6.3\times 10^{-7} \zeta_{-16}^{1/2} n_{\rm H,2}^{-1/2}$ \citep{mckee10b}.} The electrons and ions feel Lorentz forces from the magnetic field, but the neutrals do not, and thus are capable of drifting across magnetic field lines. This leads to the development of a bulk drift between the ions tied to field lines and the neutrals crossing them. At high ionization fractions, neutrals and ions collide frequently, so this bulk drift is kept small. When there are very few ions, however, neutrals may travel significant distances before colliding with an ion, leading to significant drift velocities. When the neutrals do eventually collide and transfer momentum to the ions, the result is dissipation: conversion of the bulk flow into heat.

One can define a length scale for the dissipation associated with ambipolar diffusion just as one can for viscous dissipation \citep{balsara96a, zweibel97a, zweibel02a, li06a, li08a, tilley11a},
\begin{equation}
L_{\rm AD}  = \frac{\langle B^2\rangle}{4\pi \gamma_{\rm AD} \chi_i \rho^2},
\end{equation}
where $\langle B^2\rangle$ is the root mean square magnetic field strength, $\gamma_{\rm AD}\approx 9.2\times 10^{13}$ cm$^3$ g$^{-1}$ s$^{-1}$ is the coupling coefficient describing ion-neutral collisions, $\chi_i$ is the ion mass fraction, and $\rho$ is the total density including both ionized and neutral species. Similarly, one can define an ambipolar diffusion Reynolds number ${\rm Re}_{\rm AD} = L/L_{\rm AD}$ in analogy to the ordinary fluid Reynolds number. (One can also define the more traditional magnetic Reynolds number by considering the length scale on which Ohmic dissipation will damp out motions, but under the conditions found in molecular clouds, ambipolar diffusion is usually the more important non-ideal MHD process.) Flows with ${\rm Re}_{\rm AD} \gg 1$ behave similarly to those governed by ideal MHD except on very small scales. In those with ${\rm Re}_{\rm AD} \ll 1$, the neutrals behave as a purely hydrodynamic (i.e., non-magnetized) fluid, while the ions behave as a separate, decoupled, fluid governed by ideal MHD. When ${\rm Re}_{\rm AD}\sim 1$, there can be significant dissipation.

The actual values of $L_{\rm AD}$ and ${\rm Re}_{\rm AD}$ in molecular clouds are difficult to estimate for the same reasons that $\mathcal{M}_A$ is: it is not easy to measure reliable magnetic field strengths. There are promising statistical diagnostics based on measuring the differences in spectra between ions and neutrals \citep{li08b, li10a, tilley10a, li12a, downes12a, meyer13a}, and both observations using this technique and using direct estimates of the field strength and density suggest $L_{\rm AD} \sim 0.05$ pc, with quite significant uncertainty. This implies that ${\rm Re}_{\rm AD}$ is of order tens to hundreds. This is large enough to place clouds close to the ideal MHD limit in most circumstances, though deviations from this limit on small scales have potentially interesting dynamical effects. However, it is important to note that, even if AD is the dominant dissipation mechanism, it does not mean that there is no turbulence below the scale $L_{\rm AD}$, \blue{just that} the turbulence on scales below $L_{\rm AD}$ will be hydrodynamic rather than magneto-hydrodynamic \citep[e.g.,][]{oishi06a}. \blue{\citet{li12c} find that, for typical molecular cloud conditions, $\sim 70\%$ of the dissipation occurs through AD heating on a length scale $L_{\rm AD}$, while the remaining 30\% occurs through viscosity on scales smaller than $L_{\rm AD}$.}

\subsubsection{Turbulence: Numerical Studies}

We have seen that molecular clouds are nearly-isothermal media characterized by supersonic and trans- or super-Alfv\'{e}nic turbulence. This turbulence will alter their density and velocity structures. Most work on this topic has focused on fully-developed turbulence that has reached statistical steady state, although, as I discuss below, there is some debate about whether molecular clouds actually live long enough to reach this condition. Putting this question aside for now, in this and the next subsections I review some important results about the properties of supersonically-turbulent media.

Most numerical studies of this topic are conducted using simulations of periodic boxes, within which turbulence is artificially driven by adding random velocities on large spatial scales. The large-scale turbulence then cascades down and generates structure on smaller scales, before dissipating on yet smaller scales. The largest scales are known as the driving scale, the smallest scale on which the turbulence damps out is the dissipation scale, and the intermediate scales are known as the inertial range. If a simulation has sufficient resolution to provide a large separation between the driving and dissipation scales, then the behavior in the inertial range is (hopefully) a reflection of the physics of turbulence and not the artificial manner in which it is driven or dissipated. It is worth noting, however, that no simulation even remotely approaches the actual ratio of driving to dissipation scale in molecular clouds, which is the Reynolds number, $\mbox{Re} \sim 10^9$\blue{, though it is possible to approach the ratio of driving to AD dissipation scale, since ${\rm Re}_{\rm AD} \lesssim 300$}.

Turbulence is generally characterized by the statistics of density and velocity (and magnetic field for MHD turbulence) in the turbulent medium. These can be single-point statistics, such as the probability distribution function (PDF) of density or velocity, which describes the distribution of density or velocity at a randomly chosen point in space. They can also be two-point statistics describing the correlation or PDF of density or velocity smoothed over some particular size scale. Given the vast nature of the study of turbulence, there are numerous results both numerical and analytic, and so I focus on only the few most relevant for what follows. For a more thorough introduction to the statistics of turbulence as they apply to molecular clouds, see \citet{mckee07a} and \citet{hennebelle12a}.

\subsubsection{Turbulence: Velocity Statistics}
\label{sssec:velocitystat}

The velocity field in a turbulent medium can be characterized by a number of statistics. The most basic one is the PDF of velocities at a point. In a non-magnetized supersonically-turbulent medium where the root mean square velocity is $\sigma_v$, \blue{the PDF of a single component of the velocity follows a Gaussian distribution centered on $\sigma_v$, while the PDF of the magnitude of the vector velocity follows a powerlaw below $\sigma_v$ with a Gaussian-like cutoff above $\sigma_v$ \citep{krumholz06a}}; it is unknown how the presence of a magnetic field modifies this result. However, this quantity carries no information about the spatial structure of the flow, and so it is rarely used. It is much more common to describe the spatial structure of the flow using a variety of mathematical tools including autocorrelation functions, structure functions, and power spectra. For all of these statistics, the goal is to characterize how rapidly the velocity changes as one moves between two points in a turbulent flow. The autocorrelation function of the velocity quantifies this as
\begin{equation}
A(\mathbf{r}) \equiv \left\langle\mathbf{v}(\mathbf{x})\cdot\mathbf{v}(\mathbf{x}+\mathbf{r})\right\rangle,
\end{equation}
where $\mathbf{v}(\mathbf{x})$ is the vector velocity at position $\mathbf{x}$, and the average is over all positions. Note that $A(0)$ is simply the mean square velocity in the flow. If the turbulence is isotropic, then $A(\mathbf{r})$ should depend only on $r=|\mathbf{r}|$. In a flow of constant velocity $A(r)$ is constant, but in a turbulent flow we expect $A(r)$ to decrease as $r$ increases, as the velocities at points distant from one another become less and less correlated. It is convenient to express this scale-dependence in Fourier space. If we define
\begin{equation}
\tilde{\mathbf{v}}(\mathbf{k}) \equiv \frac{1}{\sqrt{2\pi}} \int \mathbf{v}\left(\mathbf{x}\right) e^{-i\mathbf{k}\cdot\mathbf{x}}\, d^3\mathbf{x},
\end{equation}
then we can define the power spectrum $\Psi(\mathbf{k}) \equiv |\tilde{\mathbf{v}}(\mathbf{k})|^2$. The Wiener-Khinchin theorem then tells us that the power spectrum is simply the Fourier transform of the autocorrelation function:
\begin{equation}
\Psi(\mathbf{k}) = \frac{1}{(2\pi)^{3/2}}\int A(\mathbf{r}) e^{-i\mathbf{k}\cdot\mathbf{r}} \, d^3\mathbf{r}.
\end{equation}
If the turbulence is isotropic then $\Psi(\mathbf{k})$ should be a function of $k\equiv |\mathbf{k}|$ alone. In such cases it is common to normalize out the volume element by defining
\begin{equation}
P(k) \equiv 4\pi k^2 \Psi(k).
\end{equation}
The difference is that $\Psi(k)$ is the power per unit volume in $k$-space, while $P(k)$ is the total power contained in modes with wave numbers from $k$ to $k+dk$. Unfortunately both $P(k)$ and $\Psi(k)$ are commonly-referred to as the ``power spectrum" in the literature, and so in reading papers one 
must be careful to understand which quantity is being discussed!  In general we expect that $P(k)$ will be a power law in $k$ for any system where the Reynolds number is large, simply because power laws are common in any system where there is a large separation of scales, as there is in any turbulence problem where $\mbox{Re} \gg 1$.

One can relate the index of the power spectrum to the velocity dispersion $\sigma_v(\ell)$ measured over a certain size scale $\ell$ by a simple argument. Suppose that the power spectrum scales as $P(k)\propto k^{-n}$. The total kinetic energy within a region of size $\ell$ must be proportional to $\sigma(\ell)^2$, but we can also calculate it from the power spectrum:
\begin{equation}
\label{eq:sigmavl}
\sigma_v(\ell)^2 \propto \int_{2\pi/\ell}^\infty P(k) \,dk \propto \ell^{n-1}.
\end{equation}
Thus it follows immediately that the velocity dispersion $\sigma_v(\ell)\propto \ell^{(n-1)/2}$. This scaling is commonly referred to as a linewidth-size relationship, since its observable manifestation is that the linewidths of clouds depend on their size \citep{larson81a}. It is common to express this result in terms of a sonic length $\ell_s$, defined by
\begin{equation}
\label{eq:soniclength}
\sigma_v = c_s \left(\frac{\ell}{\ell_s}\right)^{(n-1)/2}.
\end{equation}
Thus the quantity $\ell_s$ is simply the length scale for which the non-thermal velocity dispersion is equal to the sound speed; one may think of it as defining the normalization of the linewidth-size relation.

For subsonic hydrodynamic turbulence, the classic theory of \citet{kolmogorov41a}\footnote{see \citep{kolmogorov91a} for an English translation} shows that $P(k) \propto k^{-5/3}$, and \citet{goldreich95a} propose a model to extend this result to subsonic MHD turbulence, where the assumption of isotropy must break down because the magnetic field introduces a preferred direction. In this model the power spectrum is no longer isotropic. It follows the Kolmogorov scaling in the direction perpendicular to the field, $P(k_\perp)\propto k_\perp^{-5/3}$, while the same level of power in the parallel direction is found at a smaller wavenumber $k_\parallel \propto k_\perp^{2/3}$. More recently, however, this analytic estimate has been called into question, and the topic remains under discussion \citep[and references therein]{hennebelle12a}.

\begin{figure*}[ht!]
\centerline{\includegraphics{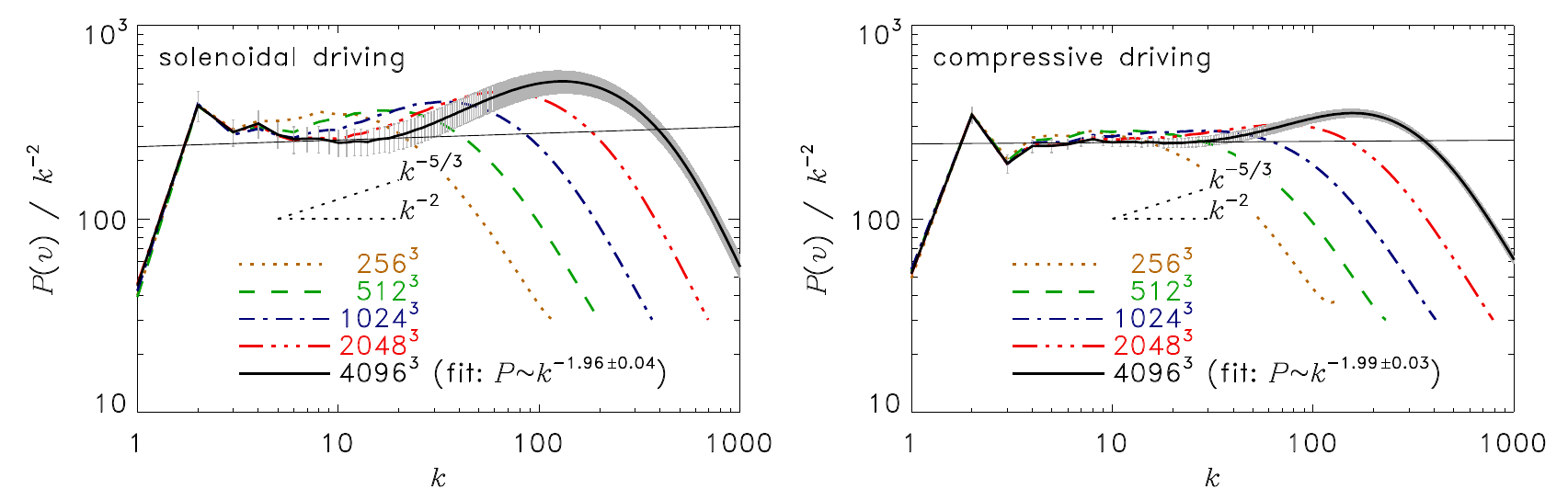}}
\caption{
\label{fig:federrath13}
Measured velocity power spectra in simulations of supersonic hydrodynamic turbulence, reprinted with permission from \citet{federrath13a}. The left panel shows the results with purely solenoidal driving, and the right panel with purely compressive driving. The quantities plotted are ``compensated" power spectra, $P(k)/k^{-2}$, so that a spectrum $P(k)\propto k^{-2}$ appears flat. The Kolmogorov slope $P(k)\propto k^{-5/3}$ is indicated. Different colors correspond to different resolution, and gray error bars show $1\sigma$ variations in time for the $4096^3$ run. The black straight line is a fit to the $4096^3$ simulation over the inertial range, $5\leq k \leq 20$ for solenoidal driving, and $10\leq k \leq 30$ for compressive driving. The drop at low $k$ shows the energy injection scale, and the bump at large $k$ is the bottleneck effect \citep{dobler03a}, which results from the transition between one-dimensional and three-dimensional turbulence that occurs at scales near the resolution limit of the computational grid.
}
\end{figure*}

Not surprisingly given its complexity, the theory of supersonic turbulence, either magnetized or unmagnetized, is much less well-developed. Partial analytic models have been proposed by \citet{boldyrev02a} and \citet{galtier11a}, and these have some support from numerical simulations \citep{kritsuk13b}, but there is no theory as well-developed as those of either Kolmogorov or Goldreich and Sridhar. In the limit of infinite $\mathcal{M}$ and $\mathrm{Re}$, 
one expects the flow to consist of a series of infinitesimally thin shocks, and thus for the velocity field to consist of a series of step functions. The power spectrum of a step function is $P(k) \propto k^{-2}$, and so heuristically one expects a power spectrum of roughly this slope for supersonic turbulence. It is important to stress, however, that this is only a heuristic argument; at present we have no rigorous derivation of the power spectrum for supersonic isothermal turbulence, with or without a magnetic field. Numerical simulations appear to show a steepening of the power spectrum from the Kolmogorov scaling $P(k) \propto k^{-5/3}$ to something closer to $P(k)\propto k^{-2}$ as the Mach number increases \citep{kritsuk07a, federrath13a}, but the exact form of the steepening appears to depend on the way that the turbulence is driven, and even the highest-resolution simulations to date ($4096^3$ for hydrodynamic turbulence \citep{federrath13a}) still achieve inertial ranges that are less than a decade wide. Figure \ref{fig:federrath13} shows a power spectrum measured from one of the most recent simulations.

\subsubsection{Turbulence: Density Statistics}
\label{sssec:densitystat}

The statistics of the density field are also of interest, and the most basic statistic is, as with velocity, the PDF. This was first studied numerically in the non-magnetic case by \citet{vazquez-semadeni94a}, and this work has since been generalized to include ideal and non-ideal MHD and self-gravity, and simulated at ever-higher resolution, by a large number of authors \citep{padoan97a, scalo98a, passot98a, ostriker99a, klessen00c, kritsuk07a, kowal07a, federrath08a, li08a, lemaster08a, schmidt09a, burkhart09a, federrath10b, price10b, price11a, kritsuk11b, collins12a, konstandin12a, molina12a, downes12a, federrath13b}. Numerical simulations have been complemented by a number of analytic studies aimed at deriving the density PDF from first principles \citep{vazquez-semadeni94a, molina12a, hopkins13b}. The general result of this work is that the density PDF for non-self-gravitating, isothermal, ideal MHD turbulence is well-approximated by a lognormal distribution
\begin{equation}
\label{eq:denpdf}
p(s)\, ds = \frac{1}{2\pi \sigma_s^2}\exp\left[-\frac{(s-s_0)^2}{2\sigma_s^2}\right] \, ds,
\end{equation}
where $s=\ln (\rho/\overline{\rho})$, $\overline{\rho}$ is the volume-averaged mean density in the region under consideration, and mass conservation requires that $s_0 = -\sigma_s^2/2$. This is the volume-weighted PDF, i.e., the PDF for the density at a random position. It is straightforward to show via a change of variables that the corresponding mass-weighted PDF (i.e., the PDF is one samples a random mass element, rather than a random position) is identical in functional form but with $s_0$ replaced by $-s_0$. The width $\sigma_s$ of the lognormal is a function of $\mathcal{M}$, $\mathcal{M}_A$, and of the mix between solenoidal and compressive modes present at the driving scale. For $\mathcal{M}_A \gg 1$, \citet{padoan11a} and \citet{molina12a} find that
\begin{equation}
\label{eq:sigmas}
\sigma_s^2 \approx \ln\left(1 + b^2 \mathcal{M}^2 \frac{\beta_0}{\beta_0+1}\right),
\end{equation}
where $\beta_0$ is the ratio of the thermal to magnetic pressure at the mean density and magnetic field strength, and $b$ is a parameter describing the compressive-solenoidal mix, with $b=1/3$ corresponding to purely solenoidal driving and $b=1$ to purely compressive. ``Natural" driving, in which the driving field contains a random mix of solenoidal and compressive components, produces $b\approx 1/2$.

That the functional form should be close to lognormal is not surprising, as in a turbulent isothermal medium the gas at any given position is subjected to a large number of random shocks and rarefactions that cause the density to be multiplied by random numbers; the central limit theorem requires that the product of a large number of independent random factors be a lognormal distribution (also see \citep{pope93a, kevlahan09a}). However, there is no formal proof that the distribution is exactly lognormal.


The introduction of physics beyond isothermal ideal MHD can alter the PDF substantially. An equation of state that is non-isothermal causes the PDF to develop a power law tail in excess of the lognormal; if ratio of specific heats $\gamma$ is smaller than unity, the tail is on the high-density side, while if it is larger than unity the tail is on the low-density side \citep{passot98a, scalo98a, audit10a}. Self-gravity also causes the PDF to develop a tail to high densities, which becomes progressively more prominent and develops a flatter and flatter slope as more of the mass collapses \citep{klessen00c, kritsuk11b, collins12a, federrath13a}. Ambipolar diffusion can produce a variety of effects, depending on ${\rm Re}_{\rm AD}$ and $\mathcal{M}_A$. In general as ${\rm Re}_{\rm AD}$ increases, the magnetic field becomes less important, which tends to broaden the PDF for neutrals -- its effects are analogous to increasing $\beta_0$ in equation (\ref{eq:sigmas}) \citep{li08a, downes12a}.

There has also been a significant amount of numerical work on higher order statistics of the density. \citet{kim05a} compute the power spectrum of density in isothermal hydrodynamic turbulence simulations with Mach numbers $\mathcal{M} = 1.2 - 12$. At low Mach numbers, they find that it is well-fit by a power law with a Kolmogorov-like exponent of $-5/3$, but that this flattens to a slope close to $-0.5$ at higher Mach numbers. \citet{beresnyak05a} and \citet{schmidt09a} report a similar flattening for MHD simulations at high Mach number. \citet{federrath09a} show that the amount of flattening is a function not only of the Mach number but also of whether the turbulence is driven compressively or solenoidally. The spectrum of the logarithm of density, on the other hand, is much less sensitive to the Mach number than the density spectrum, and retains a close-to-powerlaw form with a slope $\approx -5/3$ even at high Mach number \citep{kowal07a}. In addition to power spectra, several authors have analyzed the density structure in supersonic turbulence by examining the statistics of either randomly placed boxes or boxes centered on density peaks \citep{kritsuk07a, federrath09a}.

The numerical results on two-point density statistics have also been accompanied by a significant analytic literature, as these statistics form the basis for one class of models for the origin of the IMF (see Section \ref{sssec:imfturb}). \citet{hennebelle08b, hennebelle09a, hennebelle13a} (hereafter HC) and \citet{hopkins12e, hopkins12d, hopkins13a} have developed models for the scale-dependent density structure in turbulent media, inspired by the Press-Schechter \citep{press74a} and excursion set \citep{bond91a} formalisms first applied in the context of cosmology. The basic goal of these models is to derive a PDF for the logarithmic density $s$ after it has been smoothed to an arbitrary size scale $\ell$. To derive this quantity, we consider a volume $V$ within which the logarithmic density variation as a function of position is $s(\mathbf{x})$. We then define
\begin{equation}
\tilde{s}(\mathbf{k}) = \frac{1}{V} \int s(\mathbf{x})e^{i\mathbf{k}\cdot\mathbf{x}} \, d^3x,
\end{equation}
as its Fourier transform, and invoke Parseval's theorem:
\begin{equation}
\frac{1}{V} \int s(\mathbf{x})^2 \, d^3x = \frac{V}{(2\pi)^3}\int |\tilde{s}(\mathbf{k})|^2\,d^3k.
\end{equation}
The quantity on the left-hand size is the mean square amplitude of the logarithmic density variation per unit volume, and if $s$ is normally distributed (i.e., the density PDF is lognormal), it is easy to show that this is simply $\sigma_s^2$. The quantity inside the integral on the right-hand side is simply the power spectrum of the logarithmic density field. Thus the dispersion of the lognormal density PDF $\sigma_s$ is determined by the power spectrum of the logarithmic density field.

Following the standard approach in cosmology, one can then define the density field smoothed on some scale $\ell$ by
\begin{equation}
s_\ell(\mathbf{x}) \equiv \frac{V}{(2\pi)^3} \int \tilde{s}(\mathbf{k}) W(k,\ell) e^{-i\mathbf{k}\cdot\mathbf{x}} \, d^3k,
\end{equation}
where $W(k,\ell)$ is a window function chosen to remove power at high $k$, corresponding to small spatial scales. Many functional forms for $W(k,\ell)$ are possible, but the usual choice is a spherical top-hat function, where $W(k,\ell) = 1$ for $k<1/\ell$ and $W(k,\ell) = 0$ for $k\geq 1/\ell$, and the HC and \citeauthor{hopkins12e} models both use this approach. This window function amounts to simply truncating the power spectrum at $k=1/\ell$, so that there are no density fluctuations below that size scale. In analogy with the dispersion of the unsmoothed density field, one can then define the dispersion $\sigma_s(\ell)$ of the smoothed density field by
\begin{eqnarray}
\sigma_s(\ell)^2 & = & \frac{V}{(2\pi)^3}\int |\tilde{s}_\ell(\mathbf{k})|^2\, d^3k \\
& = & \frac{V}{(2\pi)^3}\int |\tilde{s}(\mathbf{k})|^2 W(k,\ell)^2 \,d^3k.
\label{eq:sigmasell}
\end{eqnarray}

To proceed further, HC and \citeauthor{hopkins12e} then make two additional assumptions. The first, common to both approaches, is that $s_\ell$ is, like $s$, normally distributed, in which case $\sigma_s(\ell)$ must also be the dispersion of the normal distribution for $s_\ell$. As with $s$, the assumption of a normal distribution cannot be strictly true, since it requires that the density variations on different scales be uncorrelated, which is not consistent with mass conservation. However, it appears to be approximately true.

The second assumption is somewhat different in the two models, but the results are similar. HC note that numerical results indicate that the power spectrum of density fluctuations $|\tilde{s}(\mathbf{k})|^2$ is itself a powerlaw with an index close to that of the velocity power spectrum, and based on this result they hypothesize that the two indices are in fact the same. With this assumption, it is straightforward to evaluate equation (\ref{eq:sigmasell}) to obtain
\begin{equation}
\label{eq:sigmasellhennebelle}
\sigma_s(\ell)^2 = \sigma_s(0)^2\left[1-\left(\frac{\ell}{L}\right)^{n-1}\right],
\end{equation}
where $L$ is the scale on which the turbulence is injected and $\sigma_s(0)^2$ is the dispersion $\sigma_s$ of the unsmoothed density field, given by equation (\ref{eq:sigmas}). Thus the dispersion of logarithmic densities smoothly varies from a maximum of $\sigma_s(0)$ when the density field is not smoothed at all to a minimum of 0 when it is smoothed over the size scale $L$ at which the turbulence is injected. Note that one is required to posit that there is some outer, injection scale to the turbulence beyond which there are no further density or velocity fluctuations; otherwise the integral in equation (\ref{eq:sigmasell}) diverges.

In contrast, the assumption that \citeauthor{hopkins12e} makes is that, if the Mach number $\mathcal{M}$ on some size scale $k$ is $\mathcal{M}_k$, then the dispersion $\sigma_k$ on that size scale obeys a simplified version of equation (\ref{eq:sigmas}) with $b^2=0.25$ and $\beta_0\rightarrow \infty$, as would be expected for a non-magnetized medium with a mix of solenoidal and compressive modes\footnote{Note that the coefficient of $\mathcal{M}_k^2$ in the version of equation (\ref{eq:sigmakhopkins}) in Hopkin's papers is 3/4 rather than 1/4. This difference arises from the choice of whether to define $\mathcal{M}_k$ as the 1D or 3D Mach number. My convention here is to use the 3D Mach number, while Hopkins uses the 1D Mach number.}:
\begin{equation}
\label{eq:sigmakhopkins}
\sigma_k^2 = \ln\left(1 + \frac{1}{4} \mathcal{M}_k^2\right).
\end{equation}
From the linewidth-size relationship (equation \ref{eq:sigmavl}), we have $\mathcal{M}_k\propto k^{-(n-1)/2}$. \citeauthor{hopkins12e} then further assumes that the dispersions on different scales $k$ are uncorrelated, so that the dispersion at a particular size scale $\ell$ in the smoothed density field is simply the quadrature sum of all the dispersions,
\begin{equation}
\label{eq:sigmasellhopkins}
\sigma_s^2(\ell) = \int \sigma_k^2 W(k,\ell)^2 \, d \ln k.
\end{equation}
As in \citeauthor{hennebelle08b}'s formalism, this integral will diverge if one continues to use equation (\ref{eq:sigmakhopkins}) to arbitrarily small $k$, corresponding to arbitrarily large spatial scales, because $\mathcal{M}_k$ diverges as $k\rightarrow 0$. \citeauthor{hopkins12e} handles this divergence by positing that, once $1/k$ is of order the scale height of the galactic disk, turbulent fluctuations are damped by differential rotation preventing material from collecting over large distances. He models this effect by replacing $\mathcal{M}_k^2$ with $\mathcal{M}_k^2/(1 + 2 \mathcal{M}_h^2 / 3|kh|^2)$ in equation (\ref{eq:sigmakhopkins}), where $h$ is the disk scale height and $\mathcal{M}_h$ is the 3D Mach number on that length scale. As long as $n<3$, this quantity approaches 0 as $k\rightarrow 0$, removing the divergence. Unlike equation (\ref{eq:sigmasellhennebelle}) in the HC formalism, equation (\ref{eq:sigmasellhopkins}) does not evaluate to a simple analytic function. However, the qualitative behavior is similar.

With either set of approximations, the outcome of this procedure is a model for the PDF of the smoothed logarithmic density field $s_\ell$, which is given by equation (\ref{eq:denpdf}), with $\sigma_s(\ell)$ evaluated using either equation (\ref{eq:sigmasellhennebelle}) or (\ref{eq:sigmasellhopkins}). As I discuss further in the remainder of this review, this PDF is a powerful tool, since it allows one to take advantage of the large amount of mathematical machinery that has been developed around Press-Schechter and excursion set theory for calculating mass functions, correlations, and similar statistics. 

However, it is important to emphasize that, while these models are based on plausible hypotheses about the scale-dependent behavior of the density PDF, neither of them have been derived in a truly rigorous manner, and they have only begun to be tested against simulations \citep[e.g.,][]{schmidt10a}. There is no full theory for the power spectrum of the density field in a supersonically turbulent medium, which is the central object that is required to derive the smoothed density PDF, and numerical simulations show that it only approximately follows the functional forms assumed by the analytic models \citep[e.g.,][]{kowal07a}. Indeed, the differing assumptions that \citeauthor{hennebelle08b} and \citeauthor{hopkins12e} make regarding this power spectrum lead to noticeable differences in their predictions for structure at small scales, and both approaches are plausible, suggesting the need for further investigation. In addition, neither model takes into account magnetic fields or self-gravity in calculating the density statistics (though they are included in a heuristic way when considering collapse, as I discuss later). 

There is also significant uncertainty about how to define the input parameters required for the models in a real molecular cloud. In both models, the predicted density dispersion on a particular size scale depends on the sonic length $\ell_s$, which defines the normalization of the linewidth-size relation (equation \ref{eq:soniclength}). The difficulty in the real world is that in observed molecular clouds, different parts of the same cloud do not all fall on a single linewidth-size relation with the same normalization \citep{plume97a, shirley03a}. Indeed, for fixed \blue{virial ratio $\alpha_{\rm vir}$ (see Section \ref{ssec:gmcstability})}, it is easy to show that the normalization of the linewidth-size relationship is fully determined by the surface density of the gas \citep{larson81a}, and thus the models are unambiguously-defined only for molecular clouds with constant surface density. While molecular clouds as measured by CO emission do have (roughly) constant surface densities \citep[and references therein]{dobbs14a}, regions of active star cluster formation often have vastly higher surface densities \citep[e.g.,][]{faundez04a, fontani05a, fall10a}\blue{, and in general molecular clouds display a range of distributions of area versus surface density \citep[e.g.,][]{lada13a}}. In such cases, it is unclear whether the density structure should be calculated using the surface density / linewidth-size normalization that applies to the entire cloud as traced by CO emission, or to the much smaller region of active star formation. We will encounter this issue again when we come to theoretical models for the IMF.

A further concern with the Hopkins model is its treatment of rotational stabilization from the galactic scale to suppress density variations at large scales. Observed molecular clouds in fact have negligible rotational support, and many of them are actually counter-rotating with respect to their host galaxy \citep{phillips99a, rosolowsky03a, imara11a, imara11b}. Thus the approach of treating galactic rotation as the main stabilizing agent on large scales seems dubious. Given these limitations, models of this type are powerful enough to be worth investigating further, but definitely in need of further testing.

\subsubsection{Turbulence: Decay Rates}
\label{sssec:turbdecay}

A final property of turbulence that is important for our purposes is that it decays. Subsonic flows are local in Fourier space, meaning that motions at some particular wave vector $k$ can only transfer energy to adjacent values of $k$. Thus for subsonic turbulence decay occurs through a cascade from large to small scales, finally terminating at the viscous dissipation scale \citep{kolmogorov41a, kolmogorov91a}. Supersonic turbulence can decay via this mechanism as well, but it can also dissipate directly by forming shocks, which couple together all wave vectors $k$ directly. In any event, it is clear that turbulent motions must eventually reach the scale where they can be dissipated as heat, and that, given the large radiative cooling rates calculated above, the energy will then be lost to radiation in short order.

It is natural to measure a dimensionless dissipation rate for turbulence as
\begin{equation}
\epsilon_{\rm diss} = \frac{1}{2} \frac{\dot{E}_{\rm turb}}{E_{\rm turb}} \frac{L}{V},
\end{equation}
where $E_{\rm turb}$ is the turbulent kinetic energy and $L$ and $V$ are the outer (driving) length scale of the turbulence and the characteristic velocity at that scale. A number of simulations have sought to measure $\epsilon_{\rm diss}$ for the case of supersonic MHD turbulence relevant to molecular clouds, and have generally obtained values of $\epsilon_{\rm diss}\approx 0.6-1$ \citep{stone98a, mac-low98a, mac-low99b, padoan99a, lemaster09a}. A detailed comparison of a large number of grid codes running an identical decaying turbulence problem shows that the decay rates in the different codes agree to $\sim 10\%$ \citep{kritsuk11a}, so the factor of $\sim 2$ differences in $\epsilon_{\rm diss}$ reported in the literature are likely due to variations in exact problem setup (e.g., the ratio of solenoidal to compressive modes in the driving pattern, whether and how the driving pattern is changed in time) than to numerical issues. Regardless of these small differences, it is clear from these results that isotropic supersonic MHD turbulence will decay in a time of order the flow crossing time. However, if the turbulence is anisotropic, the dissipation timescale depends only on the isotropic component of the velocity \citep{hansen11a}. Thus a highly-anisotropic velocity field will decay less rapidly than an isotropic one of the same magnitude.

\subsection{Large-Scale Force Balance}
\label{ssec:gmcstability}

Thus far we have examined the microphysical chemical and radiative processes that govern molecular clouds, and the properties of the turbulent flows within them. The final section of this review of the theoretical background addresses the large-scale force balance and stability of molecular clouds. The goal here is to understand the forces that drive molecular cloud evolution at the largest scales. The virial theorem provides an invaluable tool for this type of analysis. Depending on whether one begins from the Lagrangian or Eulerian version of the equations of motions, one can derive either Lagrangian \citep{chandrasekhar53a, mestel56a} or Eulerian \citep{mckee92a, krumholz06d, goldbaum11a} forms of the theorem. The latter is more appropriate for molecular clouds, as they can gain or lose mass to their environments and thus do not necessarily consist of fixed mass elements. The most basic Eulerian form, including hydrodynamic motions, gas pressure, magnetic fields, and self-gravity, is
\begin{eqnarray}
\frac{1}{2}\ddot{I} & = & 2 (\mathcal{T}-\mathcal{T}_0) + \mathcal{B} -\mathcal{B}_0 + \mathcal{W} \nonumber \\
& & {} - \frac{1}{2}\frac{d}{dt}\oint_{\partial V} (\rho\mathbf{v}r^2)\cdot d\mathbf{S}
\label{eq:virialthm}
\end{eqnarray}
where $V$ is the volume of interest, and the remaining terms are
\begin{eqnarray}
I & = & \int_V \rho r^2 \, dV\\
\label{eq:tdefn}
\mathcal{T} & = & \frac{1}{2} \int_V (3P + \rho v^2)\, dV\\
\mathcal{T}_0 & = & \frac{1}{2} \oint_{\partial V} \mathbf{r}\cdot\mathbf{\Pi}\cdot d\mathbf{S} \\
\mathcal{W} & = & \int_V \rho \mathbf{r}\cdot\mathbf{g} \, dV \\
\mathcal{B} & = & \frac{1}{8\pi}\int_V B^2 \, dV \\
\mathcal{B}_0 & = & -\oint_{\partial V} \mathbf{r}\cdot\mathbf{T}_M\cdot d\mathbf{S}.
\end{eqnarray}
Here $\mathbf{\Pi} = \rho \mathbf{vv} + P\mathbf{I}$ is the gas pressure tensor, $\mathbf{T}_M = (1/4\pi)[\mathbf{BB}-(B^2/2)\mathbf{I}]$ is the Maxwell stress tensor, $\mathbf{I}$ is the identity tensor, and $\mathbf{g}$ is the gravitational acceleration. This relation is an exact theorem, which follows rigorously from the equations of ideal MHD.

The terms that appear in the virial theorem have relatively straightforward physical interpretations: $I$ is the moment of inertia of the volume being studied, $\mathcal{T}$ is the sum of the kinetic and thermal energies, $\mathcal{B}$ is the total magnetic energy, and, in the absence of external gravitational accelerations, $\mathcal{W}$ is simply the gravitational binding energy. The terms $\mathcal{T}_0$ and $\mathcal{B}_0$ represent forces exerted at the surface of the volume; the former includes ordinary fluid pressure and ram pressure, while the latter includes magnetic pressure and tension. Finally, the term $(1/2)(d/dt)\oint (\rho\mathbf{v}r^2)\cdot d\mathbf{S}$ represents changes in the moment of inertia within the volume of interest due to advection of material across the bounding surface.

The sign of $\ddot{I}$ determines whether the matter in the volume is accelerating or decelerating toward the center, and so the balance between the terms on the right hand side provides an important guide to the large-scale behavior. The terms $\mathcal{T}$ and $\mathcal{B}$ are strictly positive, in the absence of an external gravitational potential $\mathcal{W}$ is strictly negative, and except for quite unusual configurations of the flow and magnetic field, $\mathcal{T}_0$ and $\mathcal{B}_0$ will be negative as well. The final time-dependent term can take on either sign, depending on the direction of the bulk flow across the surface. Three  parameters of particular importance are the Jeans number $n_J$, which describes the ratio $2\mathcal{T}/|\mathcal{W}|$ including only the thermal pressure contribution to $\mathcal{T}$, the virial ratio $\alpha_{\rm vir}$, which describes the ratio $2\mathcal{T}/|\mathcal{W}|$ including both thermal and non-thermal contributions to $\mathcal{T}$, and the dimensionless mass to flux ratio $\lambda$, which describes the ratio $\mathcal{B}/|\mathcal{W}|$. I say ``describe" here because in practice the exact quantities that appear in the virial theorem cannot be determined from observations, and so one generally defines closely related quantities that are directly measurable instead. 

The Jeans number is simply the ratio
\begin{equation}
n_J = \frac{M}{M_J}, 
\end{equation}
where $M$ is the cloud mass, the Jeans mass is
\begin{equation}
\label{eq:jeansmass}
M_J = \frac{\pi^{3/2}}{8} \frac{c_s^3}{\sqrt{G^3\rho}}
\end{equation}
and $\rho$ is the gas density\footnote{The prefactor $\pi^{3/2}/8$ in the definition of $M_J$ is somewhat arbitrary, as one could plausibly define the Jeans mass as $\rho(\lambda_J/2)^3$, $\rho\lambda_J^3$, $(4/3)\pi\rho\lambda_J^3$, or $(4/3)\pi\rho(\lambda_J/2)^3$, where $\lambda_J=(\pi c_s^2/G\rho)^{1/2}$ is the Jeans length. All four choices can be found in the literature. The prefactor given in equation (\ref{eq:jeansmass}) corresponds to the first of them. One could also define $n_J$ in terms of the Bonnor-Ebert mass \citep{ebert55a, bonnor56a}, $M_{\rm BE} = 1.18 c_s^3/(G^3\rho)^{1/2}$. This is the maximum mass for which a pressure-bounded isothermal sphere can be hydrostatic equilibrium.\label{note:jeansmass}}. For a uniform-density sphere of constant sound speed, one can easily verify that $n_J$ is, up to factors of order unity, simply $2\mathcal{T}/|\mathcal{W}|$, including only the thermal pressure $P$ when computing $\mathcal{T}$ (c.f.~equation \ref{eq:tdefn}). For the virial ratio, \citet{bertoldi92a} define
\begin{equation}
\alpha_{\rm vir} = \frac{5\sigma_{\rm 1D}^2 R}{G M},
\end{equation}
where $\sigma_{\rm 1D}$ is the one-dimensional thermal plus non-thermal velocity dispersion of a cloud of radius $R$. It is straightforward to confirm that this is quantity is simply $2\mathcal{T}/|\mathcal{W}|$ for a uniform-density sphere, including both $P$ and $\rho v^2$ when evaluating equation (\ref{eq:tdefn}). \blue{\citeauthor{bertoldi92a} also show that $\alpha_{\rm vir}$ remains close to $2\mathcal{T}/|\mathcal{W}|$ even for non-spherical and non-isochoric clouds.} Note that $\alpha_{\rm vir}$ and $n_J$ are related by the Mach number, $n_J \approx \mathcal{M}^2 \alpha_{\rm vir}$. Finally, one can define
\begin{equation}
\lambda = 2\pi\sqrt{G} \frac{M}{\Phi} = 2\pi\sqrt{G}\frac{\Sigma}{B},
\end{equation}
where $\Phi = \pi R^2 B$ is the magnetic flux threading a cloud. This is again, up to a factor of order unity, the value of \blue{$(|\mathcal{W}|/\mathcal{B})^{1/2}$} for a uniform spherical cloud. Clouds with $\lambda>1$ are said to be magnetically supercritical, while those with $\lambda < 1$ are subcritical. This distinction is particularly important because, in the ideal MHD limit where the magnetic field is frozen into the matter, $\lambda$ is invariant under overall expansions or contractions of the gas. This means that, if a cloud is magnetically subcritical and subject to ideal MHD, the magnetic term in the virial theorem will always exceed the gravitational one, and the cloud will not be able to undergo a self-gravitating collapse. It can still be accelerated inward, as the other negative terms may be larger than $\mathcal{B}$, but it can never undergo a self-gravitating collapse. Conversely, if a cloud is supercritical, its magnetic field will never be strong enough to overcome gravity.

Observations across a wide range of galactic environments, both in the Milky Way and in external galaxies, generally give $\alpha_{\rm vir} \approx 1$ for the clouds that constitute most of the molecular mass in galaxies \citep{solomon87a, fukui08a, bolatto08a, heyer09a, roman-duval10a, wong11a}, suggesting that gravity and bulk flow are roughly equally important in determining the behavior of clouds. \blue{However, these results are subject to significant systematic uncertainties, and it is important to note that, while large-scale molecular clouds defined by CO emission generally show $\alpha_{\rm vir} \approx 1$, the denser structures within them often have $\alpha_{\rm vir} \gg 1$ \citep{bertoldi92a, barnes11a}.} It is \blue{also} important to note that \blue{even a value of $\alpha_{\rm vir}$ close to unity} does \textit{not}, by itself, mean that turbulence supports clouds against gravity: a gas undergoing pressureless free-fall collapse will have $\alpha_{\rm vir} = 2$\blue{, and the difference between 1 and 2 is not large enough to be measured with confidence}. Nonetheless, \blue{the fact that $\alpha_{\rm vir} \sim 1$ does} mean that ram pressure forces are non-negligible in comparison to gravity. Moreover, the lack of \blue{clouds identified by CO emission} with $\alpha_{\rm vir} \gg 1$ \textit{does} strongly rule out the possibility that \blue{giant molecular clouds on the largest scales} are purely pressure-confined objects for which gravity is unimportant. Since observed Mach numbers $\mathcal{M}$ are far larger than unity, these same observations imply $n_J \gg 1$ except at the very smallest scales in molecular clouds, and thus thermal pressure support alone is unimportant.

The dimensionless mass to flux ratio $\lambda$ has historically been much harder to determine due to the difficulty of measuring magnetic field strengths. However, a long-term campaign over the past two decades now appears to have borne fruit, and there is an emerging observational consensus that $\lambda \approx 2-3$ \citep[and references therein]{crutcher12a}, implying that clouds are typically supercritical and that their magnetic fields are not sufficient to prevent collapse.

The relative sizes of the surface terms in the virial theorem are unfortunately much more difficult to determine from observations, but this does not mean that they are unimportant. In numerical simulations of turbulent flows, where clouds are identified simply as structures above some specified density threshold, several authors have found that the surface terms are comparable in importance to the volume ones \citep{ballesteros-paredes99a, dib07a}. One should be wary about reading too much into this result, because not all methods of observationally selecting clouds are equivalent to simple density thresholds -- indeed, as discussed in the context of molecular cloud chemistry above, the requirement for the formation of H$_2$ and CO is more akin to a column density threshold than a volume density one, and structures defined by column density are likely to be less transient than those defined by volume density. This is particularly the case as it applies to H$_2$ formation, where the relatively long equilibration time means that a structure must survive for some non-negligible length of time before it undergoes the chemical transitions required for it to be visible as a molecular cloud. However, it is clear that if there is a coherent flow across a cloud surface, for example an accretion flow, the terms in the virial theorem associated with that can be comparable to the more easily-measurable internal ones \citep{goldbaum11a}.

\section{The Star Formation Rate}
\label{sec:sfr}

With the background review complete, I now turn to the first of the three major themes of this review: what sets the star formation rate, both on galactic and sub-galactic scales? There are two classes of theoretical approach to this problem, which for the purposes of this review I will call top-down and bottom-up models. The basic assumptions of each can be summarized simply: in top-down models, the star formation rate is assumed to be set mostly by galactic-scale processes, and to be essentially independent of the rate of star formation on the scales of individual molecular clouds or smaller. In the bottom-up model, one attempts to construct the galactic-scale star formation rate as arising from the sum of star formation operating on the scales of many individual molecular clouds, coupled with a model for how those clouds are arranged on galactic scales.

\subsection{Star Formation Rates: the Top-Down Approach}

\subsubsection{Gravity-Only Models}

Historically, top-down models for the star formation rate were motivated by the need to account for two observations. The first is the strong, non-linear correlation between gas surface density and star formation rate surface density when both are averaged over the scale of entire galaxies, and the equally strong correlation between surface density of star formation and the gas surface density divided by the galaxy orbital period. Quantitatively, \citet{kennicutt98a, kennicutt98b} found $\Sigma_{\rm SFR}\propto \Sigma^{1.4}$ and $\Sigma_{\rm SFR}\propto \Sigma/t_{\rm orb}$. The second observational datum is an apparent edge to the star forming regions in galaxies, associated with a transition from the gas being Toomre \citep{toomre64a} stable to Toomre unstable \citep{quirk72a, kennicutt89a, martin01a, yang07a}. Both of these observations would seem to suggest that galaxy-scale processes are at work in regulating the star formation rate.

The first models that attempted to explain these observations invoked no physics beyond hydrodynamics and self-gravity. In these models, the primary mechanism driving star formation is the gathering of material from the disk into gravitationally-unstable clouds, and the physics of this process is what accounts for the observations. If this is the case, then a sharp cutoff in star formation when the disk ceases to be gravitationally-unstable is straightforward to explain: if the Toomre $Q$ is sufficiently large, then the gravitational instability responsible for fueling star formation shuts down \citep{kennicutt89a, li05a, li06b, li06c}.

It is also possible to produce a non-linear correlation between gas surface density and star formation rate using the same physical ingredients. Such a non-linearity is generically expected in models where gravity is the dominant player simply because the characteristic timescale associated with any gravitational phenomenon is the free-fall time, and the free-fall time should scale roughly with the gas surface density as $t_{\rm ff} \propto (\Sigma/h)^{-1/2}$, where $h$ is the galactic scale height. Thus if we generally expect star formation rates to depend on mass supply over characteristic time scale, we should expect a dependence $\Sigma_{\rm SFR} \propto \Sigma^n$ with $n>1$ but with an exact value that depends on how the scale height varies with $\Sigma$ or other parameters \citep{madore77a, elmegreen02a, shu07a}. Numerical simulations of gravitationally-unstable disks generally reproduce this scaling \citep{li05a, li06b, li06c, tasker06a}.

Several authors have also proposed that this non-linearity occurs because star formation is determined by the amount of mass in the high-density tail of the density PDF in a turbulent galactic disk (see Section \ref{sssec:densitystat}) \citep{kravtsov03a, wada07a}. The basic idea is to exploit the property of lognormals (or any similarly-shaped functions) that the fraction of the distribution that lies above some threshold that is well above the peak of the distribution is non-linearly sensitive to the location of the peak. To put it another way, if one considers a lognormal distribution of densities for which the mean density is $\overline{\rho}$, and computes the fraction of the mass $M(>\rho_{\rm th})$ that lies above a density $\rho_{\rm th} \gg \overline{\rho}$, then $M(>\rho_{\rm th})$ is a super-linear function of $\overline{\rho}$. For suitably-chosen values of the star formation threshold, one recovers a relationship $\Sigma_{\rm SFR} \propto \Sigma^n$ with values of $n$ close to the observed index of 1.4. However, the powerlaw index is \textit{not} independent of the choice of threshold (something that has important observational consequences that I discuss in Section \ref{ssec:bottomup}), and in these models no physical justification for the choice of threshold is specified.

While models of this type do seem able to reproduce the non-linear relationship between the surface densities of gas and star formation, they have traditionally not tried to reproduce the additional correlation with galaxy orbital period. However, there are a related class of models that have attempted to explain the observed correlation with orbital period in terms of a process of agglomeration or collision of pre-existing dense clouds \citep{wyse86a, wyse89a, elmegreen94b, silk97a, hunter98a, tan00a, tan10a}, or in terms of shocking by passage of spiral arms \citep{shu07a, dobbs09a, bonnell13a}. The central observation made in these models is that any process that involves collisions, agglomerations, or shock passages in a shearing disk, and where the accumulation is \textit{not} driven by local self-gravity will proceed on a timescale comparable to the orbital period, modified by a factor that depends on the shape of the rotation curve in regions where the rotation curve is not flat. If one assumes that the rate of star formation is proportional to the rate at which such collisions or agglomerations happen, then the scaling $\Sigma_{\rm SFR}\propto \Sigma/t_{\rm orb}$ naturally falls out.

In general it is difficult to distinguish observationally between models that invoke non-gravity-driven agglomeration as the central physical process and those that invoke gravitational instability (and thus gravitationally-driven agglomeration) instead. This is because galaxy disks have Toomre $Q$ parameters close to unity, and one way of writing the condition $Q\approx 1$ is that $t_{\rm ff} \sim t_{\rm orb}$ at the midplane, i.e.~the condition for marginal gravitational stability can be stated as the requirement that the local free-fall timescale and the local orbital period be comparable (omitting constant factors of order unity). Thus one generically expects the same scaling between $\Sigma_{\rm SFR}$, $\Sigma$, and $t_{\rm orb}$ regardless of whether the driving physics is gravitational collapse or random agglomeration, because Toomre stability requires that the timescale for both processes be comparable\footnote{However, note that the simulations of \citet{li06c} give $\Sigma_{\rm SFR} \propto \Sigma^{1.4}$, but also $\Sigma_{\rm SFR}\propto (\Sigma/t_{\rm orb})^{1.5}$. This is likely because the gravitational instability in reality depends on both gas and stars, something that the simple analysis given here omits.}.

While these models have a number of attractive features, the observational picture that originally motivated them has become considerably muddied. First of all, while there is a non-linear correlation between the surface densities of star formation and gas content when entire galaxies are treated as points, as discussed in Section \ref{ssec:sfrobs}, the correlation looks much closer to linear on sub-galactic scales. While this point is subject to considerable observational dispute, and some observers report super-linear correlations even on sub-galactic scales, the existence of a super-linear correlation on sub-galactic scales can no longer be taken as an established fact. This would present a problem for models where gravitational instability is the dominant physical process regulating star formation rates, as this should produce a non-linear correlation even on sub-galactic scales. On the other hand, models in which the dominant physics is spiral shock passages generally are able to reproduce the closer to linear slope \cite{dobbs09a, bonnell13a}.

The observational situation has changed even more sharply for the correlation between star formation and orbital time. On sub-galactic scales, there is no significant correlation between $\Sigma_{\rm SFR}$ and $\Sigma/t_{\rm orb}$ \citep{wong02a, leroy08a, krumholz12a}. This is difficult to reconcile with models where cloud collisions, or something else that scales with the local orbital period, is the driving physics.

Finally, the evidence for a sharp cutoff in the star formation rate at a threshold value of Toomre $Q$ has mostly evaporated. The first major piece of evidence came from \citet{hunter98a}, who showed in a sample of irregular galaxies that there was no strong evidence for a threshold at a fixed value of $Q$. For spiral galaxies, the shift began with the advent of far ultraviolet observations using \textit{Galactic Evolution Explorer (GALEX)} satellite \citep{iglesias-paramo06a}, which revealed that star formation as traced by FUV in fact continues well past the purported edge \citep{boissier07a, thilker07a}. The sharp cutoffs reported in the literature were based on ground-based H$\alpha$ observations, which, for reasons discussed in Section \ref{sssec:unresolvedimf}, cease to be faithful tracers of star formation once the star formation rate becomes too small. Moreover, as with the orbital time, sub-galactic observations reveal no correlation between the star formation rate in a given region and the local value of Toomre $Q$ \citep{leroy08a}.

A second difficulty for the gravity-driven models is that, while they may be able to explain correlations between star formation rates and various galaxy properties, they have little to say about the proportionality constants, which dictate the absolute rate of star formation and thus the value of $\epsilon_{\rm ff}$. Models that include no physics other than gravity and hydrodynamics tend to produce gas depletion times that are $\sim 2$ orders of magnitude too short \citep[e.g.,][]{tasker06a, bonnell13a}. This is not surprising, as the observed depletion time corresponds to $\epsilon_{\rm ff} \approx 0.01$, while a gas that is not supported against collapse will naturally produce stars at a rate corresponding to $\epsilon_{\rm ff} \approx 1$.\footnote{When gas is arranged in filaments whose characteristic width is smaller than or comparable to the Jeans length, $\epsilon_{\rm ff}$ can be smaller than unity by a factor that depends on the aspect ratio of the filament \citep{toala12a, pon12a}, but the fact that numerical simulations still produce star formation rates that are much higher than observed implies that this cannot be the main reason why $\epsilon_{\rm ff} \approx 0.01$ averaged over galactic scales.} The problem can be mitigated by introducing artificial pressure support, manually restricting the rate or efficiency of star formation, or adding similar subgrid models that limit the conversion of gas into stars on the smallest scales resolved in the simulation \citep[e.g.,][among many others]{springel03b, li05b, li06c}, but these subgrid models must be hand-tuned to reproduce the observed star formation rate, and of course this approach does not address the question of why $\epsilon_{\rm ff}\approx 0.01$ in the first place.

\subsubsection{Feedback-Regulated Models}
\label{sssec:feedbackmodel}

The need for models of star formation that can successfully explain the long depletion time of the neutral ISM led to a focus on star formation feedback as the key mechanism. In order to reduce the star formation rate and produce $\epsilon_{\rm ff} \ll 1$, stellar feedback needs to be able to disperse self-gravitating clouds of dense, molecular gas. As discussed in Section \ref{ssec:gmcthermo}, one of the key characteristics of this gas is that it is able to lose energy via radiation very efficiently. Thus the amount of heat delivered to molecular gas is relatively unimportant, unless it is so great as to be able to destroy the molecules and raise the temperature to the point where the radiative cooling time becomes long rather than short compared to the dynamical time. While this certainly happens to some quantity of gas (e.g., that which is hit directly by supernova ejecta traveling at $\sim 10^4$ km s$^{-1}$), even supernovae generally do not provide enough power to heat the bulk of the gas to such high temperatures. Instead, what matters is the momentum delivered to the gas, since this cannot be radiated away. If stellar feedback is able to deliver enough momentum to the gas in a molecular cloud, it may be able to unbind and disperse that cloud, halting any further star formation. Thus the key quantity in assessing the impact of stellar feedback is the amount of momentum delivered to the molecular gas.

This momentum can be characterized by either a momentum production rate per unit stellar mass $\dot{V} = (dp/dt)/M_*$ (units of velocity per time) for a given population of stars, or as a momentum production rate per unit star formation rate $V = (dp/dt)/\dot{M}_*$ (units of velocity) for a region with a specified star formation rate, and where the population is in statistical equilibrium between new stars forming and old stars dying. The former is more relevant for a regions whose characteristic dynamical times are smaller than the lifetimes of individual massive stars (the ``young stars" limit, in the terminology of \citet{krumholz10b}), while the latter is more relevant on larger size scales where the massive stellar population has time to reach statistical steady state (the ``old stars" limit). For a given mechanism to have a significant effect on the star formation rate, either $\dot{V}$ multiplied by the dynamical time of a star-forming cloud, or $V$, must be at the same order of magnitude as the cloud escape speed. For a mechanism to be capable of launching galactic winds, $V$ must be of order the galactic escape speed.

Stars can provide momentum through a number of distinct channels (see \citet{krumholz14a} for a recent review), some of which are likely important only on the scales of individual protostellar cores or star clusters, and others of which have effects felt even at galactic or super-galactic scales. Here I summarize the mechanisms that have been invoked by various authors as important regulators of the star formation rate.

\paragraph{Supernovae and Main Sequence Stellar Winds}
The classical mechanism that has been invoked to regulate star formation since the earliest simulations of galaxy formation was supernovae. For a Chabrier or Kroupa IMF, there will be $\approx 1$ supernova per 100 $\msun$ of stars formed, each of which will release $\approx 10^{51}$ erg in the form of $\approx 1$ $\msun$ of ejecta traveling at $\approx 10^4$ km s$^{-1}$. For a population of stars forming at a constant rate, the ejecta themselves carry a momentum flux $V_{\rm SN} \approx 48$ km s$^{-1}$ per unit SFR, but the actual momentum delivered is likely to be significantly larger than this. When the ejecta shock against the ISM, their high speed causes the post-shock temperature to be very high, $\sim 10^7$ K. At such high temperatures, the cooling time for the gas is very long, and as a result the over-pressured hot gas can expand adiabatically, in the process sweeping up cold gas and imparting a great deal of momentum to it. This process can plausibly increase the momentum delivered by more than an order of magnitude.

Although they have not traditionally been included in galaxy formation simulations the way supernovae have, the fast winds of early-type main sequence stars behave very similarly. At Solar metallicity, they carry instantaneous and time-integrated momentum fluxes of $\dot{V}_{\rm w} = 9$ km s$^{-1}$ \citep{krumholz14a} and $V_{\rm w} = 140$ km s$^{-1}$ \citep{dekel13a}, but they also have very high speeds and thus very high post-shock temperatures, potentially leading to an adiabatic phase where the momentum will grow substantially. However, stellar winds are also highly metallicity-dependent \citep{vink01a}, so they will not be significant in low-metallicity galaxies \citep{dib11a}. This suggests that stellar winds cannot be a dominant feedback mechanism even at Solar metallicity, because if they were we would expect to see large increases in star formation rates per unit mass in low-metallicity galaxies, quite the opposite of what is actually observed.

There are well-developed theoretical models for the interaction of both stellar winds, supernovae, or a combination of the two with a surrounding ambient medium in the case where that medium is uniform. In the case of supernova blast waves \citep[e.g.,][]{chevalier74a, mckee77a, ikeuchi84a, cioffi88a, ostriker88a, koo92a, koo92b, blondin98a, thornton98a}, the expansion passes through several distinct phases. At the earliest times the ejecta have not yet encountered a mass comparable to their own, and they expand freely. Once they encounter their own mass, they shock and their energy thermalizes, leading to a pressure-driven, adiabatic Sedov-Taylor phase. As the gas expands adiabatically, its temperature drops, and eventually radiative cooling starts to become significant, leading to a transition to a non-adiabatic but still pressure-driven expansion phase, known as the pressure-driven snowplow. Once cooling completely depletes the energy of the hot gas, the momentum becomes constant, leading to a phase known as the momentum-conserving snowplow.

Depending on the mean density of the ambient medium and on whether there are multiple supernovae within an expanding bubble, one or more of these phases may be skipped or modified \citep[e.g.,][]{dekel86a, melioli04a}, but the endpoint is always the same: an expanding shell of fixed momentum. It is the momentum in this final phase that is the relevant quantity to assess the importance of supernovae as a form of star formation feedback. Stellar winds are slightly different in that the injection of energy is continuous rather than sporadic, but the basic dynamics are the same: as long as the bubble of hot gas is kept hot (in this case by the continuing injection of wind energy), it sweeps up more material and adds more momentum to the swept-up shell, ceasing only when the driving stars turn off \citep{castor75a, weaver77a, capriotti01a, tenorio-tagle06a, arthur12a, silich13a}. 

Unfortunately the interactions of either supernovae or stellar winds with more realistic, non-uniform interstellar media are far less well-understood, and there have been far fewer either simulations or analytic models of such cases. Non-uniformity on average probably decreases the momentum imparted to the ambient medium, because it provides new mechanisms for energy to escape from the system rather than being used to drive motion. This escape can be either radiative, due to mixing between the hot and cold gas that enables faster radiative cooling \citep{mckee84a}, or it can be mechanical, with the hot gas punching holes and escaping \citep{tenorio-tagle07a, dale08a, harper-clark09a, rogers13a, creasey13a}. The relatively low X-ray luminosities observed from stellar wind bubbles provide direct observational confirmation that some form of escape must be taking place at least for stellar winds \citep{townsley03a, stevens03a, townsley06a, harper-clark09a, townsley11a, lopez11a, lopez13a}.

None of this physics is captured in most simulations that seek to understand the regulation of star formation, simply for reasons of resolution. The Sedov-Taylor phase, when much of the build-up of momentum occurs, happens when the blast wave is at radii in the range \citep[pp.~430-433]{draine11a}
\begin{eqnarray}
R_{\rm S-T,min} & \approx & 0.4 M_{\rm ej,0}^{1/3} n_{\rm H,2}^{-1/3}\mbox{ pc} \\
R_{\rm S-T,max} & \approx & 3.4 E_{51}^{0.29} n_{\rm H,2}^{-0.42}\mbox{ pc},
\end{eqnarray}
where $M_{\rm ej,0}$ is the mass of supernova ejecta in units of $M_\odot$ and $E_{51}$ is the kinetic energy of the ejecta in units of $10^{51}$ erg. Thus much of the momentum build-up occurs in the space of a few pc. Simulations that do not resolve these scales have difficulty properly capturing the momentum of supernovae, because if they simply add energy to a given computational element (cell or smoothing kernel, depending on the code type), it will be diluted by the large mass in that cell, producing a low temperature that allows the energy to be radiated away with unphysical rapidity \citep{katz92a}. There are a large variety of numerical work-arounds for this problem, including giving large kicks to individual smoothed-particle hydrodynamics (SPH) particles to ensure that the ISM receives a large amount of momentum \citep[e.g.,][]{navarro94a, springel03a, dalla-vecchia08a}, 
direct insertion of momentum into existing particles or cells based on the analytic blast wave solutions \citep[e.g.,][]{stinson06a, dubois08a, shetty08a, ostriker11a, dobbs11a, kim11a, shetty12a}, modifying the equation of state based on a subgrid models for unresolved hot gas \citep[e.g.,][]{springel03a, scannapieco06a}, disabling cooling for some period of time after supernovae occur \citep[e.g.,][]{thacker00a, stinson06a, governato07a}, and adding energy stochastically so as to ensure that, when it is added, the temperature is high enough to prevent rapid cooling \citep[e.g.,][]{dalla-vecchia12a}. A few of the highest resolution simulations can avoid the need for such tricks \citep[e.g.][]{hopkins12b}, but only if they include other forms of feedback that \textit{are} handled by subgrid models; these serve to reduce the density around young stars before supernovae go off, making the blast wave easier to resolve. \red{At high resolution and in galaxies with low star formation rates, it is also necessary to properly account for the stochastic nature of supernova feedback and its dependence on sampling from the IMF, a topic that has received scant attention thus far.}

As I discuss in more detail below, depending on how it is implemented, supernova feedback may or may not be sufficient to regulate the rate of star formation in galaxies and set $\epsilon_{\rm ff}$ on galactic scales to small values in good agreement with observation. However, this result depends on the choice of subgrid model. Moreover, even if supernovae do set $\epsilon_{\rm ff}$ to small values on the scales of entire molecular clouds, they cannot regulate $\epsilon_{\rm ff}$ on smaller scales, simply because of the $\approx 4$ Myr delay between the onset of star formation and the first supernovae. If one hypothesizes that supernovae can be effective only in clouds for which the crossing time is no more than half the lifetime of the most massive stars, then one concludes that supernovae matter only for clouds that satisfy the inequality $\Sigma_2 \lesssim 230 M_5^{1/3}$, where $\Sigma_2$ is the surface density measured in units of 100 $\msun$ pc$^{-2}$, and $M_5$ is the mass in units of $10^5$ $\msun$ \citep{fall10a}. Giant molecular clouds (marginally) satisfy this condition, but smaller star-forming structures, which also have measured values of $\epsilon_{\rm ff}\approx 0.01$, do not.

\paragraph{Radiation Pressure}
A significant fraction of the energy and momentum budget of a zero-age stellar population is emitted in far- and extreme-ultraviolet photons (FUV and EUV, $\approx 8-13.6$ and $>13.6$ eV, respectively), both of which have very large cross sections with dust grains, and the latter of which have even larger cross sections with neutral hydrogen atoms. The EUV photons carry $\approx 1/3$ of the momentum, and will be absorbed almost entirely within the ISM even for the most dust-poor galaxies. The FUV photons carry much of the rest, and a significant fraction of these will be absorbed within a galaxy of even moderate metallicity and column density. The radiation field produced by a zero-age stellar population carries a momentum per unit time per unit stellar mass $\dot{V}_L = 24$ km s$^{-1}$ Myr$^{-1}$ \citep{murray10b, krumholz14a}, and a population of stars forming at a steady rate produces a steady momentum flux per unit star formation rate $V_L = 190$ km s$^{-1}$ \citep{dekel13a}. These numbers immediately suggest that radiation pressure may be important on the scales of individual molecular clouds, which have escape speeds of $\sim 10$ km s$^{-1}$ or less. Analytic models of radiation pressure feedback indeed suggest that it is likely to be able to eject significant amounts of gas from regions of the most intense star cluster formation \citep{krumholz09d, fall10a, murray10a}, and observations appear to support this view (\citep{lopez11a}, though see \citep{pellegrini11a} for a contrasting view). However, the momentum budget of the radiation field is not large enough provide a general explanation for low values of $\epsilon_{\rm ff}$, unless more momentum can be extracted from the radiation than the above estimates suggest.

Such enhanced extraction is in principle possible, because each photon emitted by a star can be scattered, or absorbed and then re-emitted, more than once. Since the photon deposits momentum each time is it scattered or absorbed, the actual momentum delivered can be much greater than that initially input by the stars\footnote{This statement should be understood in a scalar sense. Of course the total vector momentum is conserved no matter how many times the radiation is absorbed or scattered, and in fact the total vector momentum emitted by an isotropically-radiating star is zero. However, the radial component of the momentum of the stellar radiation field is non-zero, and the radial momentum imparted to the gas can be even larger.}. 
\red{To understand the physics behind this trapping, it is helpful to examine two limiting cases. Consider a stellar source of luminosity $L$ surrounded by gas. The first case is one where every stellar photon is absorbed once, but then is scattered or re-emitted isotropically and escapes. In this case, the initial absorption deposits momentum in the radially-outward direction, while the isotropic re-emission leads to no change in the gas momentum on average. The net effect is that a stellar source of luminosity $L$ imparts momentum to the gas at a rate $L/c$.}

\red{Now consider the opposite limit, where the mean free path for the re-emitted photons is very short. In this case every photon is absorbed and re-emitted many times. One might think that this would make the radiation force small, because the radiation field would be nearly isotropic, but upon reflection one can see that this cannot be the case. If the radiation field were truly isotropic, there would be no net flux of energy. However, in equilibrium there must be a non-zero radially-outward flux $F=L/4\pi r^2$, where $r$ is the distance from the stars, in order to carry away the energy injected by the stars. Thus the photon distribution cannot be exactly isotropic, because energy balance requires that there be slightly more photons traveling radially-outward than radially-inward. This in turn means that, while every absorbing particle is hit by many photons from all sides, there are slightly more outward-moving than inward-moving photons, and this imbalance gives rise to a force per unit mass that is proportional to the net radiation flux: $f = \kappa F/c = \kappa L/4\pi r^2 c$, where $\kappa$ is the opacity per unit mass. If this force is applied to a sphere of matter of density $\rho$ and radius $R$ centered on the emitting stars, then the total rate of momentum deposition is $\kappa \rho R L/c$. We can identify the quantity $\kappa \rho R$ as simply the center-to-edge optical depth of the sphere of gas, and thus we have a momentum deposition rate $\tau L/c$. If the absorbing cloud is very optically thick, this greatly exceeds the momentum $L/c$ carried by the emitted stellar radiation field.
}

\citet{krumholz09d} parameterize \red{the behavior between these two limits} in terms of a ``trapping factor" $f_{\rm trap}$ which measures the amount of momentum deposited in the gas by the radiation field divided by the amount emitted by the stars\footnote{\red{One might justifiably ask, based on the above example, what the difference is between $f_{\rm trap}$ and the optical depth $\tau$. For a frequency-independent opacity they are in fact identical, but for a frequency-dependent opacity the optical depth cannot be defined in a frequency-independent manner, and it becomes unclear what value of $\tau$ to use. The trapping factor avoids this ambiguity.}}. Thus $f_{\rm trap}=1$ corresponds to each photon being absorbed and depositing momentum only once, while $f_{\rm trap} \gg 1$ corresponds to many absorptions per photon. The theoretical maximum value of $f_{\rm trap}$ (called the ``photon-tiring limit" in the stellar astrophysics community) is limited only by the amount of available energy, and is of order $c/v$, where $v$ is the characteristic speed of the outflowing gas \citep{dekel13a}. Very few simulations of star formation feedback in galaxies include radiative transfer in any form, let alone the ability to simulate trapping and re-emission of the radiation by dust, and so the amount of radiative trapping to include in a given simulation is left essentially as a free parameter, with authors making assumptions that range from assuming no trapping beyond the initial absorption, i.e.~$f_{\rm trap}=1$ \citep[e.g.][]{wise12a, kim13a, kim13b, ceverino13a}, to strong trapping, $f_{\rm trap} \sim 10-100$ \citep{ostriker11a, hopkins11a, hopkins12a, genel12a, aumer13a}, to a myriad of values and interpolation functions in between \citep{agertz13a, stinson13a, renaud13a, bournaud14a}. Much the same is true of analytic models, with some authors assuming strong trapping \citep{thompson05a, murray10a}, and others weak trapping \citep{krumholz09d, krumholz10b, fall10a}. The only true radiation-hydrodynamic simulations reported thus far in the literature suggest that the realistic answer is toward the low end of this range \citep{krumholz12c, krumholz13a}, due to radiation Rayleigh-Taylor instability \citep{jacquet11a, jiang13a}, which limits the ability of trapped radiation to transfer momentum to the gas. As with supernovae, the results for regulation of the star formation rate depend critically on the adopted subgrid model, as I discuss in detail below.

\paragraph{Cosmic Rays}
The fast shocks produced by stellar winds and supernovae can accelerate both electrons and ions to relativistic velocities, producing a population of non-thermal particles known as cosmic rays (CRs; see \citet{zweibel13a} for a recent review). Evidence for the relationship between star formation and CR acceleration comes in part from the strong correlation between galaxies' far infrared luminosity and their non-thermal radio emission both locally \citep[and references therein]{condon92a} and at high redshift \citep{mao11a}. The fraction of supernova kinetic energy that goes into CRs is significantly uncertain, but indirect observational constraints suggest that it is $\sim 10\%$, partitioned roughly $10:1$ between protons and electrons \citep{lacki10a}. Though they are relativistic, cosmic rays do not free-stream through the ISM. Instead, their propagation speed is restricted by scattering off Alfv\'{e}n waves \citep{wentzel74a, cesarsky80a}, and they also lose energy due to inverse Compton scattering off the ISRF and the cosmic microwave background, synchrotron emission as they spiral around magnetic field lines, and inelastic scattering off atoms in the the ISM through bremsstrahlung, photoionization, and pion production. Quantitative estimates of the rates of both diffusion and energy loss are significantly uncertain \citep[e.g.,][]{thompson06a, lacki10a}.

Observations show that, at the mid-plane of the Milky Way, CR pressure is comparable to magnetic pressure, and both make a non-negligible contribution to the overall pressure budget \citep{boulares90a, beck05a}. However, the spatial distribution of the CRs is not well known; measurements of the synchrotron emissivity suggest that the CR scale height is of order 1 kpc in the Milky Way and similar galaxies, but it may be as small as $\sim 100$ pc in starburst galaxies \citep[and references therein]{lacki10a}. If the CRs are distributed homogeneously over these scales, they may be important for driving galactic winds, but they will not be able to regulate star formation on the scale even of individual molecular clouds, since they will not provide a significant pressure gradient on the relevant size scale. On the other hand, if local star formation and subsequent CR production can drive a significant non-uniformity in the CR population on scales significantly smaller than the CR scale height, then CR pressure may be able to regulate star formation, a hypothesis first advanced by \citet{socrates08a}.

There are relatively few simulations of star formation in galaxies that include CR feedback. Those that do generally treat them as an additional fluid described by a relativistic equation of state, and with additional terms describing gain in energy due to injection by star formation, and loss in energy due to some or all of the mechanisms mentioned above. They also include a prescription for cosmic ray transport, either approximated as pure diffusion or as a slightly more sophisticated streaming model with similar effects \citep{jubelgas08a, wadepuhl11a, uhlig12a, salem13a, booth13a}. The details of the microphysical treatment vary from one implementation to another, but all the three-dimensional simulations performed thus far use relatively simple prescriptions that do not include the energy-dependence of the gain, loss, and diffusion terms, nor do they include more realistic treatments of scattering off MHD waves, features that are included in one-dimensional calculations \citep[e.g.,][]{dorfi12a}. They also do not include the dependence of the cosmic ray diffusion coefficient on the angle relative to the large-scale magnetic field (see \citet{zweibel13a} for a recent discussion). Because of these simplifications, three-dimensional simulations must choose a number of quite uncertain parameters, and the results depend on these choices. For some plausible choices CRs strongly regulate the SFR, while for others they have relatively little effect on the SFR, though they may still drive galactic winds.

\paragraph{Photoionization and Far Ultraviolet Heating}
A final form of feedback that might be important to the star formation rate is heating of the ISM by FUV and EUV photons. Since EUV photons are absorbed by neutral hydrogen atoms, they have very short mean free-paths through the neutral ISM, and thus they tend to produce localized regions of ionized gas with characteristic temperatures of $\approx 10^4$ K. FUV photons have longer mean-free paths: Solar metallicity galaxies typically have FUV attenuations of $\sim 1$ mag at total gas surface densities of 10 $\msun$ pc$^{-2}$ \citep{boissier07a}, corresponding to about half the photons being absorbed or scattered over a length scale comparable to the galactic scale height. When these photons are absorbed by dust grains, they can launch photoelectrons that heat the gas (see Section \ref{ssec:gmcthermo}).

EUV photons have long been invoked as important regulators of star formation rates and efficiencies on the size scales of individual molecular clouds \citep[and references therein]{dobbs14a, krumholz14a}, and have been proposed to set limits on $\epsilon_{\rm ff}$ in them \citep[e.g.,][]{goldbaum11a, vazquez-semadeni10a, vazquez-semadeni11a, dale12a, dale13a, zamora-aviles12a, zamora-aviles13a}. Direct observations show that the pressure of photoionized gas generally dominates H~\textsc{ii} regions \citep{lopez11a, lopez13a}. However, on galactic scales their effect on the SFR appears to be relatively modest. Several simulations of such galaxies have included photoionization heating, either via a Str\"{o}mgren volume approach in which the gas around young stars is simply set to a temperature of $10^4$ K out to a radius large enough for ionizations and recombinations to balance \citep[e.g.,][]{hopkins11a, renaud13a}, or via explicit solution of the equation of radiative transfer \citep[e.g.,][]{wise11a, kim13a, kim13b}. They all find that the effects are subdominant compared to other feedback mechanisms, mainly because ionized gas at $10^4$ K has a sound speed of $10$ km s$^{-1}$, which is not much larger than the escape speed from the largest molecular clouds \citep{dale12a, dale13a}, which, for a realistic cloud mass spectrum, contain most of the molecular mass in a galaxy \citep{dobbs14a}.

FUV heating is more complex, as it does not affect the molecular gas itself at all. As discussed in Section \ref{ssec:gmcchemistry}, the ratio of FUV radiation flux to density in the cold phase of the atomic ISM is roughly constant, and a further corollary of this result is that the pressure in the cold phase is roughly proportional to the FUV radiation flux. In two-phase equilibrium, the total pressure is as well. Thus FUV radiation production from stars can pressurize the ISM, and this may inhibit production of star-forming clouds. \citet{ostriker10a} present an analytic model in which this process regulates the star formation rate at galactic scales: the ISM is partitioned between a diffuse phase in hydrostatic equilibrium and a set of gravitationally-bound, star-forming clouds, and the balance between these phases is set by the mean pressure in the diffuse phase, which in turn is set by the rate of star formation in the gravitationally-bound phase. This forms a feedback loop in which the star formation rate self-regulates to produce a preferred partition between diffuse and bound phases. The model does a particularly good job of reproducing the radial variation of star formation rates within nearby galaxies.

There have been relatively few numerical models of the effects of FUV heating. \citet{tasker11a} finds in her simulations that the inclusion of FUV heating reduces star formation rates only modestly compared to simulations with only supernova feedback, and that the overall star formation rate is still too high compared to observations. However, these simulations adopt a FUV heating rate that is independent of the local star formation rate, so there is no true feedback. \citet{kim11a} conduct simulations of an isolated portion of a low-surface density region of a galactic disk where the FUV heating is linked to the local star formation rate. They find the FUV radiation is effective at regulating the star formation rate, but only if they also adopt a small value of $\epsilon_{\rm ff}$ on small scales, and use a subgrid model for supernova momentum feedback. Based on this result, they propose an analytic model that combines the effects of FUV and supernova feedback, following \citet{ostriker10a} for the former and \citet{kim11a} for the latter.

\paragraph{Feedback-Regulated Models: General Results}

\begin{figure}
\centerline{\includegraphics[width=3.0in]{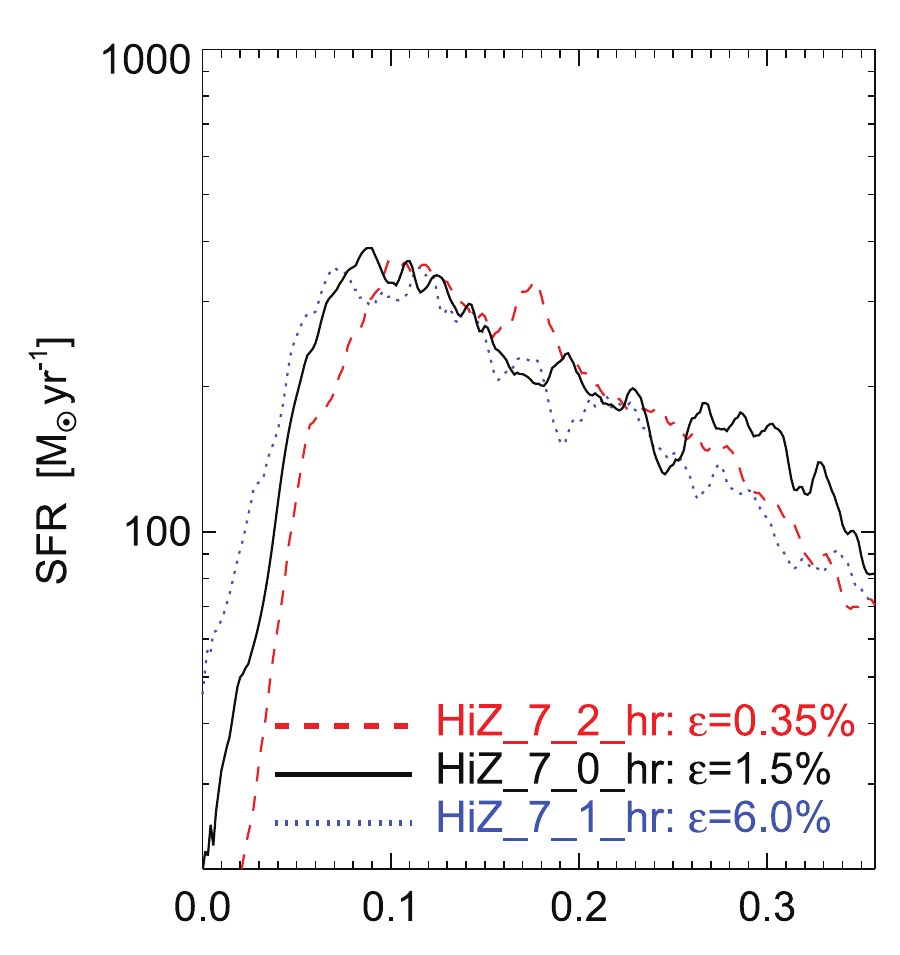}}
\centerline{\includegraphics[width=3.0in]{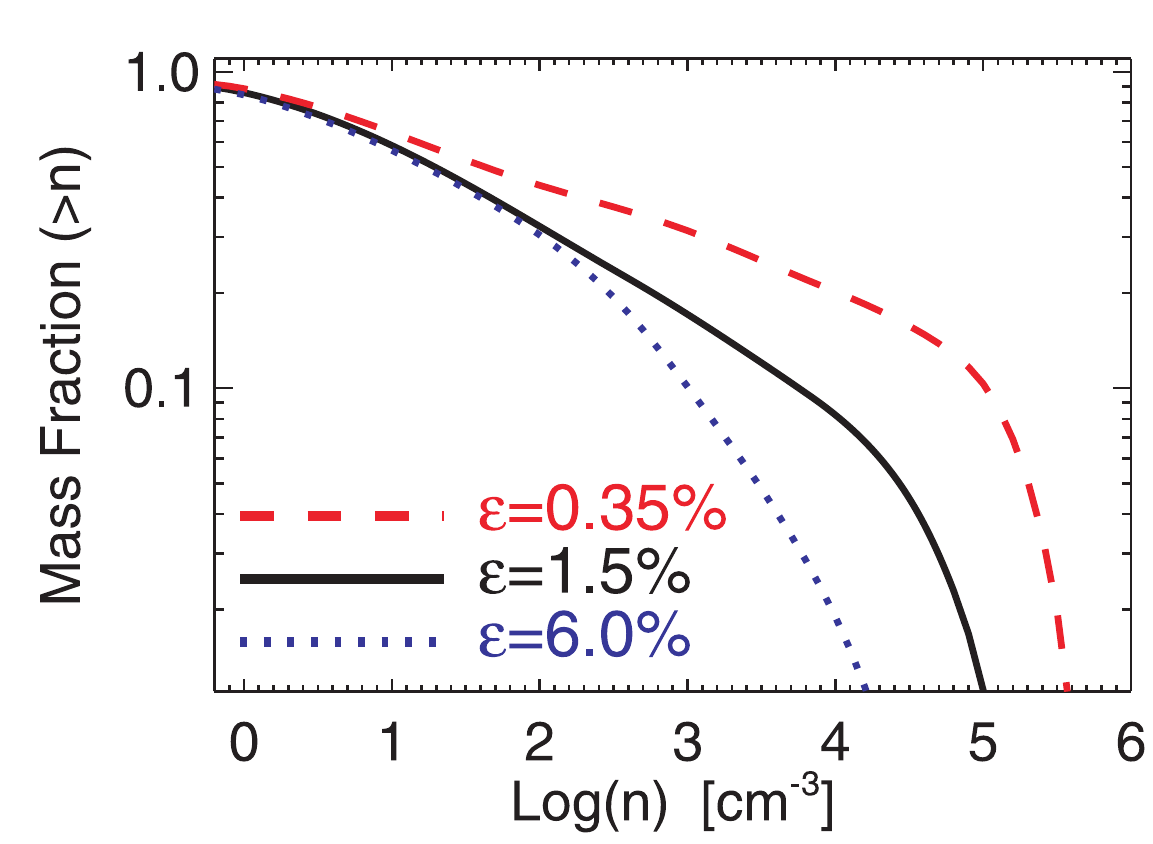}}
\caption{
\label{fig:hopkins11}
Results from three simulations of isolated disk galaxies by \citet{hopkins11a}. In these simulations, the subgrid feedback model was left fixed, but $\epsilon_{\rm ff}$ at the smallest scales resolved in the simulation was varied from $0.0035$ to $0.06$, as indicated. Top panel: star formation rate per unit time integrated over the galaxy versus simulation time (in Gyr). Bottom panel: fraction of the total gas mass $M(>n)$ at densities $n$ or more, measured from a snapshot of the simulations.
}
\end{figure}

\red{Now that we have reviewed potential feedback mechanisms, it is helpful to stand back and ask about the common features of models in which one or more of these mechanisms are the primary regulators of star formation on galactic scales. The central conceptual idea of these models is that the SFR is ultimately set by the need for the energy or momentum injected by some form of feedback to balance the weight of the ISM, the dissipation of turbulence within it, or something similar. An important corollary of this picture is that, at least when averaged over galactic scales, the SFR is insensitive to whatever rules govern the formation of stars within individual molecular clouds. Only the balance between momentum / energy injection and gravity / energy dissipation matters, and by some mechanism the SFR will self-adjust to maintain this balance.}

Figure \ref{fig:hopkins11} shows an example of this behavior. The figure shows the results of three simulations by \citet{hopkins11a}, all computed using the same subgrid model of strong radiation pressure feedback, with $f_{\rm trap}\gg 1$. In these simulations, gas that exceeds a specified density threshold (typically $n_{\rm H,2}=1$) is converted to stars at a rate $\dot{M}_* = \epsilon_{\rm ff} M_{\rm gas} / t_{\rm ff}$, where $\tff$ is the \red{free-fall time computed from the local} gas density. The three lines in the figure correspond to the results of simulations where $\epsilon_{\rm ff}$ is varied over a factor of 16 from $0.0035$ to $0.015$ to $0.06$, but all else is held fixed. The result is that the star formation rate in the simulations is essentially unchanged. Instead, what changes is the gas density distribution: if $\epsilon_{\rm ff}$ is decreased, then more gas builds up at higher density so that $M/\tff$ goes down and the star formation rate stays the same. The reverse happens if $\epsilon_{\rm ff}$ is increased. \red{This change in the density distribution is the mechanism by which the galaxy self-adjusts to achieve the SFR that produces the rate of momentum injection required to keep the ISM from collapsing.}

The simulations shown in Figure \ref{fig:hopkins11} are an example, but this behavior is generic to feedback-regulated models, and numerous other simulations find the same result if the feedback is sufficiently strong \citep[e.g.][]{ostriker11a, shetty12a, agertz13a}. The same is true in the analytic models: in the FUV-regulated star formation model of \citet{ostriker10a}, the depletion time within the star-forming gravitationally bound clouds explicitly drops out of the model prediction of the star formation rate, and instead enters only in the model prediction of the partition of the ISM between the diffuse and star-forming phases. Similarly, in the analytic models of \citet{ostriker11a} and \citet{faucher-giguere13a}, one of the free parameters is the value of $\epsilon_{\rm ff}$ within molecular clouds, but the predicted rate of star formation at the galactic scale is nearly independent of this choice.

\begin{figure}
\centerline{\includegraphics[width=2.8in]{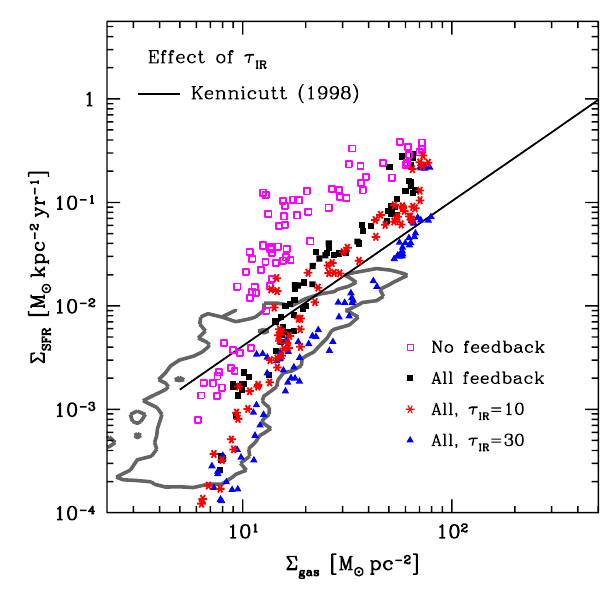}}
\caption{
\label{fig:agertz13}
Results from simulations of isolated disk galaxies by \citet{agertz13a}, reprinted with permission. The plot shows the star formation rate per unit area versus gas surface density per unit area, exactly as shown in Figure \ref{fig:sflawtot}. The black line is whole-galaxy relation of \citet{kennicutt98a}, while the gray contour shows the same \citet{bigiel08a} as the blue pixels in Figure \ref{fig:sflawtot}. The magenta, black, red and blue points show averages over $720$ pc bins measured from the simulation. The different colors correspond to different treatments of feedback in the simulations; magenta are no feedback, black is a fiducial model including supernovae and radiation pressure, where the trapping factor $f_{\rm trap}$ (called $\tau_{\rm IR}$ in the simulation legend) is estimated based on local dust temperatures and surface densities. The red and blue points show simulations with alternate, fixed values of $\tau_{\rm IR}$.
}
\end{figure}

\begin{figure}
\centerline{\includegraphics[width=2.8in]{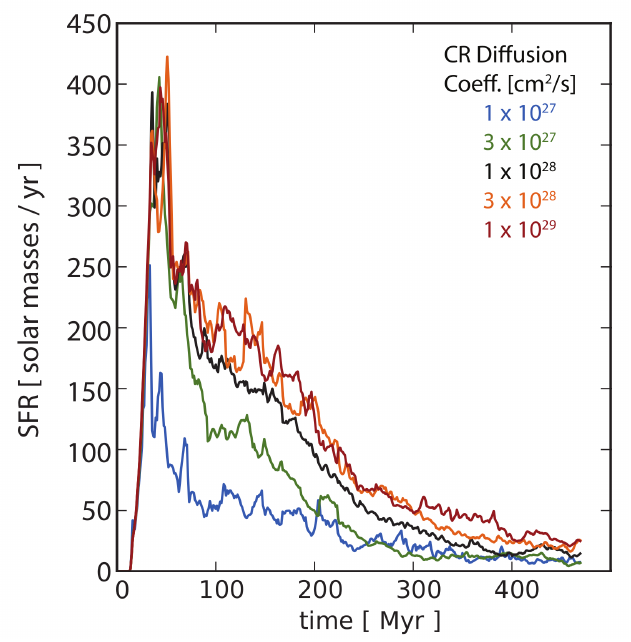}}
\caption{
\label{fig:salem13}
Results from simulations of isolated disk galaxies including cosmic ray feedback by \citet{salem13a}, reprinted with permission. The plots show star formation rate integrated over the whole galaxy versus time, and different colors correspond to simulations using different values for the cosmic ray diffusion coefficient.
}
\end{figure}

With appropriately chosen parameters for the strength of feedback, these models are capable of reproducing the observed star formation rates of galaxies, at least averaged over large scales. Figure \ref{fig:agertz13} shows an example: a simulation without feedback (magenta points in the figure) dramatically over-predicts the star formation rate compared to observations, while models with feedback (red, black and blue points) provide a much better match to the observations. However, it is important to understand that this is not a prediction that is free of subgrid model parameters. The parameters describing star formation on small scales do not affect the galaxy scale star formation rate, but those describing the feedback do.

For the analytic models this dependence is explicit: in \citet{ostriker10a}'s model, the star formation rate is close to inversely proportional to the amount of FUV heating delivered per unit star formation, while for \citet{ostriker11a}'s and \citet{faucher-giguere13a}'s model, the star formation rate is instead inversely proportional to the momentum injected per unit stellar mass formed. In \citet{kim11a}'s model the star formation rate varies inversely with a combination of those two parameters.

The dependence is less explicit in the numerical models, but is clear nonetheless, as illustrated by the examples in Figures \ref{fig:agertz13} and \ref{fig:salem13}. In Figure \ref{fig:agertz13}, the blue, red, and black points show results using three different sets of parameters in the subgrid radiation pressure feedback model. Arguments can be made for all three choices (though the radiation-hydrodynamic simulations of \citet{krumholz12c, krumholz13a} would argue for something closest to the black points), but depending on which choice is made, the star formation rate at fixed gas surface density can change by an order of magnitude. Similarly, Figure \ref{fig:salem13} shows the results of varying the assumed CR diffusion coefficient in otherwise-identical simulations of CR feedback. The black line shows the fiducial value adopted in this simulation, but the true diffusion coefficient is uncertain by at least an order of magnitude, and not all authors agree on a fiducial choice; for example, \citet{salem13a} favor the value corresponding to the black line, while \citet{booth13a} favor the value corresponding to the green line. The range of uncertainty in the star formation rate corresponding to this uncertainty in the diffusion coefficient is at minimum a factor of several, and plausibly an order of magnitude.

Thus the state of feedback-regulated models for the star formation rate can be summarized by saying that, for plausible parameterizations of various feedback processes, one can reproduce at least roughly the observed value of $\epsilon_{\rm ff}$ on kpc or larger scales, and that this value is independent of \red{the value of $\epsilon_{\rm ff}$ that one adopts on $\sim 10$ pc or smaller scales}. However, there are other, equally plausible parameterizations of feedback where neither of these statements hold: either the predicted rate of star formation is much too high or too low compared to what we observe, and/or the results are no longer independent of the value of $\epsilon_{\rm ff}$ that one adopts on small scales. Authors of feedback-regulated models generally favor feedback parameter choices where $\epsilon_{\rm ff}$ on small scales does not matter, but pending actual first-principles determinations of these parameters, these models should be treated in the same spirit as other semi-analytic models: plausibly capable of reproducing observations if one tunes the parameters appropriately, but not true first-principles predictions.

\subsubsection{Potential Problems with the Top-Down Approach}
\label{sssec:topdownproblems}

Top-down models either with or without feedback, despite their successes, also face two major objections. One is that they have trouble accounting for the observed dependence of star formation rates on the phase of the ISM. As discussed in Section \ref{sssec:sfrobsgalactic}, star formation correlates extremely well with the molecular phase of the interstellar medium, and much more poorly with either the total gas content or the atomic gas. However, if the dominant physics responsible for controlling star formation is simply the self-gravity of the gas, spiral arm shocks, stellar feedback, and similar processes operating at the galactic scale, why should the chemical phase matter? And yet observations clearly show that it does: recall Figures \ref{fig:leroy08} -- \ref{fig:sflawh2}.

One can attempt to circumvent this objection by proposing that both H$_2$ and star formation are the results of large-scale processes that drive gas to high density, so that they are correlated but causally-unrelated \citep{hartmann01a, heitsch08b, dobbs09a, bonnell13a}. However, this strategy does not seem to account for observations of low-metallicity galaxies. Recall from Section \ref{sssec:sfrobsgalactic}, and in particular Figures \ref{fig:sflawtot} and \ref{fig:sflawh2}, that changes in the metal content of a galaxy at fixed gas surface density affect both the H$_2$ content and the star formation rate, in such a way that the star formation rate per unit H$_2$ mass remains unchanged. The Small Magellanic Cloud, the outskirts of Lyman break galaxies, and damped Lyman $\alpha$ absorbers are all systematically displaced from other, higher-metallicity galaxies in the diagram of $\Sigma_{\rm SFR}$ versus $\Sigma$. While a top-down model coupled to an assumptions that H$_2$ simply forms wherever stars do could conceivably account for a constant SFR per unit H$_2$, it is hard to see how it could also account for the dependence on metallicity. If gravitational instability or spiral shocks drive or cloud collisions drive star formation, why should the metallicity of the ISM matter?

For the most part top-down models have not attempted to address this question, or have not succeeded at reproducing the observations. For example, \citet{hopkins12a} find that the star formation rate they obtain in their simulations is insensitive both to whether they include chemistry and to whether they include cooling below $10^4$ K, and this is the only part of their simulations where metallicity enters. They therefore seem to predict that metallicity has no effect on star formation, which is clearly inconsistent with the SMC, LBG, and DLA data.  \citet{bolatto11a} suggest that it might be possible to modify the \citet{ostriker10a} model to match the SMC observations by positing that the rate of FUV heating per unit star formation scales inversely with gas metallicity, but this appears to require rather unphysical assumptions about photon escape from the vicinities of molecular clouds \citep{krumholz13c}. At present there are no other proposals in the literature for how to explain the observations of the SMC or LBG outskirts in the context of any of the other feedback-regulated models.

A second issue with top-down models, though it is more than an omission than a true problem, is that they have little to say about the rate of star formation on \blue{the scales of individual molecular clouds, or even, for some models, on any scale smaller than a galaxy as a whole}. Appropriately-tuned feedback-regulated models are able to get galaxy-scale star formation rates correct regardless of the value they adopt for $\epsilon_{\rm ff}$ on the smallest scales they consider, but conversely this means that they have little to say about why $\epsilon_{\rm ff}\approx 0.01$ on scales smaller than an entire galaxy. This problem manifests in an inability to explain certain observations without adopting additional assumptions. For example, \blue{while \citet{ostriker10a} make predictions for star the depletion time including all phases of the ISM that are independent of their small-scale assumptions, they are forced to} take their value of the depletion time in molecular clouds directly from observations, and appeal to bottom-up models (see below) to explain this value. Similarly, \citet{hopkins13c} show that their simulations including radiation pressure feedback are able to reproduce the observed ratio of HCN($1\rightarrow 0$) to CO($1 \rightarrow 0$) emission in spiral galaxies, which is a proxy for the density-dependence of $\epsilon_{\rm ff}$, but only if they input the observationally-motivated value $\epsilon_{\rm ff} \approx 0.01$ at the smallest scales they can resolve. Thus diagnostics like the HCN($1\rightarrow 0$) to CO($1\rightarrow 0$) ratio and the depletion time in individual molecular clouds can be explained in the context of top-down models only by invoking additional mechanisms that set $\epsilon_{\rm ff}$ at small scales, independent of those responsible for setting it on larger scales.

The issue also arises if one considers nearby molecular clouds. Recall that \citet{evans09a} find $\epsilon_{\rm ff} \approx 0.03-0.06$ in five Solar neighborhood clouds, none of which contain any stars massive enough to produce supernovae, cosmic rays, or significant winds. The same is true for all of the Milky Way clouds shown in both panels of Figure \ref{fig:eff}, which show $\epsilon_{\rm ff} \approx 0.01$. If feedback from massive stars is the primary regulator of star formation why do these clouds not have much larger values of $\epsilon_{\rm ff}$? Conversely, given that something is responsible for setting $\epsilon_{\rm ff}\approx 0.01$ in these local clouds, why not invoke that process to regulate star formation elsewhere in the galaxy, rather than invoking massive star feedback?

\subsection{Star Formation Rates: the Bottom-Up Approach}
\label{ssec:bottomup}

The bottom-up approach is motivated in part by the need to address the two problems identified above with the top-down approach: its inability to reproduce the observed phase- and metallicity-dependence of star formation, and its silence on the question of why $\epsilon_{\rm ff}$ should be small on all scales, regardless of whether there are massive stars present or not. Both of these issues are much easier to address in a model in which one assumes that regulation of star formation is a local process rather than a galaxy-scale one. Of course this approach is not without its own problems, as I discuss below, but it provides a very useful complement to the top-down models.

\subsubsection{Which Gas is Star-Forming?}
\label{sssec:whichsf}

The first step in a bottom-up model is to ask what conditions the ISM must satisfy in order to be star-forming in the first place. As illustrated in Figures \ref{fig:sflawtot} and \ref{fig:sflawh2}, the ISM shows a $\sim 3$ order of magnitude range of depletion times even excluding starburst galaxies, with the most weakly star-forming neutral gas having depletion times an order of magnitude longer than the Hubble time. In contrast, molecular gas never seems to have a depletion time much longer than a few Gyr. What accounts for this range of behaviors?

\paragraph{\blue{Applications of the H$_2$-SFR Correlation}}
Several authors have simply taken the observed correlation between H$_2$ and star formation as a starting point for models, and have therefore modeled star formation by first determining where H$_2$ will form. The first models of this type were implemented as subgrid recipes for star formation in simulations of either isolated galaxies or cosmological galaxy formation. Since these simulations generally do not resolve the small scales of molecular clouds, or resolve them only marginally, the transition from H~\textsc{i} to H$_2$ must itself be treated with a subgrid model; this is necessary because the H$_2$ formation rate per unit volume scales as the square of density (equation \ref{eq:h2formrate}), so clumping on unresolved scales can greatly alter the total formation rate.

One method of handling this problem is to pre-compute a grid of models from a photochemistry code like \texttt{cloudy} \citep{ferland98a} and implement them as lookup tables in a simulation \citep{robertson08b}. Another is to directly solve the formation and dissociation equations, but with an increased formation rate to model unresolved clumping \citep{pelupessy06a, gnedin09a, christensen12a}. A third approach is to use an analytic model to estimate the equilibrium H$_2$ abundance \citep{kuhlen12a, jaacks13a}. The latter two methods appear to give very similar results when averaged over $\sim 100$ pc scales, except at metallicities below $\sim 1\%$ of Solar, where the low rate of H$_2$ formation renders equilibrium a poor assumption \citep{krumholz11a}. Regardless of the method used to estimate the H$_2$ fraction in the ISM, a simulation code using an H$_2$-based star formation recipe can then set the star formation rate based on the H$_2$ content. This guarantees correct reproduction of the observed relationship between star formation and ISM phase, though of course it does not address the physical question of why the two are correlated.

One can also take a similar approach analytically. The simple slab calculation shown in Section \ref{ssec:gmcchemistry} can be made more sophisticated in a variety of ways, but the generic result that there is a transition between H~\textsc{i} and H$_2$ at a characteristic column density of $\approx 10$ $M_\odot$ pc$^{-2}$ at Solar metallicity survives regardless of the method used, with a strong dependence on metallicity. \citet{krumholz08c, krumholz09a} and \citet{mckee10a} develop analytic models that predict the H$_2$ fraction of spherical clouds in terms of their mean surface densities and metallicities, and find that the transition column density scales slightly sublinearly with metallicity. \citet{krumholz13c} describes an extension of this model to H~\textsc{i}-dominated outer galaxies, where the assumption of two-phase equilibrium for the atomic ISM may not hold. Once a formula for the H$_2$ fraction is in hand, one can then formulate an analytic model for the relationship between star formation and total gas content simply by assuming that stars form only in H$_2$, and assuming or calculating (see Section \ref{sssec:effsmall}) a depletion time in the H$_2$.

\paragraph{\blue{Physical Basis of the H$_2$-SFR Correlation}}
Both the analytic and numerical approaches we have just reviewed assume that star formation follows H$_2$ (admittedly an assumption that seems very well-justified by observations), but they do not address the question of why. As discussed in Section \ref{sssec:topdownproblems}, one way of coming at this question is to posit that the relationship is simply caused by gas converting into H$_2$ wherever it gets dense enough to form stars, but this approach appears to founder when confronted with the data summarized in Figures \ref{fig:sflawtot} and \ref{fig:sflawh2} showing that changes in metallicity lead to changes in both star formation rate and H$_2$ mass at fixed gas surface density, but \textit{not} to changes in the star formation rate per unit H$_2$ mass. Thus the real challenge is to explain simultaneously why star formation correlates with H$_2$, and why metallicity affects both in the same way.

One proposal is that the onset of star formation, and the development of high H$_2$ fractions, are both caused by the development of a cold phase of the ISM \citep{elmegreen94c, schaye04a}. In these models, there is a threshold column density at which a cold atomic phase appears. This destabilizes the disk, effectively reducing the Toomre $Q$, and causes star formation. The threshold column density for formation of H$_2$ (see Section \ref{ssec:gmcchemistry}) is roughly the same at Solar metallicity, and so star formation and H$_2$ go together. However, the predicted scaling of star formation and H$_2$ fraction with metallicity in this model is that the threshold at which both H$_2$ and star formation turn on should vary as roughly $Z^{-1/2}$, which seems to be too shallow compared to what is observed \citep{erkal12a}.

Another possible explanation, first posited based on analytic models by \citet{krumholz11b} and independently discovered numerically by \citet{glover12a}, comes from examining the thermodynamics of the gas and the relationship between dust shielding, H$_2$ content, and photoelectric heating. As discussed in Section \ref{ssec:gmcthermo}, there are two main heating mechanisms in the dense ISM: photoelectric and cosmic ray heating. FUV radiation is exponentially attenuated by dust absorption, and at the relatively high visual extinctions seen in Galactic star-forming regions, is probably-subdominant compared to cosmic rays, which are attenuated by the gas either not at all or much more weakly. Under this condition, the equilibrium temperature is $\approx 10$ K. However, now consider fixing the gas column density but changing the metallicity, and thus the amount of dust and the total extinction. For sufficiently low dust content, photoelectric heating will become dominant, and the equilibrium temperature will be much larger, raising the Jeans mass (equation \ref{eq:jeansmass}). If the gas temperature rises from 10 K to 100 K, this will produce a factor of $30$ increase in the mass that can be supported against collapse by thermal pressure, and this increase in thermal support could plausibly suppress star formation.

\begin{figure}[ht!]
\centerline{\includegraphics[width=2.8in]{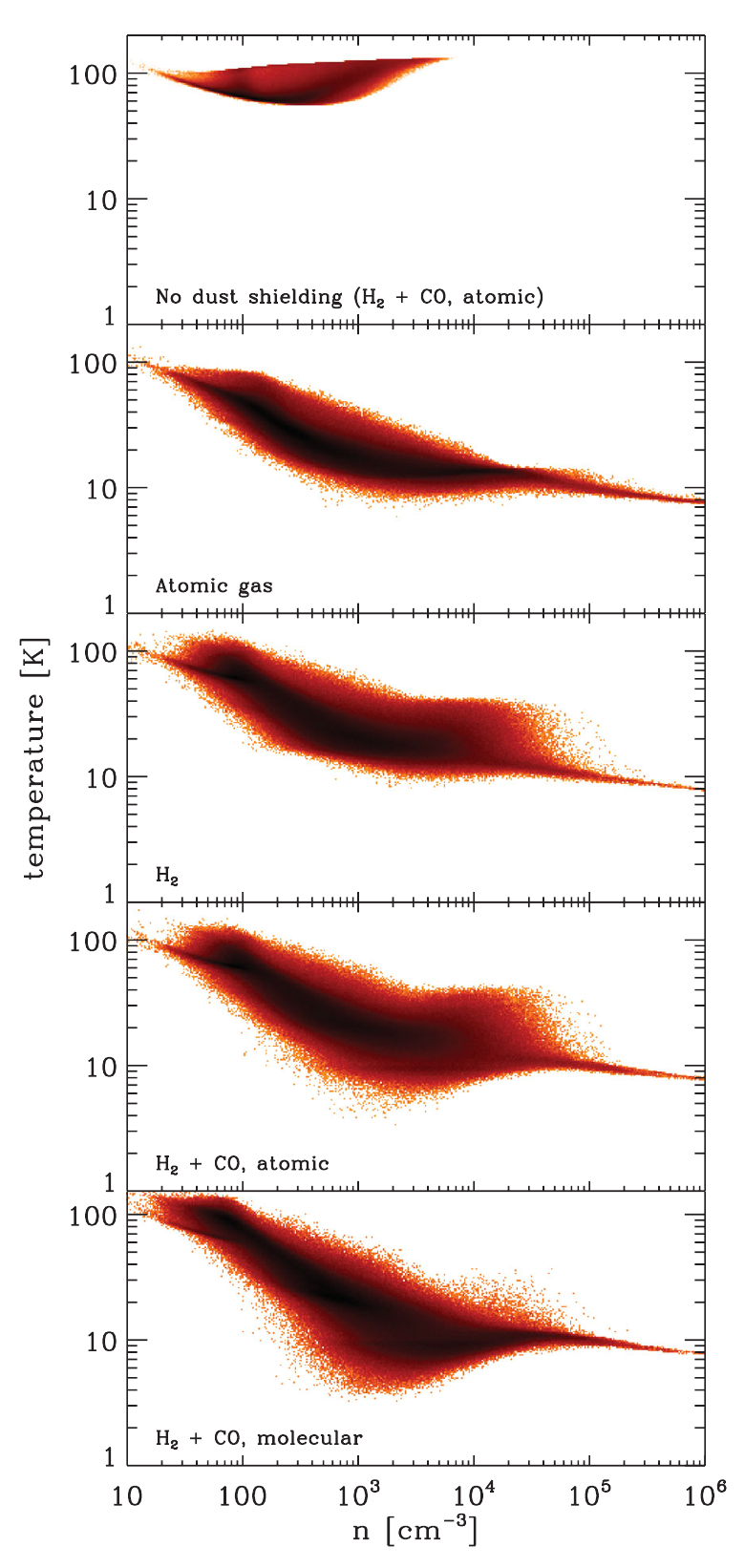}}
\caption{
\label{fig:glover12}
Density and temperature distributions from four simulations of turbulent clouds, using different treatments of thermodynamics and chemistry. \blue{The clouds have an initial column density of $0.018$ g cm$^{-3}$ ($85$ $M_\odot$ pc$^{-2}$).} In all panels, colors indicate the distribution of mass in the density-temperature plane at the same evolutionary time in each simulation. The bottom two panels show runs including all thermal and chemical processes (those discussed in Section \ref{ssec:gmcchemistry} and \ref{ssec:gmcthermo}, and several others as well); the bottom panel starts with fully-molecular initial conditions, and the second from the bottom starts from fully atomic ones. In the top three panels, one or more processes are disabled. In the top panel, chemical reactions and radiative heating and cooling proceed as normal, but dust attenuation of the FUV radiation field is set to zero. In the second panel, attenuation, heating, and cooling are normal, but reactions leading to the formation of H$_2$ and CO are suppressed. In the third panel, H$_2$-forming reactions are allowed, but CO-forming ones are not. Figure taken from \citet{glover12a}, reprinted by permission.
}
\end{figure}

Figure \ref{fig:glover12} illustrates this effect, showing the results of a series of otherwise-identical simulations in which various thermal and chemical processes are disabled to explore their effects. For now, compare the top and bottom panels; the bottom panel shows the distribution of gas temperature and density produced in the simulation when all physical processes are included, and the top panel shows the results when dust attenuation of the ISRF is omitted. In the simulation with no shielding, there is a marked lack of the cold, dense gas that is present in the other simulation, and as a result there is no star formation at the time shown in the simulation. However, the effect on star formation is more a delay than complete prevention. \citet{glover12b} find that, while lowering the metallicity does delay the onset of star formation, the rate of star formation once it begins is only weakly dependent on metallicity.

The observed correlation with H$_2$ comes about because the same $\sim 1000$ \AA~FUV photons that are responsible for heating the gas are also responsible for dissociating H$_2$. Thus in regions where the attenuation is high enough to let the gas reach low temperatures, it is also high enough to suppress H$_2$ photodissociation and let the ISM become H$_2$-dominated. Conversely, in regions with little dust attenuation, the ISM will be both molecule-poor and too warm to form stars. In retrospect this result is not terribly surprising, as the equations describing chemical and thermal equilibrium take remarkably similar forms. When photoelectric heating dominates, thermal equilibrium balances a heating term, for which the energy added per unit volume is proportional to $\chi_{\rm FUV} n_{\rm H} e^{-\tau}$, against a cooling term proportional to $n_{\rm H}^2$. Similarly, chemical equilibrium balances a photodissociation term, in which the H$_2$ destruction rate per unit volume is proportional to $\chi_{\rm FUV} n_{\rm H} e^{-\tau}$, against a formation term proportional to $n_{\rm H}^2$. The analogy is not perfect, because the optical depth $\tau$ that affects photoelectric heating is mostly the dust continuum one, while $\tau$ for the dissociation rate also contains a contribution from line shielding. Similarly, both the cooling and formation rates depend on temperature, and not in the same way. Nonetheless, the analogy is close enough that, in practice, there is an excellent correlation between temperature and H$_2$ abundance.

Importantly, the formation of H$_2$ or CO is not actually important for the thermodynamics. The middle three panels of Figure \ref{fig:glover12} show the results of simulations that are similar to the fiducial one shown in the bottom panel, but with various modifications to the treatment of chemistry. In the second panel from the top, all molecule formation is suppressed, in the middle panel formation of CO is suppressed, and in the panel below that H$_2$ and CO formation are allowed, but the initial conditions are purely atomic. Qualitatively the results are no different than for the fiducial run, showing that molecule formation is not important for the dynamics. Molecules are simply a tracer of dust shielding.

A strong prediction of this picture, again made independently by the two groups, is that the correlation between H$_2$ and star formation should break down at sufficiently low metallicities \citep{krumholz12e, glover12b}. The H$_2$ equilibration time in the ISM is much longer than the thermal equilibration time, and both scale inversely with the metallicity. At sufficiently low metallicity, the gas should be able to cool to low temperature and proceed to star formation faster than H$_2$ can form, breaking the correlation; the effect should be large enough to be detectable at metallicities below $\approx 10\%$ of Solar. However, we lack an easy way of measuring H$_2$ masses at such small metallicities, as CO breaks down as a tracer of molecular gas at low metallicity for the reasons discussed in Section \ref{ssec:gmcchemistry}. The observations required to test this prediction will therefore be challenging, though there are a number of possible strategies \citep{krumholz12e}.

\subsubsection{The Star Formation Rate at Small Scales}
\label{sssec:effsmall}

The discussion in the previous section helps clarify why the depletion time in H~\textsc{i}-dominated regions is so long compared to that in molecular ones, but it still leaves unanswered the question of why $\epsilon_{\rm ff}$ is so small even in molecular gas. There are a number of possible explanations, with varying degrees of theoretical and observational support.

\paragraph{Threshold Models}
Several authors have noted that, in resolved observations of local clouds, the correlation between gas mass and star formation rate becomes increasingly tight as one considers only gas at higher and higher column densities \citep{goldsmith08a, lada10a, heiderman10a, evans14a}. For unresolved observations, the same tightening in the correlation is seen between star formation rates and gas mass as we proceed to masses traced by molecules of increasing critical volume density \citep{gao04a, wu05a}. This has led some authors to posit that the value of $\epsilon_{\rm ff}$ is low because only gas above a certain (volume or column) density threshold is able to form stars efficiently, and the fraction of mass in a typical molecular cloud that satisfies this condition is quite small \citep{heiderman10a, lada12a}. Gas that is above the threshold, referred to as ``dense" gas, forms stars on some specified timescale $t_{\rm dense}$.

There is undeniably an observational correlation between the presence of gas at high densities and the star formation rate. However, as a model the threshold picture is at best a skeleton. The threshold may be related to dust shielding against FUV radiation \citep{clark13a}, but in this case it is not clear why the column density requirement is any different than the requirement that the gas be molecular. In particular, why does it need to be dense enough to be HCN-emitting? Other than this proposal, there are no other reasons given in the literature for a physical origin for the proposed threshold. Perhaps more importantly, the timescale $t_{\rm dense}$ in the dense gas is still far longer than the free-fall time using the density estimated either from the critical density or from resolved observations of the regions above the threshold. \blue{Only once once proceeds to densities so high that the typical objects have masses comparable to that of a single star do we find star formation and free-fall timescales that are comparable, indicating $\epsilon_{\rm ff}$ approaching unity.} Thus \blue{in a threshold model,} the low value of $\epsilon_{\rm ff}$ in regions \blue{above the density threshold} remans unexplained. The situation is analogous to that in the top-down models, where a low value of $\epsilon_{\rm ff}$ at galactic scales is to be explained by feedback, while the similarly low value at sub-galactic scales is left unexplained. Here the low value of $\epsilon_{\rm ff}$ at molecular cloud scales is explained by invoking a threshold, but the low value at smaller scales is unexplained. 

\paragraph{Turbulence-Regulated Models}
A second class of models to explain the value of $\epsilon_{\rm ff}$ is based on statistics of turbulence in molecular gas. Recall that, in the virial theorem that describes the large-scale force balance of molecular clouds (Section \ref{ssec:gmcstability}), turbulence enters as a term that tends to increase $\ddot{I}$, preventing large-scale contraction. Thus turbulence opposes collapse on average. However, turbulence is intermittent, and there are always regions where the turbulent flow tends to raise the gas density and thereby promote gravitational collapse. Thus turbulence plays a dual role, preventing collapse on large-scales while encouraging it in unusual, local regions \citep{mac-low04a}.

In its simplest form, one can understand this phenomenon simply using the virial theorem coupled to the statistics of turbulence. Consider a molecular cloud of characteristic size $L$ in large-scale force balance between gravity and turbulence, so that the virial ratio $\alpha_{\rm vir}\approx 1$. Now consider a sub-region within the cloud of size $\ell < L$. For a randomly-chosen region, the mass contained within the region will scale as $M(\ell)\propto \ell^3$, and thus the gravitational binding energy of the region will scale as $|\mathcal{W}|\propto M(\ell)^2/\ell \propto \ell^5$. On the other hand, the linewidth-size relationship for turbulence (Section \ref{sssec:velocitystat}) implies that the velocity dispersion on size scale $\ell$ should vary as $\sigma(\ell)\propto \ell^{(n-1)/2}$ with $n\approx 2$ in the highly-supersonic limit. Thus the kinetic energy should vary as $\mathcal{T}\propto M\sigma(\ell)^2 \propto \ell^4$. It therefore follows that the virial ratio $\alpha_{\rm vir}(\ell)\propto \mathcal{T}/|\mathcal{W}|\propto \ell^{-1}$. If clouds are roughly virial on the largest scales, then on all smaller scales they are on average highly-supervirial, with $\alpha_{\rm vir} \gg 1$. The virial theorem then implies that sub-regions inside molecular clouds should have $\ddot{I} > 0$, in which case they should not collapse to form stars.\footnote{This argument ignores the fractal structure of molecular clouds. Using a realistic fractal dimension gives a weaker scaling of mass, binding energy, and kinetic energy with length scale -- see \citet{kritsuk13a}. The basic conclusion is, however, the same as for this overly simple analysis: the typical sub-region within a molecular cloud is gravitationally-unbound.}

However, turbulence also creates a distribution of densities, with some regions much denser than the mean. These regions will also have higher gravitational binding energies than the mean. On the smallest turbulent size scales, those where the non-thermal velocity dispersion becomes comparable to the sound speed, the condition for a region to have $|\mathcal{W}| \gtrsim \mathcal{T}$ can be written as a density threshold $\rho > x_{\rm crit} \overline{\rho}$, where $\overline{\rho}$ is the mean density and $x_{\rm crit}$ is a dimensionless number that depends on the same parameters as the density PDF.  Given the PDF of densities produced by the turbulence (Section \ref{sssec:densitystat}), one can compute the mass above this threshold. If this mass then collapses on some timescale $t_{\rm coll}$, which is \textit{not} much longer than $t_{\rm ff}$, then this gives an estimate of the dimensionless star formation rate $\epsilon_{\rm ff}$.

The first quantitative analysis of the star formation rate that results from this competition was developed by \citet{krumholz05c}, who posited that $t_{\rm coll}$ would be of order the mean density free-fall time $t_{\rm ff}(\overline{\rho})$, and who calibrated various fudge factors based on the simulations of \citet{vazquez-semadeni03a}. This calculation yields $\epsilon_{\rm ff}$ as a function of $\alpha_{\rm vir}$, the Mach number $\mathcal{M}$ of the cloud, and the fraction $\epsilon_{\rm core}$ of a collapsing gas core that ends up in a star rather than being ejected by the protostellar outflow. The numerical result was that $\epsilon_{\rm ff} \approx 0.01$ over a very broad range of $\mathcal{M}$ as long as $\alpha_{\rm vir}$ is of order unity.

Subsequent authors have generalized and improved this model in by introducing magnetic fields \citep{padoan11a}, which adds the Alfv\'{e}n Mach number $\mathcal{M}_A$ or equivalently the plasma $\beta$ as a further parameter, by considering variations in $t_{\rm coll}$ with density \citep{hennebelle11b, hopkins12e, hopkins13a, hennebelle13a}, and by considering a dependence on the compressive-to-solenoidal ratio of the turbulence \citep{federrath12a}, parameterized by $b$ (Section \ref{sssec:densitystat}). \citet{padoan11a}, \citet{padoan12a}, and \citet{federrath12a} present the most complete set of simulations published to date, compare them to a range of analytic models, and give calibrated estimates for $\epsilon_{\rm ff}(\alpha_{\rm vir}, \mathcal{M}, \mathcal{M}_A, b, \epsilon_{\rm wind})$.

The numerical value of $\epsilon_{\rm ff}$ that comes out of these models depends on whether the turbulence is magnetized or not, on the virial parameter, and on the compressive / solenoidal mix of the driving. However, it is clear that the \citet{krumholz05c} model overestimates the ability of turbulence alone to inhibit star formation. In the new models and the simulations used to calibrate them, $\epsilon_{\rm ff} \approx 0.01$ can be achieved only through some combination of magnetic fields, mostly solenoidal driving, and slightly super-virial turbulence. Non-magnetic models, models with more compressive turbulence, and models with precisely virial turbulence tend to give $\epsilon_{\rm ff} \approx 0.1$ for realistic parameter choices. This is the steady-state value, but the simulations also show that star formation rates tend to accelerate in time. They begin at $\epsilon_{\rm ff} \approx 0.01$ even in non-magnetic cases with virialized, mixed compressive-solenoidal turbulence, but accelerate to $\epsilon_{\rm ff} \approx 0.1$ after $\sim 1-2$ free-fall times. This acceleration appears to be related to the density PDF developing a high-density powerlaw tail of self-gravitating gas significantly in excess of a lognormal \citep{federrath13a}.

These results mean that, once steady-state conditions are reached, the original \citet{krumholz05c} proposal that virialized turbulence alone regulates star formation is probably not correct. Virialized turbulence does lower $\epsilon_{\rm ff}$ by an order of magnitude, but not by the two orders of magnitude required to match the observations. There are a number of possible ways to get the remainder of the way, however. First, $\epsilon_{\rm ff}$ is proportional to $\epsilon_{\rm core}$; the simulations used to calibrate $\epsilon_{\rm ff}$ do not include outflows, so this factor is inserted ex post facto. \citet{krumholz05c} adopted $\epsilon_{\rm core} = 0.5$ based on the analytic models of \citet{matzner00a}, and subsequent authors have retained this value. However, more recent observational and theoretical work has suggested that it is likely closer to $0.2-0.33$ \citep[e.g.,][]{alves07a, nutter07a, enoch08a, rathborne09a, konyves10a, hansen12a}, and this will bring $\epsilon_{\rm ff}$ down by a factor of $\sim 2$, a non-negligible fraction of the distance between typical simulation values and observed ones. Another possible important effect is that feedback localized to high-density peaks where stars are forming reduces $\epsilon_{\rm ff}$, both by directly expelling matter and by disrupting the powerlaw tail of the density PDF, thereby keeping the turbulence in a state more like that found early in self-gravitating turbulence simulations, when $\epsilon_{\rm ff}$ is low. Simulations including one obvious local feedback, protostellar outflows, do show that it reduces $\epsilon_{\rm ff}$ even compared to cases where there is turbulence but no local feedback \citep{wang10a}.

Local feedback may also be required to remedy another omission in turbulence-regulated models. As discussed in Section \ref{sssec:turbdecay}, turbulence decays in roughly a crossing time if it is not driven, and in turbulence-regulated models the driving mechanism is not specified. There are numerous proposals for the origin of the turbulence in molecular clouds, including the energy of accretion flows \citep{klessen10a, goldbaum11a}, a cascade of energy from galactic-scale shear, cloud-cloud collisions, \blue{and external supernova shocks} \citep{tasker09a, tasker11a, dobbs11a, dobbs11b, dobbs12a, dobbs13a, hopkins12e, van-loo13a}, photoionization feedback from star formation \citep{matzner02a, krumholz06d, gritschneder09a, goldbaum11a, walch12a}, radiation pressure from star formation \citep{krumholz09d, fall10a, murray10a, hopkins11a}, and momentum injection by protostellar outflows \citep{norman80a, mckee89a, li06b, nakamura07a, matzner07a, wang10a}. Essentially any of the mechanisms that have been invoked to destroy molecular clouds or limit star formation efficiency could also be responsible for driving turbulence within them, with the possible exception of supernovae, which may be delayed by too long to prevent the turbulence from decaying before the first supernova occurs \citep{fall10a}. However, it is an open question whether these mechanisms are capable of driving turbulence for an extended period, or whether they either fail to do so strongly enough to prevent collapse, or, conversely, inject so much energy that clouds suffer rapid disruption before they ever achieve statistically-saturated turbulence. In the latter case, one would return to something closer to the top-down picture.

\paragraph{Magnetically-Regulated Models}
The classical explanation for the low value of $\epsilon_{\rm ff}$ is to appeal not to the turbulence term in the virial theorem, but instead to the magnetic one \citep[e.g.,][]{shu87a}. As discussed in Section \ref{ssec:gmcstability}, in the ideal MHD limit, the dimensionless quantity $\lambda$ that measures the mass to flux ratio relative to the critical value is invariant under large-scale expansions or contractions of a cloud. Thus if $\lambda < 1$ and the cloud is magnetically subcritical, and ideal MHD applies, then gravity can never win against magnetic support and induce collapse. Before roughly the 1990s, the standard model of star formation was that clouds generally had $\lambda < 1$, and $\epsilon_{\rm ff}$ was small because it required many free-fall times for non-ideal MHD processes to allow enough magnetic flux to leak out of clouds \citep{mouschovias76a, shu87a}. However, observations have cast serious doubt on this model. As mentioned above, direct measurements of magnetic fields in $\sim 100$ clouds using the Zeeman effect now seem to show that $\lambda \approx 2-3$ is more typical \citep{crutcher99a, crutcher10a, crutcher12a}. Moreover, mass to flux ratios are generally observed to be higher in the centers of dense cores than in their envelopes, exactly the opposite of what would be expected if star formation proceeded via magnetic flux leaking out of dense cores and into their envelopes \citep{crutcher09a}, though there has been some debate on this result \citep{mouschovias09a, mouschovias10a, crutcher10b}. These observations seem to rule out the possibility that $\epsilon_{\rm ff}$ is low due to magnetic support.

\subsubsection{Bottom-Up Models: General Results}

Given models for both where in the ISM star formation will take place, and for the rate of star-formation in those parts, one can compute the relationship between star formation and gas content simply as
\begin{equation}
\Sigma_{\rm SFR} = f_{\rm SF} \epsilon_{\rm ff} \frac{\Sigma}{t_{\rm ff}},
\end{equation}
where $f_{\rm SF}$ is the fraction of the ISM that is star-forming (generally equivalent to the fraction that is molecular). In a three-dimensional simulation, this areal law can be replaced by an equivalent volumetric one. The quantity $f_{\rm SF}$ is computed from the models described in Section \ref{sssec:whichsf}, and the quantity $\epsilon_{\rm ff}$ can be either set to a fixed value based on observations, or computed based on the models described in Section \ref{sssec:effsmall}. The result is a prediction for the relationship between gas and star formation in the ISM, with possible additional dependencies on any other factors that affect $f_{\rm SF}$, such as metallicity.

\begin{figure}[ht!]
\centerline{\includegraphics[width=2.8in]{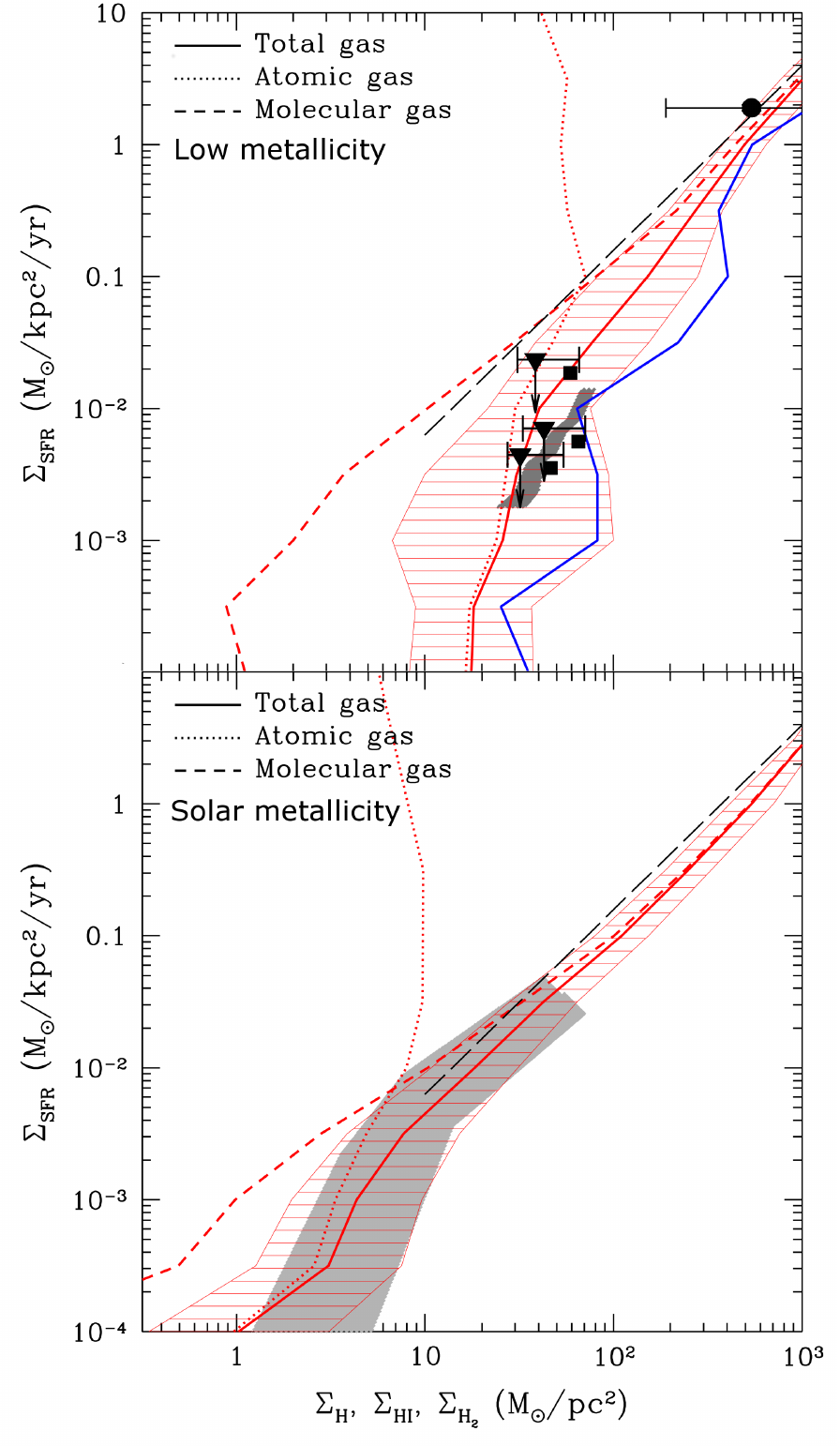}}
\caption{
\label{fig:gnedin10}
Star formation rate versus gas surface density for galaxies at $z\approx 3$ in cosmological simulations of galaxy formation, from \citet{gnedin10a}, reprinted by permission of the AAS. The top panel shows simulations using self-consistently computed metallicities (which are sub-Solar), while the bottom one shows simulations where the metallicity is artificially set to Solar. In each panel, the thick red line shows the relationship between surface formation $\Sigma_{\rm SFR}$ and total gas surface density $\Sigma_{\rm H}$, with the hatched region showing the range of variation. The dashed red lines show the $\Sigma_{\rm SFR} - \Sigma_{\rm H_2}$ relation, and the dotted red lines show the $\Sigma_{\rm SFR}-\Sigma_{\rm HI}$ relation. The gray bands and black points show observations. The the top panel, the data are from $z\approx 3$ surveys of (presumably) low-metallicity systems, while in the bottom panel gray band is from \citet{bigiel08a}, an earlier version of the same data set shown in blue in Figure \ref{fig:sflawtot}. Black dashed lines in both panels are the \citet{kennicutt98a} relation. Notice the difference between the Solar- and low-metallicity relations for both the data and the simulations.
}
\end{figure}

\begin{figure}[ht!]
\centerline{\includegraphics[width=2.8in]{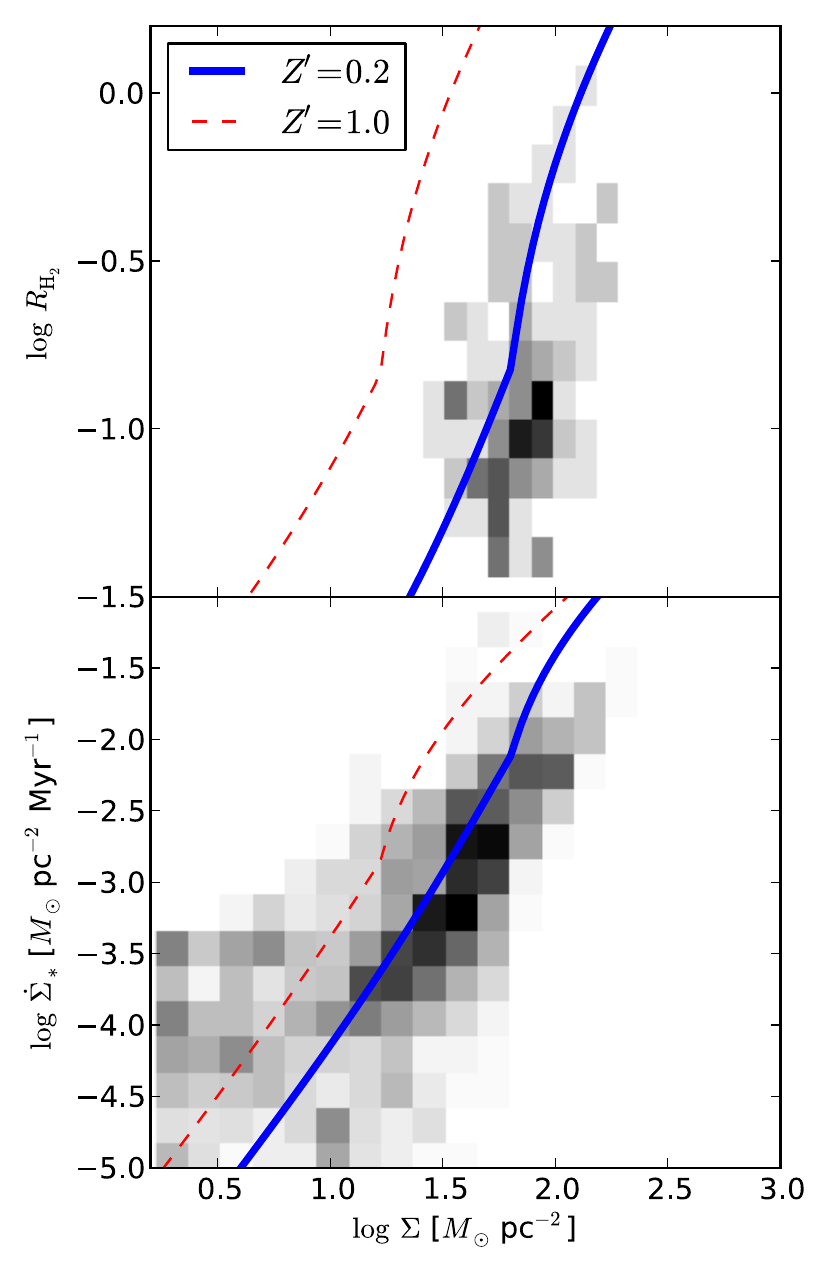}}
\caption{
\label{fig:krumholz13}
Top panel: ratio of molecular and atomic gas surface densities $R_{\rm H_2} = \Sigma_{\rm H_2}/\Sigma_{\rm HI}$ versus total gas surface density $\Sigma = \Sigma_{\rm H_2} + \Sigma_{\rm HI}$. Bottom panel: star formation rate per unit area $\dot{\Sigma}_*$ versus $\Sigma$. The black pixels in each panel show observations of the Small Magellanic Cloud by \citet{bolatto11a} (also shown as the green pixels in Figure \ref{fig:sflawtot}), with the color of the pixel indicating the density of individual lines of sight at the indicated values of $(\Sigma,R_{\rm H_2})$ and $(\Sigma, \dot{\Sigma}_*)$. The blue lines are the analytic model of \citet{krumholz13c}, evaluated using the SMC's metallicity of $Z/Z_\odot = 0.2$. The red dashed lines are the values the model would predict for $Z=Z_\odot$. The model correctly predicts the displacement of the SMC's molecular abundance and star formation rate from the values seen in higher metallicity galaxies.
}
\end{figure}

Compared to the top-down models, models of this form have two major successes. First, models in which star formation is assumed to trace H$_2$ provide a much better match to the observed phase- and metallicity-dependence of star formation rates than do models where feedback is assumed to be the dominant process. In particular, they are able to reproduce the break in the $\Sigma_{\rm SFR} - \Sigma$ relation at $\sim 10$ $\msun$ pc$^{-2}$ at Solar metallicity, and the observation that the location of the break is metallicity-dependent, as illustrated in Figure \ref{fig:sflawtot}. This result has now been reproduced independently by several groups, using a variety of techniques: simulations where the H$_2$ fraction is determined from a time-dependent chemical model \citep{gnedin09a, gnedin10a, gnedin11a}, simulations where the H$_2$ fraction is calculated from a time-independent numerical or analytic subgrid model \citep{robertson08a, kuhlen12a}, and purely analytic models \citep{krumholz09b, krumholz13c}. Figures \ref{fig:gnedin10} and \ref{fig:krumholz13} shows examples from two recent papers demonstrating this agreement. It is worth pointing out that, while the figures show recent work, these models actually predicted the metallicity-dependence of the relationship between star formation and gas surface density before it was established by observation, so this constitutes a genuine prediction.

A second success of the bottom-up approach is that, at least in the turbulence-regulated model, it can successfully explain why $\epsilon_{\rm ff}$ is \blue{roughly independent of size or density scale, rather than rising dramatically in denser gas}. The basic explanation is simple: a constant value of $\epsilon_{\rm ff}$ means that the mass $M(>\rho)$ above some specified density threshold $\rho$ must decrease as roughly $\rho^{-1/2}$, so that the ratio $M(>\rho)/t_{\rm ff}(\rho)$ remains about constant. The lognormal PDF produced by supersonic turbulence has approximately this property over a a broad range of Mach numbers \citep{krumholz07g, mckee07a, narayanan08a, narayanan08b}. The ability of these models to give roughly constant $\epsilon_{\rm ff}$ manifests directly in their ability to explain molecular line observations. The infrared and line luminosities of galaxies are often observed to obey scaling relations
\begin{equation}
L_{\rm IR} = c L_{\rm line}^p
\end{equation}
where the constant of proportionality $c$ and index $p$ depend on the particular line being observed \citep[e.g.,][]{gao04a, greve05a, riechers06a, wu05a, wu10a, bussmann08a, narayanan08a, bayet09a, juneau09a, schenck11a, garcia-burillo12a}. Two groups, one using analytic models \citep{krumholz07g} and one using numerical ones \citep{narayanan08b}, have used turbulence-regulated models to predict $c$ and $p$ for a range of lines, including much of the rotational ladders of CO, HCO$^+$, and HCN. A generic prediction of these models is that the index $p$ depends on the critical density of the line in question, and this prediction is in generally good agreement with observations \citep{bussmann08a, narayanan08a, juneau09a, schenck11a}, though it seems to fail for the highest $J$ lines \citep{bayet09a}, likely because these models do not correctly model the temperature-dependence of the emission. Nonetheless, the ability to reproduce the (approximate) density-independence of $\epsilon_{\rm ff}$ and its observational manifestation in molecular line data constitutes a major success of bottom-up models, which top-down models, as noted above, are able to match only by adding an ad hoc assumption of low $\epsilon_{\rm ff}$ on small scales. Again, this is a prediction, not a post-diction, of the models.

\subsubsection{Potential Problems with the Bottom-Up Approach, and the Need for a Synthesis}

While the bottom-up approach provides a very good match to a number of observations, the models also contain significant holes and assumptions. One has to do with the exact mechanism by which the correlation between star formation and molecular gas, and the metallicity-dependence it implies, is established. As discussed above, suppression of star formation in gas that is insufficiently shielded from the grain photoelectric heating by the ISRF appears to provide a possible explanation. However, much work still remains to flesh out this hypothesis. For example, \citet{krumholz11b} show that regions with low H$_2$ fractions will generally have Bonnor-Ebert (or Jeans) masses of $\gtrsim 10^2-10^3$ $M_\odot$, much larger than the mass of a typical star. However, these masses are still much less than the mass of an entire giant molecular cloud, and it is not clear exactly how high the Jeans mass must be in order for star formation to be suppressed by a certain amount.

Similarly, \citet{glover12a, glover12b} show that star formation is significantly delayed in clouds that are not shielded, but it is not suppressed completely. The $10^4$ $M_\odot$ clouds they consider are too massive to be supported solely by thermal pressure, and in the absence of internal feedback or external perturbations, neither of which are included in the simulations, collapse eventually occurs even in unshielded clouds. \blue{In such clouds the gas would still become molecular before fully collapsing, since the collapse time scales with density $n^{-1/2}$ and the molecular formation time varies as $n^{-1}$. However, if only the collapsing gas became molecular, this would presumably be reflected in a very high value of $\epsilon_{\rm ff}$ for this gas, which is not observed in low-metallicity systems like the SMC.}

A second gap in the bottom-up models is an account for exactly how the turbulence is maintained in molecular gas so as to keep $\epsilon_{\rm ff}$ low. There are numerical simulations demonstrating that this is possible on the scales of individual star clusters \citep{li06b, nakamura07a, wang10a}, and analytic models suggesting that it should work on large scales as well \citep[e.g.,][]{krumholz06d, klessen10a, goldbaum11a}, but not yet a full simulation that shows turbulence being maintained and keeping $\epsilon_{\rm ff}$ at the size scales of molecular clouds. In part this is an issue of dynamic range and numerical limitations. Simulations of entire galaxies, which are required to model processes like driving turbulence by accretion and cloud-cloud collisions, generally have resolutions too small to model the turbulence within molecular cloud, while simulations including feedback either have to resort to subgrid models, or have difficulty achieving enough dynamic range to simulate both the small scales where feedback occurs and the large ones that describe entire molecular clouds. Until such a high-dynamic range simulation can be performed, the maintenance of turbulence by either external or internal driving will remain a conjecture.

Both of these gaps suggest that the way forward in questions of the star formation rate has to be a synthesis between the top-down and bottom-up approaches. To address the question of how metallicity really determines both the phase balance and the rate of star formation, we require simulations that start from galactic scales but also resolve the small scales where the gas transitions from warmer and atomic to cold and molecular, including realistic treatments of FUV heating and dust shielding that are missing from almost all present-day galaxy-scale models. We also need these simulations so that we can simultaneously resolve the environments and interiors of molecular clouds, without resorting to parameterized feedback models where the choice of input parameters essentially determines the outcome. Such simulations would also help resolve the question of how large-scale feedback interacts with small scale efficiency, and with the need for the ISM to form a shielded phase before proceeding to star formation. At present, these questions depend crucially on the parameters one uses to describe feedback. For example, \citet{christensen12a} and \citet{kuhlen12a, kuhlen13a} both conduct cosmological simulations of low-metallicity dwarf galaxies using recipes that restrict star formation to molecule-rich gas. However, they reach very different conclusions about whether the metallicity-dependence of star formation actually changes the mean star formation rate or efficiency of galaxies over cosmological times, likely because they make different assumptions about the strength of stellar feedback.

Simulations with enough dynamic range to span from galactic to sub-pc scales, without relying on subgrid models that put in the feedback by hand, will be extraordinarily technically-challenging, and may be some distance in the future. In their absence, another viable approach will be to tune the star formation and feedback parameters needed in the large-scale models using small-scale simulations that do not rely on subgrid models. Efforts to formulate such subgrid recipes for feedback by radiation pressure \citep{krumholz12c, krumholz13a} and supernovae \citep{creasey13a} are already underway, but more work of this type is clearly needed.

\section{Stellar Clustering}
\label{sec:clustering}

The origins of stellar clustering have received significantly less theoretical attention than the star formation rate, but, as I discuss in this section, the major theoretical questions in both fields are tightly bound. The way that stars cluster is strongly related to the question of how $\epsilon_{\rm ff}$ is regulated and on what scales. To review, the central questions that interest us in this section revolve around the process by which stars that are initially born in high-density, highly-clustered regions disperse into their final configuration consisting of some fraction, usually the great majority, in a field population that is not bound to any structure smaller than the galaxy as a whole, while a minority end up in gravitationally-bound clusters with a mass function at birth $dN/dM \propto M^{-\beta}$ with $\beta\approx 2$. The theoretical problem of explaining these observations can be roughly broken into two parts. First, one must ask about the initial conditions for the problem: how are newborn stars arranged, both in space and kinematically, and both relative to each other and to the gas? Second, one must ask how this structure evolves as star formation comes to an end, gas is removed, and the evolution transitions from gas-dynamical to collisionless.

\subsection{Spatial and Kinematic Structure of Newborn Stars}
\label{ssec:starstructure}

The qualitative result that young stars are highly clustered is straightforward to understand from theory: the gas from which these stars form is moving highly supersonically, leading to very large density contrasts. Since the timescale for gravitational collapse scales as $\rho^{-1/2}$, the densest regions tend to run away and form stars first, thus leading to a stellar configuration that reflects or even further amplifies the highly structured nature of the gas. There have been relatively few attempts to go beyond this rough qualitative agreement to check for quantitative matches between theory and observations. \citet{klessen00a} and \citet{hansen12a} both calculated two-point correlation functions from their simulations of star cluster formation, and found reasonable matches to observed two-point correlation functions. In particular, both sets of simulations reproduce the observed behavior that the two-point correlation function behaves as a powerlaw over a large range of scales, but has a break at small scales. \red{The small-scale break} corresponds, perhaps equivalently, to either the Jeans length in the initial cloud or to the transition from the regime at small separation where most correlated stars are bound binaries to the regime at large separation where the stars are not individually bound to one another, but only to the larger cluster. Both simulations also have the property that the two-point correlation functions are nearly time-invariant.

\citet{hopkins13d} proposes an analytic model for the spatial clustering of stars, situated within the larger framework of the \citet{hopkins12e} excursion set model for turbulence. In this model the clustering of stars is dictated by the hierarchical nature of supersonic turbulence, and is calculated with precisely the same techniques as for the hierarchical clustering of galaxies. The results are also generally consistent with observations in the regime above the binary / Jeans length scale, but they fail to recover the break at small scales. This analytic model is a plausible explanation for what occurs in the \citet{hansen12a} simulations, since these include turbulence, but it is hard to see why they would describe the \citet{klessen00a} simulations, which start with essentially no turbulence ($\alpha_{\rm vir} = 0.01$). The simulations also differ in other respects as well: \citeauthor{klessen00a} include no stellar feedback, while \citeauthor{hansen12a}~include radiative transfer and protostellar outflows. The fact that these wildly disparate setups both produce qualitatively the same result, and that these results are time invariant, suggests that the spatial distribution of stars probably results from the simple property that cold, self-gravitating gas tends to collect into hierarchical structures. Turbulence, the basis of the Hopkins model, is one way of establishing this hierarchy, but not the only possible way.

A number of authors have also examined the distribution of stars in simulations not in an attempt to reproduce observations, but instead with the goal of understanding the early phases of star cluster evolution that occur while the stars are still hidden by large quantities of dust. \citet{maschberger10a} and \citet{kruijssen12c} analyze a simulation by \citet{bonnell08a}, and find that a final star cluster is assembled hierarchically, with a number of small bound sub-clusters forming first, and then later merging to produce fewer larger ones. The individual sub-clusters stop growing when they have depleted most of their gas, so the structure that results resembles a gaseous cloud with small stellar-dominated regions inside it. As time progresses, the stellar-dominated regions grow and merge, consuming more and more of the cloud. However, it is unclear how general these results are: the development of stellar-dominated regions occurs because $\epsilon_{\rm ff} \approx 1$ on the $\sim 0.1$ pc scales of star clusters in the simulations, and is close to that large even on the $\sim 10$ pc scale of the entire molecular cloud\footnote{\citet{bonnell11a} argue for a closely-related simulation that the time-averaged value of $\epsilon_{\rm ff}$ in the simulation is only $0.04$, which would match observations much better. However, they arrive at this estimate by averaging over a long initial phase of their simulations before star formation begins, and this phase is long because their simulations start with no initial density perturbations. Simulations that begin with initial turbulent structure show much shorter delays in the onset of star formation \citep[e.g.,][]{federrath12a, padoan12a}. Observations strongly rule out the possibility that real molecular clouds experience a long quiescent phase before the onset of star formation \citep{hartmann01a, ballesteros-paredes07a, tamburro08a}, suggesting that a more sensible procedure would be to ignore the initial transient when the star formation rate is low, and instead measure $\epsilon_{\rm ff}$ from the onset of star formation rather than from the start of the simulation; doing so yields $\epsilon_{\rm ff}\approx 0.3$ for \citeauthor{bonnell11a}'s simulation. Measuring $\epsilon_{\rm ff}$ only after star formation has begun is also the standard procedure in all other published measurements of $\epsilon_{\rm ff}$ from simulations. \label{note:bonnell11a}}. \red{This high value of $\epsilon_{\rm ff}$ does not agree well with observations, and is probably an artifact of the limited physics included in the simulations, which omit} magnetic fields, feedback, or anything else to slow down the star formation. Stellar-dominated regions also form in other simulations with similarly-high values of $\epsilon_{\rm ff}$ \citep{girichidis11a, girichidis12b, moeckel12a}, but do not appear to form in simulations that include feedback and magnetic fields and thus have lower $\epsilon_{\rm ff}$ \citep[e.g.,][]{li06b, nakamura07a, wang10a}.

Several authors have also investigated the kinematic properties of newborn stars, and their relationship with the gas. A generic finding of these simulations is that, while the overall velocities of the stars remain well-correlated with those of the surrounding gas, the stars also have velocity dispersions significantly smaller than that of the gas  \citep{offner09a, kruijssen12c, girichidis12b}. This appears to occur because stars are formed at the nodes of shocks, where the bulk velocity found in the cloud has mostly cancelled. Indeed, the same phenomenon applies to the gas cores even before stars form \citep{offner08a}, and detailed comparisons between the kinematics of the cores and the stars produced in simulations and those actually observed in nearby regions appear to produce reasonable agreement \citep{offner08a, offner09b}. Interestingly, the result that simulated stellar velocities are well-correlated with gas ones but with lower velocity dispersion, and the good agreement between simulations and observations, appears to hold over a wide variety of simulation methods and assumptions; for example, it matters little whether the turbulence is driven or not. This suggests that the stellar kinematics have a very simple and universal origin, the most likely candidate for which is the general phenomenon that densities and velocities tend to be anti-correlated in supersonically-turbulent flows \citep{padoan01a}.

As a result of their lower velocity dispersions, stars tend to be sub-virial with respect to the gas, but roughly virial if one considers only the stellar mass \citep{offner09a, kruijssen12c, girichidis12b}. \citet{kruijssen12c} attribute this to the stellar-dominated nature of the sub-clusters, but \citet{offner09a} find the same result in simulations that do not proceed to high star formation efficiency and do not produce stellar-dominated regions. Regardless of its origin, the fact that stars have much lower velocity dispersions than gas has important implications for the evolution of star clusters once gas is removed. However, as with the result regarding the existence of gas depleted regions, we should treat this result with caution because the simulations from which it is derived do not include any form of feedback, and form stars too rapidly and efficiently compared to what we observe.

\subsection{Gas Removal}

\subsubsection{General Theory}

At some point in the star formation process, the gas is removed, either because it has all been converted into stars, or because some stellar feedback process ejects it. The classical theory for how the stars will respond, first described by \citet{hills80a}, is quite simple, though more sophisticated analytic models exist \citep{adams00a, boily03a}. If one starts with a virialized system of gas and stars with negligible support from magnetic fields, the kinetic and potential energy are related by
\begin{equation}
-2 \mathcal{T} = \mathcal{W},
\end{equation}
and the two terms individually scale with the mass $M$ of the system as $\mathcal{T}\propto M$ and $\mathcal{W}\propto M^2$. If one rapidly removes some of the mass, leaving a mass $\epsilon M$ behind in the form of stars\footnote{It is critical to distinguish between $\epsilon$ and $\epsilon_{\rm ff}$. The latter describes the instantaneous rate of conversion from gas to stars, while the former is the integrated fraction of the gas that is converted into stars over the entire lifetime of the star-forming system. The two are identical only if the system lives exactly one free-fall time, and if there is no gain or loss of gas mass through any process other than star formation during that interval.}, then the total energy of the resulting system is
\begin{equation}
E = \mathcal{T}' + \mathcal{W}' = \epsilon \mathcal{T} + \epsilon^2 \mathcal{W} = \epsilon \left(1-2\epsilon\right)\mathcal{T},
\end{equation}
where $\mathcal{T}'$ and $\mathcal{W}'$ are the new kinetic and potential energy after gas removal. The total energy $E$ is negative, indicating that the system is bound, only if $\epsilon > 1/2$. \red{(Note that the exact same calculation implies that binary companions to stars that go supernova will in general become unbound, unless the asymmetric kick of the supernova happens to push the neutron star in exactly the right direction to keep the system together.)} This process of star clusters disrupting due to rapid gas expulsion goes by the somewhat macabre name ``infant mortality". On the other hand, if the mass loss is slow compared to a dynamical time, then the system remains in virial equilibrium at all times, and it is straightforward to show that in this case the system always remains bound, but that its radius increases from an initial value $R$ to a final value $R' = R /\epsilon$.

Numerous authors have studied this process with N-body simulations as well. The most common procedure is to start with a star cluster in a gas potential, whose depth relative to the potential produced by the stars is specified by the star formation efficiency. The stars themselves can be either smoothly distributed and in virial equilibrium with the gas \citep{tutukov78a, lada84a, goodwin97a, kroupa01b, geyer01a, boily03b, bastian06a, goodwin06a, baumgardt07a, pelupessy12a}, smooth and sub-virial \citep{chen09a, goodwin09a}, distributed in a fractal or other sub-structured distribution \citep{scally02a, goodwin04b}, or taken directly from the output of gas-dynamical simulations \citep{smith11a, smith11b, smith13b}. The cluster potential is then removed over some time scale, either via a prescribed analytic formula, or by running a fluid-dynamics simulation together with the N-body one and causing the gas to disperse using a simple prescription for the effects of stellar feedback \citep{geyer01a, pelupessy12a}. The primary free parameters in this approach are the star formation efficiency $\epsilon$, the timescale over which the gas is removed, and the virial ratio of the stars at the time the simulations begin.

The simulations generally agree with the simple analytic argument given above, but with some important differences. First, even for an initially-virialized stellar population and instantaneous gas removal, $\epsilon = 0.5$ does not represent a hard line for cluster survival or disruption. Instead, at least some bound remnants will be left even with values of $\epsilon \approx 0.33$, mainly because the kinetic energy is not uniformly distributed among the stars; instead, when the potential is removed, those stars on the high-energy tail of the Maxwellian distribution carry away a disproportionate share of the energy, while those with lower energies remain behind. However, at values of $\epsilon < 0.5$, clusters do suffer increasing mass loss, which becomes total at $\epsilon \lesssim 0.3$. Conversely, even if mass removal is as slow at $\sim 10$ crossing times, for low values of $\epsilon$ substantial mass loss can still occur, thanks to the presence of the galactic tidal field. This tends to strip stars that wander too far from the cluster during mass removal, even if that removal is slow. 

Second, for a cluster that is smooth but not initially virialized, clusters are much more likely to survive than if the initial conditions is virialized, and $\epsilon$ alone is not a good predictor of the outcome. Instead, the fraction of the stars that remain bound is determined primarily by the effective star formation efficiency $\epsilon_{\rm eff}$, defined simply as the virial ratio of the stars immediately after gas removal. \blue{(Note that $\epsilon_{\rm eff}$ should not be confused with $\epsilon_{\rm ff}$, the dimensionless star formation rate per free-fall time, which represents an entirely different concept -- unfortunately the letter $\epsilon$ is used for too many different things in this field.)} Thus if the stars are sub-virial with respect to the gas, while it is present, gas removal will result in an effective star formation efficiency that is larger than the true star formation efficiency $\epsilon$, and the stellar cluster will be correspondingly more difficult to disrupt.

Third, in cases where the initial conditions are highly sub-structured, either through a specified structure model or by taking the results from fluid simulations, the results are highly stochastic, and can change wildly from one realization of the structure to another, even when all the parameters used to generate those realizations (e.g., the star formation efficiency and the initial virial ratio) are held fixed. Thus the amount of mass that remains in bound clusters in this case is highly random, and can only realistically be determined from very large statistical ensembles of simulations.

\subsubsection{The Cluster Mass Function}
\label{ssec:icmf}

The relationship between the star formation efficiency $\epsilon$ and cluster survival provides a tool to investigate the origin of the mass function of clusters that do survive.\red{\footnote{\red{Recall from the previous section that, for stars that are not in virial equilibrium when the gas is removed, the survival of the star cluster is determined not by the actual efficiency $\epsilon$, but by effective efficiency $\epsilon_{\rm eff}$. For simplicity in this section I will ignore this complication, but the results may be extended to the non-virialized case in a straightforward manner.}}} Suppose that we observe cluster-forming gas clouds with a mass function $dN_{\rm obs}/dM_g$. \red{The distribution of masses present at any time is proportional to the rate at which clouds of a given mass form, $dN_{\rm form}/dM_g$, multiplied by the (potentially mass-dependent) lifetime $t_l(M_g)$, which implies that the mass function of forming clouds is distributed as}
\begin{equation}
\frac{dN_{\rm form}}{dM_g} \propto \frac{1}{t_l(M_g)}\frac{dN_{\rm obs}}{dM_g}.
\end{equation}

\red{We wish} to relate this to the mass spectrum of the newly-formed star clusters. \red{The mass of stars formed in a cloud of mass $M_g$ is simply $\epsilon M_g$, where $\epsilon$ is the star formation efficiency, which may itself be a function of $M_g$. However, not all of the stars formed necessarily remain as part of a star cluster; some may disperse into the galactic field. To account for this,} let $f_{\rm cl}(\epsilon)$ be the fraction of the stellar mass remains part of a final star cluster; \red{Thus a gas cloud of mass $M_g$ forms a mass $\epsilon M_g$ of stars, and a final star cluster of mass $M_c = f_{\rm cl} \epsilon M_g$.} If one is observing very young clusters that have not had time to disperse even if they are unbound (as expected for ages $\lesssim 10$ Myr), then $f_{\rm cl}(\epsilon)$ is simply unity\footnote{Recall from Section \ref{ssec:obscluster} that some authors define clusters only as objects that are more than a dynamical time old, and by this definition there are no such things as ``unbound clusters". My usage of the generic term ``cluster" to include both things that are bound and old and things that are younger than a dynamical time is not an attempt to take sides in this definitional debate so much as it is an attempt to avoid clumsy language such as ``unrelaxed stellar agglomerates".}. \red{Given these definitions,} one can then compute the mass spectrum of newly formed clusters simply from application of the chain rule:
\begin{eqnarray}
\frac{dN_{\rm form}}{dM_c} & = & 
\left( \frac{dM_c}{dM_g}\right)^{-1} \frac{dN_{\rm form}}{dM_g}
\nonumber \\
& \propto & \left[\epsilon f_{\rm cl} + f_{\rm cl} \frac{d\epsilon}{d\ln M_g}
+ \epsilon \frac{df_{\rm cl}}{d\epsilon} \frac{d\epsilon}{d\ln M_g}
\right]^{-1}
\nonumber \\
& & \quad{} \cdot
\frac{1}{t_l(M_g)}\frac{dN_{\rm obs}}{dM_g}.
\end{eqnarray}
This relation encodes the ingredients by which one can translate between an observed mass function of gaseous objects $dN_{\rm obs}/dM_g$ and the mass function of the star clusters they produce. The necessary ingredients are the mass-dependent lifetime $t_l(M_g)$, the functional form of $f_{\rm cl}(\epsilon)$ (which can be determined from the N-body simulations or analytic theory), and the dependence of $\epsilon$ on $M_g$. This relationship simplifies in various cases. If one is observing young clusters, such that there has been no time for unbound stars to disperse and one can set $f_{\rm cl} = 1$, then the term in square brackets reduces to simply $\epsilon + d\epsilon/d\ln M_g$.

The mass function of proto-cluster gas clouds is somewhat ambiguous, given that it depends on how one defines such clouds and on the observational tracer used to measure the gas (more on this below), but as discussed in Section \ref{ssec:obscluster}, for a wide range of reasonable choices the \blue{observed} cloud mass function $dN_{\rm obs}/dM_g$ is roughly a powerlaw $M_g^{-\beta_g}$ of index $\beta_g$ in the range $1.5$ to $2$. The lifetime of molecular clouds is hotly disputed \citep[and references therein]{dobbs14a}, but it is reasonable to expect that it should scale with the crossing time or the free-fall time. \blue{It is important to caution that there is no observational evidence one way or another on this point, but in the absence of evidence, this hypothesis is a reasonable starting point.} For a population of clouds with fixed surface density, which appears to roughly describe molecular clouds on large scales \citep[and references therein]{dobbs14a} and also gas clumps selected by a variety of methods \citep{fall10a}, this implies $t_l\propto M_g^{1/4}$; for a population of clouds with fixed radius or density instead, we have $t_l\propto M_g^{-1/2}$ or $t_l \propto M_g^0$, respectively. Thus a reasonable estimate for $t_l^{-1} (dN_{\rm obs}/dM_g)$ is that it is a powerlaw with an index in the range $-1.75$ to $-2.25$, though one could obtain values outside this range under strong assumptions (e.g., fixed radius and an observed mass function at the high end of the observed range, which would give an index of $-2.5$).

The largest uncertainty in the above analysis is the dependence of $\epsilon$ on $M_g$. However, several authors have made theoretical arguments for these dependences in an attempt to derive the mass function of star clusters; conversely, some authors have used the observed cluster mass function in attempt to constrain the process of star formation. \citet{kroupa02a} argue that $\epsilon$ should have a local minimum at mass $M_g \sim 10^4$ $M_\odot$, corresponding to clusters massive enough to host O stars but not so massive that ionized gas cannot escape from them, and that this should induce features in the cluster mass function in the range $10^3 - 10^4$ $M_\odot$. \citet{baumgardt08a} and \citet{parmentier08a} argue for a similar features at $\sim 10^5$ $M_\odot$ induced by supernova-driven mass loss, and propose that this explains the observed turnover in the globular cluster mass function. \citet{fall10a} focus on the early phase when $f_{\rm cl} = 1$, and argue that the observed $\beta\approx 2$ slope for young clusters can be reproduced naturally if the gas clouds have roughly constant densities, and if the process responsible for stellar feedback is a momentum-driven mechanism such as radiation pressure. \citet{parmentier11a} and \citet{parmentier12a} argue that a constant density initial condition provides a better explanation for the observed similarity between the cloud and cluster mass functions.

Finally, it is worth noting that all of the models discussed here are of the ``spherical cow" variety: one starts with a gas cloud of defined mass, which converts a defined fraction of its gas to stars and expels the rest over some time scale. In reality, of course, star-forming regions do not have well-defined edges. The mass and surface density vary continuously with position, and the star formation efficiency is a function of the scale over which it is averaged, with smaller regions (probably) achieving higher efficiencies than larger ones. However, at present there is no model for the origin of the cluster mass function that properly accounts for this hierarchical structure. As I discuss in the next section, however, there have been attempts to take this structure into account when calculating the fraction of star formation occurring in long-lived structures.

\subsubsection{The Fraction of Stars Forming in Clusters}
\label{sssec:clusterfrac}

The mass function is only half of the story; the other part is what fraction of the star formation is in stellar clusters. Before addressing this question theoretically, it is important to sharpen it a bit. The very influential review of \citet{lada03a} argued that the great majority of stars form in clusters, but subsequent work using larger and more complete catalogues of embedded stars has shown that the question itself is somewhat ill-posed, and depends on what one means by ``forming in a cluster". \blue{\citeauthor{lada03a} define a cluster as a collection of stars whose density is much greater than that of the galactic field, and that are dense enough to survive tidal disruption and numerous enough to avoid immediate dissolution by two-body relaxation. By this definition, essentially all stars are born in clusters. However, this definition admits as clusters objects that have much lower densities than typically observed for older, open clusters, and if one sets a higher surface or volume density threshold, then the fraction of stars born in clusters, and the properties of the clusters that one identified, depends critically on the threshold one adopts \citep{bressert10a}. This may be because density thresholds are an insufficiently precise tool to identify clusters even if distinct clusters are present \citep{pfalzner12a}, but no more precise tool is available at the moment.}

Imposing criteria for what constitutes a cluster \blue{beyond simple density cuts} can introduce yet more variation. For example, \citet{chandar10b} and \citet{bastian12a} produce wildly-conflicting cluster catalogs for the same portion of M83, with the majority of the difference stemming from the fact that \citeauthor{chandar10b}~use an automated selection based (roughly) on surface brightness alone, while \citeauthor{bastian12a}~start with a surface brightness-selected automated catalog but then perform an additional by-eye inspection and remove from the catalog any regions that do not have round, symmetric morphologies.

This review of the observations suggests that the question ``what fraction of stars form in clusters" does not have a meaningful answer that can be specified independent of a definition of cluster. However, one can ask a related question that is physically meaningful: after the gas has been expelled from a star-forming cloud and the dynamics become purely N-body, what fraction of stars are parts of gravitationally-bound structures smaller than the galaxy as a whole (excluding binaries or similar small-N star systems)? This in turn connects to the observational question of why stars forming in bound gas clouds at densities much higher than that of the galactic field manage nonetheless to find themselves at far lower densities $\sim 10-100$ Myr later. 

\blue{One might think that the answer to this question might turn on the question of whether molecular clouds are gravitationally bound or not, but simulations suggest otherwise. For example, \citet{clark05b} simulate a molecular cloud with a mass of $10^5$ $M_\odot$ and a virial ratio $\alpha_{\rm vir}=4$, rendering it moderately unbound. Without any stellar feedback in the simulations, they find that most of the stars end up in bound stellar structures. The effect of the high virial ratio is that, rather than a single cluster, the cloud forms a modest number of sub-clusters that are not mutually bound to one another, but are each bound internally. This strongly suggests that even moderately unbound clouds will still produce bound stellar structures in the absence of feedback, and so any successful model of stellar clustering must invoke stellar feedback somehow.}

The models used to explore \blue{how stellar feedback affects} the cluster mass function could in principle be used to explore this question as well. In practice, however, there are numerous obstacles to doing so: in the vicinity of $\epsilon = 0.5$ and expulsion times of order the crossing time, the fraction of stars that remain bound becomes extremely sensitive to small changes in either the efficiency or the gas expulsion timescale, making predictions highly uncertain. Moreover, as noted above, when the stars in question are not dynamically-relaxed, the results are highly stochastic and appear to depend on the details of the initial conditions in unpredictable ways. Clearly a more statistical approach is needed.

\citet{kruijssen12a} proposes one approach to the problem, based on the density PDF of supersonic turbulence. In this model, one estimates the characteristic star formation efficiency $\epsilon(\rho)$ achieved by gas of density $\rho$ as $\epsilon(\rho) = \epsilon_{\rm ff} [t_{\rm fb}/t_{\rm ff}(\rho)]$, where $t_{\rm fb}$ is a characteristic timescale required for feedback (assumed to be dominated by supernovae in the model) to halt further star formation\red{, and the density $\rho$ is averaged over an infinitesimally-small volume, not over any finite size scale}. This in turn is calculated via a simple argument balancing the pressure of the ISM against the pressure of supernova-heated gas. Once $\epsilon(\rho)$ is known, one can then estimate the fraction of mass that remains bound as a function of $\epsilon$ from the results of numerical simulations, and calculate the overall bound fraction simply by integrating over the density PDF. High-density regions put most of their stars into bound structures because they have large $\epsilon$ due to their short free-fall times, while lower-density regions make little contribution to bound structure because most of their mass is still gaseous, and is expelled by feedback.

This model predicts that the fraction of stars born in bound structures is an increasing function of the ISM pressure (parameterized via the total gas surface density), which enters by determining how long star formation can continue before being halted by feedback; there is some observational evidence in favor of such a trend \citep{larsen00a, goddard10a, bastian11a, bastian12a, silva-villa13a}, though, as with much relating to star cluster observations, this claim is disputed \citep{fall09a, chandar10a, chandar10b, chandar11a, fall12a}. More generally, we should recognize that this model depends on a large-number of poorly-constrained parameters, including the characteristic timescale on which feedback halts star formation, and the mapping between local star formation efficiency and fraction of stars that remain bound. One could arrive at significantly different results for plausible choices of these parameters, and, unfortunately, these parameters are much harder to calibrate from simulations than are the parameters that enter models of turbulence-regulated star formation: it is much harder to simulate the full suite of stellar feedback processes discussed in Section \ref{sssec:feedbackmodel} than to simulate boxes of driven turbulence! Nonetheless, the model represents a first attempt to to build a theory for star cluster formation that grapples with the continuous and hierarchical nature of molecular clouds.

A final complication in getting from a theoretical prediction of star formation occurring in bound structures to the actual observable of bound stellar systems of a certain age is that there are processes that can unbind star clusters after they become gas-free, but before they have time to spread apart significantly and thus become distinguishable from those stars that were unbound thanks to gas expulsion. Star clusters can be disrupted due to tidal shocking by nearby gas clouds \citep{kruijssen12c, kruijssen12b} and, for small clusters (mass $\lesssim 10^3$ $M_\odot$), by two-body evaporation, mass loss through stellar evolution, energy released during core collapse, and related dynamical effects, all of which are accelerated when the stars are highly sub-structured to begin with \citep{moeckel12a}.

\subsection{Stellar Clustering: Ways Forward}

More complete theoretical explanations for how and why stars cluster will require progress on two major fronts. The first of these is numerical. As should be apparent from the preceding discussion, much of the work that has been done on both the ``initial conditions" for the problem of stellar clustering (i.e., the distribution of newborn stars relative to each other and to the gas) and the response of the system to gas removal have been based on simulations that include no stellar feedback. Given that we know that such simulations produce star formation rates that are far too high compared to observed values, and that the star formation efficiency appears to be an important parameter in determining the outcome of gas removal, this is obviously a concern. As discussed in Section \ref{sssec:effsmall}, more complete simulations that include local feedback, turbulence, or some combination of both are now able to produce star formation rates in much better agreement with observations, and it seems urgent to analyze these simulations using the same techniques that have been applied to simulations without feedback, in order to determine which results are robust and which are not.

A more ambitious numerical goal would be to perform simulations that calculate star formation and gas removal self-consistently within a single simulation, so that the fraction of stars in the final, bound structure could be computed directly. At present no simulations quite achieve this goal, though some are close. \citet{wang10a} simulate star cluster formation including outflow feedback and magnetic fields (which appear to be crucial to getting the effects of feedback right -- see the discussion in that paper and also in \citet{gendelev12a} and \citet{krumholz14a}), but not the feedback from massive stars that could eject the bulk of the gas. \citet{vazquez-semadeni10a, vazquez-semadeni11a} simulate molecular cloud formation and, via a simple subgrid model, gas removal by H~\textsc{ii} regions, but not with enough resolution to resolve individual stars, as would be necessary to follow their N-body evolution. \citet{rogers13a} extend this work with a much more realistic treatment of the effects of stellar wind and supernova feedback, but do not follow the dynamics of the stars. \citet{pelupessy12a} also simulate gas removal with a simple subgrid model, and do track the trajectories of individual stars, but they insert the stars by hand rather than following their formation self-consistently. \citet{dale12a, dale13a} simulate both star formation and subsequent disruption of protoclusters by ionizing radiation, and do so with enough resolution to follow individual stars and with a much better approximation to ionizing radiation feedback than has been used by other authors. However, they include neither magnetic fields nor any other form of feedback. As this list should make clear, most of the ingredients needed to solve this problem exist, but thus far not all in a single simulation or a single code. Solving this problem will require putting the necessary ingredients into a single simulation and running it end to end.\footnote{An alternative approach is to chain together existing codes using a high-level software control structure \citep[e.g.,][]{portegies-zwart13a}, though this does not obviate the need to have low-level codes that actually implement all the requisite physics.}

On the analytic side, progress will require a move toward theories that grapple with the continuous and hierarchical structure of the ISM. The excursion set formalism used by \citet{hopkins13d} to predict the two-point correlation function of cores and young stars, and the density PDF model of stellar clustering by \citet{kruijssen12a} represent steps in this direction, but each has major holes left to be filled. The excursion set approach needs to be supplemented with analytic models for feedback in order to say anything meaningful about the final state of stellar clustering rather than just the initial condition, and a model based on single-point statistics such as the PDF is for obvious reasons unable to make any predictions about cluster mass functions or similar non-point properties. A useful starting point for theoretical progress might be to explore a synthesis of the two approaches. Any theory of this form will have significant unknown parameters that will have to be calibrated from simulations, and in the future it is imperative that the simulations used for this purpose be as realistic as possible, in particular in their treatment of stellar feedback.

\section{The Initial Mass Function}
\label{sec:imf}

The origin of the initial mass function forms the final topic of this review. As discussed in Section \ref{ssec:imfintro}, the IMF has two particularly important features that are robustly measured over a wide range of environments, and that demand explanation. First, there is a powerlaw tail at high masses, with a slope that appears to be universal or nearly so. Second, there is a distinct peak at a mass between $0.1$ and $1$ $M_\odot$. The location of this peak does not seem to vary in local environments, while there is tentative but increasingly-convincing evidence for variation in the most massive present-day early-type galaxies. The behavior of the IMF at masses below the peak might constitute a third subject for theoretical exploration, but this is by far the least observationally-constrained part of the IMF, and, moreover, it has relatively little significance on galactic scales, since objects of such low mass produce negligible light, and below the IMF peak they also contribute relatively little mass. I structure this section around potential explanations for the slope of the IMF first, and then potential explanations for the peak.

\subsection{The High-Mass Slope}

As discussed in Section \ref{ssec:imfintro}, the high-mass end of the IMF always appears to follow a powerlaw $dN/dm \propto m^{-\alpha}$ with $\alpha \approx 2.3-2.4$. Such a scale-free powerlaw dependence naturally calls for a scale-free phenomenon to explain it, and there are two obvious candidates: gravity-driven accretion, and turbulence.

\subsubsection{Competitive Accretion Models}
\label{sssec:compacc}

One way of producing a scale-free powerlaw is to rely on self-similar growth. Models of this class are generically known as ``competitive accretion" models, and were recently reviewed by \citet{bonnell07a}. The central idea, first proposed qualitatively by \citet{larson78a, larson82a}, and first quantitatively worked out by \citet{zinnecker82a}, is to consider what happens in a cluster when a series of stars are born with a relatively small range of masses, but then accrete in a manner such that the accretion rate is an increasing function of the current mass, i.e.,
\begin{equation}
\frac{dm}{dt} = f(m),
\end{equation}
with $f(m)$ an increasing function. For the specific case of a powerlaw function $f(m) \propto m^\eta$, this equation has the analytic solution
\begin{equation}
\frac{m(t)}{m_0} =
\left\{
\begin{array}{ll}
\left[1-\left(\eta-1\right) \tau \right]^{1/(1-\eta)}, 
\, & \eta\neq 1 \\
\exp\left(\tau\right), &
\eta=1
\end{array}
\right.,
\end{equation}
where $m_0$ and $\dot{m}_0$ are the mass and accretion rate at $t=0$, and $\tau = \dot{m}_0 t/m_0$ is the dimensionless time. For $\eta>0$, these functions have the common feature that they produce runaway growth on a timescale $m_0/\dot{m}_0$; for $\eta>1$ the growth is super-exponential, and reaches infinite mass at time $\tau=1/(\eta-1)$. Now consider a group of stars that accrete following this rule, but that have a range of values of $\tau$ at which they stop accreting. This can be because there is a wide range of actual accretion times $t$, because there is a range of initial masses $m_0$, because there is a range of initial accretion rates $\dot{m}_0$, or some combination of all three. If the distribution of $\tau$ values is $dN / d\tau$, and all the stars start with about the same value of $m_0$,\footnote{I assume fixed $m_0$ only for algebraic simplicity, as it is straightforward to generalize the argument to the case of a range of $m_0$ values.} then the distribution of final masses is
\begin{equation}
\frac{dN}{dm} \propto \left(\frac{dN}{d\tau}\right)\left(\frac{dm}{d\tau}\right)^{-1} \propto \left(\frac{dN}{d\tau}\right)m(\tau)^{-\eta}.
\end{equation}
The implication is that, if one starts with a distribution of $\tau$ values that is (for example) uniform over the range $\tau_{\rm min}$ to $\tau_{\rm max}$, the final distribution of masses will follow a powerlaw distribution from $m(\tau_{\rm min})$ to $m(\tau_{\rm max})$. In this manner, a relatively narrow and flat distribution of accretion durations, accretion times, and/or initial masses can be broadened to a powerlaw distribution by the action of mass-dependent accretion. The broadening occurs because, when accretion rates increase with mass, growth is highly non-linear. Thus a small difference in the number of dimensionless accretion times is amplified to a very large difference in final mass. This competitive behavior is what gives its name to this class of models.

There have been an extremely large number of both analytic models and simulations exploring models of this sort. In his original proposal, \citet{zinnecker82a} suggested that stars would accrete via a Bondi-Hoyle process \citep{hoyle39a, bondi52a}, which corresponds to $\dot{m}\propto m^2$ (i.e., $\eta=2$), and pointed out that this would give rise to a mass spectrum $dN/dm\propto m^{-2}$, fairly similar to the observed index of $\alpha = 2.35$. Subsequently, \citet{bonnell97a, bonnell01a, bonnell01b} found that the spectrum would be steepened slightly by mass segregation: more massive stars tend to sink to the center of a growing cluster, so they find themselves in higher density gas, and on average this makes the accretion rate grow with stellar mass faster than $m^2$, as would be expected for Bondi-Hoyle accretion in a uniform-density medium. A number of authors have also studied more complex accretion laws motivated by a range of physical scenarios, and have also considered a range of possible physical origins for a distribution in $\tau$ values \citep[e.g.,][]{adams96a, basu04a, bate05a, myers08a, myers09b, myers11b, myers12a}. Possibilities include random start times coupled with simultaneous truncation due to stellar feedback, a range of mass loss rates due to stellar outflows, ejection of stars due to dynamical interactions with other stars, and exhaustion of gas from a star's immediate vicinity due to the local star formation efficiency reaching 100\%.

\begin{figure}[ht!]
\centerline{\includegraphics[width=2.6in]{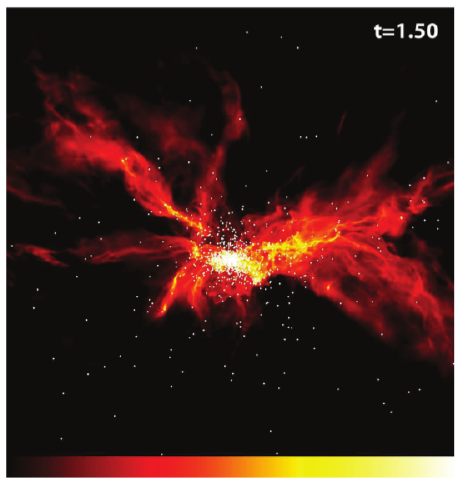}}
\centerline{\includegraphics[width=2.8in]{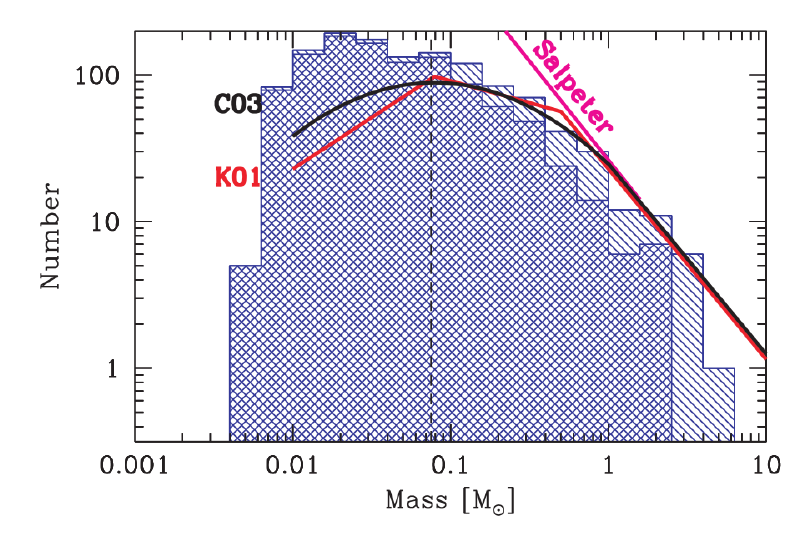}}
\caption{
\label{fig:bate09}
Results from a simulation by \citet{bate09b}, reproduced by permission. The top panel shows the logarithm of projected density, with the color scale running from $10^{-1.4} - 10^1$ g cm$^{-2}$, after $1.5$ free-fall times of evolution. The region shown is 0.8 pc on a side. The bottom panel shows the distribution of the stellar masses in the simulation at the time shown in the upper panel. The single-hatched region shows all objects, and the double-hatched one shows only those objects that have stopped accreting. For comparison, the magenta line is the \citet{salpeter55a} slope $\alpha=-2.35$, and the red and black lines are the functional forms for the IMF proposed by \citet{kroupa01a} and \citet{chabrier03a}, respectively.
}
\end{figure}

The only physical processes invoked in the above argument are hydrodynamics and gravity, and simulations containing only these two ingredients indeed appear capable of producing a powerlaw distribution of masses with a slope similar to the observed one \citep{bonnell03a, bate03a, bonnell04a, bate09b}. Including radiative heating due to protostars does not appear to change this fundamental result \citep{bate09a}. Figure \ref{fig:bate09} shows an example result from a simulation of this type. \red{The simulation shown, and numerous other ones like it, begin with a gas cloud $\sim 1$ pc in size, which is seeded with an initial turbulent velocity field such that the initial virial ratio $\alpha_{\rm vir} \sim 1$. The cloud is then allowed to evolve in response to hydrodynamic and gravitational forces.\footnote{\red{There is one complication that applies to the simulation shown in Figure \ref{fig:bate09}, and to all other simulations I discuss in this section. None of these simulations have the dynamic range required to resolve the stars themselves. Instead, they represent stars using ``sink particles" \citep{bate95a, krumholz04a, federrath10a, gong13a}: in regions where the gas density exceeds as specified threshold, and the gas is bound and gravitationally unstable, the code converts some of the gas into a Lagrangian point particle that continues to interact gravitationally with the remainder of the simulation, and can accrete gas, but no longer feels or exerts pressure forces. The exact recipe used to determine when and where to insert sink particles, and how to accrete mass onto them once they are created, varies from code to code. In the remainder of this section, when I refer to the mass distribution of stars in the context of numerical simulations, readers should understand that I am really referring to the mass distributions of sink particles that are intended to represent stars.}}}

A common feature of these models is that there is essentially no correlation between the properties of the gas around a star at the time of its formation and its final mass \citep{bonnell04a, smith09b}. (However see \citet{chabrier10a} for a counter-argument.) Instead, the mass that ends up in stars, at least more massive ones, is drawn from a region comparable to the size of the entire initial cloud, and one cannot identify any bound structures with masses comparable to the masses of more massive stars and smaller than that of the cloud as a whole.

However, not all simulations based on hydrodynamics and gravity produce the observed IMF powerlaw slope. The models that are successful begin with a uniform density and initial turbulence that is at virial levels (i.e., $\alpha_{\rm vir} \approx 1$), but no mechanical feedback or cascade from larger size scales to maintain the turbulence or keep $\epsilon_{\rm ff}$ small. In contrast, simulations that start with little initial turbulence produce flatter mass spectra $dN/dm\propto m^{\alpha}$ with $\alpha\approx -1.5$ \citep{klessen98a, klessen00a, klessen01b}. This appears to occur because in this case the stars pack close together along filaments, and their tidal fields inhibit accretion onto their neighbors, resulting in an accretion law roughly following $\dot{m}\propto m^{2/3}$. One can also produce similar variations in the mass spectrum by starting with supervirial conditions \citep{clark08a}, or even by starting with virialized turbulence but varying the initial density distribution, with steeper density distributions leading to progressively flatter and more top-heavy mass functions \citep{girichidis11a, girichidis12a}\blue{, while mass distributions that are closer to uniform produce few or no massive objects \citep{martel06a, urban10a}}. In contrast, simulations that begin from conditions of saturated turbulence also seem able to produce powerlaw mass distributions whose slope matches that of the observed IMF, but via a mechanism that is fundamentally different than competitive accretion and bears much more resemblance to the turbulence-based models described in Section \ref{sssec:imfturb} \citep{padoan07a, schmidt10a, krumholz12b}.

The need for carefully-selected initial conditions to match the observed mass spectrum is a symptom of a more general potential problem with the model. Competitive accretion only appears to operate in the presence of decaying turbulence \citep{krumholz05e, bonnell06c}, since it relies on a coherent supersonic collapse to bring both gas and stars into a small arena where the competition can take place. If there is significant feedback or the turbulence is not allowed to decay to low levels, this inhibits the coherent collapse on which competitive accretion depends \citep{wang10a, krumholz12b}. The difficulty arises in the fact that the same mechanisms that seem to be required to reproduce the observed star formation rate, and possibly also the stellar clustering fraction, may also prevent competitive accretion from occurring. Indeed, there are no published simulations\footnote{As mentioned in Section \ref{ssec:starstructure}, \citet{bonnell11a} report a value of $\epsilon_{\rm ff} = 0.04$ for one of their simulations that does show competitive accretion. However, this reported value is computed via a methodology that is at odds with both observational constraints and with the standard used in other numerical work. See footnote \ref{note:bonnell11a}.} that show competitive accretion but also have $\epsilon_{\rm ff} \lesssim 0.1$ \citep[see also][]{krumholz07e}, and it is far from clear that it is possible for competitive accretion to occur in an environment with low $\epsilon_{\rm ff}$.

\subsubsection{Turbulence-Based Models}
\label{sssec:imfturb}

The other scale-free process that could naturally explain a powerlaw distribution of stellar masses is turbulence. As discussed in Section \ref{ssec:gmcflows}, turbulence produces powerlaw velocity spectra and nearly-powerlaw spectra of logarithmic density, so it is \textit{a priori} plausible that it might also produce powerlaw distributions of stellar mass. Models based on this premise proceed in three stages. First, one must estimate the distribution of fragment masses without regard to whether they are gravitationally-bound or not (the so-called ``\blue{core}" mass spectrum). Second, one must ask which of these \blue{cores} are gravitationally bound (\blue{making them ``bound cores"}). Third, one must ask how the masses of stars related to the masses of \blue{bound cores}.

\paragraph{The \blue{Core} Mass Function}

The first turbulence-based model for the IMF was proposed by \citet{padoan97a} and extended by \citet{padoan02a} (hereafter PN). They make a simple heuristic argument based on two premises. The first is that stars form from cores of dense gas that are produced by shock passages. The typical mass of these cores is determined by magnetized isothermal shock jump conditions: if we consider a region of initial size $\ell_0$ and mean density $\rho_0$, then when it is hit by a shock of Alfv\'{e}n Mach number $\mathcal{M}_A$, the shock jump conditions dictate that it will be compressed to a characteristic size $\ell_1 = \ell_0/\mathcal{M}_A$ and density $\rho_1 = \rho_0 \mathcal{M}_A$. If cores are produced in the post-shock region, then one might expect their typical mass to be
\begin{equation}
\label{eq:pneqn1}
m_1 \sim \rho_1 \ell_1^3 = \frac{\rho_0 \ell_0^3}{\mathcal{M}_A^2}.
\end{equation}
However $\mathcal{M}_A$ is proportional to the shock velocity, and the characteristic velocity is itself scale dependent; as shown in Section \ref{sssec:velocitystat} (equation \ref{eq:sigmavl}), the velocity dispersion on size scale $\ell$ varies as $\sigma_v(\ell) \propto \ell^{(n-1)/2}$, where the velocity power spectrum is follows a powerlaw $P(k) \propto k^{-n}$. Thus we can write the characteristic Mach number $\mathcal{M}_{A}$ on size scale $\ell$ as
\begin{equation}
\mathcal{M}_{A} = \mathcal{M}_{A,0} \left(\frac{\ell}{\ell_0}\right)^{(n-1)/2}.
\end{equation}
Substituting this into equation (\ref{eq:pneqn1}) gives
\begin{equation}
\label{eq:pneqn2}
m_1 \sim \frac{\rho_0 \ell_0^3}{\mathcal{M}_{A,0}^2} \left(\frac{\ell_1}{\ell_0}\right)^{4-n}.
\end{equation}
One can choose $\ell_0$ and $\rho_0$ to be the outer scale of the turbulence and the mean density of a cloud, respectively, in which case this expression simply reduces to the characteristic mass of cores measured on some size scale $\ell$ smaller than the size of the cloud. The second premise of the argument is that the turbulent flow is self-similar in the sense that the number of cores present due to turbulent fluctuations on size scale $\ell$ is simply proportional to the available volume at that size scale, i.e., $N(\ell)\propto \ell^{-3}$. Combining this scaling with equation (\ref{eq:pneqn2}), PN arrive at
\begin{eqnarray}
\label{eq:pneqn3}
\frac{dN}{d\log m} & \propto & \ell(m)^{-3} \propto m^{-3/(4-n)} \qquad\Rightarrow
\nonumber  \\
\frac{dN}{dm} & \propto & m^{-(7-n)/(4-n)},
\end{eqnarray}
where $\ell(m) \propto m^{1/(4-n)}$ is simply the inverse of the mapping between $m$ and $\ell$ given by equation (\ref{eq:pneqn2}). For powerlaw indices $n=5/3 - 2$, as expected for high Mach number turbulence, this gives an index in the range $\alpha=2.3$ to $2.5$ for the IMF, in good agreement with the observed value.

It is worth pointing out that there are several significant sleights of hand in this argument. First, equation (\ref{eq:pneqn2}) follows from equation (\ref{eq:pneqn1}) only if the Alfv\'{e}n Mach number that one substitutes into the denominator of equation (\ref{eq:pneqn1}) is that on scale $\ell_1$, and \textit{not} that on scale $\ell_0$. However, in the shock jump conditions used to write equation (\ref{eq:pneqn1}), the Alfv\'{e}n Mach number that enters is the pre-shock one, which one would have thought would be simply $\mathcal{M}_{A,0}$. Moreover, the shock jump conditions themselves used in deriving equation (\ref{eq:pneqn1}) are appropriate only for shocks where the flow velocity is perpendicular to the magnetic field, and it is not clear why this is the appropriate assumption to make. Finally, the argument regarding the number of objects $N(\ell)$ as a function of scale $\ell$, and the way this result is used to derive the mass function, are both suspect. For example, it is not clear in deriving equation (\ref{eq:pneqn3}) why one should take $dN/d\log m\propto \ell(m)^{-3}$ rather than $dN/dm$. This uncertainty stems from a lack of a rigorous definition of what an ``object" to be counted at size scale $\ell$ really is.

Despite these ambiguities, however, there is some numerical support for a picture like that proposed by PN. One prediction of this model is that the mass spectrum depends on the powerlaw index $n$ of the turbulence. A second prediction comes from considering the case of a non-magnetic flow, where the post-shock size-scale and density scale with the Mach number as $\ell_1 = \ell_0/\mathcal{M}^2$ and density $\rho_1 = \rho_0 \mathcal{M}^2$. Using these scalings and following the same argument as above, one arrives at a prediction $dN/d\log m\propto m^{3/(5-2n)}$. \citet{padoan07a} conduct HD and MHD simulations of driven turbulence with no gravity, and calculate the mass spectrum of overdensities identified using a clump-finding algorithm. Depending on whether the simulation uses HD or MHD, and depending on the code they use, they find different powerlaw indices for the turbulent power spectrum, ranging from $n=1.9 - 2.2$, and they find that the resulting clump mass spectrum in each simulation is consistent with the predicted value given the measured index $n$ and the presence or absence of magnetic fields.

\citet{hennebelle08b, hennebelle09a, hennebelle13a} (HC) and \citet{hopkins12e, hopkins12d, hopkins13a} helped clarify some of the ambiguities in \citeauthor{padoan02a}'s model by embedding it within, respectively, the Press-Schechter \citep{press74a} and excursion set \citep{bond91a} formalisms for describing the statistics of random fields with structure on many scales. These models for the density distribution in a supersonically-turbulent medium were introduced briefly in Section \ref{sssec:densitystat}. To remind readers, both models hypothesize that the PDF of the log of density, smoothed on size scale $\ell$, is a lognormal. They then provide a formalism for calculating the scale-dependent width $\sigma_s(\ell)$ as a function of the length scale at which the turbulence is injected, $L$, and the Mach number $\mathcal{M}_L$ on that length scale. The Hopkins model goes on to identify this outer scale with the scale height of the disk, and to compute the Mach number on that scale from the requirements of Toomre stability of the disk.

Given such a model, one can use the standard techniques of Press-Schecher or excursion set theory to count the number of objects above some particular density threshold. The procedure is fairly simple. In the Press-Schecter approach, one first considers a particular size scale $\ell$. The mass of an object of density $\rho$ at that size scale is $M(\rho)\sim \rho \ell^3$, and the total mass of objects with density between $\rho$ and $\rho+d\rho$ is simply $\rho p(\rho) d\rho$, where $p(\rho)$ is the PDF of the density, which is trivially related to the PDF $p(s)$ of the log of density. Dividing the total mass of objects at density $\rho$ by the mass $M(\rho)$ of each object gives the number of objects at that mass at that particular size scale $\ell$. To obtain the overall mass distribution, one must then simply integrate over all size scales $\ell$. This final step gives rise to the well-known ``cloud-in-cloud" problem: some objects that are counted at a particular size scale $\ell$ will also be part of a larger object on size scale $\ell' > \ell$, and the integral double-counts them. One can attempt to correct for this approximately. The excursion set approach allows one to perform the correction precisely, by performing a random walk starting at a very large size scale and then adding up the randomly-chosen density fluctuations as the walk proceeds to progressively smaller size scales. The random walk ends when the density reaches the specified threshold value.

\paragraph{From \blue{All Cores to Bound Cores}}

The \blue{core} mass function specifies objects by a particular density threshold, but does not determine whether these objects are bound. \blue{Filtering the cores to identify the bound ones is therefore the next step in the argument.} In the PN model, boundedness is determined by comparing the mass of an object to the Jeans mass, which is determined from the density PDF. The procedure is that one draws a density from the lognormal PDF given by equation (\ref{eq:denpdf}), then evaluates the Jeans mass (equation \ref{eq:jeansmass}) using that density. If the Jeans mass is smaller than the mass of the object, it is considered gravitationally bound and liable to collapse to a star. This does not alter the mass function for high mass objects, because it is extremely unlikely that a draw from the density PDF will produce a density so small that such an object is rendered stable. Thus the slopes of the mass functions \blue{for all cores and for bound cores} are identical. However, the \blue{bound} core mass function does become truncated at low masses, something I discuss further in Section \ref{ssec:imfpeak}.

The effects of the boundedness criterion are more significant for the HC and Hopkins models, because they consider non-thermal support as well as thermal support. In these models, one uses the linewidth-size relation (equation \ref{eq:sigmavl}) to specify the scale-dependent velocity dispersion $\sigma_v(\ell)$. With some additional assumptions one can also make a guess at the magnetic field strength. In the Hopkins formalism, one also knows the degree of rotational support, though this is unimportant except on large scales.

These various sources of support can be combined with thermal pressure to define a boundedness criterion that is, up to factors of order unity, simply an application of the virial theorem (equation \ref{eq:virialthm}). One considers a region bound if the virial theorem gives $\ddot{I} \leq 0$, i.e., if the gravitational potential energy exceeds the sum of the thermal energy, the turbulent kinetic energy, the magnetic energy, and the kinetic energy in ordered rotation. All of these energies are fully determined by the size scale $\ell$ under consideration and the local density $\rho$ smoothed on this size scale. Thus one can rewrite the condition for boundedness as simply a condition on the density; note that this is almost exactly the same as the procedure used in \citet{krumholz05c} and similar models for the turbulence-regulated star formation rate. It is also subject to some of the same weaknesses, the most notable of which is that this procedure assumes that density and velocity are uncorrelated, while it is well known that in a turbulent medium density and velocity are in fact highly-correlated \citep[e.g.,][]{padoan01a}.

Armed with the scale-dependent threshold density $\rho_{\rm th}(\ell)$, one can derive the mass function using much the same procedure as is used to compute the mass function with a fixed, scale-independent density threshold. In the HC Press-Schecter-based model the procedure is exactly the same, while in the excursion set formalism one again performs a random walk to compute the scale-dependent density, but one identifies cores with the smallest scale at which the density exceeds the threshold, rather than the largest. The qualitative effect of the scale-dependent threshold is to steepen the mass function compared to a fixed threshold. This steepening occurs because many of the most massive objects tend to be found on relatively large size scales, where the turbulent support is also greatest, and this leads to a suppression of the high end of the mass function. As in the PN model, thermal support at the low-mass end leads to a turnover in the core mass function below a certain mass.

\paragraph{From \blue{Bound} Cores to Stars}

The final assumption in the turbulence-based models for the origin of the IMF is that there is a self-similar mapping between \blue{bound} core and star masses, so that the CMF and IMF have the same functional form. This is far from an obvious conclusion, particularly for cores on size scales $\ell$ such that $\sigma_v(\ell) \gg c_s$, i.e., where the cores are expected to have significant internal turbulence. When such a self-gravitating structure forms and decouples from the surrounding turbulent flow, why does it not fragment further? Simulations of turbulent cores that include only the physics included in the PN, HC, and Hopkins models (i.e., hydrodynamics plus gravity) show that this is exactly what happens, with a result that the final stellar masses bear no resemblance to the initial core mass \citep{dobbs05a}. This problem applies even to cores with $\sigma_v(\ell) \sim c_s$, because as these collapse, the collapse will drive supersonic turbulence within them \citep{huff07a, klessen10a, robertson12a}, and hydrodynamic simulations confirm that even cores with little initial turbulence can still fragment \citep{goodwin04a, walch12b}.

One can also ask the question another way: once a self-gravitating turbulent structure forms in a turbulent flow, why should should one not apply something like the PN, HC, or Hopkins formalism to it, and calculate its sub-fragmentation into yet smaller pieces? The analogy with cosmology is instructive. When one uses something like the Press-Schecter or excursion set formalism to calculate that a dark matter halo has collapsed, one does not expect the dark matter in that halo to further sub-fragment. However, this is only because the dark matter is collisionless and lacks a way to dissipate energy, so that when it collapses its velocity dispersion rises and stabilizes it against further fragmentation. The baryons may also undergo initial shock heating as they fall into a dark matter halo, but if they are able to cool rapidly, they will not stabilize. They will instead collapse and fragment to smaller and smaller scales, so that the final mass of the stable baryonic structures bears no resemblance to the mass of the structure into which they initially collapsed. It would seem that molecular gas should behave like the baryons in this cosmological example, but the PN, HC, and Hopkins formalisms amount to treating them like the dark matter instead. \blue{Indeed, \citet{hopkins13a} realizes this problem and notes that, in the absence of some support that halts the collapse, application of his formalism indeed suggests that fragmentation should continue indefinitely to ever-smaller scales.}

There are a number of possible ways around this problem. Simulations including physics other than simply hydrodynamics and gravity shows that there are a number of physical effects that strongly suppress fragmentation. Radiation feedback from embedded massive stars raises the Jeans mass, and is capable of causing a cores to collapse more or less monolithically rather than undergoing sub-fragmentation \citep{krumholz06b, krumholz07a, krumholz08a, bate09a, offner09a, krumholz10a, offner10a, krumholz12b, bate12a}. Magnetic fields, particularly in combination with radiation feedback, are also effective at suppression fragmentation \citep{hennebelle11a, commercon11c, myers13a}. However, none of these processes are currently included in the analytic models, and there is no compelling reason to believe that they allow a simple one-to-one mapping from the CMF to the IMF in all circumstances and environments. Instead, it seems likely that fragmentation or non-fragmentation within cores will modify the mapping from CMF to IMF in non-trivial ways that models have not yet begun to explore.

\paragraph{\blue{Observational Implications of Turbulence-Based Models}}

\blue{
Turbulence-based models make a strong prediction that there should be a correspondence between the mass functions of cores (CMFs) and stars. Depending on the details of the model, this correspondence can apply only for bound cores, or for all cores regardless of their boundedness. Such a comparison is not quite as trivial as it might first seems, however, due to questions of timescale. A one-to-one mapping from bound cores to stars means that formation rate of stars matches the formation rate of the cores that produce them, but the observed CMF is the product of the formation rate distribution and the lifetime distribution.\footnote{Precisely the same issue arises in comparing the mass functions of star clusters and the gas clouds in which they form -- see Section \ref{ssec:icmf}.} How this should influence the relationship between the observed CMF and the IMF depends on ones models for how cores evolve in time. For example, \citet{clark07a} point out that, if all cores are one thermal Jeans mass in size, then their collapse times should be proportional to their masses, in which case we should find $dN_{\rm core}/dM_{\rm core} \propto M_{\rm core}(dN_*/dM_*)$. In contrast, if one assumes that cores have masses of one turbulent Jeans mass, or equivalently that they have constant bounding pressures, then, the difference between core and star mass function is much reduced: $dN_{\rm core}/dM \propto M_{\rm core}^{0.25}(dN_*/dM)$ \citet{hennebelle09a}. \citet{padoan11b} point out that one must consider not only core collapse timescale, but also assembly timescales. They argue that, because cores should be collapsing and putting mass into stars at the same time as they are assembling, the observable gas mass in a core will generally be less than that of the stars they produce. The difference in mass is particularly large for the cores producing the most massive stars, such that very massive bound cores should be far rarer than massive stars. 
} 

\blue{The observational situation is only slightly clearer than the theoretical one. Observations of the CMF and its relationship to the IMF} are reviewed in \citet{offner14a}, and I refer readers there for a more thorough discussion. A short summary is that the mass function of cores in many regions, and measured using many techniques, looks remarkably like the stellar IMF. There is a powerlaw tail at high masses with a slope of roughly $dN/dM\propto M^{-\alpha}$ with $\alpha \approx 2$ to $2.5$ \citep{motte98a, testi98a, johnstone00a, reid05a, reid06a, enoch07a, alves07a, enoch08a, roman-zuniga10a, nutter07a, andre10a, konyves10a}. \blue{This result is at least qualitatively consistent with the idea that the CMF and the IMF should be similar, and that there is a relative straightforward mapping between the two, without a very large difference in evolution timescales between low and high mass cores. However, the data only extend to core masses of $\sim 10$ $M_\odot$.}

In some observational studies the authors also report the detection of a break in the CMF at a mass $\sim 1$ $M_\odot$, giving it a distinct peak similar to the stellar IMF but shifted to higher mass by a factor of $\sim 3-4$. This detection may be correct, but one can also justifiably treat it with skepticism, because the detected peak is always within a factor of a few of the completeness limit of the survey, and does not appear to be invariant when this completeness limit varies For example, \citet{reid06b} collect a large number of published CMF determinations, including ones for distant regions where the mass completeness limit is much higher than in the closer regions. They find that all the CMFs follow a common shape, with a powerlaw of slope $\alpha\approx 2.3$ at high masses and a break at lower masses, but that the mass at which the break takes place is not common from one region to another. Instead, the break is at lower masses in more nearby regions, and at higher masses in more distant regions.

More recently, \citet{andre10a} found that the CMF in the Aquila region, with a completeness limit of $\approx 0.3$ $M_\odot$, peaks at $\approx 0.6$ $M_\odot$, while the CMF in the Polaris region, with a completeness limit of $\approx 10^{-2}$ $M_\odot$, peaks at $\approx 2\times 10^{-2}$ $M_\odot$. The Polaris cloud is much less dense than Aquila, and it is possible that the difference in the CMF is a real reflection of the physical conditions in the cloud, but the fact that in both regions $M_{\rm peak} / M_{\rm completeness} \approx 2$ is highly suggestive that something observational rather than physical is at work. \blue{\citet{andre10a} report that most of the cores they identify in Polaris region are not gravitationally bound, and thus it is possible that this discrepancy might be reduced if one selected cores based on a physical boundedness criterion, rather than one of the more commonly-used clump-finding or thresholding algorithms that effectively identify structures based on contrast in position-position or position-position-velocity data. However, this has yet to be demonstrated.}

\subsection{The IMF Peak}
\label{ssec:imfpeak}

\subsubsection{The Problem of the Mass Scale}

Explaining the peak of the IMF is a somewhat distinct challenge from explaining the high mass slope. A powerlaw is scale-free, but the peak singles out a particular mass scale. The general challenge that one faces in explaining the peak is that the physical mechanisms most often invoked to explain the slope -- isothermal ideal magnetohydrodynamics and gravity -- are provably scale-free \citep{mckee10b, krumholz11e}. One can demonstrate this result simply by non-dimensionalizing the equations governing such a system, which are
\begin{eqnarray}
\frac{\partial \rho}{\partial t} & = & -\nabla\cdot (\rho\mathbf{v}) \\
\frac{\partial}{\partial t}(\rho \mathbf{v}) & = & -\nabla \cdot \left(\rho\mathbf{vv}\right) - c_s^2 \nabla \rho \nonumber \\
& & {} + \frac{1}{4\pi} (\nabla\times\mathbf{B})\times\mathbf{B} - \rho \nabla \phi \\
\frac{\partial\mathbf{B}}{\partial t} & = & -\nabla\times\left(\mathbf{B}\times\mathbf{v}\right) \\
\nabla^2 \phi & = & 4\pi G \rho
\end{eqnarray}
Here $\rho$ is the density, $\mathbf{v}$ is the velocity, $\mathbf{B}$ is the magnetic field, $c_s$ is the (constant) sound speed, and $\phi$ is the gravitational potential. The first two equations express mass and momentum conservation, the third expresses magnetic flux freezing, and the final equation is the Poisson equation for the gravitational potential. One can non-dimensionalize these equations by choosing a characteristic length scale $L$, velocity scale $V$, density scale $\rho_0$, and magnetic field scale $B_0$, and making a change of variables $\mathbf{x} = \mathbf{x}'L$, $t = t' L/V$, $\rho = r \rho_0$, $\mathbf{B} = \mathbf{b} B_0$, $\mathbf{v} = \mathbf{u} V$, and $\phi = \psi G\rho_0 L^2$. With fairly minimal algebra, the equations then reduce to
\begin{eqnarray}
\frac{\partial r}{\partial t'} & = & -\nabla'\cdot (r\mathbf{u}) \\
\frac{\partial}{\partial t'}(r \mathbf{u}) & = & -\nabla' \cdot \left(r\mathbf{uu}\right) - \frac{1}{\mathcal{M}^2}\nabla' r \nonumber \\
& & {} + \frac{1}{\mathcal{M}_A^2} (\nabla'\times\mathbf{b})\times\mathbf{b} - \frac{1}{\alpha_{\rm vir}} \nabla' \psi \\
\frac{\partial\mathbf{b}}{\partial t'} & = & -\nabla'\times\left(\mathbf{b}\times\mathbf{u}\right) \\
\nabla'^2 \psi & = & 4\pi r,
\end{eqnarray}
where $\nabla'$ indicates differentiation with respect to $x'$. The dimensionless ratios appearing in these equations are
\begin{eqnarray}
\mathcal{M} & = & \frac{V}{c_s} \\
\mathcal{M}_A & = & \frac{V}{V_A} = V\frac{\sqrt{4\pi \rho_0}}{B_0} \\
\alpha_{\rm vir} & = & \frac{V^2}{G \rho_0 L^2},
\end{eqnarray}
which are simply the Mach number, Alfv\'{e}n Mach number, and virial ratio for the system. These three dimensionless numbers fully characterize the evolution of the system.

The reason to write out the non-dimensionalization explicitly is to point out that $\mathcal{M}$, $\mathcal{M}_A$, and $\alpha_{\rm vir}$ are all left invariant if we change the density, length, and magnetic field scales by $\rho_0' = f \rho_0$, $L' = f^{-1/2} L$, and $B_0' = f^{1/2} B_0$, where $f$ is an arbitrary positive number. However, under such a rescaling, the masses of all structures change by a factor $(\rho_0'/\rho_0)(L'/L)^3 = f^{-1/2}$. One implication of this for numerical simulations is that it is impossible to claim that a simulation containing only ideal isothermal MHD produces objects of a particular characteristic mass, as the simulation could always be rescaled to change that mass to an arbitrary value; indeed, many isothermal simulations are performed in dimensionless units, and are scaled to dimensional units only after the simulation is run.

A much broader implication is that \textit{any} theory based solely on ideal isothermal MHD and gravity cannot possibly explain the origin of the IMF without adding some additional assumption or argument to specify the values of the dimensional quantities, at least to the point that one is no longer free to rescale the combination $\rho_0 L^3$ to arbitrary values. These additional assumptions or arguments will then contain the key physics that determines the stellar mass scale. Theories for the origin of the peak of the IMF can be divided into two groups based on what additional piece of physics they choose to add to assign a definite mass scale. One approach is to invoke some galaxy-scale physics to set an outer scale for molecular clouds or the turbulence in them, and to derive the characteristic mass scale from that, thereby linking the characteristic mass of stars to some property of galaxies. The other is to invoke deviations from purely isothermal behavior.\footnote{A third potential approach, which has not been explored thus far and that I will not discuss further, would be to invoke a non-ideal MHD process to set the characteristic mass scale. However, one can show that the most obvious candidate process, ambipolar diffusion, does not add a characteristic mass scale as long as the ionization fraction behaves as a powerlaw function of density \citep{mckee10b}. Thus a more complex non-ideal MHD mechanism would be required.}

\subsubsection{Setting the IMF Peak from Galactic Properties}
\label{sssec:imfpeakgal}

\paragraph{The Jeans Mass Hypothesis}

The simplest hypothesis for the origin of the peak of the IMF is that it simply reflects the mean-density Jeans mass (equation \ref{eq:jeansmass}) in the star-forming cloud \citep[e.g.,][]{larson92a, bate05a}, and a number of authors have applied this hypothesis to cosmological models \citep[e.g.,][]{tumlinson07a, narayanan12b, narayanan13a}. This hypothesis was also included in many of the early competitive accretion models (see Section \ref{sssec:compacc}) to provide the characteristic value for the narrow spectrum of initial fragment masses that competitive accretion would broaden, though in principle one could also couple the competitive accretion explanation for the IMF's high mass tail with any model for the origin of the IMF peak. 

There is some numerical support for the proposition that characteristic fragment masses trace the mean-density Jeans mass, as this behavior is seen in a number of simulations performed with an isothermal equation of state \citep{klessen98a, klessen00a, klessen01a, bate03a, clark05c, bonnell06a}. In these simulations, the initial density and sound speed, the two quantities that enter the Jeans mass, are simply set by hand. As a theory for the origin of the IMF more generally, this approach amounts to the hypothesis that star-forming molecular clouds have some characteristic mean density and temperature, or range of these values, and that the peak of the IMF is determined by whatever processes select them.

This model has some serious shortcomings. The first is that, in the real world as opposed to idealized simulations, it is not obvious what should be counted as the ``cloud" whose mean density determines the location of the IMF peak. As discussed in Section \ref{sssec:whichsf}, there is a phase transition from the atomic to the molecular ISM that appears to be important in allowing star formation to begin, and so one can justifiably limit the material under consideration to molecular gas. However, even within molecular clouds the range of densities is immense, ranging from $\sim 100$ cm$^{-3}$ if one considers all the material traced by CO to $>10^6$ cm$^{-3}$ or more if one considers only very high-density tracers. There is no obvious physical justification for choosing one scale or another, but the predicted characteristic mass can vary by orders of magnitude depending on how this choice is made.

A second major difficulty is that this model would seem to predict that the IMF should vary strongly from one star-forming cloud to another even within the Milky Way, let alone in external galaxies where the range of star-forming environments is much broader. As discussed in Section \ref{sssec:imfresolved}, the data on all young clusters with spatially-resolved stellar populations appears consistent with a universal IMF peak, despite the fact that the gas and stellar densities of the clusters in question varies by many orders of magnitude. \blue{While the data certainly do not rule out variations in the IMF peak mass at the factor of a few level, and at the extreme perhaps even the factor of ten level, if the Jeans mass hypothesis were correct than the variations should be substantially larger than that. For example, \citet{hartmann02a} finds a mean stellar separation of $\sim 0.25$ pc$^{-3}$ in Taurus, corresponding to a density $\sim 60$ stars pc$^{-3}$.\footnote{\blue{\citet{hartmann02a}} notes that the stars are mostly strung out along linear filaments, so the true volume density is in fact probably lower than this.} In comparison, the central density in the Orion Nebula Clusters is roughly $2\times 10^4$ stars pc$^{-3}$ \citep{hillenbrand98a}. If the characteristic stellar mass really varies as the square root of density, one would expect a factor of $\sim 20$ variation between these regions, far too large to be consistent with observations. This variation may be partly offset if the temperatures of the gas increase with density \citep{padoan07a, elmegreen08a}, but it would require something of a coincidence for this to cancel out the density variation precisely enough to leave the IMF peak unchanged.}

\paragraph{The Turbulent Jeans Mass Hypothesis}

As discussed in Section \ref{sssec:imfturb}, simulations in which the turbulence is initially saturated, or in which it is driven to that state by stellar feedback, appear to fragment in a manner that is fundamentally different than simulations starting from smooth initial conditions. This affects the peak of the IMF as well as the powerlaw tail. The PN, HC, and Hopkins models that predict the slope of the IMF in a turbulent medium also predict a location for its peak, and, as with the simple Jeans hypothesis, this location is ultimately set by galactic-scale processes.

In the PN model, one starts with a powerlaw distribution of fragment masses and then converts that to the IMF by scaling the number of objects at a given mass by the probability that an object of that mass exceeds the Jeans mass. One determines this probability by drawing a random density from a lognormal PDF, calculating the associated Jeans mass, and comparing that to the mass of the fragment. For a low-mass fragment, the randomly-drawn density must be quite high for the fragment mass to exceed the Jeans mass, and this greatly reduces the number of low-mass fragments that are counted as eventually forming stars. The net effect is to impose a lognormal-shaped cutoff for masses below the Jeans mass, evaluated at the peak of the density PDF.

Though the exact implementation is different, the HC and Hopkins models rely on essentially the same mechanism: small-scale, low-mass structures in the turbulent density field are extremely unlikely to yield stars, because thermal support renders it improbable that they will be gravitationally-unstable. \red{In these models, if one defines structures as connected regions where the density exceeds a specified, constant threshold, and counts the number of such structures as a function of their mass, the result is a pure powerlaw. However, with a constant density threshold, the lowest mass of these objects will not be gravitationally bound. If, instead of using a constant density threshold, one requires that the object be gravitationally bound, then the threshold density rises rapidly at small mass or length scales. This rising threshold} gives rise to a lognormal suppression of the number of fragments at low mass.

In all of these models, thermal support induces a break in the stellar mass function at a mass corresponding to the Jeans mass evaluated at the median density, which is much higher than the mean density due to turbulent compression. From equation (\ref{eq:denpdf}), the median density corresponds to $s_0 = \ln(\rho/\overline{\rho}) = \sigma_s^2/2$, and with the aid of equation (\ref{eq:sigmas}) for $\sigma_s$, this gives
\begin{equation}
\rho_{\rm med} = \overline{\rho} \exp\left[\frac{1}{2}\ln\left(1+b^2\mathcal{M}^2\frac{\beta_0}{\beta_0+1}\right)\right].
\end{equation}
Thus the median density is larger than the mean density by a factor that, in the limit $\mathcal{M}\gg 1$, approaches $\mathcal{M}$ times a factor of order unity that depends on $b$ and $\beta_0$. If we define $M_{J,0} = (\pi^{3/2}/8) c_s^3/(G^3\overline{\rho})^{1/2}$ as the Jeans mass evaluated at the mean density (equation \ref{eq:jeansmass}), then when $\mathcal{M}\gg 1$ we see that the characteristic peak of the IMF will be at a mass
\begin{equation}
\label{eq:mpeak1}
M_{\rm peak} \approx \frac{M_{J,0}}{\mathcal{M}}.
\end{equation}
One may think of this as the turbulent Jeans mass, since it is simply the Jeans mass evaluated using the median density that results from turbulent compression rather than the mean density that would prevail without turbulence. This result would seem to suggest that the IMF peak should depend on three quantities set by the large-scale properties of molecular clouds, $\rho_0$, $c_s$, and $\mathcal{M}$. However, two of these can be combined in a simple way. With a little algebra, one can show that for turbulence following a Burgers' spectrum (powerlaw index $n=2$), the peak mass can also be expressed as (up to factors of order unity)
\begin{equation}
\label{eq:mpeak2}
M_{\rm peak} \approx  \frac{M_{J,0}}{\mathcal{M}} \approx \alpha_{\rm vir}^{1/2} \ell_s \frac{c_s^2}{G}.
\end{equation}
Even if the powerlaw index $n$ is not precisely $2$, this results holds approximately for any reasonable value of $n$. Thus to the extent that clouds are roughly virialized, $\alpha_{\rm vir}\approx 1$, under this hypothesis the IMF peak mass depends only on the normalization of the linewidth-size relation $\ell_s$ and the sound speed $c_s$. Because it is linked to the sonic length, this mass is sometimes referred to as the sonic mass. \blue{Alternately, for a virialized cloud, one may write this mass in terms of the cloud column density \citep{padoan07a}, relying on the one-to-one mapping between column density and $\ell_s$ that holds for virialized clouds \citep{larson81a}.}

As with the proposal that the IMF peak is set by the Jeans mass alone, a model in which it is set by the turbulent Jeans mass ultimately links the characteristic mass of stars to whatever galactic-scale processes are responsible for setting the characteristic temperatures and linewidth-size relations of molecular clouds. In the PN and HC models, these galactic scale processes are left unspecified. \citet{hopkins13e} assumes that the phase transition between atomic and molecular gas has negligible dynamical effects and simply treats the molecular clouds as part of a continuous turbulent cascade starting at galactic scales. This allows him to compute the velocity dispersion at size scales of a galactic scale height by requiring that this be sufficient to put the disk in a state of marginal gravitational stability. Under these assumptions, one can express the mass of the IMF peak as
\begin{equation}
M_{\rm peak} \approx \frac{c_s^4}{Q G^2 \Sigma},
\end{equation}
where $Q\approx 1$ is the \citep{toomre64a} stability parameter for the disk, and $\Sigma$ is the gas surface density. Thus Hopkins predicts that the parameters controlling the IMF are the gas sound speed and the gas surface density of galactic disks. If one assumes that the bulk of the stars in present-day elliptical galaxies were formed in starbursts characterized by very high surface densities, then for plausible values of $\Sigma$ and $c_s$ this dependence may explain the shift in IMF peak to lower masses seen in elliptical galaxies (see \ref{sssec:unresolvedimf}). However, the parameters must be chosen by hand to recover the correct sense of variation, since there are two effects pushing in opposite directions: galaxies undergoing intense star formation will tend to have higher $c_s$, pushing to higher masses, but also higher $\Sigma$, pushing the peak to lower masses. It is not \textit{a priori} obvious which of these effects should dominate, and there is a great deal of freedom in how to choose the parameters because the temperature and thus the sound speed within starburst galaxies is highly non-uniform. Thus one could also choose plausible parameters that would yield an increase rather than a decrease in $M_{\rm peak}$ in such environments.

The turbulent Jeans mass model has a significant advantage over the simple Jeans mass hypothesis, in that the normalization of the linewidth-size relation is observed to be essentially the same in all molecular clouds traced by CO within the Milky Way \citep{heyer09a}, and also does not appear to vary much for CO clouds in other local galaxies with similar properties \citep{bolatto08a}. This is in line with Hopkins's prediction that the normalization of the linewidth-size relation is related to the gas surface density of galaxies, which does not vary tremendously over the most nearby galaxies. This could help explain why the IMF does not vary much. On the other hand, as discussed briefly in Section \ref{sssec:densitystat}, it is not the case that the linewidth-size relation is actually constant within clouds. Regions of active star cluster formation tend to show linewidths that are a factor of $\sim 5$ higher than would be predicted using the linewidth-size relation measured for CO clouds \citep{plume97a, shirley03a}. This corresponds to a factor of $\sim 25$ difference in the predicted value of $M_{\rm peak}$. \blue{Similarly, in M51, which is substantially more molecule-rich than the Milky Way, \citet{hughes13a} find that linewidths are a factor of $\sim 2$ larger than in M33 or the LMC for CO clouds of the same size, corresponding to a factor of $\sim 4$ predicted difference in $M_{\rm peak}$.}

Thus one is left with precisely the same problem as in the simplest Jeans mass hypothesis: the answer depends sensitively on whether the sonic length used in evaluating the model is the one that applies to molecular clouds at large, low-density scales, or the one that applies at small, high-density scales. There is no obvious reason in the theoretical models to prefer one or the other, which means that the prediction is highly ambiguous. \blue{Furthermore, the model predicts that the IMF should vary systematically in galaxies like M51 with higher surface densities. As with the simple Jeans mass hypothesis, variations in temperature that correlate with variations in density may cancel out some of these effects, but there is no obvious reason why the cancellation should work as well as it would need to in order to explain the observed lack of IMF variation.}

\paragraph{Is There a Characteristic Fragment Mass of Isothermal Turbulence?}

A final concern for both the simple and turbulent Jeans mass hypotheses, which might at first seem technical but that actually exposes a deep physical issue, is that it is not clear that any of the simulations used to investigate these models are converged. It is noteworthy that no published simulation of isothermal fragmentation of initially-turbulent gas has \textit{ever} demonstrated numerical convergence, and the one published convergence study in the literature, performed by \citet{martel06a}, found that the problem does not have a converged solution. Instead, they found that, as the resolution increases, the characteristic fragment mass decreases, apparently without limit. Similar behavior is seen under isothermal conditions in more idealized settings. \citet{kratter10a} find that the fragment mass does not converge in simulations of isothermal gravitationally-unstable accretion disks.

One possible explanation for this non-convergence comes from a particularly simple problem first studied by \citet{boss91a} and now widely used as a code test: the collapse of a cloud with a Gaussian initial density profile in solid body rotation, and with a small-amplitude $m=2$ initial density perturbation. The first simulations of such a system found that it fragmented into multiple objects, but \citet{inutsuka92a} showed analytically that a filament of the type formed in the simulation should not fragment at all, but instead should collapse to infinite line density before it fragmented into singular point masses. The disagreement between simulation and analytic theory remained unresolved until \citet{truelove97a} were able to recover the analytic result numerically by using an adaptive mesh code that progressively increased the resolution as needed to follow the collapse. They found that, as long as they increased the resolution, the filament would not fragment. It did so only if they either changed the equation of state away from isothermal \citep[see also][]{boss00a} or stopped increasing the resolution to follow the rising density. In the latter case, the characteristic mass of the fragments was a decreasing function of the numerical resolution, in precisely the same manner found by \citet{martel06a}. 

The conclusion to draw from this numerical result is that the question ``what characteristic fragment mass is produced by the collapse of a Gaussian cloud in solid-body rotation?" is ill-posed. There is no characteristic fragment mass, because the collapse does not produce point-like fragments. While this is an idealized case, it is noteworthy that filamentary structures of the type formed in a rotating Gaussian cloud are ubiquitous in simulations of isothermal turbulence as well. Thus it seems at least plausible that the non-convergence observed by \citet{martel06a} is, like that found by \citet{truelove97a}, not a failure of numerics but a result of a real physical phenomenon: that the correct answer to the problem of how an isothermal, turbulent, self-gravitating medium collapses is a system of singular filaments rather than singular points, implying that questions about the characteristic mass spectrum produced by such media are ill-posed.\red{\footnote{\red{There appears to be an analogous issue in N-body simulations of warm/hot dark matter, where cosmological filaments that should be stable instead fragment into clumps due to finite numerical resolution \citep{wang07a, breysse14a}.}}} If this is the case, then it would be a fatal blow to all models based on isothermal fragmentation, because the characteristic fragment mass in such an isothermal system would be undefined. Thus there is an urgent need to reinvestigate and settle the question of whether isothermal fragmentation actually produces a well-defined mass scale under any circumstances.

\subsubsection{The IMF Peak from Non-Isothermal Fragmentation}
\label{sssec:imfpeaknoniso}

The alternative approach to explaining the peak of the IMF is to invoke deviations from isothermality. Recall from the discussion in Section \ref{ssec:gmcthermo} that isothermality in molecular clouds holds only approximately, and some of these deviations from isothermality may be important to setting the characteristic mass of stars. Several possible mechanisms have been proposed for how non-isothermality could determine the location of the IMF peak.

\paragraph{Barotropic Equation of State Models}

One class of models for how the IMF could be set involves giving up on the assumption that gas is isothermal, but continuing to treat the gas as barotropic, meaning that the pressure and temperature are determined solely by the gas density. In order to understand what such a barotropic equation of state should look like, one must review the important heating and cooling processes that take place in star-forming clouds. This topic has been explored by a large number of authors \citep{larson73a, larson85a, masunaga98a, masunaga00a, goldsmith01a, omukai00a, omukai05a, commercon11b, vaytet13a}, and what I present here is merely a short summary of the results of these papers.

As discussed in Section \ref{ssec:gmcthermo}, the dominant heating source in clouds depends on their column density, with grain photoelectric heating dominating at low column density, and cosmic ray heating dominating at higher column density when photoelectric heating is blocked. If gas is collapsing, adiabatic compression that occurs as the gas density rises can also be a significant source of heating. The heating rate due to this process is highly density-dependent, since the gas cannot compress on a timescale shorter than the free-fall time, which varies as density to the $-1/2$ power.

The competing cooling processes are line emission and collisional coupling with cold dust. Line cooling can be dominated by C$^+$ or CO lines, depending on the chemical state of the gas, and the cooling rate per unit volume can vary with density as either $n^2$, if the density is below the critical density of the cooling line, or $n$, if the density is above the critical density. The rate of thermal exchange with the dust also varies as $n^2$, and at high densities thermal exchange with dust is always dominant. At Solar metallicity, tight dust-gas coupling is established at a density of $\sim 10^{4-5}$ cm$^{-3}$. In this case the gas and dust temperatures become locked together. At the lower-density end of this regime of tight dust-gas coupling, the dust is nearly isothermal because it is able to reach thermal equilibrium with the ambient radiation field very quickly; thus the gas is close to isothermal too. However, if the rate of gas heating due to adiabatic compression becomes too high, the dust will not be able to keep up, and the temperature of both dust and gas will rise. The same thing will happen if the column density rises high enough for the dust to become optically thick to its own cooling radiation. 

The net effect of combining these processes depends on a number of free parameters, such as the column density of the region under consideration, the strength of the external radiation field in both the infrared and ultraviolet parts of the spectrum, and the collapse rate of the gas. Most numerical models of these processes are calculated for one particular physical scenario, most often an isolated, $\sim 1$ $M_\odot$ cloud of gas undergoing free-fall collapse, and subjected to no external radiation except a low-level galactic background. In this scenario, one can map out the behavior of temperature versus density and write this as an approximate equation of state.

The most widely used barotropic equations of state are due to \citet{masunaga98a} and \citet{larson05a}. \citeauthor{masunaga98a}~focus on the regime after tight dust-gas coupling has been established. They find that the gas is essentially isothermal over the density range $n_{\rm H} \sim 10^5-10^{10}$ cm$^{-3}$. At even higher densities, the dust can no longer keep up with adiabatic contraction heating, and the gas becomes approximately adiabatic, such that $T \propto \rho^{\gamma-1}$ with ratio of specific heats $\gamma\approx 1.4-1.7.$\footnote{The value of $\gamma$ is a complex question because, under interstellar conditions, H$_2$ does not behave like either an ideal monatomic or diatomic gas. Deviations from ideal behavior occur because the level spacings are large enough that quantum mechanical effects are non-negligible, and because the ratio of ortho-H$_2$ to para-H$_2$ does not reach thermal equilibrium. As a result, $\gamma$ varies in a complex manner that can depend not only on the instantaneous density and temperature, but also on the thermal history of a particular fluid element, and the conditions under which the H$_2$ molecules formed \citep{black75a, boley07a, tomida13a, krumholz14b}.} This is generally implemented as an equation of state of the form
\begin{eqnarray}
\frac{P(\rho)}{\rho k_B/\mu} & =  &T(\rho) 
\nonumber \\
& =  &T_{\rm min} \left\{
\begin{array}{ll}
1, & n_{\rm H} < n_{\rm bar} \\
(n_{\rm H}/n_{\rm bar})^{\gamma-1}, & n_{\rm H} \geq n_{\rm bar},
\end{array}
\right.
\label{eq:masunagaeos}
\end{eqnarray}
where $\mu=3.9\times 10^{-24}$ g is the mean mass particle in fully molecular gas. Different authors have used different values of the parameters $T_{\rm min}$, $n_{\rm bar}$, and $\gamma$, but typical values are $T_{\rm min} = 10$ K, $n_{\rm bar} \sim 10^8-10^{11}$ cm$^{-3}$ and $\gamma \approx 5/3 - 7/5$. Some authors add an additional branch where $\gamma$ switches from $5/3$ to $7/5$ at some density, or even perform a more detailed calculation to follow the run of $\gamma$ versus $n_{\rm H}$.

In contrast, \citet{larson05a} focuses on the lower-density regime that straddles the density where dust-gas coupling becomes strong. He finds that, at densities below $\sim 10^{5}$ cm$^{-3}$, the temperature is a slightly decreasing function of the density, because the heating rate is dominated by photoelectric and cosmic ray heating, which are both linear in density, while cooling is dominated by C$^{+}$ emission, which has a super-linear density dependence because the density is below the critical density for the C$^+$ 158 $\mu$m line\blue{\footnote{\blue{Note that \citeauthor{larson05a}'s model assumes that the carbon remains C$^+$ rather than CO even at densities up to $10^5$ cm$^{-3}$, so that CO cooling is never important.}}}. Above $\sim 10^5$ cm$^{-3}$, dust-gas thermal exchange takes over as the dominant cooling mechanism, and adiabatic compression as the dominant heating mechanism. The former gives a rate that is linear in density, and the rate for the latter is superlinear, so the net effect is that the gas begins to become warmer as the density increases. \citeauthor{larson05a} parameterizes this behavior via a barotropic equation of state
\begin{equation}
\label{eq:larsoneos}
T(\rho) =
T_{\rm min}\left\{
\begin{array}{ll}
(n_{\rm H}/n_{\rm bar})^{-0.27}, & n_{\rm H} < n_{\rm bar} \\
(n_{\rm H}/n_{\rm bar})^{0.07}, & n_{\rm H} \geq n_{\rm bar}
\end{array}
\right.,
\end{equation}
with $T_{\rm min} = 4.4$ K and $n_{\rm bar} = 4.3\times 10^5$ cm$^{-3}$. This equation of state is roughly consistent with observations, but these observations are limited to starless cores in nearby, low-mass, low-density star-forming regions, and it is not clear how general they are.

The equation of state is significant for determining the location of the IMF peak because idealized experiments show that the way gas fragments is highly sensitive to $\gamma$. As discussed in Section \ref{sssec:densitystat}, supersonically-turbulent media with $\gamma < 1$ have density PDFs with a powerlaw tail on the high side, while those with $\gamma > 1$ have a powerlaw tail at low density. Thus the value of $\gamma$ affects the amount of mass that turbulence drives to densities high enough to potentially become self-gravitating. Moreover, the condition for gravitational instability also depends on $\gamma$. For spherical structures we have the standard result that polytropes become unstable for $\gamma \leq 4/3$, while for cylindrical structures the analogous boundary is at $\gamma = 1$ \citep{ostriker64a, ostriker64b, mestel65a}; \citet{larson05a} reviews results for a variety of other geometries. Numerical simulations of self-gravitating turbulence confirm the amount of fragmentation is highly sensitive to $\gamma$, with values $<1$ producing much more fragmentation than values $>1$ \citep{li03a}. \citet{jappsen05a} simulate turbulent gas with an equation of state given by equation (\ref{eq:larsoneos}) but with varying values of $n_{\rm bar}$. They find that the characteristic fragment mass varies as $n_{\rm bar}^{-1/2}$. Similarly, \citet{bonnell06a} find that, when they use something similar to equation (\ref{eq:larsoneos}) as their equation of state, the characteristic mass no longer varies with the initial density and temperature, as it does for an isothermal equation of state.

Based on these results, \citet{whitworth98a} and \citet{larson05a} have posited that the characteristic mass of stars is determined by the Jeans mass evaluated at the density and temperature where the barotropic equation of state changes from $\gamma \leq 1$ to $\gamma > 1$. For the equation of state given by equation (\ref{eq:larsoneos}), the Jeans mass at this transition is $0.04$ $M_\odot$, close enough to the observed peak of the IMF to make a connection plausible.\footnote{\citet{larson05a} reports the Jeans mass as $0.3$ $M_\odot$. This difference is due to the ambiguity in defining the Jeans mass (see footnote \ref{note:jeansmass}). Larson's definition uses a coefficient of $\pi^{3/2}$ rather than $\pi^{3/2}/8$ in equation (\ref{eq:jeansmass}). This should serve as a caution about putting too much weight on close numerical agreement between theoretically predicted and observed masses when there are uncertain numerical factors present.}

This model for the origin of the IMF has several advantages over the hypothesis that characteristic masses are set from galactic scales. First,  it avoids the issue of whether isothermal turbulence has a well-defined mass scale at all. Simulations with non-isothermal equations of state manifestly do pick out a definite mass scale, and can converge on its value \citep{bonnell06a, bate09a, kratter10a, bate12a}. Second, it avoids the problem of ambiguities in defining the ``cloud" for which the mean density or the sonic length is to be measured. Third, it avoids the related problem of predicting large variations on the IMF from one star-forming region to another, and, by linking the IMF to microphysical processes rather than galactic properties, provides a plausible explanation for why the IMF is apparently so invariant. Indeed, \citet{elmegreen08a} argue that the Jeans mass at the grain-gas coupling density will vary very little in response to a number of possible variations in the galactic environment.

The major worry about barotropic equation of state models is that the equation of state itself is highly uncertain. Recall that barotropic equations of state, such as equations (\ref{eq:masunagaeos}) and (\ref{eq:larsoneos}), are derived for a particular choice of parameters for the background radiation field and column density, and a particular dynamical scenario. Other, equally plausible scenarios can generate quite different results, which in turn would lead to very different predicted IMF peak masses. For example, consider the temperature-density relationships shown in Figures \ref{fig:despotic} and \ref{fig:glover12}. Neither one shows a run of temperature versus density that looks much like that predicted by equation (\ref{eq:larsoneos}). For the calculation shown in Figure \ref{fig:despotic} this is likely because the calculation assumes a higher gas column density \blue{than that adopted by \citet{larson05a}}. This allows the transition from C$^+$ to CO composition, and a concomitant drop in temperature, at lower density than predicted by equation (\ref{eq:larsoneos}). The density-dependence of temperature is also different because cooling is through CO lines with low critical densities rather than the high-critical density C$^+$ line. Conversely, for the calculation shown in Figure \ref{fig:glover12}, collapse at high densities is not as coherent or rapid as assumed in the \citeauthor{larson05a} model, and so adiabatic heating is less important. As a result, the gas remains sub-isothermal to significantly higher densities than equation (\ref{eq:larsoneos}) predicts. The point is that, in a real galaxy, there seems to be no good reason to assume that all star-forming regions will obey the same barotropic equation of state, and thus an explanation of IMF universality based on an equation of state is much less secure than it might appear at first. Indeed, as I discuss below, a barotropic equation of state may be a very poor approximation even within a single star-forming region.

\paragraph{Radiative Feedback Models}

The realization that the gas equation of state is important for controlling fragmentation has led to a renewed investigation of the processes that control it. It has been known for a very long time that, once stars appear, they can radically alter the temperature structure around them \citep[e.g.,][]{masunaga00a, whitehouse06a}, but the importance of this for fragmentation was not widely realized until the work of \citet{larson05a} and other authors on barotropic equation of state models for the IMF peak. The point that radiative feedback is important for the IMF was first made in the context of massive star formation \citep{krumholz06b, krumholz07a, krumholz08a}, but subsequent work has shown that its effects are important in low-mass star-forming regions too \citep{bate09a, bate12a, offner09a, urban10a}. This is because even low mass protostars can have very high luminosities early in their lives when they are undergoing rapid accretion. For a star of mass $m$ and radius $r$ accreting at a rate $\dot{m}$, the accretion luminosity is
\begin{equation}
L_{\rm acc} \approx \frac{G \dot{m} m}{r} 
= 31 \dot{m}_{-5} m_0 r_{1}^{-1}L_\odot,
\end{equation}
where $m_{-5} = \dot{m}/10^{-5}$ $M_\odot$ yr$^{-1}$, $m_0 = m/1$ $M_\odot$, and $r_1=r/10$ $R_\odot$. \blue{This picture is an oversimplification, because it ignores the likelihood that the accretion rate is highly time-variable \citep[and references therein]{dunham14a}, but it nonetheless suggests that radiation is a non-negligible effect.}

\begin{figure}[ht!]
\centerline{\includegraphics[width=2.9in]{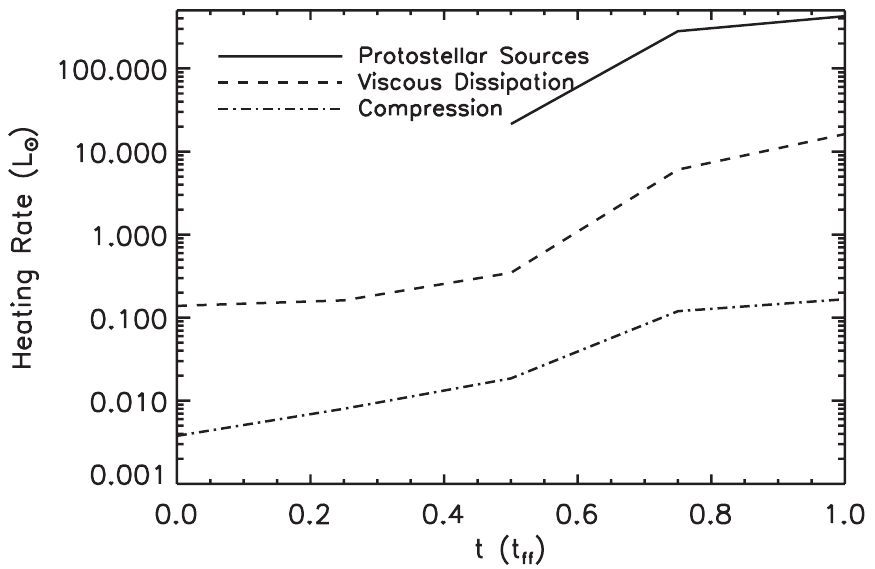}}
\centerline{\includegraphics[width=3in]{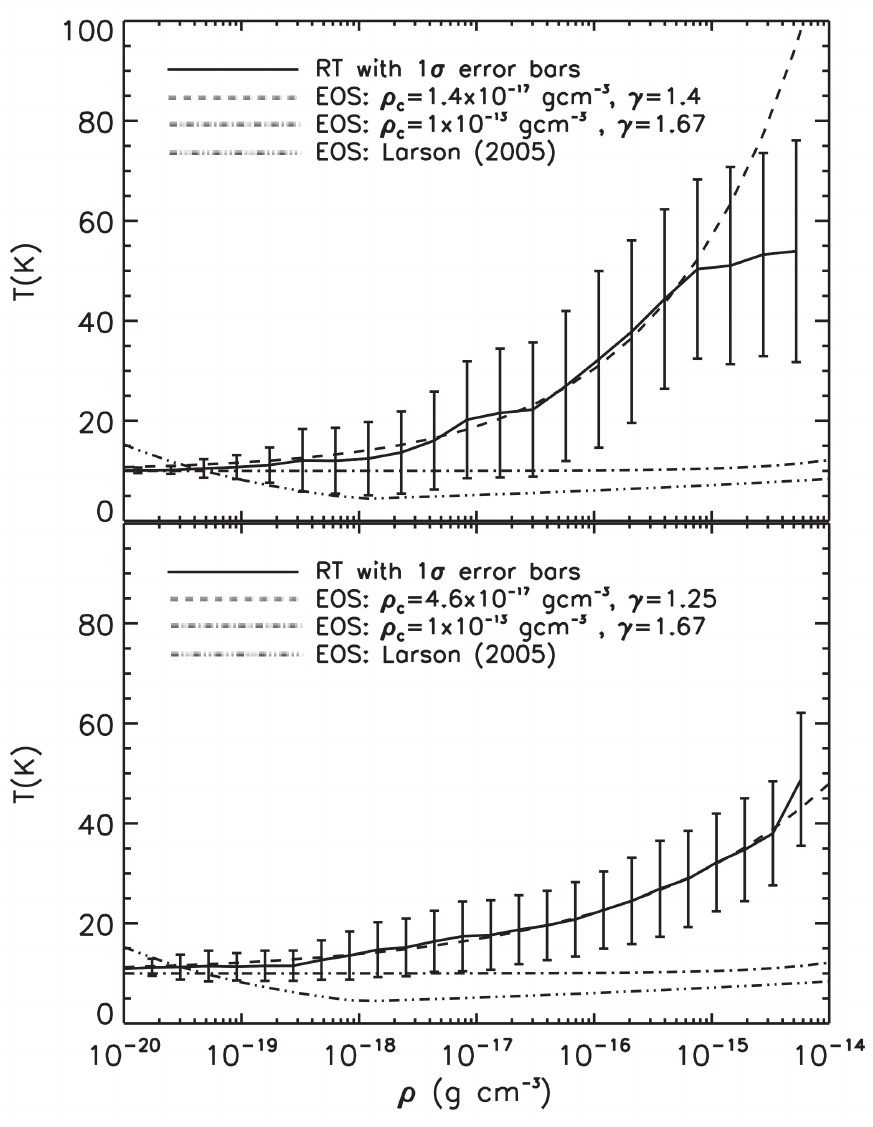}}
\caption{
\label{fig:offner09}
Results from a radiation-hydrodynamic simulation by \citet{offner09a}, reproduced by permission. The top panel shows the time-dependent contribution of various sources to the heating rate of the gas in the simulation: accretion luminosity from protostars, viscous dissipation, and adiabatic compression. The line for accretion appears when the first stars form. The bottom two panels show the temperature-density distribution in the simulation at two different times. The solid line shows the simulation results, with error bars showing the $1\sigma$ range of variation. The dashed line shows a best-fit barotropic equation of state of the form given by equation (\ref{eq:masunagaeos}), with $n_{\rm bar}$ and $\gamma$ allowed to vary. The dot-dashed line shows equation (\ref{eq:masunagaeos}) with some of the standard parameters often adopted in the literature, while the triple dot-dashed line shows the \citet{larson05a} equation of state, equation (\ref{eq:larsoneos}).
}
\end{figure}

The upper panel of Figure \ref{fig:offner09} shows a comparison between protostellar heating and other sources of energy in a radiation-hydrodynamic simulation. As illustrated in the Figure, accretion luminosity is high enough that it strongly dominates the energetics of a star-forming region. The energy release by accretion luminosity will initially heat the dust, but if the dust is hot enough then thermal transfer to the gas can become a dominant heating source even well before the density is high enough to achieve tight coupling between the two \citep{urban10a}. Moreover, the effects of radiative heating cannot be described by a barotropic equation of state. Radiative heating depends on local sources whose positions and luminosities change with time, making the temperature a function of position and time as well as density. The lower two panels of Figure \ref{fig:offner09} illustrate the problems of trying to model radiative heating with a barotropic equation of state. One can find a reasonable fit to the mean density-temperature relationship at a given time, but it is not time-invariant, and even at a single time there is a very large dispersion.

While this creates problems for the explanation of the IMF peak as arising from a universal barotropic equation of state, it provides another possible explanation: stars determine their own masses via radiative feedback. Analytic models for how this might happen have been developed by \citet{bate09a} and \citet{krumholz11e}. These models begin by analytically estimating the temperature as a function of distance from an accreting protostar, which is a function of the escape speed from the stellar surface and the gas density (which also sets the accretion rate, through the density-dependence of the free-fall time). They then follow a procedure somewhat analogous to that in the HC and Hopkins IMF models. They consider regions of size $R$ around the star, and compare the Jeans mass within that region $M_J(R)$, as determined by the density and temperature, with the enclosed mass $M_{\rm enc}(R)$. At small radii, $M_J(R)$ is large because the temperature is high, while $M_{\rm enc}(R)$ is small, giving $M_J(R) \ll M_{\rm enc}(R)$. This means that the enclosed mass is less than Jeans mass, and is unable to fragment to form a new star. At large radii, the temperature and Jeans mass fall and the enclosed mass rises, so that the inequality reverses and $M_J(R) \gg M_{\rm enc}(R)$. This means that the enclosed mass is able to fragment into new stars.

Bate and Krumholz hypothesize that the characteristic mass that goes into a single star, and thus the peak of the IMF, is given by the mass over which stellar radiative feedback is just marginally able to suppress its ability to fragment:
\begin{equation}
M_{\rm peak} = M_J(R) = M_{\rm enc}(R).
\end{equation}
The resulting value of $M_{\rm peak}$ is nearly independent of both the absolute value of the ambient density and its distribution around the star due to a cancellation: raising the density raises the enclosed mass $M_{\rm enc}(R)$ at fixed resolution, but it also tends to raise $M_J(R)$ by a nearly equal amount. Although a rising density at fixed temperature leads to a lower Jeans mass, when radiative feedback is considered this is more than outweighed by the increase in temperature due to higher accretion rates and more effective trapping of the stellar radiation. Thus the final dependence of $M_{\rm peak}$ on any interstellar parameter is quite weak: $M_{\rm peak} \propto \rho^{-1/9}$ \citep{bate09a} or $M_{\rm peak} \propto P^{-1/18}$ \citep{krumholz11e}, where $\rho$ and $P$ are the density and pressure, respectively.  The sense of variation -- that higher density and pressure regions form an IMF with a lower characteristic mass -- is consistent with what is required to explain the observations of elliptical galaxies discussed in Section \ref{sssec:unresolvedimf}. Moreover, the IMF does not depend on metallicity, as long as the dust abundance is high enough to render the gas optically thick to stellar radiation \citep{myers11a, krumholz11e}.

The IMF peak mass does depend on the escape speed from the stellar surface, but \citet{krumholz11e} points out that this is not a free parameter, because for much of their main accretion phase protostars are burning deuterium in their cores, and this fixes their core temperatures to $\approx 10^6$ K. This core temperature determines the escape speed.\footnote{The statement that the core temperature fully determines the surface escape speed is exactly true for a polytrope of fixed index \citep{chandrasekhar39a}. If the polytropic index is allowed to vary, or the star is not precisely a polytrope, the core temperature does not have a one-to-one mapping with the escape speed, but the variation is generally no more than a factor of $\sim 2$.} The characteristic temperature of deuterium burning can be written in terms of fundamental nuclear physics constants, and this makes it possible to write the characteristic mass of stars almost entirely in terms of fundamental constants. An attractive feature of this approach is that it also helps resolve a question that goes back to a controversy between Eddington and Jeans \citep{eddington26a, jeans26a, eddington26b, jeans26b}.\footnote{It is well worth the time to go back and read these exchanges, which are available online through the NASA Astrophysics Data System, if only for a demonstration that nasty letters complaining about not being cited are not unique to the modern era.} The mystery is that the stellar mass scale is actually quite special, in that it is in the right range to make sustained nuclear reactions possible. If stellar masses were two orders of magnitude smaller, then stars would be supported by degeneracy pressure without ever igniting fusion, and if it were three orders of magnitude larger, they would likely collapse to black holes via pair instability before spending an appreciable time on the main sequence. However, there is no obvious reason why the interstellar processes that form stars should know anything about nuclear physics. The importance of deuterium burning for determining the strength of stellar radiative feedback provides a link between nuclear and interstellar scales that may explain what would otherwise be the sort of coincidence that would force one to resort to the anthropic principle. \blue{Note that the deuterium is important not because of the luminosity it produces (which is much smaller than the accretion luminosity), but because of its role in fixing the stellar radius, and thus the yield of luminous energy per unit mass accreted.}

Numerical simulations provide significant support to this model, but also suggest complications. \citet{offner09a} find that radiation feedback stabilizes protostellar disks against fragmentation, preventing spurious formation of brown dwarfs as had been seen in earlier, non-radiative calculations. \citet{bate09a, bate12a} perform radiation-hydrodynamic simulations and find that they produce stars whose mass spectrum agrees well with the observed peak of the IMF. Moreover, as with barotropic equation of state simulations, the location of the IMF peak does not depend on the initial conditions. \citet{krumholz12b} include protostellar outflows as well as radiation in their calculations, and find that they too reproduce the observed IMF very well, including matching the powerlaw tail out to $\sim 10$ $M_\odot$. On the other hand, \citet{hansen12a} find that, in regions of low density, protostellar outflows serve to reduce the luminosity of accreting protostars enough to render protostellar radiation unimportant for fragmentation. \citet{krumholz11c} \blue{and \citet{martel12a}} find the opposite problem: in simulations with high $\epsilon_{\rm ff}$, protostellar radiation heating is so effective that it eventually suppresses all fragmentation, leading to a top-heavy IMF. \citet{krumholz12b} show that the problem can be fixed by setting up simulations with more realistic initial conditions and more feedback, such that $\epsilon_{\rm ff}$ is closer to observed values, but more investigation is clearly needed.

\subsection{The Origin of the IMF: Ways Forward}

As the bulk of this section should make clear, there are promising models for explaining many features of the IMF, but not yet a full theory that satisfactorily explains all of it. Analytically, the turbulent fragmentation models of PN, HC, and Hopkins seem very promising explanations for the origin of the IMF slope, but they have difficulty with issues of sub-fragmentation of massive cores, and more generally with explaining the peak of the IMF in a way that is independent of arbitrary choices about what parts of a cloud to use to measure input parameters. Explanations for the peak of the IMF based on deviations from isothermality on small scales, on the other hand, avoid such ambiguities, have promising support from numerical simulations, and provide a compelling explanation for the near universality of the IMF peak. However, they have nothing to say about the powerlaw tail.

Given this state of affairs, the most promising avenue for immediate investigation would seem to be to generalize some of the models for turbulent fragmentation so that they can handle non-isothermal gas, potentially providing a full theory for both the slope and peak of the IMF. While HC and Hopkins have generalized their models for barotropic equations of state \citep{hennebelle09a, hopkins13a}, we have seen that the barotropic approach is a very crude approximation to what really happens inside a molecular cloud once stellar feedback becomes significant. More work is needed to incorporate position-dependent analytic feedback models like those of \citet{bate09a} and \citet{krumholz11e} into the larger framework of the turbulent fragmentation models.

Numerically, the work that has been done thus far convincingly shows that a credible simulation that seeks to explain the origin of the IMF must include stellar feedback, in particular radiative transfer. However, this opens up a significant can of worms: since radiative feedback is mostly accretion-driven, anything that could potentially affect either the accretion rate or the gas temperature structure can be important to determining the IMF. The list of potentially important effects includes things such as magnetic fields and protostellar outflows that are only starting to be incorporated into simulations. It includes effects such as imperfect dust-gas coupling, photoelectric heating, and optically thick line cooling that have been included in simulations, but not in tandem with radiative feedback from stars. It also includes initial conditions, since the star formation rate or history in a particular region might well depend on the density or velocity structure of the molecular cloud from which it is born. Thus there is a need to include more physics in simulations and see how those processes interact with radiative heating, and there is also a need to conduct simulations starting from more realistic initial conditions, which means following the assembly of clouds as well as their subsequent collapse and fragmentation. \blue{Sanity-checking the results against observations at several points along this road seems highly advisable.}

\section{Discussion and Conclusions}
\label{sec:discussion}

\subsection{From the SFR to Clustering to the IMF: Linking the Scales}

I hope that those who have made it this far through the review will have drawn a few conclusions of their own. One that should be obvious is that the big three problems of the SFR, stellar clustering, and the IMF are inextricably linked, and a successful theory to explain one of them is likely to either be a theory for all three, or at a minimum to require certain assumptions about all three. These linkages of the galactic to the stellar scale come about in several ways, and which of them are important depends on the theory. At a minimum, the formation of molecular clouds at large scales provides initial conditions for the subsequent formation of stellar clusters and fragmentation of gas into individual stars. The feedback produced by those stars then provides a boundary condition for the star formation on galactic scales. Realistically, the linkages go beyond that. For example, radiative feedback may well be important in setting the IMF, but its strength is affected both by the star formation rate, which determines the accretion luminosity, and by the spatial clustering of accreting stars, which determines how strongly the radiation from one star can influence its neighbors.

Trying to build theories that address all these scales at once has been somewhat at odds with the traditional approach to star formation theory, and even with many of the current models. For example, the ``top-down" approach to building a theory for the star formation rate amounts to hypothesizing that everything important is determined by galactic scale processes, while everything on the cluster and smaller scale hidden behind one or two parameters. Conversely, models that predict the peak of the IMF in terms of a (turbulent) Jeans mass pre-suppose the presence of a cloud with a well-defined density or linewidth-size relation, and treat these as single numbers that are set at the galactic scale, and can be regarded as fixed parameters. While such simplifications can be powerful, one of the lessons that I hope readers take from this review is that we are starting to bump up against their limitations. Models that treat all the small-scale physics in the star formation rate as a single number seems to founder on how to handle the observed metallicity-dependence of the star formation rate, and models for the IMF that treat the galactic scale via a single density or sonic length have trouble explaining why the IMF does not vary more than we observe.

These and other examples suggest that future progress will depend on linking the scales by developing theories that are naturally multi-scale in nature. On the analytic side, the most promising venue for this appears to be through statistical theories of turbulence, which naturally embrace the multi-scale nature of the cold interstellar medium. Their power is illustrated by the fact that approaches of this sort have been featured prominently in the discussions above on each of the three big problems, sometimes even within the same theoretical framework. The major limitation of these models is that they are scale-free and at present have difficulty incorporating the necessarily messier physics of stellar feedback, which is decidedly non-scale free. We have seen, however, that turbulence alone is probably not sufficient to explain the star formation rate, explain the fraction of stars that form in bound clusters, or determine the position of the IMF peak. Some other ingredient is required, and stellar feedback is the natural suspect.

The numerical challenge is to build simulations that include enough physics to go persuasively from the galactic scale down to the point of resolving individual stars. Due to the huge dynamic range of the problem, a simulation that does this is likely to be able to run for only a very small fraction of a galactic dynamical time, but even such a short-duration run is likely to yield useful results. The real challenge will be building in the requisite physics. At present, galaxy-scale simulations tend to assume optically-thin heating and cooling and use either no chemistry or only highly-simplified chemical networks. They generally do not include radiative transfer to treat stellar radiation feedback, or indeed any other feedback processes except via a parameterized energy or momentum injection recipe. In contrast, small-scale simulations often include much more detailed treatments of stellar feedback that do not rely on recipes, but make a number of assumptions that are clearly inappropriate on the galactic scale -- for example that dust and gas are tightly coupled, or that the gas is fully molecular. There are also intermediate-scale simulations that treat the atomic to molecular transition, but generally do not include stellar feedback. At present, there are no simulation codes that include all the physical processes that are necessary to credibly simulate the ISM on scales from the galactic to the stellar. Building such a code must be a major goal of theoretical work on star formation in the next five years.

As an adjunct to this, there will also need to be significant work on connecting the output to observations. Some of the observational indicators discussed in this review are fairly straightforward, such as maps of the locations and numbers of protostars as compared to maps of dust mass and column density. Some, however, require significantly more interpretation, such as observations of optically thick molecular lines. Proper comparison with observations, and successful mapping from observed to physical quantities, will require post-processing simulation results with radiative transfer and chemistry codes.

\subsection{Problems for the Next Few Years}

It seems appropriate to end this review with some suggestions for ``low-hanging fruit": problems that, unlike the grand challenges we have just discussed, should be solvable in the next one to two years, and that would probably generate a large number of citations should an ambitious young postdoc or grad student come up with a solid answer. Most of these ideas have been mentioned before in the review, but hopefully summarizing them here will provide a quick cheat sheet of inspiration for projects. 

\begin{itemize}
\item Supernova feedback in a turbulent medium. The feedback-regulated model of star formation discussed in Section \ref{sssec:feedbackmodel} has as one if its crucial unknowns the amount of momentum that is produced by expanding supernova blast waves. Present simulations for the most part do not reach the resolution required to determine this quantity in a reliable manner. Using an AMR code or similar, it should be possible to simulate star formation at a galactic scale and follow the collapse down to the point where supernova blast waves would be resolved, and then to set off supernovae (perhaps using a subgrid model for the star formation itself) and measure the amount of momentum they impart. Such a simulation would probably be too expensive to run for multiple galactic dynamical times, but just following it far enough for the supernova remnant to reach the momentum-conserving phase would provide a very useful estimate for a quantity whose current value is very poorly known.
\item Stellar clustering with gas removal. The turbulence models of Hennebelle and Chabrier and Hopkins make predictions for the spatial distribution of dense gas cores and young stars that appear at least roughly consistent with observations (Section \ref{ssec:starstructure}), but say nothing about how gas is removed and how those stars evolve once feedback begins to remove gas. In contrast, Kruijssen presents a model of the response of a hierarchically-structured stellar population to gas removal, but, since his model relies only on the single-point density PDF, his treatment of the spatial arrangement of stars before gas removal is very crude, and he cannot easily make predictions for the cluster mass function (Section \ref{sssec:clusterfrac}). It ought to be possible to put these elements together to come up with an improved theory that can simultaneously predict both the cluster mass function and the fraction of stars forming in clusters.
\item Isothermal collapse: points or filaments? As discussed in Section \ref{sssec:imfpeakgal}, one major class of theories for the origin of the peak of the initial mass function is that it is set by turbulence. However, it has yet to be demonstrated that turbulence plus self-gravity produces a set of point masses with a definite mass scale. We know of at least one example of an isothermal fragmentation problem where the mode of collapse is to a singular filament rather than singular points, in which case there is no well-defined answer to the question of what is the characteristic mass of the fragments. Is this just a special, singular case, or does isothermal collapse generally produce networks of singular filaments rather than singular points? By the same token, is it possible to obtain a numerically-converged result for the gravitational fragmentation of an isothermal turbulent medium, or do such simulations invariably fail to converge and produce a mass spectrum that extends down to the resolution limit?
\item Localized radiative feedback in turbulent fragmentation models. As discussed in Section \ref{sssec:imfpeaknoniso}, radiative feedback appears capable of modifying the way gas fragments, but this effect is not included in any of the broader theories of turbulent fragmentation. While the models can handle non-isothermal equations of state that apply everywhere, they cannot yet handle localized feedback. Extending them to do so, and checking how this affects their predictions for the IMF, would be very valuable.
\end{itemize}

Of course (as often happens) some of these problems may be much deeper and more difficult than they appear. As with this entire review, this list also reflects the biases of the author as to what is worth doing. Caveat emptor.

\section*{Acknowledgements} 

I thank all the authors who contributed figures or data for this review: K.~Alatalo, M.~Bate, F.~Bigiel, A.~Bolatto, G.~Bryan, N.~Da Rio, T.~Davis, N.~Evans, N.~Gnedin, M.~Fall, C.~Federrath, S.~Glover,  R.~Gutermuth, P.~Hopkins, C.~Lada, A.~Leroy, M.~Lombardi, S.~Offner, M.~Rafelski, M.~Salem, A.~Schruba, A.~Wolfe, and M.~Wolfire. I also thank the co-authors on these works, who are too numerous to list here. I thank the following people for helpful discussions and/or comments on the manuscript: D.~Balsara, P.~Clark, C.~Conroy, N.~Evans, C.-A.~Faucher-Gigu\`{e}re, C.~Federrath, A.~G.~Kritsuk, P.~Kroupa, M.-M.~Mac Low, C.~McKee, P.~Padoan, and D.~Weisz. During the writing of this review I was supported by an Alfred P.~Sloan Fellowship, NSF CAREER grant AST-0955300, NASA ATP grant NNX13AB84G, and NASA TCAN grant NNX14AB52G. I also thank the Aspen Center for Physics, which is supported by NSF Grant PHY-1066293, for hospitality during the writing of this review.





\bibliographystyle{apj}
\bibliography{refs}







\end{document}